\documentclass[prd,tightenlines,preprint,preprintnumbers,aps,eqsecnum,nofootinbib,%
superscriptaddress,showpacs,floatfix]{revtex4-1}
\pdfoutput=1

\usepackage{graphicx}
\usepackage{amsmath,amsfonts,amssymb}
\usepackage[sort&compress]{natbib}
\usepackage{array,longtable}
\usepackage{bm}
\usepackage{multirow}
\usepackage[pdftex]{color}
\usepackage[colorlinks=true,linkcolor=blue,filecolor=blue,urlcolor=blue,citecolor=blue,pdftex,plainpages=false]{hyperref}

\definecolor{gray}     {rgb}{0.6,0.6,0.6}

\newcommand{\BR}{\ensuremath{\mathcal{B}}}            
\newcommand{\CR}{\nonumber\\}
\newcommand{\f}{\frac}
\newcommand{\FerMILC}{Fermilab/MILC}
\newcommand{\gev}{\text{GeV}}
\newcommand{\mev}{\text{MeV}}
\newcommand{\mph}{\ensuremath{\phantom{-}}}
\newcommand{\MSbar}{\ensuremath{\overline{\rm MS}}}   
\newcommand{\order}[1]{\ensuremath{\mathcal{O}\left( #1 \right)}} 
\newcommand{\Nc}{\ensuremath{N_c}}                    
\newcommand{\sh}{\ensuremath{\hat{s}}}
\newcommand{\spp}{\vphantom{\Big(}}

\renewcommand{\Re}{\mathop{\text{Re}}}

\newcommand{\BtoK}{\ensuremath{B\to K}}
\newcommand{\BtoKll}{\ensuremath{B\to K\ell^+\ell^-}}

\newcommand{\BtoKmumu}{\ensuremath{B\to K\mu^+\mu^-}}

\newcommand{\BptoKpll}{\ensuremath{B^+\to K^+\ell^+\ell^-}}
\newcommand{\BptoKpmumu}{\ensuremath{B^+\to K^+\mu^+\mu^-}}
\newcommand{\BptoKpee}{\ensuremath{B^+\to K^+e^+e^-}}

\newcommand{\BntoKnll}{\ensuremath{B^0\to K^0\ell^+\ell^-}}
\newcommand{\BntoKnmumu}{\ensuremath{B^0\to K^0\mu^+\mu^-}}
\newcommand{\BntoKnee}{\ensuremath{B^0\to K^0e^+e^-}}
\newcommand{\BtoKnunu}{\ensuremath{B\to K\nu\bar{\nu}}}

\newcommand{\Btopi}{\ensuremath{B\to\pi}}
\newcommand{\Btopill}{\ensuremath{B\to\pi\ell^+\ell^-}}

\newcommand{\Btopimumu}{\ensuremath{B\to\pi\mu^+\mu^-}}

\newcommand{\Bptopipll}{\ensuremath{B^+\to\pi^+\ell^+\ell^-}}
\newcommand{\Bptopipmumu}{\ensuremath{B^+\to\pi^+\mu^+\mu^-}}

\newcommand{\Bntopinll}{\ensuremath{B^0\to\pi^0\ell^+\ell^-}}

\newcommand{\Btopinunu}{\ensuremath{B\to\pi\nu\bar{\nu}}}

\begin{document}
\allowdisplaybreaks

\title{Phenomenology of semileptonic \texorpdfstring{\boldmath$B$-meson}{B-meson} decays with \\
form factors from lattice QCD}

\author{Daping Du}
\email[]{dadu@syr.edu}
\affiliation{Department of Physics, Syracuse University, Syracuse, New York, 13244, USA}

\author{A.X.~El-Khadra}
\affiliation{Department of Physics, University of Illinois, Urbana, Illinois, 61801, USA}

\author{Steven~Gottlieb}
\affiliation{Department of Physics, Indiana University, Bloomington, Indiana, 47405, USA}

\author{A.S.~Kronfeld}
\affiliation{Theoretical Physics Department, Fermi National Accelerator Laboratory, Batavia, Illinois, 60510, USA}
\affiliation{Institute for Advanced Study, Technische Universit\"at M\"unchen, 85748 Garching, Germany}

\author{J.~Laiho}
\affiliation{Department of Physics, Syracuse University, Syracuse, New York, 13244, USA}

\author{E.~Lunghi}
\email[]{elunghi@indiana.edu}
\affiliation{Department of Physics, Indiana University, Bloomington, Indiana, 47405, USA}

\author{R.S.~\surname{Van de Water}}
\email[]{ruthv@fnal.gov}
\affiliation{Theoretical Physics Department, Fermi National Accelerator Laboratory, Batavia, Illinois, 60510, USA}

\author{Ran~Zhou}
\email[]{zhouran@fnal.gov}
\affiliation{Theoretical Physics Department, Fermi National Accelerator Laboratory, Batavia, Illinois, 60510, USA}

\collaboration{Fermilab Lattice and MILC Collaborations}
\noaffiliation

\preprint{FERMILAB-PUB-15/425-T}
\preprint{NSF-KITP-15-134}

\begin{abstract}
We study the exclusive semileptonic $B$-meson decays $\BtoK(\pi)\ell^+\ell^-$, $\BtoK(\pi)\nu\bar\nu$, and
$B\to\pi\tau\nu$, computing observables in the Standard model using the recent lattice-QCD results for the
underlying form factors from the Fermilab Lattice and MILC Collaborations.
These processes provide theoretically clean windows into physics beyond the Standard Model because the
hadronic uncertainties are now under good control for suitably binned observables.
For example, the resulting partially integrated branching fractions for $\Btopi\mu^+\mu^-$ and $\BtoK\mu^+\mu^-$ outside
the charmonium resonance region are 1--2$\sigma$ higher than the LHCb Collaboration's recent measurements,
where the theoretical and experimental errors are commensurate.
The combined tension is 1.7$\sigma$.
Combining the Standard-Model rates with LHCb's measurements yields values for the Cabibbo-Kobayashi-Maskawa
(CKM) matrix elements $|V_{td}|=7.45{(69)}\times10^{-3}$, $|V_{ts}|=35.7(1.5)\times10^{-3}$, and
$|V_{td}/V_{ts}|=0.201{(20)}$, which are compatible with the values obtained from neutral $B_{(s)}$-meson
oscillations and have competitive uncertainties.
Alternatively, taking the CKM matrix elements from unitarity, we constrain new-physics contributions at the
electroweak scale.
The constraints on the Wilson coefficients ${\rm Re}(C_9)$ and ${\rm Re}(C_{10})$ from $\Btopi\mu^+\mu^-$
and $\BtoK\mu^+\mu^-$ are competitive with those from $B\to K^* \mu^+\mu^-$, and display a 2.0$\sigma$
tension with the Standard Model.
Our predictions for $\BtoK(\pi)\nu\bar\nu$ and $B\to\pi\tau\nu$ are close to the current experimental limits.
\end{abstract}


\date{\today} 

\maketitle

\section{Introduction and Motivation}
\label{sec:Intro}

The experimental high-energy physics community is searching for virtual effects of new heavy particles that
would give rise to deviations from Standard-Model predictions via a broad range of precision
measurements~\cite{Hewett:2012ns}.
Because the masses and couplings of the new particles are not known \emph{a priori}, indirect searches are
being pursued in many areas of particle physics, including the charged-lepton
sector~\cite{Albrecht:2013wet}, the Higgs sector~\cite{Dawson:2013bba}, and the quark-flavor
sector~\cite{Butler:2013kdw}.
Within heavy-quark physics, $B$-meson semileptonic decays provide numerous observables such as decay rates,
angular distributions, and asymmetries that are expected to be sensitive to different new-physics scenarios.
For example, the rare decays $\BtoKll$, $\BtoKnunu$, $\Btopill$, and $\Btopinunu$ proceed via $b\to s$ and
$b \to d$ flavor-changing neutral currents (FCNCs) and are sensitive to the effects of new heavy particles
that can arise in a wide range of models.
These include supersymmetry~\cite{Bobeth:2001sq,Demir:2002cj,Choudhury:2002fk,Wang:2007sp},
leptoquarks~\cite{Hiller:2014yaa,Gripaios:2014tna,Sahoo:2015wya}, and a fourth
generation~\cite{Hou:2013btm}; models with flavor-changing $Z'$ gauge bosons~\cite{Gauld:2013qba,%
Buras:2013qja,Gauld:2013qja,Buras:2013dea,Altmannshofer:2014cfa,Buras:2014fpa,Crivellin:2015mga,%
Crivellin:2015lwa}; and models with
extended~\cite{Aliev:1998sk,Iltan:1998ra,Bobeth:2001sq,Erkol:2002nw,Erkol:2004me,Song:2008zzc} or
composite~\cite{Gripaios:2014tna} Higgs sectors.
Decays to $\tau$-lepton final states such as $\Btopi\tau\nu$ are especially sensitive to charged scalars
that couple preferentially to heavier
particles~\cite{Chen:2006nua,Nierste:2008qe,Tanaka:2010se,Crivellin:2012ye,Sakaki:2012ft,Celis:2012dk}, such
as those that occur in two-Higgs-doublet models.
Tree-level CKM-favored $b \to u$ charged-current processes can be modified due to the presence of new
right-handed currents~\cite{Crivellin:2009sd,Feger:2010qc,Bernlochner:2014ova}.
If deviations from the Standard Model are observed in $B$-meson semileptonic decays, correlations between
measurements can provide information on the underlying masses and couplings of the new-physics scenario that
is realized in Nature.
(See, \emph{e.g.}, Refs.~\cite{Buras:2013ooa,Altmannshofer:2015sma} for recent reviews.)

Several tensions between theory and experiment have recently been observed in $B$-meson semileptonic decays.
The BaBar experiment found excesses in both $R(D) \equiv {\BR(B\to D \tau \nu)}/{\BR(B \to D \ell \nu)}$ and
$R(D^*) \equiv \BR(B\to D^* \tau \nu)/\BR(B \to D^* \ell \nu)$ with a combined significance of
3.4$\sigma$~\cite{Lees:2012xj,Bailey:2012jg}.
These results were subsequently confirmed by Belle~\cite{Huschle:2015rga} and LHCb~\cite{Aaij:2015yra},
albeit with somewhat lower significance; a recent HFAG average of these measurements quotes a combined
significance of 3.9$\sigma$ \cite{Amhis:2014hma}.
The LHCb experiment recently reported a measurement of the ratio of $\BptoKpmumu$ over $\BptoKpee$ branching
fractions (denoted $R_{K^+}^{\mu e}$ below) in the range $1~\textrm{GeV}^2\le q^2\le6~\textrm{GeV}^2$ that
is 2.6$\sigma$ lower than Standard-Model expectations~\cite{Aaij:2014ora}.
The Standard-Model predictions for the $B \to K^{(*)}\mu^+\mu^-$ differential decay rates are slightly, but
systematically, higher than experimental measurements by
LHCb~\cite{Bouchard:2013mia,Horgan:2013pva,Aaij:2014pli}.
Discrepancies of 2--3$\sigma$ between theory and experiment have also been observed for several $B\to
K^*\ell\ell$ angular observables~\cite{Aaij:2013qta,ChristophLangenbruchonbehalfoftheLHCb:2015iha}.
The long-standing $\approx 3\sigma$ tensions between determinations of the CKM matrix elements $|V_{ub}|$
and $|V_{cb}|$ obtained from inclusive and exclusive tree-level semileptonic $B$-meson decays were recently
confirmed with a new high-precision lattice-QCD calculation of the $B \to \pi \ell \nu$ form
factors~\cite{Lattice:2015tia}, the first unquenched lattice-QCD calculations of the $B\to D\ell\nu$ form
factors at nonzero recoil~\cite{Lattice:2015rga,Na:2015kha}, and the first unquenched lattice-QCD
calculation of the ratio of $\Lambda_b \to p \ell\nu$ to $\Lambda_b \to \Lambda_c \ell\nu$ form
factors~\cite{Detmold:2015aaa}.%
\footnote{Note, however, that the Belle experiment's preliminary measurement of the $B\to D\ell\nu$
differential decay rate~\cite{GlattauerEPS2015}, when combined with lattice-QCD form-factor
calculations~\cite{Lattice:2015rga,Na:2015kha}, yields a value of $|V_{cb}|$~\cite{GambinoEPS2015} that is
in better agreement with the inclusive determination~\cite{Alberti:2014yda}.}

Experimental progress on $B$-meson semileptonic decays has also been, and will continue to be, significant.
The LHCb experiment recently announced the first measurement of the \Bptopipmumu\ differential
decay rate~\cite{Aaij:2015nea}, as well as for the ratio of \Bptopipmumu\ to \BptoKpmumu\ rates.
This enables a more stringent test of the Standard Model via comparison of the shape to the theoretical
prediction.
The Belle experiment recently presented an upper limit on the total rate for $B^0 \to \pi^- \tau \nu$
decay~\cite{Hamer:2015jsa} from their first search for this process that is less than an order of magnitude
above that of the Standard-Model prediction.
The upcoming Belle~II experiment expects to observe $B\to\pi\tau\nu$ and other heretofore unseen processes
such as $B^0\to\pi^0\nu\bar\nu$~\cite{Aushev:2010bq}.
(The charged counterpart $B^+\to\pi^+\nu\bar\nu$ is part of the analysis chain $B^+\to\tau^+\nu$,
$\tau^+\to\pi^+\bar{\nu}$~\cite{Aubert:2009wt,Lees:2012ju,Adachi:2012mm,Kronenbitter:2015kls}.) %
Given the several observed tensions in semileptonic $B$-meson decays enumerated above and the recent and
anticipated improvement in experimental measurements, it is important and timely to critically examine the
assumptions entering the Standard-Model predictions for semileptonic $B$-decay observables and to provide
reliable estimates of the theoretical uncertainties.

The Fermilab Lattice and MILC Collaborations (\FerMILC) recently completed calculations of the form factors
for $\BtoK$~\cite{Bailey:2015dka} and $\Btopi$~\cite{Lattice:2015tia,Bailey:2015nbd} transitions with
lattice QCD using ensembles of gauge configurations with three dynamical quark flavors.
For $\BtoK$, the errors are commensurate with earlier lattice-QCD results~\cite{Bouchard:2013eph}.
For $\Btopi$, the results of Refs.~\cite{Lattice:2015tia,Bailey:2015nbd} are the most precise form factors
to date, with errors less than half the size of previous ones~\cite{Bailey:2008wp,Flynn:2015mha}.
Reference~\cite{Lattice:2015tia} also contains a joint fit of lattice-QCD form factors with experimental
measurements of the differential decay rate from BaBar and
Belle~\cite{delAmoSanchez:2010af,Ha:2010rf,Lees:2012vv,Sibidanov:2013rkk} to obtain the most precise
exclusive determination to date of the CKM matrix element $|V_{ub}|=3.72(16)\times10^{-3}$.
This fit also improves the determination of the vector and scalar form factors $f_+$ and $f_0$, compared to
those from lattice-QCD alone, provided that new physics does not contribute significantly to tree-level
$B\to \pi\ell\nu\; (\ell=e, \mu)$ transitions.

Given the landscape of quark-flavor physics described above, it is timely to use the form factors from
Refs.~\cite{Lattice:2015tia,Bailey:2015nbd,Bailey:2015dka} to obtain Standard-Model predictions for various
$B$-meson semileptonic-decay observables.
(For brevity, the rest of this paper refers to these results as the ``\FerMILC\ form factors.'') %
The new \emph{ab initio} QCD information on the hadronic matrix elements allows us to obtain theoretical
predictions of the observables with fewer assumptions than previously possible.
In this work, we consider the processes $\BtoKll$, $\BtoKnunu$, $\Btopill$, $\Btopinunu$, and $B\to
\pi\tau\nu$.
We present the following observables: differential decay rates, asymmetries, combinations of $\Btopi$ and
$\BtoK$ observables, and lepton-universality-violating ratios.
For partially integrated quantities, we include the correlations between bins of momentum transfer $q^2$.
Where possible, we make comparisons with existing experimental measurements.
We also combine our predictions for the $B\to K(\pi) \ell^+\ell^-$ Standard-Model rates with the most recent
experimental measurements to constrain the associated combinations of CKM matrix elements
$|V_{tb}V_{td}^*|$, $|V_{tb}V_{ts}^*|$, and $|V_{td}/V_{ts}|$.
For the $\Btopi$ vector and scalar form factors, we use the more precise Standard-Model determinations,
which use experimental shape information from $\Btopi\ell\nu$ decay.

We do not consider $B\to K^*$ processes in this paper, although there is extensive experimental and
theoretical work.
Lattice-QCD calculations of the hadronic form factors are available~\cite{Horgan:2013pva,Horgan:2013hoa},
albeit without complete accounting for the $K^*\to K\pi$ decay~\cite{Briceno:2014uqa}.
The phenomenology of these processes~\cite{Grinstein:2004vb,Becher:2005fg,Ali:2007sj,Bobeth:2010wg,%
Beylich:2011aq,Bobeth:2011nj,Hou:2014dza} often assumes various relations deduced from flavor symmetries.
Here we use the \BtoK\ and \Btopi\ form factors obtained directly from lattice QCD~\cite{Lattice:2015tia,%
Bailey:2015nbd,Bailey:2015dka} to test some of the symmetry relations employed in the literature.

The semileptonic form factors suffice to parametrize the factorizable hadronic 
contributions to $\Btopi$ and $\BtoK$ decays in all extensions of the Standard Model.
New heavy particles above the electroweak scale only modify the short-distance Wilson coefficients of the
effective Hamiltonian~\cite{Grinstein:1988me,Buras:1993xp,Huber:2005ig,Altmannshofer:2008dz}.
Here we use the \FerMILC\ form factors to obtain model-independent constraints on the Wilson coefficients
for the effective operators that govern $b\to d(s)$ FCNC transitions.
To facilitate the use of these form factors for additional phenomenological studies, the original
papers~\cite{Lattice:2015tia,Bailey:2015nbd,Bailey:2015dka} provided complete parametrizations of the
$\Btopi$ and $\BtoK$ form factors as coefficients of the $z$ expansions and their correlations.
To enable the combined analysis of both modes, this paper supplements that information by providing the
correlations between the $\Btopi$ and $\BtoK$ form-factor coefficients.


This paper is organized as follows.
We first provide an overview of the theoretical framework for the semileptonic decay processes studied in
this work in Sec.~\ref{sec:Theory}.
Next, in Sec.~\ref{sec:SymTests}, we summarize the calculations behind the \FerMILC\ \Btopi\ and \BtoK\ form
factors~\cite{Lattice:2015tia,Bailey:2015nbd,Bailey:2015dka}, providing a table of correlations among all
form factors.
Here, we also use the form factors to directly test heavy-quark and SU(3) symmetry relations that have been
used in previous Standard-Model predictions for rare semileptonic $B$-meson decay observables.
In Sec.~\ref{sec:SMPheno}, we present our main results for Standard-Model predictions for $\Btopill$,
$\BtoKll$, $B\to K (\pi) \nu \bar{\nu}$, and $B\to \pi\tau\nu$ observables using the \FerMILC\ form factors,
discussing each process in a separate subsection.
Then, in Sec.~\ref{sec:Implications} we use our predictions for the partially integrated branching fractions
together with experimental rate measurements to constrain the associated CKM matrix elements
(Sec.~\ref{subsec:Vtx}) and relevant Wilson coefficients (Sec.~\ref{subsec:WilsonConstraints}).
To aid the reader in digesting the information presented in Secs.~\ref{sec:SMPheno}
and~\ref{sec:Implications}, we summarize our main results in Sec.~\ref{sec:Summary}.
Finally, we give an outlook for future improvements and concluding remarks in Sec.~\ref{sec:Conclusions}.

Three Appendices provide detailed, supplementary information.
In Appendix~\ref{app:Results}, we tabulate our numerical results for \Btopill and \BtoKll
observables in the Standard Model integrated over different
$q^2$ intervals.
We present the complete theoretical expressions for the $B\to K(\pi)\ell^+\ell^-$ differential decays rates
in the Standard Model, including nonfactorizable terms, in Appendix~\ref{app:Formulae}.
The numerical values of the parametric inputs used for our calculations are provided in
Appendix~\ref{app:Inputs}.

\section{Theoretical background}
\label{sec:Theory}

Here we summarize the Standard-Model theory for the semileptonic decay processes considered in this work.
First, Sec.~\ref{subsec:ff} provides the standard definitions of the form factors.
Next, in Sec.~\ref{subsec:ll}, we discuss the theoretical framework for rare processes with a charged-lepton
pair final state, $b\to q\ell\ell \, (q=d,s)$.
Then we briefly summarize the formulae for rare decays with a neutrino pair final state $b\to q \nu
\bar{\nu} \, (q=d,s)$ in Sec.~\ref{subsec:nunubar} and for tree-level $b\to u \ell \nu_\ell$ semileptonic
decays in Sec.~\ref{subsec:ellnu}.
The latter two processes are theoretically much simpler, being mediated by a single operator in the
electroweak effective Hamiltonian.

\subsection{Form-factor definitions}
\label{subsec:ff} 

The pseudoscalar-to-pseudoscalar transitions considered in this paper can be mediated by vector, scalar, and
tensor currents.
It is conventional to decompose the matrix elements into Lorentz-invariant forms built from the
pseudoscalar- and $B$-meson momenta $p_P$ and $p_B$, multiplied by form factors that depend on the Lorentz
invariant $q^2$, where $q=p_B-p_P$ is the momentum carried off by the leptons.
For the vector current,
\begin{align}
    \langle P(p_P)|\bar{q}\gamma^\mu b|B(p_B)\rangle&=
        f_+(q^2) \left[ (p_B+p_P)^\mu - q^\mu\frac{M_B^2-M_P^2}{q^2}\right] +
        f_0(q^2) q^\mu\frac{M_B^2-M_P^2}{q^2},
    \label{eq:vector-ff} \\
        &=  f_+(q^2) (p_B+p_P)^\mu + f_-(q^2) (p_B-p_P)^\mu .
\end{align}
The form factors $f_+(q^2)$ and $f_0(q^2)$ couple to $J^P=1^-$ and~$0^+$, respectively, and therefore enter
expressions for differential decay rates in a straightforward way.
Because the terms proportional to $f_0(q^2)$ carry a factor of $q^\mu$, their contributions to differential
decay rates are weighted by the lepton mass, $m_\ell$, and is therefore significant only in the case of
$\tau$-lepton final states.
The form factor $f_-(q^2)$ is useful for a test of heavy-quark symmetry, discussed in
Sec.~\ref{subsec:HQSTests}.
Partial conservation of the vector current implies that $f_0$ also parametrizes the matrix element of the
scalar current:
\begin{equation}
     \langle P(p_P)|\bar{q}b|B(p_B)\rangle  = \frac{M_B^2-M_P^2}{m_b-m_q} f_0(q^2).
    \label{eq:scalar-ff}
\end{equation}
Finally, the matrix element of the tensor current is
\begin{equation}
    \langle P(p_P)|i\bar{q}\sigma^{\mu\nu}b|B(p_B)\rangle  = \frac{2}{M_B+M_P}
        (p_B^\mu p_P^{\nu}-p_B^\nu p_P^{\mu})  f_T(q^2),
    \label{eq:tensor-ff}
\end{equation}
where $\sigma^{\mu\nu}=i[\gamma^\mu,\gamma^\nu]/2$.

These form factors suffice to parametrize the hadronic transition when the leptonic part of the reaction
factorizes.
Particularly important corrections arise in the penguin decays $\Btopi\ell\ell$ and $\BtoK\ell\ell$
studied in this work, as discussed in the next subsection and in Appendix~\ref{app:Formulae}.

\subsection{Rare \texorpdfstring{\boldmath$b\to q \ell\ell \, (q=d,s)$}{b to qll} decay processes}
\label{subsec:ll} 

In this subsection, we first present the effective Hamiltonian for this case in Sec.~\ref{subsec:EffHam},
followed by a description of how we obtain the short-distance Wilson coefficients of the effective
Hamiltonian at the relevant low scale in Sec.~\ref{subsec:WCs}.
To obtain physical observables, one also needs the on-shell $b\to d(s)\ell\ell$ matrix elements of the
operators in the effective Hamiltonian.
As discussed in Sec.~\ref{subsec:MEs}, for decays into light charged leptons, $\ell = e, \mu$, it is
necessary to treat the different kinematic regions within different frameworks.
In Sec.~\ref{subsec:DDR} we present the general structure of the double differential decay rate.
Details of the calculations at high and low $q^2$ are relegated to Appendix~\ref{app:Formulae}.

\subsubsection{Effective Hamiltonian} \label{subsec:EffHam}

The starting point for the description of $b\to q\ell\ell \, (q=d,s)$ transitions is the effective
Lagrangian~\cite{Huber:2005ig}: 
\begin{align}
    \mathcal{L}_\text{eff} = 
        & + \frac{4 G_F}{\sqrt{2}} V^*_{tq} V_{tb}^{} \left[ {\sum_{i=1}^{8} C_i(\mu) Q_i + \frac{\alpha_e (\mu)}{4\pi}
        \sum_{i=9}^{10} C_i(\mu) Q_i} +
            \sum_{i=3}^{6} C_{iQ}(\mu) Q_{iQ} + C_b(\mu) Q_b \right] \nonumber \\
        & + \frac{4 G_F}{\sqrt{2}} V^*_{uq} V_{ub}^{} \sum_{i=1}^{2} C_i(\mu) \left[ Q_i - Q_i^u \right] 
           + \mathcal{L}_{\text{QCD}\times\text{QED}} .
\end{align}
Throughout this paper, as in the literature, we refer to $\mathcal{H}_\text{eff}=-\mathcal{L}_\text{eff}$ as
the (electroweak) effective Hamiltonian.

At leading order in the electroweak interaction, there are twelve independent operators, which we take to be
\begin{align}
Q_1^u   &= (\bar{q}_L \gamma_{\mu} T^a u_L) (\bar{u}_L \gamma^{\mu} T^a b_L),
\nonumber \\[0.5em]
Q_2^u   &= (\bar{q}_L \gamma_{\mu}     u_L) (\bar{u}_L \gamma^{\mu}     b_L),
\nonumber \\[0.5em]
Q_1   &= (\bar{q}_L \gamma_{\mu} T^a c_L) (\bar{c}_L \gamma^{\mu} T^a b_L),
\nonumber \\[0.5em]
Q_2   &= (\bar{q}_L \gamma_{\mu}     c_L) (\bar{c}_L \gamma^{\mu}     b_L),
\nonumber \\[0.5em]
Q_3   &= (\bar{q}_L \gamma_{\mu}     b_L) \sum_{q^\prime} (\bar{q}^\prime\gamma^{\mu}     q^\prime),
\nonumber \\[0.5em]
Q_4   &= (\bar{q}_L \gamma_{\mu} T^a b_L) \sum_{q^\prime} (\bar{q}^\prime\gamma^{\mu} T^a q^\prime),    
\nonumber \\[0.5em]
Q_5   &= ({\bar{q}_L} \gamma_{\mu_1}
                     \gamma_{\mu_2}
                     \gamma_{\mu_3}    b_L)\sum_{q^\prime} (\bar{q}^\prime \gamma^{\mu_1} 
                                                         \gamma^{\mu_2}
                                                         \gamma^{\mu_3}     q^\prime),     
\label{eq:OPE} \\[0.5em]
Q_6   &= ({\bar{q}_L} \gamma_{\mu_1}
                     \gamma_{\mu_2}
                     \gamma_{\mu_3} T^a b_L)\sum_{q^\prime} (\bar{q}^\prime \gamma^{\mu_1} 
                                                            \gamma^{\mu_2}
                                                            \gamma^{\mu_3} T^a q^\prime),
\nonumber \\[0.5em]
Q_7   &=  \f{e}{16 \pi^2} m_b (\bar{q}_L \sigma^{\mu \nu}     b_R) F_{\mu \nu},
\nonumber \\[0.5em]
Q_8   &=  \f{g}{16 \pi^2} m_b (\bar{q}_L \sigma^{\mu \nu} T^a b_R) G_{\mu \nu}^a, 
\nonumber \\[0.5em]
Q_9      &= (\bar{q}_L \gamma_{\mu} b_L) \sum_\ell (\bar{\ell}\gamma^{\mu}\ell),
\nonumber \\[0.5em]
Q_{10}   &= (\bar{q}_L \gamma_{\mu}     b_L) \sum_\ell (\bar{\ell}\gamma^{\mu} \gamma_5\ell).
\nonumber
\end{align}
Because the top-quark mass is above the electroweak scale, only the five lightest quark flavors
$q'=u,d,s,c,b$ are included in operators $Q_3$ through $Q_6$.
All three lepton flavors $\ell=e,\mu,\tau$ appear in operators $Q_9$ and $Q_{10}$.

Once QED corrections are considered, five more operators must be included, which we choose to be
\begin{equation}
\begin{array}{rl}
Q_{3Q} = & (\bar{q}_L \gamma_{\mu}     b_L)
\sum_{q^\prime} e_{q^\prime} (\bar{q}^\prime\gamma^{\mu} q^\prime), \\[0.5em]
Q_{4Q} = & (\bar{q}_L \gamma_{\mu} T^a b_L)
\sum_{q^\prime} e_{q^\prime} (\bar{q}^\prime\gamma^{\mu} T^a q^\prime), \\[0.5em]
Q_{5Q} = & (\bar{q}_L \gamma_{\mu_1}
                     \gamma_{\mu_2}
                     \gamma_{\mu_3}    b_L)\sum_{q^\prime} e_{q^\prime} (\bar{q}^\prime \gamma^{\mu_1} 
                                                               \gamma^{\mu_2}
                                                               \gamma^{\mu_3}     q^\prime), \\[0.5em]
Q_{6Q} = & (\bar{q}_L \gamma_{\mu_1}
                     \gamma_{\mu_2}
                     \gamma_{\mu_3} T^a b_L)\sum_{q^\prime} e_{q^\prime} (\bar{q}^\prime \gamma^{\mu_1} 
                                                                \gamma^{\mu_2}
                                                                \gamma^{\mu_3} T^a q^\prime), \\[0.5em]
Q_b = & \f{1}{12} \left[ 
          (\bar{q}_L \gamma_{\mu_1}
                     \gamma_{\mu_2}
                     \gamma_{\mu_3}    b_L)            (\bar{b} \gamma^{\mu_1} 
                                                                \gamma^{\mu_2}
                                                                \gamma^{\mu_3}     b)
             -4 (\bar{q}_L \gamma_{\mu} b_L) (\bar{b} \gamma^{\mu} b) \right].
\end{array} 
\label{eq:QEDOps}
\end{equation}
where $e_{q'}$ are the electric charges of the corresponding quarks ($\f{2}{3}$ or $-\f{1}{3}$).

\subsubsection{Wilson coefficients}
\label{subsec:WCs}

In the calculation of any $b\to q$ ($q=d,s$) transition, large logarithms of the ratio
$\mu_\text{high}/\mu_\text{low}$ arise, where $\mu_\text{high} \sim m_t, m_W, m_Z$ is a scale associated
with virtual heavy-particle exchanges and $\mu_\text{low} \sim p_\text{ext} \sim m_b$ is a scale associated
with the typical momenta of the final state on-shell particles.

The standard procedure to resum these large logarithms is based on the factorization of short- and
long-distance physics, \emph{i.e.}, writing $\ln(\mu_\text{high}/\mu_\text{low}) = \ln(\mu_\text{high}/\mu)
+ \ln(\mu/\mu_\text{low})$ and absorbing the first logarithm into the Wilson coefficients and the second
into the matrix elements of the local operators.
The independence of the overall amplitude on the factorization scale $\mu$ leads to renormalization group
equations for the Wilson coefficients whose solution resums terms of the type $\left[\alpha_s^L \ln
(\mu_\text{high}/\mu)\right]^n$ to all orders in perturbation theory.
($L=0,1$ are known as leading and next-to-leading log approximations.) %
The scale $\mu$ can then be chosen close to $\mu_\text{low}$ (typically $\mu \simeq m_b$), thus eliminating
all large logarithms from the calculation of the amplitude.
Any residual dependence on the scales $\mu_\text{high}$ and $\mu_\text{low}$ is taken as an uncertainty from
missing higher order perturbative corrections.
We follow the standard practice of varying these scales by a factor of two around some nominal central
values, which we choose to be $\mu_\text{high} = 120~\text{GeV}$ and $\mu_\text{low} = 5~\text{GeV}$.

The $b\to d(s)\ell\ell$ case is complicated by the fact that the Wilson coefficients for the leading
semileptonic operators $Q_9$ and $Q_{10}$ carry explicit factors of $\alpha_e$, in addition to the common
factor $4G_F/\sqrt{2}$.
Moreover, the current-current operators $Q_1$ and $Q_2$ mix with the semileptonic operators at one loop in
QED and at two loops in mixed QED-QCD.
These complications can all be straightforwardly addressed in a double expansion in $\alpha_s$ and
$\alpha_e/\alpha_s$.
We refer the reader to Ref.~\cite{Huber:2005ig} for a detailed account of this double expansion as well as a
complete collection of all anomalous dimension matrices required for the running of the Wilson coefficients
and of the QED and QCD couplings.
In contrast to earlier analyses~\cite{Misiak:1992bc,Buras:1994dj}, we do not include the gauge couplings in
the normalization of $Q_9$ and $Q_{10}$, precisely to simplify the mixed QCD-QED renormalization group
equations.
Finally, note that the operator $Q_b$ contributes to the transition amplitude only via mixing with the other
operators.

\subsubsection{Matrix elements}
\label{subsec:MEs}

The calculation of exclusive $b \to s(d) \ell\ell$ matrix elements for the operators $Q_{7,9,10}$ is
relatively simple.
Because these operators contain an explicit photon or charged-lepton pair, the $B\to K(\pi) \ell\ell$ matrix
element trivially factorizes in QCD into the product of a charged-lepton current and a form factor.
The matrix element of $Q_7$ is proportional to the tensor form factor $f_T$, while those of $Q_{9,10}$ only
get contributions from the vector-current operator because of parity conservation and the fact that the
incoming and outgoing mesons are both pseudoscalars.
The vector-current matrix element leads to the form factors $f_+$ and $f_0$.
Note that $f_+$ and $f_0$, being matrix elements of a partially conserved vector current, do not renormalize
and have no scale dependence.
On the other hand, the $\mu$ dependence of $f_T$ is canceled by that of the quark mass and the Wilson
coefficient for $Q_7$.

The calculation of $B\to P\ell\ell$ ($P =K, \pi$) matrix elements of operators that do not involve an
explicit photon or charged-lepton pair is more complicated.
Schematically,
\begin{align}
    \langle P \ell\ell | Q_i (y) |\bar B\rangle \sim
    (\bar u_\ell \gamma_\mu v_\ell) \; \int d^4 x \, e^{i q \cdot (x-y)}
    \langle P | T\,  J_\text{em}^\mu (x) Q_i(y) |\bar B\rangle ,
    \label{eq:longdistance}
\end{align}
where $u_\ell$ and $v_\ell$ are the lepton spinors and $J_\text{em}^\mu$ is the electromagnetic current.
The matrix elements of the $T$-product include long-distance contributions that are difficult to calculate,
even with lattice QCD.
In certain kinematic regions, however, these complex matrix elements can be expressed in terms of simpler
objects, namely the form factors defined in Sec.~\ref{subsec:ff} plus the light-cone distribution
amplitudes, up to power corrections of order~$\Lambda_\text{QCD}/m_b$.

Before discussing the effective theories used to simplify the matrix elements in
Eq.~(\ref{eq:longdistance}), let us comment on the role of $c\bar c$ and $u\bar u$ states.
The processes $B\to K(\pi)\ell\ell$ can proceed through the following intermediate resonances: %
$B\to K(\pi)\psi_{uu,cc}\to K(\pi)\ell\ell$ where $\psi_{uu}=\rho,\omega$ and %
$\psi_{cc} = \psi(1S, 2S, 3770, 4040, 4160, 4415)$.
In the language of Eq.~(\ref{eq:longdistance}), contributions of intermediate $\psi_{cc}$ and $\psi_{uu}$
states stem from matrix elements involving the operators $Q_{1,2}$ and $Q_{1,2}^u$, respectively.
The two lowest charmonium states have masses below the open charm threshold ($D\bar D$) and have very small
widths, implying very strong violations of quark-hadron duality; consequently the regions including the
$\psi(1S)$ and $\psi(2S)$ masses (also known as $J/\psi$ and $\psi^\prime$) are routinely cut from
theoretical and experimental analyses alike.
Above the $\psi(2S)$, a resonance compatible with the $\psi(4160)$ has been observed in $B\to K\mu^+\mu^-$
decay~\cite{Aaij:2013pta}; the $\psi(3770)$ is also seen, but the signal for the $\psi(4040)$ and higher
resonances is not significant.
Because the four higher charmonium resonances are broad and spread throughout the high-$q^2$ region, in this
region quark-hadron-duality violation is estimated to be small~\cite{Beylich:2011aq} for observables
integrated over the full high-$q^2$ range.
The kinematic region where the light resonances $(\rho, \omega, \phi)$ contribute is typically not excluded
from experimental analyses.
Although their effects on branching fractions and other observables can be substantial, their contributions
cannot be calculated in a fully model-independent manner.
References~\cite{Khodjamirian:2012rm,Hambrock:2015wka} estimate the size of nonlocal contributions to $B\to
K(\pi)\ell\ell$ decays from the $\rho$ and $\omega$ using hadronic dispersion
relations~\cite{Khodjamirian:2010vf}.
They predict an enhancement of the $B^+ \to \pi^+ \mu^+ \mu^-$ differential branching fraction at low $q^2$
in good agreement with the $q^2$ spectrum measured by LHCb~\cite{Aaij:2015nea}.

At high $q^2$ the final-state meson is nearly at rest, and the two leptons carry half the energy of the
$B$~meson each.
As first discussed by Grinstein and Pirjol~\cite{Grinstein:2004vb}, the photon that produces them has
$q^2\sim M_B^2$ and the $T$-product in Eq.~(\ref{eq:longdistance}) can be evaluated using an operator
product expansion (OPE) in $1/M_B$~\cite{Beylich:2011aq,Bobeth:2011gi,Bobeth:2011nj,Bobeth:2012vn}.
The resulting matrix elements can be parametrized in terms of the three form factors $f_{+,0,T}$.
In the literature $f_T$ is usually replaced by $f_+$ using heavy quark relations~\cite{Grinstein:2002cz,
Grinstein:2004vb, Hewett:2004tv, Bobeth:2011nj}, whereas in this paper we use the lattice-QCD results for
$f_T$.
Within this framework, the high-$q^2$ rate is described entirely in terms of the form factors $f_{+,0,T}$ up
to corrections of order~$\Lambda/M_B$.
It is important to realize that the high-$q^2$ OPE requires $(x-y)^2 \sim 1/m_b^2$ implying that all matrix
elements should be expanded in $1/q^2 \sim 1/m_b^2$.
In Refs.~\cite{Grinstein:2004vb,Bobeth:2011gi}, the authors treat $m_c \ll m_b$ and expand $Q_{1,2}$ in
powers of $m_c^2/q^2$.
Here we instead follow Ref.~\cite{Beylich:2011aq} by integrating out the charm quark at the $m_b$ scale and
including the full $m_c$ dependence of the $Q_{1,2}$ matrix elements.
This approach simplifies the operator basis without introducing any loss of accuracy.
We include a 2\% uncertainty to account for quark-hadron duality violations~\cite{Beylich:2011aq}.

At low~$q^2$ the two leptons are nearly collinear, and the daughter meson recoils with large energy
$E_P\sim m_b/2$.
In this kinematic configuration, three scales play an important role: the hard scale $\sim m_b^2$, the
hard-collinear scale $\sim\Lambda_\text{QCD}m_b$ stemming from interactions of the energetic final state
quarks with the light quarks and gluons in the $B$ meson, and the purely nonperturbative scale
$\sim\Lambda_\text{QCD}^2$.
Note that the ratio of any two scales vanishes in the $m_b\to\infty$ limit.
In the soft-collinear effective theory (SCET)~\cite{Beneke:1999br,Beneke:2000ry,Bauer:2000yr,Bauer:2001yt},
an expansion in $\Lambda_\text{QCD}/m_b$ exploits this hierarchy such that all contributions stemming from
physics above $\Lambda_\text{QCD}$ can be calculated in perturbative QCD.
At leading power, the remaining nonperturbative objects are standard form factors and $B$, $\pi$ and $K$
mesons light-cone distribution amplitudes.
SCET is therefore a double expansion in $\alpha_s$ and $\Lambda_\text{QCD}/m_b$.
The application of soft-collinear factorization to exclusive $b\to s \ell\ell$ decays was pioneered in
Refs.~\cite{Beneke:2000wa, Beneke:2001at} and subsequently employed in many phenomenological analyses; see,
for instance, Refs.~\cite{Bobeth:2007dw, Altmannshofer:2008dz}.

The structure of the low-$q^2$ SCET expansion (also known as QCD
factorization~\cite{Beneke:1999br,Beneke:2000ry}) for $B\to P\ell\ell$ ($P=K,\pi$) is (omitting prefactors)
\begin{align}
    C_i \; \langle P \ell\ell| Q_i | \bar B\rangle  & \sim C_i \; 
        \Big[(1+\alpha_s)  f_T + (1+\alpha_s) f_+ + \phi_B \star T \star\phi_P \Big], \quad i=1,\ldots,6, \\
    C_7 \; \langle P \ell\ell| Q_7 | \bar B\rangle &\sim C_7 \; f_T ,\\
    C_8 \; \langle P \ell\ell| Q_8 | \bar B\rangle &\sim C_8 \; 
        \Big[ \alpha_s  f_T + \alpha_s f_+ + \phi_B \star T \star \phi_P \Big]  ,\\
    C_9 \; \langle P \ell\ell| Q_9 | \bar B\rangle &\sim C_9 \; f_+ ,\\
    C_{10} \; \langle P \ell\ell| Q_{10} | \bar B\rangle &\sim C_{10} \; f_+ ,
\end{align}
where the coefficients of $f_+$ and $f_T$ originate from hard interactions, and $\phi_B\star T\star\phi_P$
denotes a convolution of a short-distance kernel $T$, originating from hard-collinear interactions, with the
$B$-meson and final-state meson light-cone distribution amplitudes $\phi_B$ and $\phi_P$, respectively.
As explained in detail in Appendix~\ref{app:Formulae}, it is customary to collect all terms proportional to
the form factors and introduce effective Wilson coefficients $C_7^\text{eff}$ and $C_9^\text{eff}$.
The structure of the whole amplitude is then
\begin{equation}
    A(B\to P  \ell\ell) \sim C_7^\text{eff} f_T + \left(C_9^\text{eff} + C_{10}\right) f_+ +
        \phi_B \star T \star \phi_P.
\end{equation}
Further, some terms in $\langle Q_3 \rangle$ through $\langle Q_6 \rangle$ are proportional to 
$\langle Q_8\rangle$ and are usually taken into account with the introduction of the effective Wilson
coefficient~$C_8^\text{eff}$.

Within the SCET approach it is also possible to express $f_T$ in terms of $f_+$---schematically
$f_T\sim(1+\alpha_s)f_++\phi_B\star T\star\phi_P$.
Because we have direct access to the lattice-QCD calculation of $f_T$, this step would only result in the
unnecessary introduction of additional uncertainties.

\subsubsection{Differential decay rates} 
\label{subsec:DDR}

The double differential $B\to K (\pi) \ell\ell$ rate can be written as
\begin{equation}
    \frac{d^2\Gamma}{dq^2 \, d\cos\theta} = a + b \cos\theta + c \cos^2\theta ,
\end{equation}
where $\theta$ is the angle between the $B$ meson and $\ell^-$ in the dilepton rest frame, and $a,b,c$ are
functions of $q^2$ that depend on the form factors and Wilson coefficients.
The three main observables considered in the literature are the differential rate
\begin{equation}
        \frac{d\Gamma}{dq^2} = 2 \left( a + \frac{c}{3}\right) ,
\end{equation}
the forward-backward asymmetry, and the flat term~\cite{Bobeth:2007dw}.
There are two forms of the last two, either evaluated at a single value of
$q^2$~\cite{Bouchard:2013eph}
\begin{align}
    A_\text{FB}(q^2) &= \frac{b}{d\Gamma/dq^2} , \\
    F_H(q^2)         &= \frac{2 (a+c)}{d\Gamma/dq^2} ,
    \label{eq:FH_def}
\end{align}
which is useful for plotting, or a binned form~\cite{Bobeth:2007dw}
\begin{align}    
    A_\text{FB}(q^2_\text{min},q^2_\text{max}) &= \int_{q^2_\text{min}}^{q^2_\text{max}} b\,dq^2
        \left[\int_{q^2_\text{min}}^{q^2_\text{max}} 2 \left( a + \frac{c}{3}\right) dq^2\right]^{-1}, \\
    F_H(q^2_\text{min},q^2_\text{max})         &= \int_{q^2_\text{min}}^{q^2_\text{max}}(a+c)\,dq^2
        \left[\int_{q^2_\text{min}}^{q^2_\text{max}} \left( a + \frac{c}{3}\right) dq^2\right]^{-1},
    \label{eq:FH_def_bin}
\end{align}
which can be compared with experimental measurements.
In the Standard Model $b=0$, \emph{i.e.}, the forward-backward asymmetry vanishes (neglecting tiny QED
effects).
Further, in the $m_\ell = 0$ limit (an excellent approximation for $\ell=e,\mu$), one finds $c = -a$,
implying a very small flat term of order $m_\ell^2/M_B^2$.
Thus both the forward-backward asymmetry and flat term are potentially sensitive to contributions beyond the
Standard Model.
In addition, it is possible to consider the isospin and $CP$ asymmetries of the differential $d\Gamma/dq^2$
rate.
Appendix~\ref{app:Formulae} provides explicit expressions for $a$ and $c$ in the Standard Model.

\subsection{Rare \texorpdfstring{\boldmath$b\to q\,\nu\bar{\nu}\;(q=d,s)$}{b to q nu nubar} decay processes}
\label{subsec:nunubar} 

In the Standard Model, the effective Hamiltonian for the rare decay process $b\to q\nu\bar{\nu}\; (q=d,s)$
is given by
\begin{equation} 
    \mathcal{H}_\text{eff} = - \frac{4 G_F}{\sqrt{2}} V_{tb}V^*_{tq} \, C_L Q_L, 
\end{equation} 
where 
\begin{equation}
    Q_L = \frac{e^2}{16 \pi^2} \left(\bar{q}_L \gamma_\mu b_L \right)
        \sum_\nu \left( \bar{\nu}_L \gamma^\mu \nu_L
\right) , 
\end{equation} 
summing over $\nu = \nu_e, \nu_\mu, \nu_\tau$, and 
\begin{equation}
    C_L = - X_t / \sin^2 \theta_W .
\end{equation}
The function $X_t$ parametrizes top-quark-loop effects and includes next-to-leading-order QCD
contributions~\cite{Buchalla:1993bv,Misiak:1999yg,Buchalla:1998ba} and two-loop electroweak
corrections~\cite{Brod:2010hi}. We take the numerical value $X_t=1.469(17)$ from Ref.~\cite{Brod:2010hi}.

The neutrino-pair final state ensures that the complications discussed for $\BtoK(\pi)\ell^+\ell^-$ decays
in Sec.~\ref{subsec:ll} do not arise in the calculation of the decay rate for this process.
In particular, the decay rate receives no contributions from $u\bar{u}$ or $c\bar{c}$ resonances or
nonfactorizable terms.
Thus, the systematic uncertainties associated with power corrections, resonances, and duality violations are
absent~\cite{Buras:2014fpa}.
In summary, the short-distance flavor-changing-neutral-current-induced contribution to the 
Standard Model decay rate for $B \to P \nu \bar{\nu} \; (P=K,\pi)$,
which proceeds via the the flavor-changing-neutral-current interaction, depends only on the vector form
factor $f_+(q^2)$ and can be calculated over the entire kinematic range with full control over the
theoretical errors.
The differential branching fraction takes the form~\cite{Altmannshofer:2009ma,Buras:2014fpa}
\begin{equation}
    \frac{d\BR (B \to P \nu \bar{\nu})_\text{SD}}{dq^2} = C_P \tau_{B} \left| V_{tb}V_{ts(d)}^* \right|^2
        \frac{G_F^2 \alpha^2 }{32 \pi^5} \frac{X_t^2}{{\rm sin}^4 \theta_W} |\bm{p}_P|^3 f_+^2(q^2).
    \label{eq:Btonunubar_rate}
\end{equation}
where $|\bm{p}_P|$ is the magnitude of the final-state meson three-momentum in the $B$-meson rest frame.
The isospin factor $C_P = 1$ for decays to kaons and charged pions ($K^{\pm}, K^0, \pi^{\pm}$), while
$C_P=\frac{1}{2}$ for decays to neutral pions ($\pi^0$).

For the neutral modes $B^0\to K^0(\pi^0)\nu\bar{\nu}$, Eq.~(\ref{eq:Btonunubar_rate}) provides a full
Standard-Model description.
For the charged modes $B^+ \to K^+(\pi^+) \nu \bar{\nu}$, however, a tree-level amplitude arises via an
intermediate lepton between two charged interactions~\cite{Kamenik:2009kc}.
First the $B^+$ meson decays leptonically, \emph{i.e.}, $B^+ \to \ell^+ \nu$; subsequently, the charged
lepton decays as $\ell^+ \to P^+ \bar{\nu}$.
For $\ell=\tau$, the intermediate lepton can be on shell, leading to a long-distance contribution.
Interference between the long- and short-distance amplitudes is negligible~\cite{Kamenik:2009kc}, leaving the
following long-distance contribution to the rate
\begin{equation}
    \BR(B^+ \to P^+\nu_\tau \bar{\nu}_\tau)_{\mathrm{LD}}  =
        \frac{\left|  G_F^2 V_{ub} V_{us(d)}^\ast f_B f_{P} \right|^2 }{256 \pi^3 M_B^3}
        \frac{2\pi m_{\tau}(M_B^{2}-m_\tau^{2})^{2}(M_P^{2}-m_\tau^{2})^{2}}{\Gamma_\tau\Gamma_B} .
    \label{eq:Btonunubar_tree}
\end{equation}
Superficially, Eq.~(\ref{eq:Btonunubar_tree}) is suppressed relative to the loop-induced rate in
Eq.~(\ref{eq:Btonunubar_rate}) by~$G_F^2$, but the $\tau$ width $\Gamma_\tau$ is of order $G_F^2$,
canceling this suppression.
The long-distance contribution is also numerically significant because the $\tau$ mass is large.
For $B^+ \to \pi^+ \nu_\tau \bar{\nu}_\tau$, it is further enhanced relative to the short-distance
contribution by the CKM factor $|V_{ud}/V_{td}|^2$.

Taking the CKM matrix element $|V_{ub}| = 3.72(16) \times 10^{-3}$ from \FerMILC~\cite{Lattice:2015tia}, the
combinations $|V_{ud}| f_{\pi^-} = 127.13(2)(13)$~MeV and $|V_{us}| f_{K^+} = 35.09(4)(4)$~MeV from
experiment~\cite{Rosner:2015wva}, and all other inputs from Table~\ref{tab:inputs}, we obtain for the
$\nu_\tau$-pair final state:
\begin{eqnarray}
    \BR(B^+ \to \pi^+\nu_\tau \bar{\nu}_\tau)_{\mathrm{LD}}  & = & 9.48(92) \times 10^{-6} ,
    \label{eq:BtoPinunubar_LD_result} \\
    \BR(B^+ \to K^+\nu_\tau \bar{\nu}_\tau)_{\mathrm{LD}}  & = & 6.22(60) \times 10^{-7} ,
    \label{eq:BtoKnunubar_LD_result}
\end{eqnarray}
where the errors stem from the uncertainties on $f_B$ and $|V_{ub}|$, and other parametric errors are
negligible.
The long-distance contributions to the $B^+ \to K^+(\pi^+) \nu_\ell \bar{\nu_\ell}$ rate for $\ell = e, \mu$
are of order $10^{-17}$--$10^{-18}$~\cite{Kamenik:2009kc}.
Because lattice QCD provides reliable determinations of the hadronic inputs
$f_B$~\cite{Bazavov:2011aa,Na:2012kp,Dowdall:2013tga,Carrasco:2013naa,Christ:2014uea,Aoki:2014nga} and
$f_{\pi(K)}$~\cite{Follana:2007uv,Durr:2010hr,Bazavov:2010hj,Laiho:2011np,Arthur:2012yc,Dowdall:2013rya,%
Bazavov:2014wgs,Carrasco:2014poa}, the long-distance contributions to the $B^+ \to K^+(\pi^+) \nu \bar{\nu}$
decay rates are under good theoretical control.

\subsection{Tree-level \texorpdfstring{\boldmath$b\to u \ell \nu$}{b to u l nu} decay processes}
\label{subsec:ellnu} 

The tree-level semileptonic decay $B\to\pi\ell\nu_\ell\;(\ell=e,\mu,\tau)$ is mediated in the Standard Model
by the charged current interaction, and the resulting Standard-Model differential decay rate is
\begin{align}
    \frac{d\Gamma(B \to \pi \ell \nu_\ell)}{dq^2} &= C_P \frac{G_F^2 |V_{ub}|^2}{24 \pi^3}
        \frac{(q^2-m_\ell^2)^2 |\bm{p}_P|}{q^4 M_{B}^2} \left[ \left(1+\frac{m_\ell^2}{2q^2}\right)
        M_B^2 |\bm{p}_P|^2 |f_+(q^2)|^2
    \right. \nonumber \\ & + \left.
        \frac{3m_\ell^2}{8q^2}(M_{B}^2-M_\pi^2)^2|f_0(q^2)|^2\right] ,
\label{eq:B_semileptonic_rate}
\end{align}
where the isospin factor $C_P$ is the same as in Eq.~(\ref{eq:Btonunubar_rate}) above.
The decay rate depends upon both the vector ($f_+$) and scalar ($f_0$) form factors.
For decays to light charged leptons ($\ell=e,\mu$), the contribution from the scalar form factor is
suppressed by $m_\ell^2$ and hence negligibly small.
In contrast, the scalar form-factor contribution to decays into $\tau$ leptons is numerically significant.
While $B\to \pi \tau \nu_\tau$ decay is not a rare, loop-suppressed process in the Standard Model, the large
$\tau$-lepton mass makes it particularly sensitive to contributions mediated by charged Higgs bosons.

\section{Lattice-QCD form factors and symmetry tests}
\label{sec:SymTests} 

The first \emph{ab-initio} lattice-QCD results for the $\BtoK$ form factors and for the $\Btopi$ tensor form
factor became available only recently~\cite{Bouchard:2013eph,Lattice:2015tia,Bailey:2015nbd,Bailey:2015dka}.
Consequently, previous theoretical calculations of $B\to K (\pi) \ell^+\ell^-$ observables have sometimes
used expectations from heavy-quark and/or SU(3)-flavor symmetries to relate the unknown form factors to
others that can be constrained from experiment or computed with QCD models (see, {\it e.g.},\
Refs.~\cite{Bobeth:2011nj,Ali:2013zfa}).

In this section, we directly test these symmetry relations, at both high and low $q^2$, using the complete
set of \FerMILC\ \BtoK\ and \Btopi\ form factors~\cite{Lattice:2015tia,Bailey:2015nbd,Bailey:2015dka}.
For the $\Btopi$ case, we use the vector and scalar form factors $f_+$ and $f_0$ obtained from a combined
fit of lattice-QCD data with experiment.
This combination improves the precision on the form factors at low $q^2$, but assumes that no significant
new physics contributes to the tree-level $B\to\pi\ell\nu$ decays for $\ell=\mu,e$.

First, in Sec.~\ref{subsec:LatFFs}, we briefly summarize the lattice form-factor calculations, highlighting
the properties of the simulations and analysis that enable controlled systematic errors and high precision.
Then, in Sec.~\ref{subsec:HQSTests}, we present tests of heavy-quark symmetry relations for $\Btopi$ and
$\BtoK$ form factors that were not already presented in Refs.~\cite{Lattice:2015tia}
and~\cite{Bailey:2015dka}.
Finally, we calculate the size of SU(3)-flavor-breaking effects between the $\BtoK$ and $\Btopi$ form
factors and compare with power-counting expectations in Sec.~\ref{subsec:SU3Tests}.

\subsection{Lattice-QCD form-factor calculations}
\label{subsec:LatFFs}

The Fermilab Lattice and MILC Collaborations carried out the numerical lattice-QCD calculations in
Refs.~\cite{Lattice:2015tia,Bailey:2015nbd,Bailey:2015dka} in parallel.
Here we summarize the features of the work that enabled both high precision and controlled uncertainties.
Below we give the correlations between the $\Btopi$ and $\BtoK$ form factors, which have not appeared
elsewhere.
To put this new information in context, we summarize the similarities and slight differences between the
$\Btopi$~\cite{Lattice:2015tia,Bailey:2015nbd} and $\BtoK$~\cite{Bailey:2015dka} \FerMILC\ lattice-QCD
calculations.

The calculations~\cite{Lattice:2015tia,Bailey:2015nbd,Bailey:2015dka} employed the MILC asqtad
ensembles~\cite{Bernard:2001av,Aubin:2004wf,Bazavov:2009bb} at four lattice spacings from approximately
0.12~fm down to 0.045~fm; physical volumes with linear size $L\gtrsim3.8~\text{fm}$; and several choices for
the masses of the sea-quarks, corresponding to pions with mass as low as $175~\text{MeV}$.
The strange sea-quark mass was chosen close to the physical strange-quark mass, but varied a bit with
lattice spacing, allowing for adjustment of this mass \emph{a posteriori}.
The Fermilab method was used for the lattice $b$ quark~\cite{ElKhadra:1996mp}.
As in several other calculations, starting with Ref.~\cite{ElKhadra:2001rv}, the matching of the currents
from the lattice to the continuum was mostly nonperturbative, with a residual matching factor close to unity
computed in one-loop perturbation theory, with matching scale $\mu=m_b$ for the tensor current.
Because the one-loop calculation was separate from the Monte Carlo calculation of correlation functions, it
was exploited to introduce a multiplicative ``blinding'' offset.

The matrix elements for the form factors were obtained from fits to two- and three-point correlation
functions, including one excited $B$ meson in the fit.
After matching these lattice-QCD data to the continuum, as described above, two further analysis steps are
crucial for the present paper.
First, the form factors calculated in the kinematic region $q^2\gtrsim17~\text{GeV}^2$ were extrapolated to
zero lattice spacing and to the physical light-quark masses with a form of chiral perturbation theory
($\chi$PT) for semileptonic decays~\cite{Becirevic:2002sc} adapted to staggered fermions~\cite{Aubin:2007mc}.
Because the final-state pion and kaon energies can become large in the context of standard $\chi$PT, the
analyses found better fits with SU(2) hard-pion and hard-kaon $\chi$PT~\cite{Bijnens:2010ws}.
This chiral-continuum extrapolation included terms for heavy-quark discretization effects, with a functional
form taken from heavy-quark effective theory~\cite{Kronfeld:2000ck,Harada:2001fi,Oktay:2008ex}, as in
Ref.~\cite{Bazavov:2011aa}.

Next, to extend the form-factor results to the whole kinematically allowed region,
Refs.~\cite{Lattice:2015tia,Bailey:2015nbd,Bailey:2015dka} used the model-independent $z$ expansion based on
the analytic structure of the form factors.
In the present paper, we rely on the output of these fits, including correlations, so we repeat the most
pertinent details.
Following Refs.~\cite{Boyd:1994tt,Bourrely:2008za}, the complex $q^2$ plane is mapped to
\begin{equation}
    z(q^2,t_0) = \frac{\sqrt{t_+-q^2}-\sqrt{t_+-t_0}}{\sqrt{t_+-q^2}+\sqrt{t_+-t_0}},
    \label{eq:z}
\end{equation}
which maps a cut at $q^2>t_+=(M_B+M_P)^2$ to the unit circle and maps the semileptonic region to an interval
in $z$ on the real axis.
The extent of the interval can be minimized by choosing $t_0=(M_B+M_P)(\sqrt{M_B}-\sqrt{M_P})^2$, where
$M_P=M_{\pi}$ or $M_K$.
Unitarity implies that a power series in $z$ converges for $|z|<1$.
In Refs.~\cite{Lattice:2015tia,Bailey:2015nbd,Bailey:2015dka}, these series were used~\cite{Bourrely:2008za}:
\begin{align}
    f_+(q^2) &= \frac{1}{P_+(q^2)}\sum_{n=0}^{K-1} b_n^+ \left[z^n - (-1)^{K-n}\frac{n}{K}z^K \right],
    \label{eq:f+z} \\
    f_0(q^2) &= \frac{1}{P_0(q^2)}\sum_{n=0}^{K-1} b_n^0 z^n,
    \label{eq:f0z}
\end{align}
and the same for $f_T$ as for $f_+$ (with coefficients $b_n^T$).
The pole factor $P_{+,0,T}(q^2)=1-q^2/M_{+,0,T}^2$, with $M_{+,0,T}$ chosen as follows: for \Btopi,
$M_+=M_T=M_{B^*}=5.3252~\text{GeV}$ from experiment~\cite{Agashe:2014kda}, $M_0\to\infty$ (\emph{i.e.}, no
pole); for \BtoK, $M_+=M_T=M_{B_s^*}=5.4154~\text{GeV}$ from experiment~\cite{Agashe:2014kda},
$M_0=5.711~\text{GeV}$ from lattice QCD~\cite{Lang:2015hza}.
The output of the chiral-continuum extrapolation was propagated to Eqs.~(\ref{eq:f+z}) and~(\ref{eq:f0z})
using either synthetic data~\cite{Bailey:2015dka} or a functional fitting
procedure~\cite{Lattice:2015tia,Bailey:2015nbd}.

Reference~\cite{Lattice:2015tia} also presented determinations of the \Btopi\ form factors $f_+$ and $f_0$
from a combined $z$ fit to the lattice-QCD form factors and experimental measurements of the $\Btopi\ell\nu$
differential decay rate from the $B$
factories~\cite{delAmoSanchez:2010af,Ha:2010rf,Lees:2012vv,Sibidanov:2013rkk}.
This fit employed the same $z$~expansions as above.
The experimental data provides information on the shape of $f_+(q^2)$ at low $q^2$ beyond the direct reach
of lattice-QCD simulations, thereby reducing the form-factor errors at low $q^2$.
In this paper, we use these more precise \Btopi\ vector and scalar form factors for all calculations of
$\Btopi$ observables, thereby improving the precision of the Standard-Model results at the expense of the
assumption that new physics does not significantly alter the rate of this tree-level transition.

In Sec.~\ref{subsec:SU3Tests}, we present predictions for combinations of $\Btopi$ and $\BtoK$ observables,
which require the correlations between the two channels, not provided
before~\cite{Lattice:2015tia,Bailey:2015nbd,Bailey:2015dka}.
As co-authors of these papers, we have access to the relevant information.
To enable others to study both modes together, we provide the correlation coefficients in
Table~\ref{tab:Btopi-BtoK-correlations}.
With Eqs.~(\ref{eq:z})--(\ref{eq:f0z}) and the information contained in Table~XIX of
Ref.~\cite{Lattice:2015tia}, Table~XII of Ref.~\cite{Bailey:2015dka}, and Table~III of
Ref.~\cite{Bailey:2015nbd}, the supplementary information provided in
Table~\ref{tab:Btopi-BtoK-correlations} enables the reader to reproduce the form factors and combinations of
them.

\begin{table}
    \caption{Correlations between the $z$-expansion coefficients of the $\Btopi$ and $\BtoK$ vector,
        scalar, and tensor form factors, where the $\Btopi$ vector and scalar form factors include 
        experimental shape information from $\Btopi\ell\nu$ decay.
        These should be combined with Table~XIX of Ref.~\cite{Lattice:2015tia}, Table~III of
        Ref.~\cite{Bailey:2015nbd}, and Table~XII of Ref.~\cite{Bailey:2015dka}, which give the central 
        values of the coefficients as well as the remaining correlation information.}
    \label{tab:Btopi-BtoK-correlations}
    \begin{tabular}{cc@{\quad}ccccccccc}
        \hline\hline
        & & \multicolumn{9}{c}{$B\to K \ell\ell$} \\
        &             &  $b_{0}^{+}$ & $b_{1}^{+}$ & $b_{2}^{+}$ & $b_{0}^{0}$ & $b_{1}^{0}$ & $b_{2}^{0}$ & $b_{0}^{T} $& $b_{1}^{T}$ & $b_{2}^{T}$ \\
        \hline
        \parbox[t]{6mm}{\multirow{12}{*}{\rotatebox[origin=c]{90}{ $B\to \pi \ell\ell$}}}
        & $b_{0}^{+}$ & $\mph0.273$ &    $-0.002$ &    $-0.029$ & $\mph0.227$ & $\mph0.063$ & $\mph0.034$ & $\mph0.333$ &    $-0.001$ &    $-0.005$ \\
        & $b_{1}^{+}$ & $\mph0.016$ & $\mph0.085$ &    $-0.006$ &    $-0.003$ & $\mph0.061$ & $\mph0.067$ &    $-0.011$ & $\mph0.075$ & $\mph0.017$ \\
        & $b_{2}^{+}$ &    $-0.133$ &    $-0.069$ & $\mph0.024$ &    $-0.094$ &    $-0.077$ &    $-0.064$ &    $-0.124$ &    $-0.053$ & $\mph0.006$ \\
        & $b_{3}^{+}$ &    $-0.077$ &    $-0.033$ & $\mph0.060$ &    $-0.030$ &    $-0.028$ &    $-0.023$ &    $-0.062$ &    $-0.021$ & $\mph0.031$ \\
        & $b_{0}^{0}$ & $\mph0.278$ & $\mph0.098$ & $\mph0.091$ & $\mph0.299$ & $\mph0.160$ & $\mph0.124$ & $\mph0.285$ &    $-0.005$ &    $-0.005$ \\
        & $b_{1}^{0}$ &    $-0.004$ & $\mph0.225$ & $\mph0.155$ & $\mph0.065$ & $\mph0.197$ & $\mph0.171$ &    $-0.079$ & $\mph0.153$ & $\mph0.092$ \\
        & $b_{2}^{0}$ &    $-0.120$ &    $-0.231$ &    $-0.163$ &    $-0.194$ &    $-0.232$ &    $-0.183$ &    $-0.020$ &    $-0.058$ &    $-0.006$ \\
        & $b_{3}^{0}$ &    $-0.085$ &    $-0.192$ &    $-0.155$ &    $-0.144$ &    $-0.195$ &    $-0.171$ &    $-0.041$ &    $-0.121$ &    $-0.079$ \\
        & $b_{0}^{T}$ & $\mph0.319$ & $\mph0.051$ &    $-0.005$ & $\mph0.279$ & $\mph0.115$ & $\mph0.088$ & $\mph0.392$ & $\mph0.037$ & $\mph0.008$ \\
        & $b_{1}^{T}$ & $\mph0.056$ & $\mph0.080$ & $\mph0.012$ & $\mph0.051$ & $\mph0.072$ & $\mph0.063$ & $\mph0.067$ & $\mph0.097$ & $\mph0.048$ \\
        & $b_{2}^{T}$ & $\mph0.014$ & $\mph0.022$ & $\mph0.029$ & $\mph0.030$ & $\mph0.026$ & $\mph0.019$ & $\mph0.018$ & $\mph0.014$ & $\mph0.025$ \\
        & $b_{3}^{T}$ & $\mph0.005$ & $\mph0.010$ & $\mph0.026$ & $\mph0.023$ & $\mph0.015$ & $\mph0.008$ & $\mph0.010$ & $\mph0.003$ & $\mph0.022$ \\
        \hline\hline
    \end{tabular}
\end{table}

The dominant correlations between the two sets of form factors are statistical, because both calculations
used the same gauge-field ensembles.
In practice, however, both the statistical and systematic correlations are diluted in the chiral-continuum
extrapolations for $\Btopi$ and $\BtoK$, which were performed independently.
The same holds for several important systematic uncertainties, namely from the chiral-continuum
extrapolations, the uncertainty in the $B^*$-$B$-$\pi$ coupling, and the heavy-quark discretization errors.
The correlations became even smaller once the experimental $\Btopi\ell\nu$ data were used to constrain the
shape of the $\Btopi$ vector and scalar form factors.
In the end, the only significant correlations are among the leading coefficients $b_0$, which correspond
essentially to the normalization, and are therefore well determined by the data.
Even these are typically only $\sim0.3$, with the largest being $\sim0.4$.
Smaller correlations between the leading coefficients and the higher-order coefficients of $\sim0.1$--0.2
arise from the kinematic constraint $f_+(0) = f_0(0)$ enforced in the $z$-expansion fits.
The use of experimental $\Btopi\ell\nu$ data to constrain the shape of the $\Btopi$ vector and scalar form
factors, when combined with the kinematic constraint, leads to the negative entries in the correlation
matrix, which correspond to anticorrelations between $b_i^+$ and $b_i^0$ for $\Btopi$ and the other
coefficients.

\subsection{Tests of heavy-quark symmetry}
\label{subsec:HQSTests}

Several tests of heavy-quark-symmetry relations for \Btopi\ and \BtoK\ using the \FerMILC\ form factors were
already presented in Refs.~\cite{Lattice:2015tia,Bailey:2015dka}.
Figure~16 of Ref.~\cite{Bailey:2015dka} plots the $\BtoK$ form-factor ratios $f_0/f_+$ and $f_T/f_+$ at high
$q^2$ obtained from lattice QCD, and compares them with expectations from heavy-quark symmetry.
Similarly, Fig.~25 of Ref.~\cite{Lattice:2015tia} compares the $\Btopi$ form factor ratio $f_0/f_+$ with
heavy-quark-symmetry expectations.
Here we examine heavy-quark-symmetry tests of the $\Btopi$ tensor form factor.

\begin{figure}[t]
    \includegraphics[width=0.49\linewidth,trim=10pt 2pt 8pt 10pt,clip]{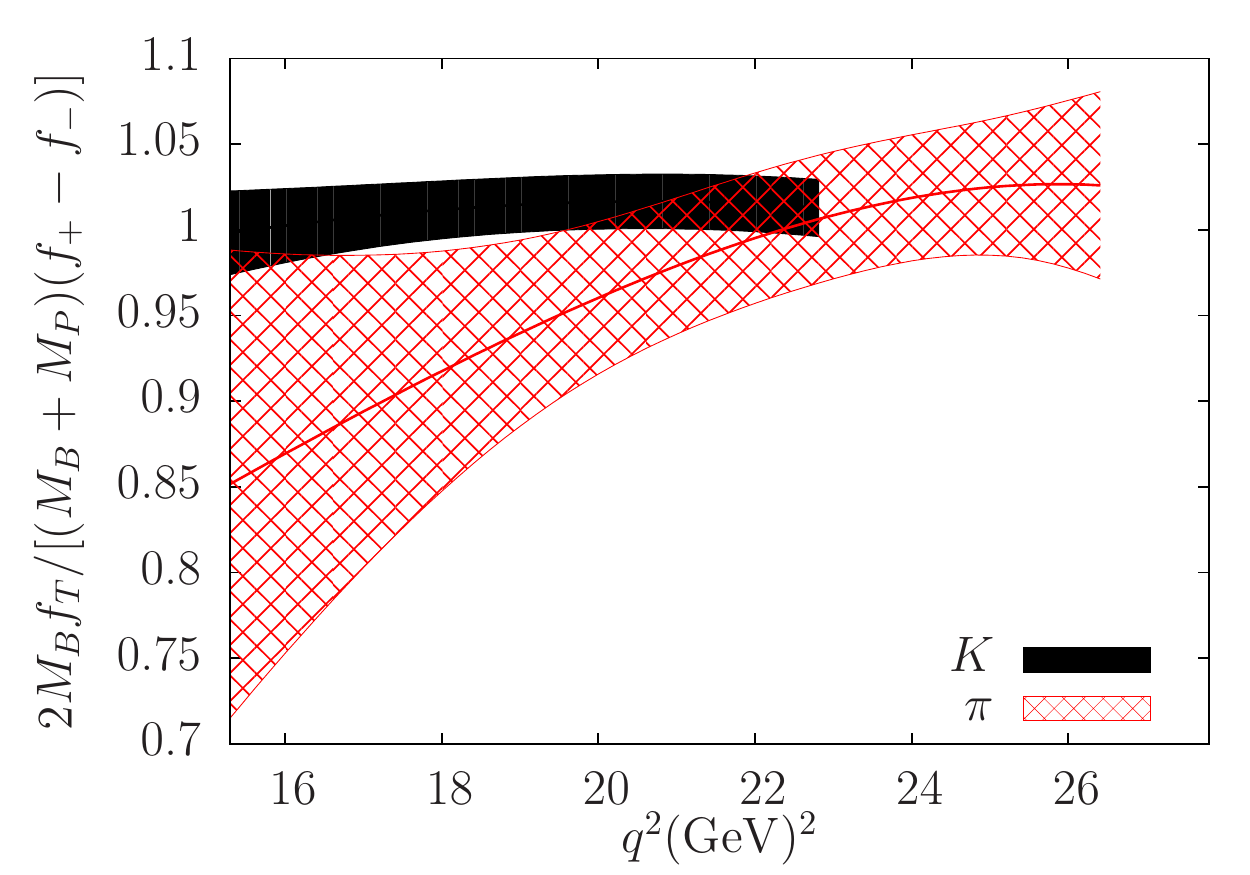} \hfill
    \includegraphics[width=0.49\linewidth,trim=10pt 2pt 8pt 10pt,clip]{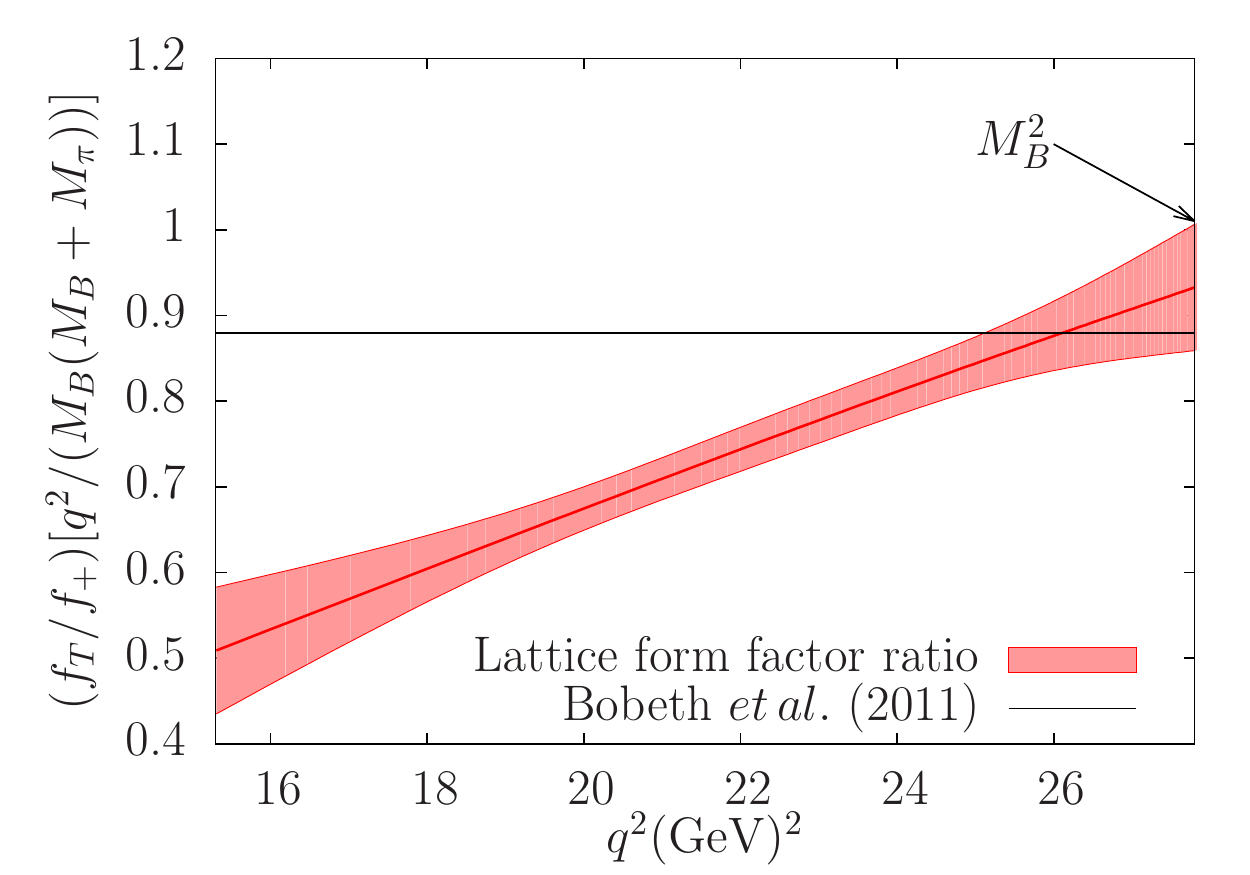}
    \caption{Tests of the heavy-quark symmetry relations for the tensor form factor using the \FerMILC\ 
        form factors~\cite{Lattice:2015tia,Bailey:2015nbd,Bailey:2015dka}.
        The left plot shows the ratio in Eq.~(\ref{eq:iwbd}) at low recoil for \Btopi\ (red hatched band) 
        and \BtoK\ (black solid band), which would become unity as 
        $m_b\to\infty$~\cite{Isgur:1990kf,Burdman:1992hg}.
        The right plot compares $(f_T/f_+)\left[q^2/(M_B(M_B + M_\pi))\right]$ at low recoil (red curve 
        with error band), for \Btopi\ with the theoretical expectation for 
        $\kappa(\mu)$~\cite{Bobeth:2011nj} (black horizontal line).}
    \label{fig:HQS}
\end{figure}

The simplest heavy-quark-symmetry relation between $f_T$ and the other form factors
is~\cite{Isgur:1990kf,Burdman:1992hg}
\begin{equation}
    f_T(q^2) = \frac{M_B+M_P}{2M_B} \left[f_+(q^2) - f_-(q^2)\right] + \order{\alpha_s, \Lambda/m_b},
    \label{eq:hqsft}
\end{equation}
which follows because the right-hand side is proportional to the matrix element
$\langle{}P|\bar{q}\gamma^ib|B\rangle$, while the left-hand side is proportional to
$\langle{}P|\bar{q}\gamma^i\gamma^0b|B\rangle$.
In Fig.~\ref{fig:HQS}, left, we plot the ratio
\begin{equation}
    \frac{2M_B}{M_B+M_P} \frac{f_T(q^2)}{f_+(q^2)-f_-(q^2)}
    \label{eq:iwbd}
\end{equation}
\emph{vs}.\ $q^2$, calculated directly from the \FerMILC\ form
factors~\cite{Lattice:2015tia,Bailey:2015nbd,Bailey:2015dka}.
We show the range $q^2\gtrsim15~\text{GeV}^2$ where the error remains small enough to provide a meaningful
test of the relation in Eq.~(\ref{eq:hqsft}).
In this region, the approximation in Eq.~(\ref{eq:hqsft}) fares very well, especially for \BtoK.

A refinement of this idea uses heavy-quark symmetry and $m_b$-scaling to eliminate
$f_-$~\cite{Grinstein:2004vb,Hewett:2004tv,Bobeth:2011nj}:
\begin{equation}
    \lim_{q^2 \to M_B^2}\frac{f_T(q^2,\mu)}{f_+(q^2)} = \kappa(\mu) \frac{M_B(M_B + M_\pi)}{q^2}
        + \order{\Lambda/m_b},
    \label{eq:fTf+lowrecoil} 
\end{equation}
in which the scale-dependent coefficient $\kappa(\mu)$ incorporates radiative corrections and is given
explicitly through order~$\alpha_s^2$ in Eq.~(\ref{eq:kappa}).
Figure~\ref{fig:HQS}, right, compares the quantity $(f_T/f_+)\left[q^2/(M_B(M_B + M_\pi))\right]$ obtained
from the \FerMILC\ \Btopi\ form factors \cite{Lattice:2015tia,Bailey:2015nbd} with the theoretical
prediction from Eq.~(\ref{eq:fTf+lowrecoil}).
For the theoretical estimate, we take $m_b = 4.18$~GeV and ${\alpha_s}^{(4)}_{\overline{\rm MS}} (m_b) =
0.2268$, giving $\kappa(m_b) \approx 0.88$~\cite{Grinstein:2004vb,Bobeth:2011nj}.

As observed in Ref.~\cite{Bailey:2015dka} for the $B\to K$ form factors, the ratio $f_T/f_+$ calculated
directly from lattice QCD agrees well with the expectation from Eq.~(\ref{eq:fTf+lowrecoil}) for
$q^2 \approx M_B^2$.
Although we do not show any errors on the theoretical prediction, we can estimate the size of higher-order
corrections in the heavy-quark expansion from power counting.
Taking $\Lambda=500$~MeV gives $\Lambda/m_b \sim 12\%$.
Equation~(\ref{eq:fTf+lowrecoil}) also receives corrections from the pion recoil energy of order~$E_\pi/m_b$.
This ratio grows rapidly from $E_\pi/m_b \sim 3\%$ at $q^2_{\rm max}$ to $E_\pi/m_b \sim 30\%$ at $q^2
\approx 14 {\rm~GeV}^2$.
The observed size of deviations from the leading heavy-quark-symmetry prediction are somewhat larger than
the rough estimate based on power counting.
Although form factors from {\it ab-initio} QCD are now available for $\Btopi$ and $\BtoK$, other analyses of
semileptonic decay processes might still use heavy-quark-symmetry relations.
Figure~\ref{fig:HQS} provides quantitative, empirical guides for estimating the associated systematic
uncertainty introduced by their use.

In the limit $q^2\ll M_B^2$, a collinear spin symmetry emerges for the energetic daughter quark, and the
vector, scalar, and tensor form factors are related to a universal $M_B$-independent form
factor~\cite{Charles:1998dr,Beneke:2000wa}:
\begin{equation}
    f_+(q^2) = \frac{M_B}{2E_P}f_0(q^2) = \frac{M_B}{M_B+M_P}f_T(q^2), \quad (q^2\ll M_B^2).
\end{equation}
The first relation merely recovers the kinematic constraint, $f_+(0)=f_0(0)$.
Unfortunately, as $q^2$ decreases, so do the correlations between the lattice-QCD determinations of $f_+$
and $f_T$.
The error on the ratio $f_T/f_+$ therefore increases, reaching 100\% at $q^2=0$.
Thus, we are unable to quantitatively test the predicted relationship between $f_T$ and $f_+$ at large
recoil.

\subsection{Tests of SU(3)-flavor symmetry}
\label{subsec:SU3Tests}

The $\Btopi$ and $\BtoK$ form factors would be equal in the SU(3)-flavor limit $m_u=m_d=m_s$ and, thus,
differ due to corrections that are suppressed by the factor $(m_s-m_{ud})/\Lambda$, where $\Lambda$ is a
typical QCD scale inside heavy-light mesons.
Indeed, approximate SU(3) symmetry implies relations among all matrix elements of the form
\begin{equation}
    \langle P_{q\bar{r}}(p_P)|\bar{q}\Gamma b|B_r(p_B)\rangle  ,
    \label{eq:su3}
\end{equation}
where $r$ denotes the flavor of the spectator quark; $\Gamma$ is $\gamma^\mu$, $1$, or $i\sigma^{\mu\nu}$;
and the subscript on the final-state pseudoscalar denotes its flavor content.

A rule of thumb~\cite{Bajc:1996py} for SU(3) breaking is that large effects can be traced to the
pseudoscalar masses---$M_K^2\gg M_\pi^2$---while SU(3)-breaking effects in matrix elements per se are small.
In considering the matrix elements in Eq.~(\ref{eq:su3}), the final-state four-momentum~$p_P$ cannot be the
same for all $P_{q\bar{r}}$ mesons, because the mass shells differ.
The masses affect the kinematic variables $q^2$, $E_P$, and $\bm{p}_P$ in different ways.

\begin{figure}
    \includegraphics[width=0.49\linewidth,trim=10pt 2pt 8pt 10pt,clip]{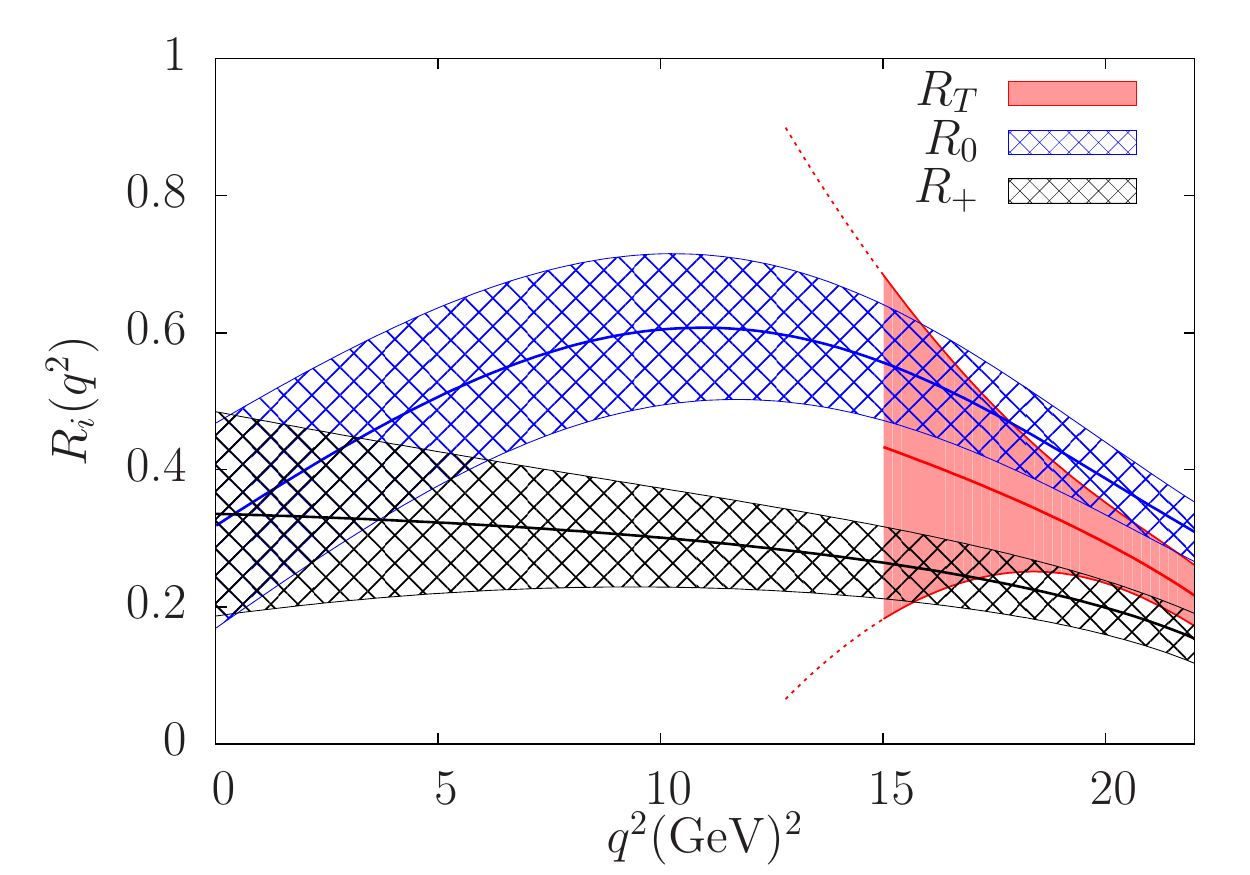} \hfill
    \includegraphics[width=0.49\linewidth,trim=10pt 2pt 8pt 10pt,clip]{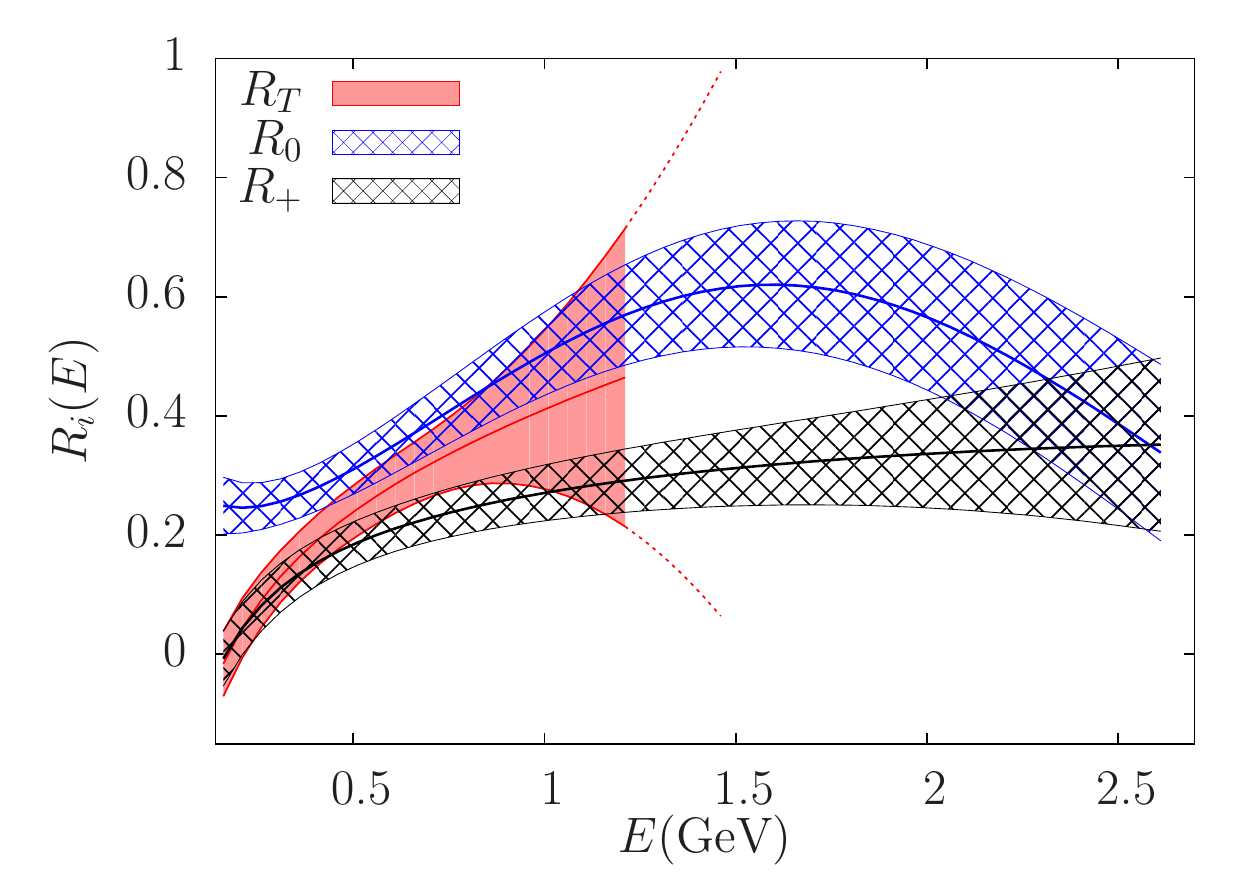}
    \caption{SU(3)-flavor-breaking ratios using the \FerMILC\ form 
        factors~\cite{Lattice:2015tia,Bailey:2015nbd,Bailey:2015dka} (solid and hatched curves with error 
        bands).   
        Left: ratios $R_{+,0,T}(q^2)$ [Eq.~(\ref{eq:SU3_Ri})].   
        Right: ratios $R_{+,0,T}(E)$ [Eq.~(\ref{eq:SU3_Ri-E})].
        We do not show $R_T$ when the error becomes too large to draw any useful inferences, although the 
        trend of the error band is shown by the thin lines extending from the $R_T$ error bands.}
    \label{fig:RT}
\end{figure}

In Ref.~\cite{Ali:2013zfa} (see also Ref.~\cite{Bajc:1996py}), SU(3)-breaking was considered as a 
function of $q^2$, with the quantities $R_{+,0,T}(q^2)$ defined as
\begin{equation}
    R_i(q^2) = \frac{f_i^{BK}(q^2)}{f_i^{B\pi}(q^2)}-1,
    \label{eq:SU3_Ri}
\end{equation}
where $i=+,0,T$.
In Ref.~\cite{Ali:2013zfa}, the ratios $R_{+,0}(q^2)$ were calculated using lattice $\Btopi$ and $\BtoK$
form factors from Refs.~\cite{Dalgic:2006dt} and~\cite{Bouchard:2013mia,Bouchard:2013pna}, respectively.
Here we repeat the tests in Ref.~\cite{Ali:2013zfa} using the more precise \FerMILC\ \Btopi\ vector and
scalar form factors~\cite{Lattice:2015tia}, and include an additional test using the \FerMILC\ \Btopi\
tensor form factor~\cite{Bailey:2015nbd}, taking the $\BtoK$ form factors from
\FerMILC~\cite{Bailey:2015dka} as well.
Figure~\ref{fig:RT}, left, plots the quantity $R_i(q^2)$, including correlations between the $\Btopi$ and
$\BtoK$ form factors from statistics as well as the dominant systematic
errors~\cite{Lattice:2015tia,Bailey:2015nbd,Bailey:2015dka}.
We find that the sizes of $R_i(q^2)$ are between 20\% and 60\%, ranging from commensurate with
$(m_s-m_{ud})/\Lambda$ (for $m_s\sim100$~MeV, $\Lambda\sim500$~MeV) to uncharacteristically large.
In the region where the error on the tensor form factors remain manageable, we find that
$R_T(q^2)\approx[R_+(q^2)+R_0(q^2)]/2$, as assumed in Ref.~\cite{Ali:2013zfa}.
With the more precise \FerMILC\ \Btopi\ form factors~\cite{Lattice:2015tia,Bailey:2015nbd}, the resulting
SU(3) breaking is larger than that deduced and employed in that work.

As an alternative to Eq.~(\ref{eq:SU3_Ri}), we consider the analogous ratio with fixed final-state energy 
(in the $B$ rest frame):
\begin{equation}
    R_i(E) = \frac{f_i^{BK}(E)}{f_i^{B\pi}(E)}-1.
    \label{eq:SU3_Ri-E}
\end{equation}
As shown in Fig.~\ref{fig:RT}, right, the SU(3) breaking in $R_i(E)$ is similar to that in~$R_i(q^2)$.

\begin{figure}
    \includegraphics[width=0.49\linewidth]{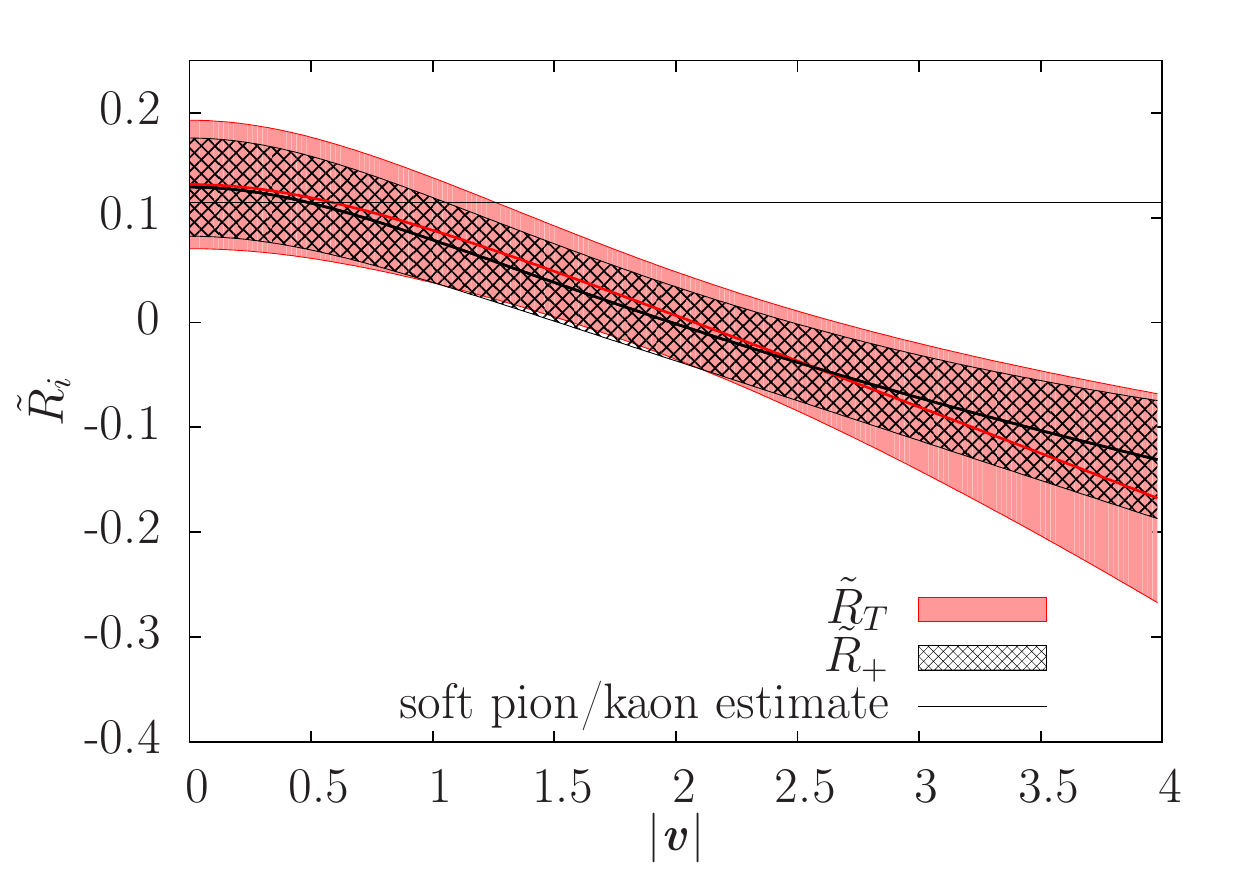} \hfill
    \includegraphics[width=0.49\linewidth]{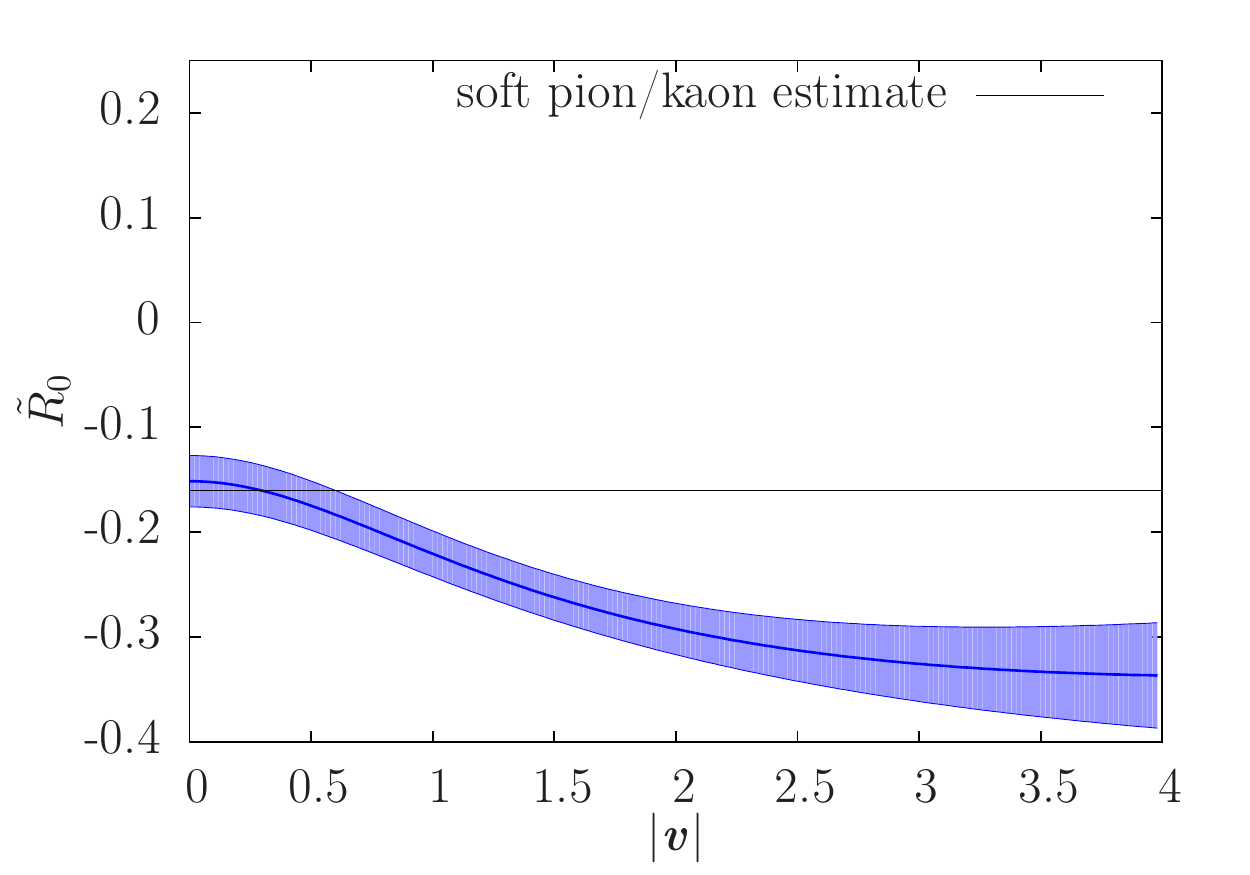}
    \caption{Velocity-based SU(3)-flavor-breaking ratios $\tilde{R}_{+,T}(|\bm{v}|)$ (left) and
        $\tilde{R}_{0}(|\bm{v}|)$ (right) using the lattice-QCD form factors from 
        Refs.~\cite{Lattice:2015tia,Bailey:2015nbd,Bailey:2015dka}.}
    \label{fig:RT-extra}
\end{figure}

A further alternative is to examine
\begin{equation}
    \tilde{R}_i(|\bm{v}|) = \frac{f_i^{BK}(|\bm{v}|)P_i^{BK}}{f_i^{B\pi}(|\bm{v}|)P_i^{B\pi}}-1,
    \label{eq:SU3_Ri-v}
\end{equation}
where $\bm{v}=\bm{p}_P/M_P$ is the final-state three-velocity (in HQET conventions) in the $B$-meson rest
frame.
The factors $P_0^{B\pi}=P_0^{BK}=1$, $P_{+,T}^{B\pi}=1-q^2/M_{B^*}^2$, and $P_{+,T}^{BK}=1-q^2/M_{B_s^*}^2$
are introduced to remove the kinematically important vector-meson pole from the vector and tensor form
factors.
As shown in Fig.~\ref{fig:RT-extra}, this measure of SU(3) breaking is under 20\% for $f_+$ and $f_T$, and
under 35\% for $f_0$ (in the momentum range shown).
The result for $\tilde{R}_0$ can be understood in the soft-pion (soft-kaon) limit,
where $f_0\propto f_B/f_P$, so
\begin{equation}
    \tilde{R}_0(|\bm{v}|\to0) = \frac{f_\pi}{f_K}-1\approx-0.16.
\end{equation}
which agrees very well with our result in Fig.~\ref{fig:RT-extra} (right). 
Similarly, for $\tilde{R}_+$
\begin{equation}
    \tilde{R}_+(|\bm{v}|\to0) = \frac{f_\pi g_{B_s^*BK}}{f_K g_{B^*B\pi}}-1.
\end{equation}
The SU(3)-breaking effects in the couplings can be estimated from the chiral extrapolation of a recent
calculation of~$g_{B^*B\pi}$~\cite{Flynn:2015xna}.
The expression of the chiral extrapolation of $g_{B^*B\pi}$ was given in Eq.~(28) of
Ref.~\cite{Flynn:2015xna}.
Replacing the pions in the loops with kaons, we estimate $g_{B_s^*BK}/g_{B^*B\pi}\approx1.33$ and,
consequently, $R_+ \approx 0.11$.
Because of the heavy-quark relation between $f_T$ and $f_+$ discussed above, $\tilde{R}_T$ should be close
to $\tilde{R}_+$, and one can see in Fig.~\ref{fig:RT-extra} (left) that this is indeed the case.

\section{Standard-Model results}
\label{sec:SMPheno}

We now use the \FerMILC\ \Btopi\ and \BtoK\ form factors~\cite{Lattice:2015tia,Bailey:2015nbd,%
Bailey:2015dka} to predict $B \to K(\pi)\ell^+\ell^-$, $B\to K(\pi)\nu\bar{\nu}$, and $B\to\pi\tau\nu$
observables (and their ratios) in the Standard Model.
For predictions of $\Btopi$ decay observables, as in the previous section, we use the more precise $\Btopi$
vector and scalar form factors obtained using the measured $\Btopi\ell\nu$ $q^2$ spectrum to constrain the
shape.
We present results for rare decays with a charged-lepton pair final state, $b\to q\ell\ell \; (q=d,s)$ in
Sec.~\ref{subsec:results_ll}, for rare decays with a neutrino pair final state $b\to q \nu \bar{\nu} \;
(q=d,s)$ in Sec.~\ref{subsec:results_nunubar}, and for tree-level $b\to u \tau\nu_\tau$ semileptonic decays
in Sec.~\ref{subsec:results_ellnu}.
Where possible, we compare our results with experimental measurements.

We compile our numerical results for the partially integrated $B \to K (\pi) \ell^+ \ell^-$ observables over
different $q^2$ intervals in Tables~\ref{tab:B2pi_FH_charged}--\ref{tab:B2P+_ratio_dBdq2_Other_correlations}
of Appendix~\ref{app:Results}.
To enable comparison with the recent experimental measurements of $B \to K (\pi) \ell^+ \ell^-$ from LHCb,
we provide the matrix of correlations between our Standard-Model predictions for the binned branching
fractions (and the ratio of $\Btopi$-to-$\BtoK$ binned branching fractions) for the same wide $q^2$ bins
below and above the charmonium resonances employed by LHCb~\cite{Aaij:2014pli,Aaij:2015nea}.

Appendix~\ref{app:Formulae} provides the complete expressions for the Standard-Model 
$B \to K (\pi)\ell^+ \ell^-$ $(\ell= e,\mu, \tau$) differential decay rates.
The simpler expressions for the $B \to K (\pi) \nu \bar{\nu}$ and $B\to \pi\tau\nu$ decay rates are
presented in the main text of Secs.~\ref{subsec:nunubar} and~\ref{subsec:ellnu}, respectively.
The Wilson coefficients and other numerical inputs used for all of the phenomenological analyses in this
work are given in Appendix~\ref{app:Inputs}.

\subsection{Rare \texorpdfstring{\boldmath$b\to q\, \ell\ell \; (q=d,s)$}{b to qll} decay observables}
\label{subsec:results_ll}

\subsubsection{\texorpdfstring{$B \to \pi \ell^+\ell^-$}{B to pi ll} observables}   
\label{subsec:B2piobs}


The Fermilab Lattice and MILC Collaborations already presented some Standard-Model predictions for
$B\to\pi\ell^+\ell^-$~\cite{Bailey:2015nbd}.
Figure~4 of that work plots the differential branching fractions for $\ell=\mu,\tau$, while Table~IV gives
the partial branching fractions in selected intervals of $q^2$ below and above the charmonium resonances,
and for the full kinematic range.
To enable correlated analyses of the partially integrated branching fractions for $B \to \pi$ from
Ref.~\cite{Bailey:2015nbd} and those for $B\to K$ presented in Sec.~\ref{subsec:B2Kobs}, we update the
large-bin numerical results from Table~IV of Ref.~\cite{Bailey:2015nbd} in Table~\ref{tab:B+2pi_dBdq2} here
by adding digits to the quoted uncertainties and combining the scale uncertainty (quoted separately in
Ref.~\cite{Bailey:2015nbd}) with the ``other" error.
In addition, we extend the phenomenological analysis of $B\to\pi \ell^+\ell^-$ by providing predictions for
the flat term of the angular distribution, cf.\ Eqs.~(\ref{eq:FH_def}) and~(\ref{eq:FH_def_bin}).

\begin{figure}
    \includegraphics[width=0.49\textwidth]{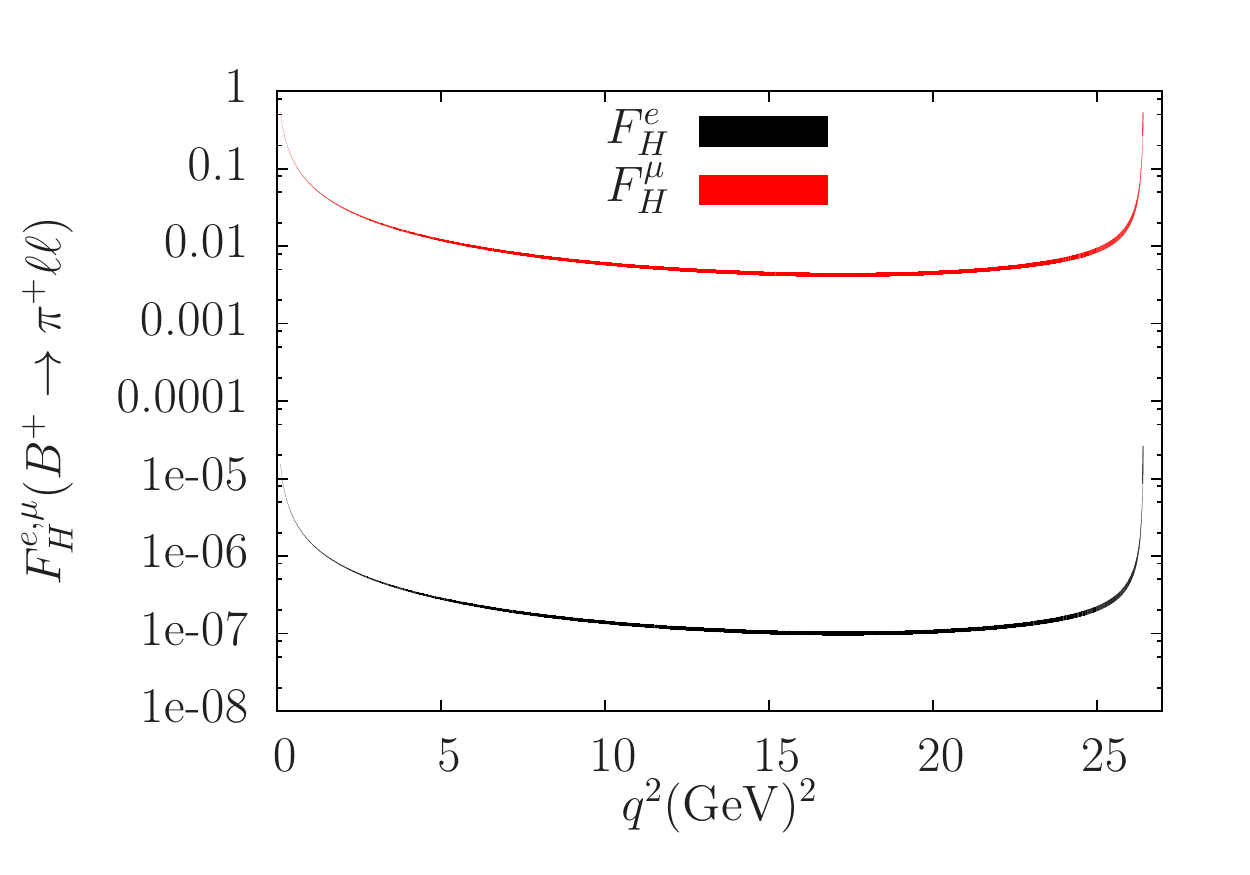} \hfill
    \includegraphics[width=0.49\textwidth]{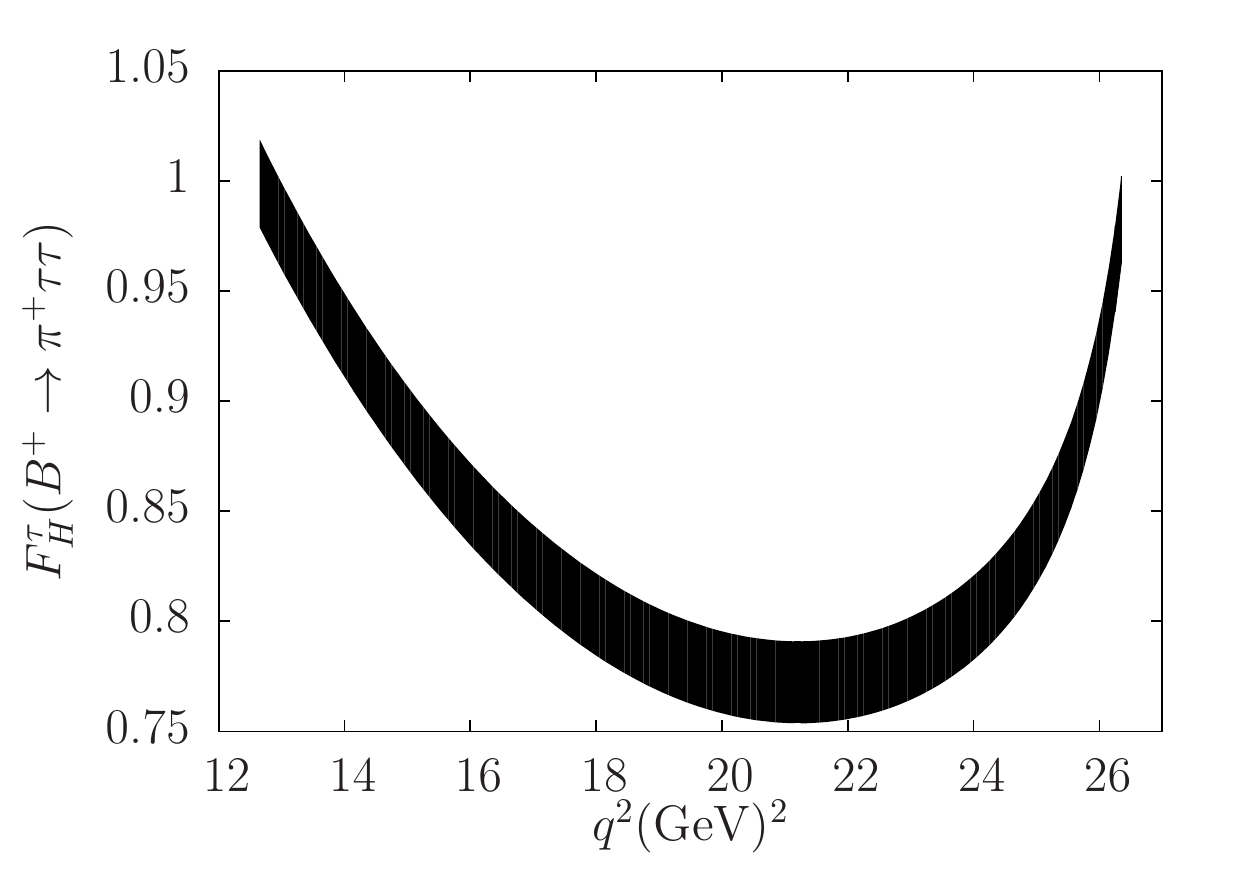} 
    \caption{Standard-Model predictions for the $B^+ \to \pi^+ \ell^+\ell^-$ flat term $F_H^\ell(q^2)$ 
        for $\ell=e,\mu$ (left) and $\ell=\tau$ (right) using the \FerMILC\ form
        factors~\cite{Lattice:2015tia,Bailey:2015nbd}.} 
    \label{fig:B2pi_FH}
\end{figure}

Figure~\ref{fig:B2pi_FH} plots our Standard-Model predictions for the $B^+\to\pi^+ \ell^+\ell^-$ flat term
$F_H^{\ell}(q^2)$, $\ell=e,\mu,\tau$, while Tables~\ref{tab:B2pi_FH_charged} and~\ref{tab:B2pi_FH_neutral}
report numerical values for the binned version $F_H^{\ell}(q^2_\text{min},q^2_\text{max})$ for the charged
and neutral decay modes, respectively.
For the dimuon final state, we find $F_H^\mu(q^2)\sim1$--2\% for most of the kinematic range, which is large
enough to be measured in future experiments.
For the electron-positron final state, $F_H^e(q^2)$ is so small---$10^{-6}$ or smaller---that any
foreseeable nonzero measurement would indicate the presence of new physics.

After the Fermilab Lattice and MILC Collaborations submitted Ref.~\cite{Bailey:2015nbd} for publication, the
LHCb experiment announced~\cite{TekampeDPF2015} a new measurement of the differential decay rate for
\Btopimumu\ decay, which is now finalized~\cite{Aaij:2015nea}.
Here we repeat the main numerical results of Ref.~\cite{Bailey:2015nbd} and compare them to the LHCb
measurement.
The Standard-Model predictions for the partially integrated branching ratio in the wide high-$q^2$ and
low-$q^2$ bins are~\cite{Bailey:2015nbd}
\begin{equation}
    \Delta \BR(B^+\to \pi^+ \mu^+\mu^-)^\text{SM} \times 10^9 = \left\{
    \begin{array}{lrcr}
      4.78(29)(54)(15)(6) &  1~\text{GeV}^2\leq & q^2 \leq &  6~\text{GeV}^2 , \\
      5.05(30)(34)(7)(15) & 15~\text{GeV}^2\leq & q^2 \leq & 22~\text{GeV}^2 ,
    \end{array} \right.
    \label{eq:BpillBins}
\end{equation}
where the errors are from the CKM matrix elements, form factors, the variation of the high and low matching
scales, and the quadrature sum of all other contributions, respectively.
LHCb quotes measured values for binned differential branching fractions~\cite{Aaij:2015nea}, which we
convert to partially integrated branching fractions for ease of comparison with Eq.~(\ref{eq:BpillBins}):
\begin{equation}
    \Delta\BR(B^+\to\pi^+\mu^+\mu^-)^\text{exp} \times 10^{9}~\text{GeV}^{2} = \left\{
    \begin{array}{lrcr}
        4.55 \left(^{+1.05}_{-1.00}\right)(0.15)  &  1~\text{GeV}^2 \leq & q^2 \leq &  6~\text{GeV}^2 , \\
        3.29 \left(^{+0.84}_{-0.70}\right)(0.07)  & 15~\text{GeV}^2 \leq & q^2 \leq & 22~\text{GeV}^2 ,
    \end{array} \right.
    \label{eq:LHCb_B2pi}
\end{equation}
where the two errors are statistical and systematic. 

\begin{figure}
    \includegraphics[width=0.49\textwidth]{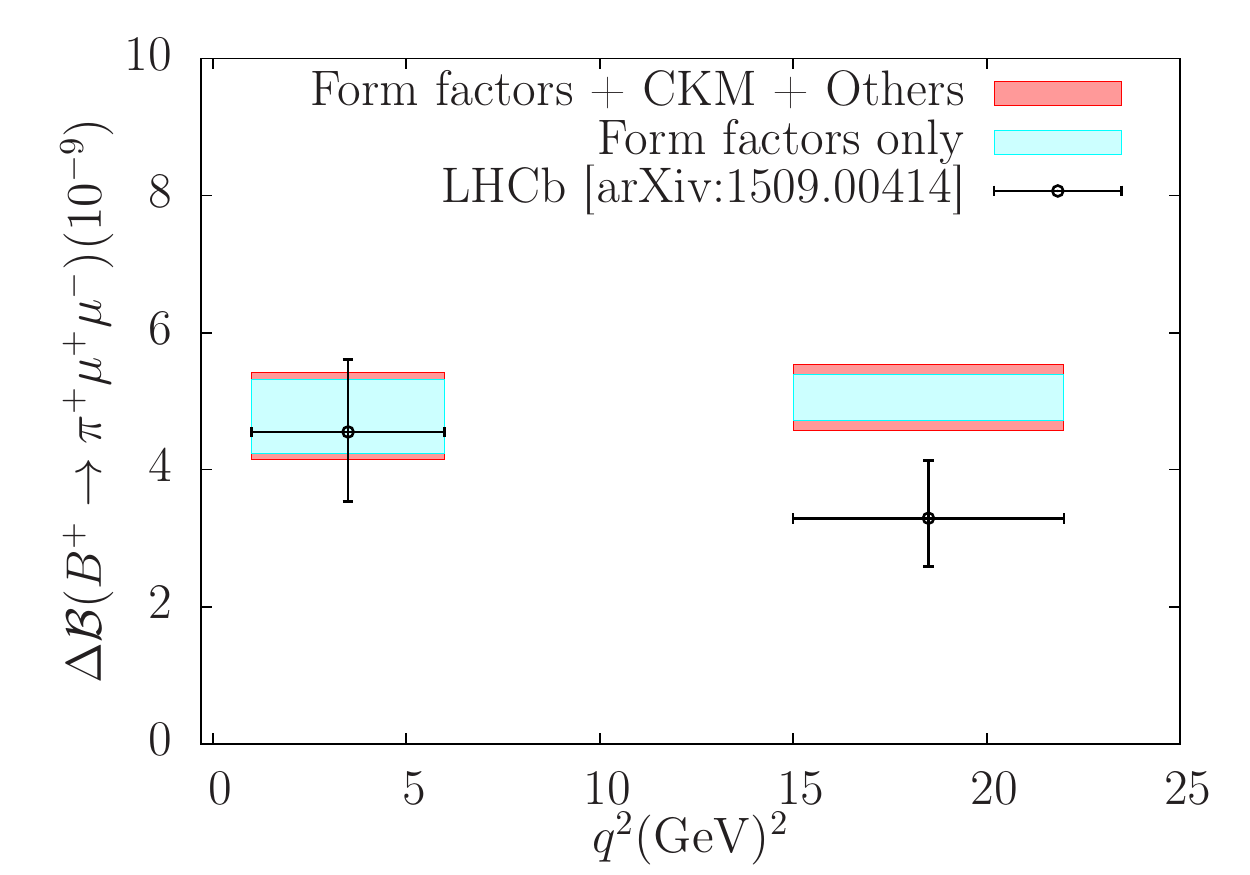}  \hfill
    \includegraphics[width=0.49\textwidth]{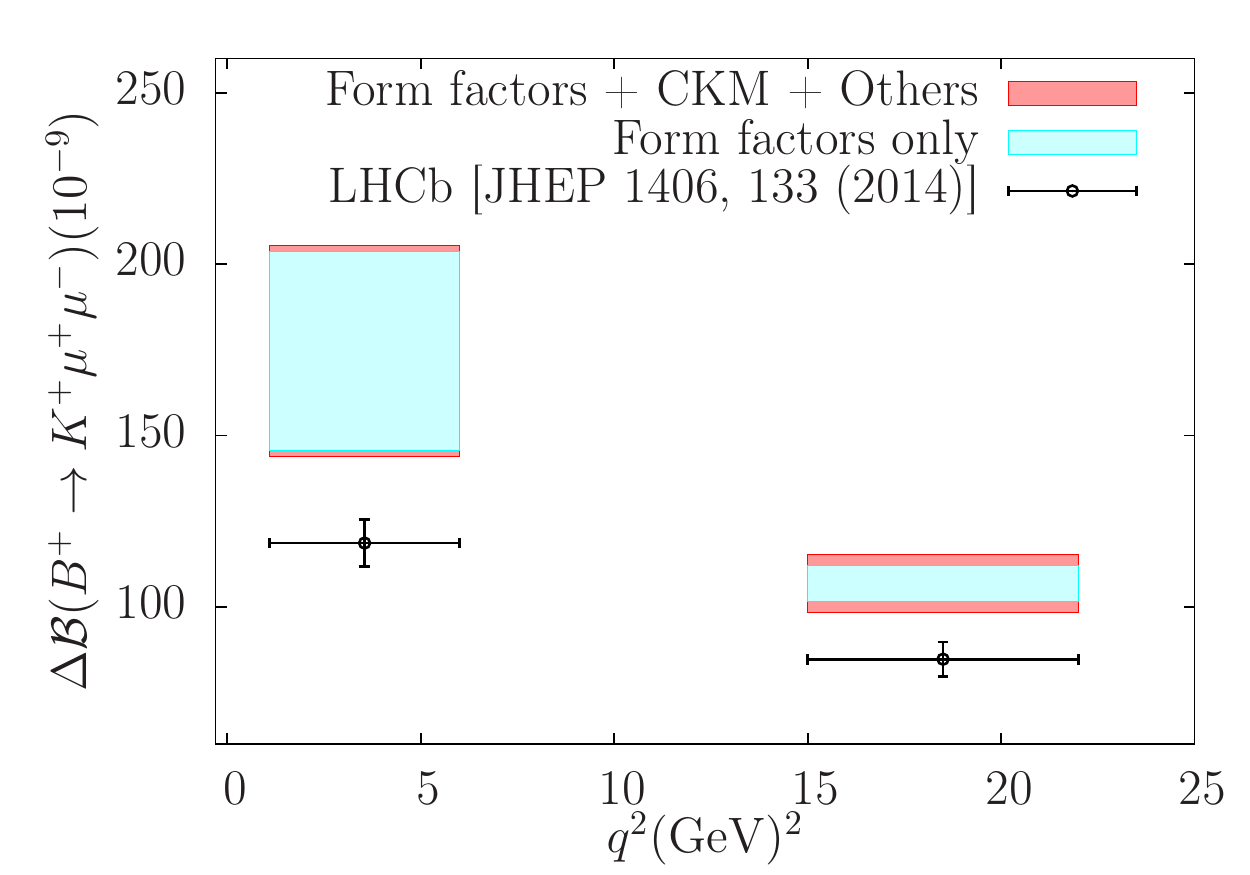}
    \caption{Standard-Model partially integrated branching ratios for $B^+\to \pi^+\mu^+\mu^-$ decay  
        (left) and $B^+\to K^+\mu^+\mu^-$ decay (right) using the \FerMILC\ form factors%
        ~\cite{Bailey:2015dka,Lattice:2015tia,Bailey:2015nbd} compared with experimental measurements 
        from LHCb~\cite{Aaij:2014pli,Aaij:2015nea} for the wide $q^2$ bins above and below the charmonium 
        resonances.}
    \label{fig:B2K_dBdq2_integrated}
\end{figure}

Figure~\ref{fig:B2K_dBdq2_integrated} (left panel) compares the Standard-Model predictions from
Ref.~\cite{Bailey:2015nbd} and LHCb for the wide bins.
The result for the low $q^2$ interval below the charm resonances agrees with the experimental measurement,
but that for the high $q^2$ interval differs at the $1.9\sigma$ level.
The combination of the two bins, including the theoretical correlations from
Tables~\ref{tab:B2pi+_B2K+_dBdq2_FF_correlations}, and~\ref{tab:B2pi+_B2K+_dBdq2_Other_correlations} and
treating the experimental bins as uncorrelated, yields a $\chi^2/\text{dof} = 3.7/2$ ($p=0.15$), and thus
disfavors the Standard-Model hypothesis at 1.4$\sigma$ confidence level.

Although LHCb's recent measurement of the $\Btopi\ell^+\ell^+$ differential decay rate~\cite{Aaij:2015nea}
is compatible with the Standard-Model predictions, the uncertainties leave room for sizable new-physics
contributions.
In the high-$q^2$ interval, $15~\text{GeV}^2\leq q^2 \leq 22~\text{GeV}^2$, the theoretical and experimental
errors are commensurate.
Future, more precise measurements after the LHCb upgrade will refine the comparison, thereby strengthening
the test of the Standard Model.

\subsubsection{\texorpdfstring{$B \to K \ell^+\ell^-$}{B to K ll} observables}
\label{subsec:B2Kobs}


Here we present results for $B\to K \ell^+\ell^-$ ($\ell = \mu, \tau$) observables in the Standard Model
using the \FerMILC\ \BtoK\ form factors~\cite{Bailey:2015dka}.
Many previous phenomenological analyses of $\BtoKll$ related the tensor form factor $f_T$ to the
vector form factor $f_+$ based on approximate symmetries~\cite{Bobeth:2007dw,Bobeth:2011nj}.
The HPQCD Collaboration has also presented results for $\BtoK$ observables using their own lattice-QCD
form-factor determinations~\cite{Bouchard:2013mia}.
We improve upon the Standard-Model predictions in that work and in Ref.~\cite{Bailey:2015dka} by
incorporating hard-scattering contributions at low $q^2$ and by using Wilson coefficients that include
logarithmically enhanced QED corrections.

\begin{figure}
    \includegraphics[width=0.49\textwidth]{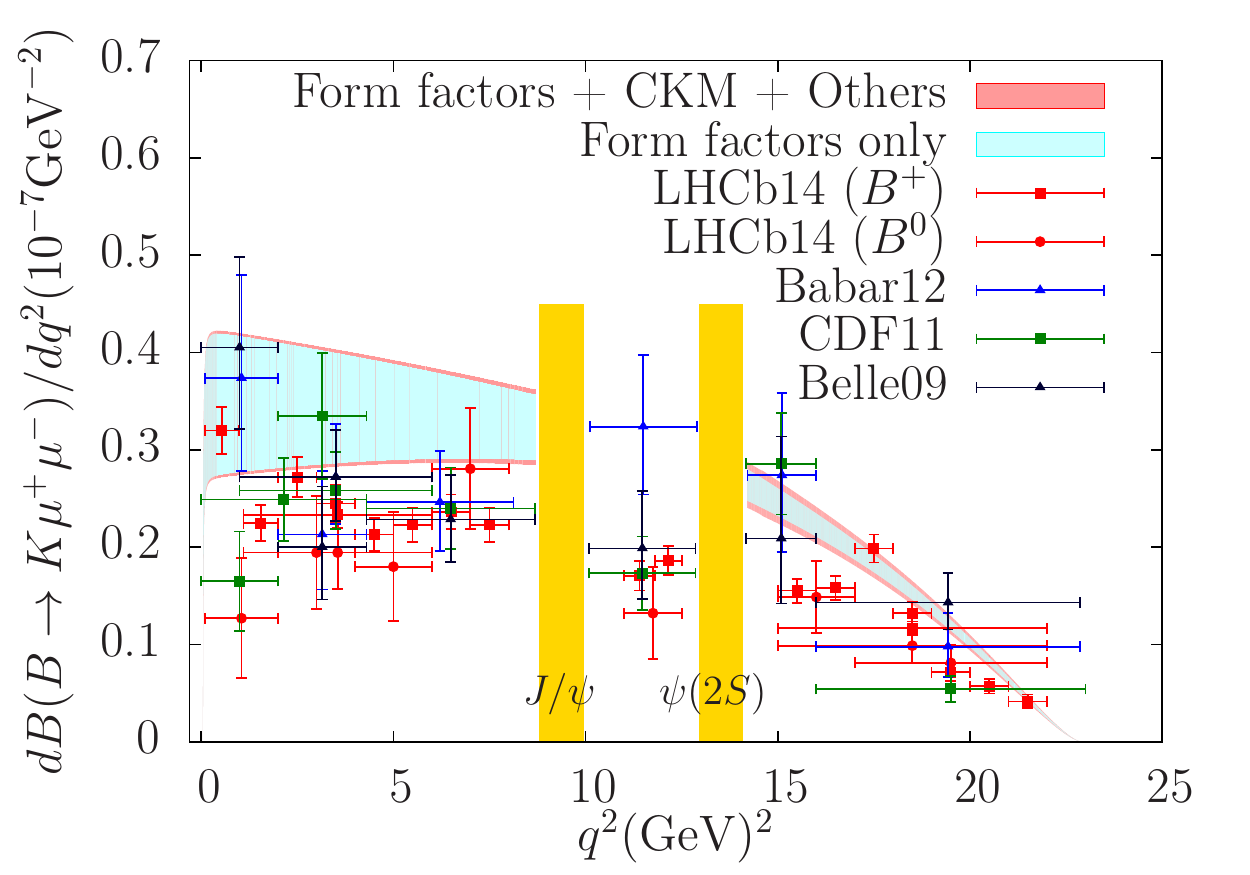} \hfill
    \includegraphics[width=0.49\textwidth]{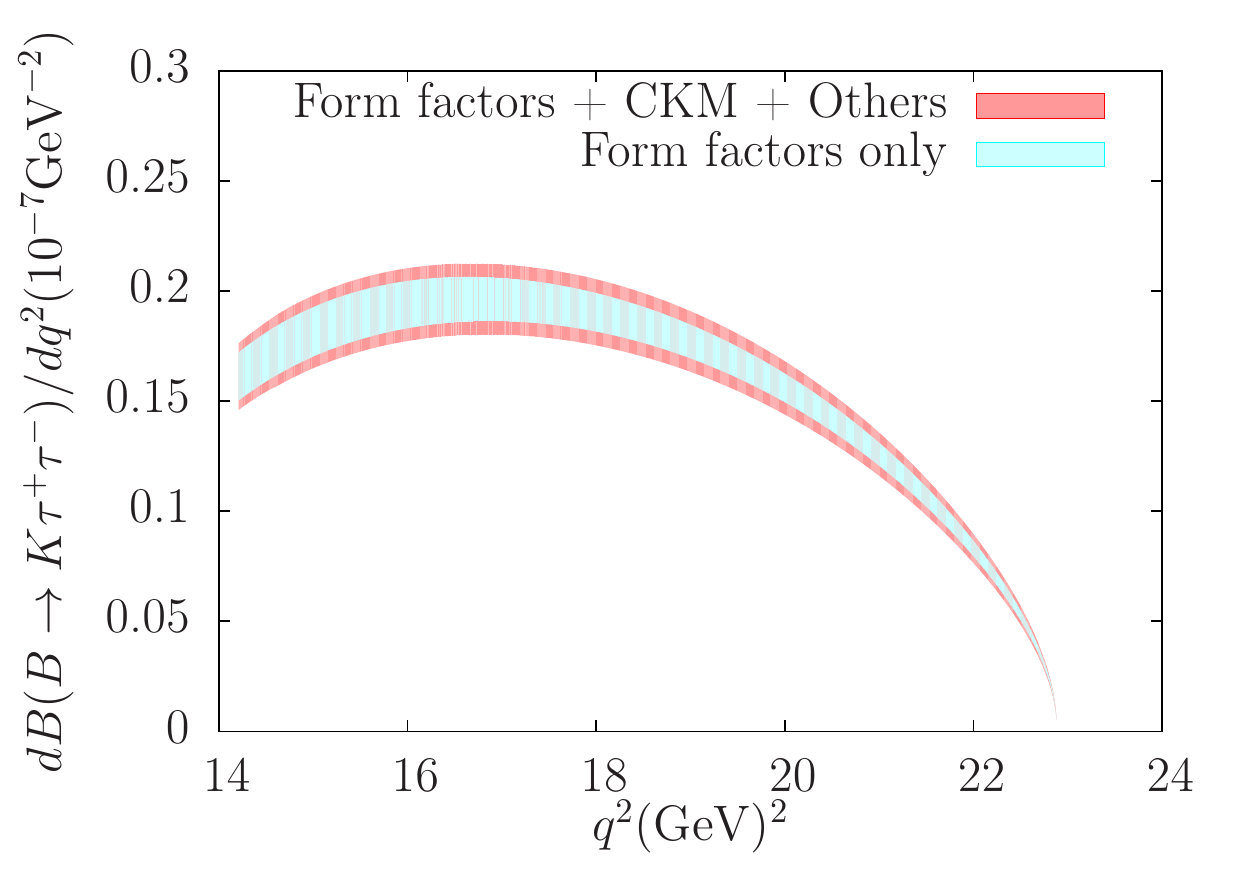}
    \caption{Standard-Model differential branching fraction (gray band) for \BtoKmumu\ decay (left) 
        and $B\to K\tau^+\tau^-$ (right), where $B$ denotes the isospin average, using the \FerMILC\ form 
        factors~\cite{Bailey:2015dka}.
        Experimental results for $B\to K\mu^+\mu^-$ are from
        Refs.~\cite{Wei:2009zv,Aaltonen:2011qs,Lees:2012tva,Aaij:2014pli}.
        The BaBar, Belle, and CDF experiments report isospin-averaged measurements.}
    \label{fig:B2K_dBdq2_mu}
\end{figure}

Figure~\ref{fig:B2K_dBdq2_mu} plots the isospin-averaged Standard-Model differential branching fractions for
$B\to K\mu^+\mu^-$ and $B\to K\tau^+\tau^-$.
For $B\to K\mu^+\mu^-$ decay, we compare our results with the latest measurements by
BaBar~\cite{Lees:2012tva}, Belle~\cite{Wei:2009zv}, CDF~\cite{Aaltonen:2011qs}, and LHCb~\cite{Aaij:2014pli}.
Tables~\ref{tab:B+2K_dBdq2} and~\ref{tab:B02K_dBdq2} give the partially integrated branching fractions for
the charged ($B^+$) and neutral ($B^0$) meson decays, respectively, for the same $q^2$ bins used by LHCb in
Ref.~\cite{Aaij:2014pli}.
In the regions $q^2\lesssim1~\text{GeV}^2$ and $6~\text{GeV}^2\lesssim q^2\lesssim 14~\text{GeV}^2$, $u\bar
u$ and $c\bar c$ resonances dominate the rate.
To estimate the total branching ratio, we simply disregard them and interpolate linearly in $q^2$ between
the QCD-factorization result at $q^2 \approx 8.5~\text{GeV}^2$ and the OPE result at $q^2 \approx
13~\text{GeV}^2$.
Although this treatment does not yield the full branching ratio, it enables a comparison with the
quoted experimental totals, which are obtained from a similar treatment of these regions.
Away from the charmonium resonances, the Standard-Model calculation is under good theoretical control, and
the partially integrated branching ratios in the wide high-$q^2$ and low-$q^2$ bins are our main results:
\begin{align}
    \Delta\BR(B^+\to K^+ \mu^+\mu^-)^\text{SM} \times 10^9 &= \left\{
    \begin{array}{lrcr}
      174.7(9.5)(29.1)(3.2)(2.2), & 1.1~\text{GeV}^2 \leq & q^2 \leq & 6~\text{GeV}^2 ,  \\
      106.8(5.8)(5.2)(1.7)(3.1),  & 15~\text{GeV}^2 \leq & q^2 \leq & 22~\text{GeV}^2 , 
    \end{array} \right.  \label{eq:SM_B2K+} \\
    \Delta\BR(B^0\to K^0 \mu^+\mu^-)^\text{SM} \times 10^9 &= \left\{
    \begin{array}{lrcr}
      160.8(8.8)(26.6)(3.0)(1.9), & 1.1~\text{GeV}^2 \leq & q^2 \leq & 6~\text{GeV}^2 ,  \\
       98.5(5.4)(4.8)(1.6)(2.8),  & 15~\text{GeV}^2 \leq & q^2 \leq & 22~\text{GeV}^2 ,
    \end{array} \right.  \label{eq:SM_B2K0}
\end{align}
where the errors are from the CKM elements, form factors, variations of the high and low matching
scales, and the quadrature sum of all other contributions, respectively.
LHCb's measurements for the same wide bins are~\cite{Aaij:2014pli}
\begin{align}
    \Delta \BR (B^+\to K^+\mu^+\mu^-)^\text{exp} \times 10^{9}~\text{GeV}^{2} &= \left\{
    \begin{array}{lrcr}
        118.6(3.4)(5.9)  & 1.1~\text{GeV}^2\leq & q^2 \leq & 6~\text{GeV}^2 , \\
        84.7(2.8)(4.2)  & 15~\text{GeV}^2\leq & q^2 \leq & 22~\text{GeV}^2 , \\ 
    \end{array} \right. 
    \label{eq:LHCb_B2K+} \\
    \Delta \BR (B^0\to K^0\mu^+\mu^-)^\text{exp} \times 10^{9}~\text{GeV}^{2} &= \left\{
    \begin{array}{lrcr}
        91.6 \left(^{+17.2}_{-15.7}\right)(4.4) & 1.1~\text{GeV}^2\leq & q^2 \leq &  6~\text{GeV}^2 , \\
        66.5 \left(^{+11.2}_{-10.5}\right)(3.5) & 15~\text{GeV}^2\leq  & q^2 \leq & 22~\text{GeV}^2 , \\ 
    \end{array} \right.  
    \label{eq:LHCb_B2K0}
\end{align}
where the errors are statistical and systematic, respectively, and again we convert the quoted differential
branching fractions to partially integrated branching fractions for direct comparison with
Eqs.~(\ref{eq:SM_B2K+}) and~(\ref{eq:SM_B2K0}).
Figure~\ref{fig:B2K_dBdq2_integrated}, right, shows the comparison between the Standard Model and the
experimental measurements.
The Standard-Model values are higher than the measurements by 1.8$\sigma$ and 2.2$\sigma$ for the low- and
high-$q^2$ bins, respectively.
The combination of the two bins, including the theoretical correlations from
Tables~\ref{tab:B2pi+_B2K+_dBdq2_FF_correlations} and~\ref{tab:B2pi+_B2K+_dBdq2_Other_correlations} of
Appendix~\ref{app:Results} and treating the experimental bins as uncorrelated, yields
$\chi^2/\text{dof}=5.7/2$, $p=0.06$, thus disfavoring the Standard-Model hypothesis with 1.9$\sigma$
significance.
Note, however, that the structures observed in the LHCb data above the $\psi(2S)$ (red points with small 
errors bars in Fig.~\ref{fig:B2K_dBdq2_mu}) warrant a great deal of caution in comparing theory and 
experiment for narrow bins~\cite{Lyon:2014hpa}.

\begin{figure}
    \includegraphics[width=0.49\textwidth]{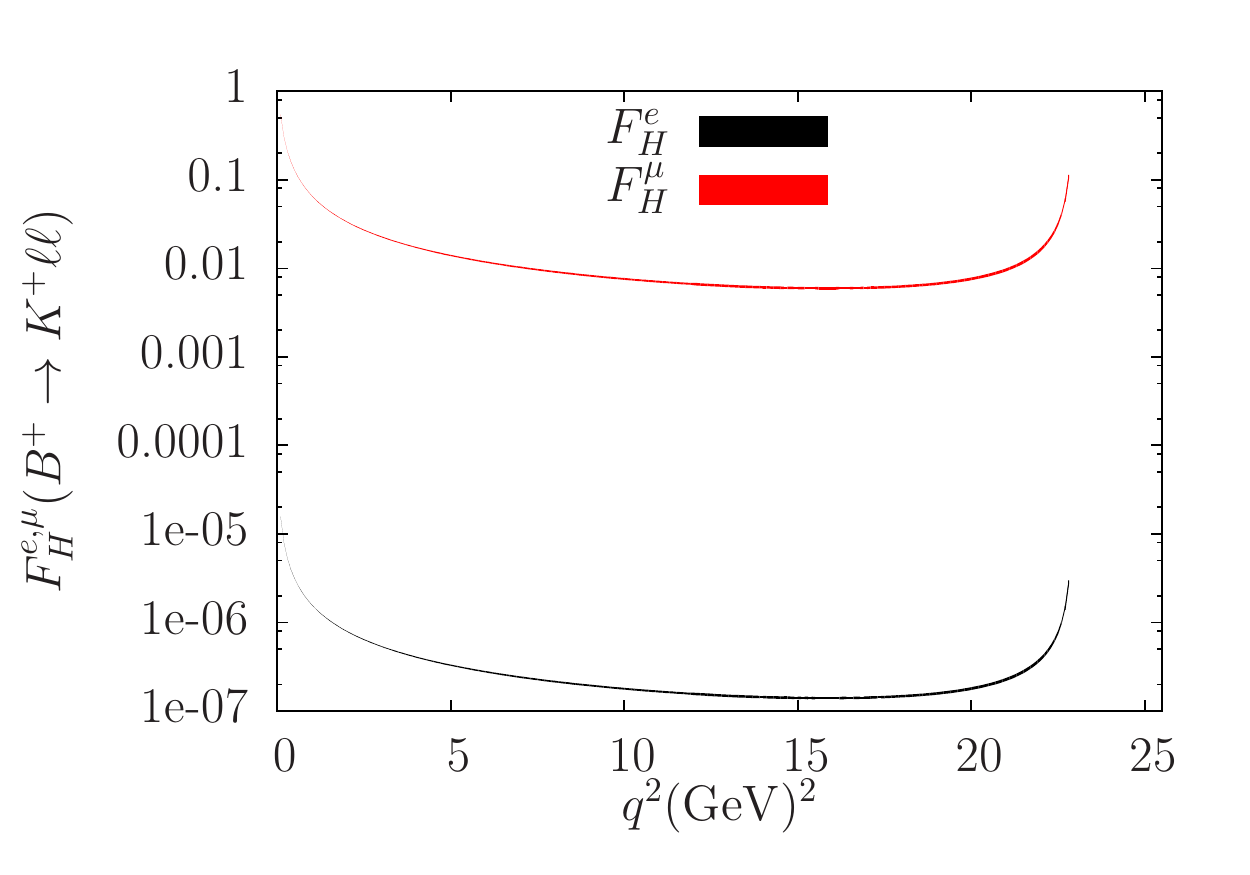}\hfill
    \includegraphics[width=0.49\textwidth]{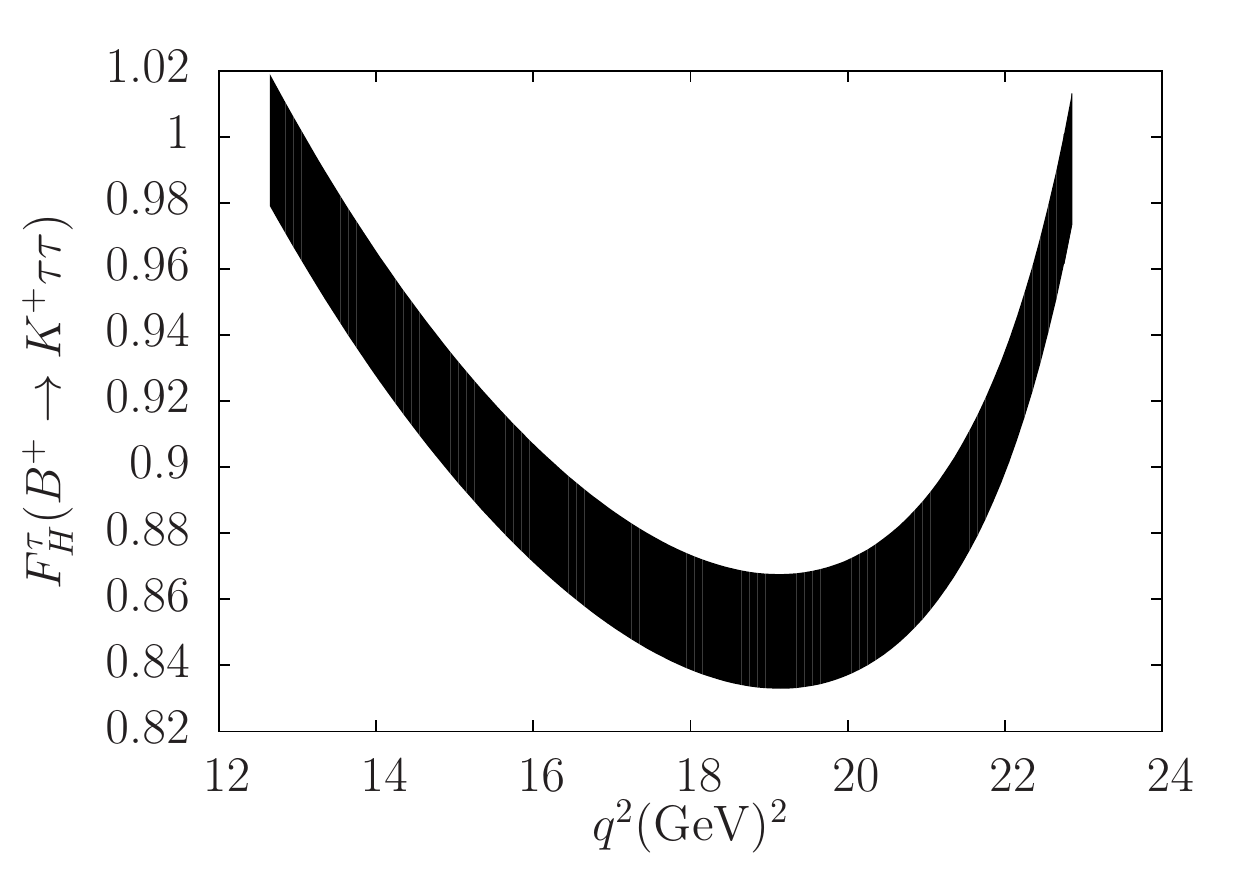}
    \caption{Standard-Model predictions for the $B^+\to K^+\ell^+\ell^-$ flat term $F_H^\ell(q^2)$ for
        $\ell=e,\mu$ (left) and $\ell=\tau$ (right) using the \FerMILC\ form factors~\cite{Bailey:2015dka}.}
    \label{fig:B2K_FH}
\end{figure}

We also calculate the Standard-Model flat term $F_H^\ell(q^2)$ with the \FerMILC\ form
factors~\cite{Bailey:2015dka} and plot it for $B^+\to K^+ \ell^+\ell^-\; (\ell = e, \mu, \tau)$ in
Fig.~\ref{fig:B2K_FH}.
LHCb reported results for the binned flat term
$F_H^\mu(q^2_\text{min},q^2_\text{max})$~\cite{Aaij:2012vr}
with uncertainties greater than 100\% in every bin.
The Standard-Model result for $F_H^\mu$ agrees with LHCb's measurement, but the comparison is limited by the
large experimental errors, which will improve with new measurements after the LHCb upgrade.
For future comparisons, Tables~\ref{tab:B2K_FH_charged} and~\ref{tab:B2K_FH_neutral} of
Appendix~\ref{app:Results} provide results for the binned $F_H^\ell(q^2_\text{min},q^2_\text{max})$ in
both the charged and neutral decay modes, respectively.

\subsubsection{Combinations of \texorpdfstring{\Btopi}{B2pi} and \texorpdfstring{\BtoK}{B2K} observables}
\label{subsec:B2Kpicombos}


Figure~\ref{fig:B2K_dBdq2_integrated} shows that the \Btopimumu\ and \BtoKmumu\ Standard-Model
partially integrated branching ratios are larger than the experimental results for all four wide $q^2$ bins.
The four Standard-Model values are, however, highly correlated.
Thus the combination of all four measurements disfavors the Standard-Model hypothesis at the
1.7$\sigma$ level ($\chi^2/\text{dof} = 7.8/4$, $p=0.10$). This significance lies in between the individual
exclusions from the two $\Btopi\mu^+\mu^-$ bins and the two \BtoKmumu\ bins.

The first observation of $\Btopill$ by LHCb implies that the ratio of branching fractions ${\cal
B}(B^+\to\pi^+\mu^+\mu^-)/{\cal B}(B^+\to K^+ \mu^+\mu^-) = 0.053(14)(1)$, where the errors are statistical
and systematic, respectively~\cite{LHCb:2012de}.
More recently, LHCb reported first results for the partially integrated branching fractions for
$B^+\to\pi^+\mu^+\mu^-$~\cite{Aaij:2015nea}, and also provided the ratio of
$\Delta\BR(B^+\to\pi^+\mu^+\mu^-)$-to-$\Delta\BR(B^+\to K^+\mu^+\mu^-)$ for the same $q^2$ intervals above
and below the charmonium resonances.
This ratio probes new-physics scenarios that would affect the shape of the $q^2$ distribution
differently for $\Btopi\ell^+\ell^-$ and $\BtoK\ell^+\ell^-$.
On the other hand, it is not sensitive to new physics that would affect the overall $\Btopi\ell^+\ell^-$ and
$\BtoK\ell^+\ell^-$ rates in the same way.

\begin{figure}
	\includegraphics[width=0.52\textwidth]{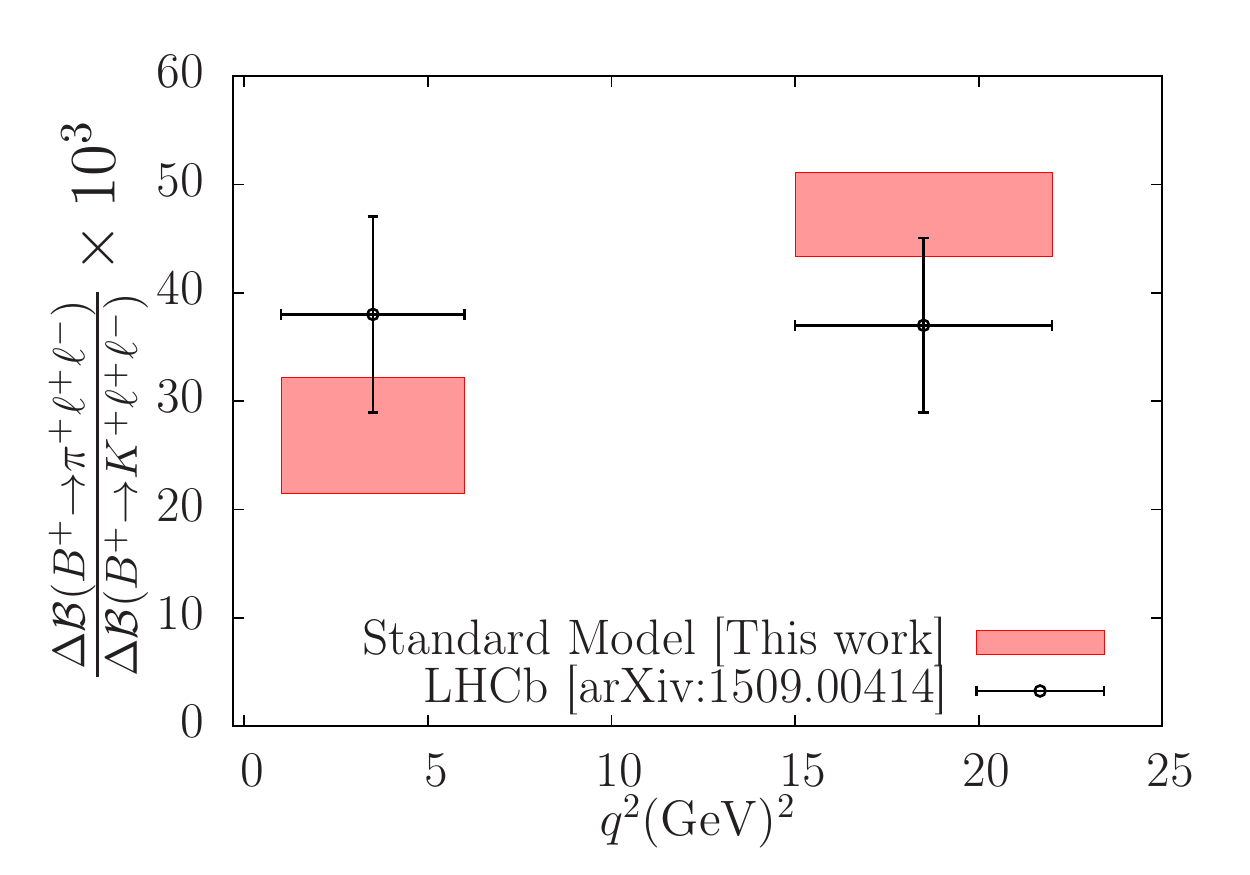} 
    \caption{Ratio of partially integrated branching ratios 
        $\Delta\BR(B^+\to\pi^+\mu^+\mu^-)$/$\Delta\BR(B^+\to K^+ \mu^+\mu^-)$ in the Standard Model using 
        the \FerMILC\ form factors~\cite{Lattice:2015tia,Bailey:2015nbd,Bailey:2015dka}, compared with 
        experimental measurements from LHCb~\cite{Aaij:2015nea}
        The errors in the Standard-Model results are dominated by the form-factor uncertainties; the 
        others are too small to be shown separately.}
	\label{fig:B2pi_B2K_ratio}
\end{figure}

The ratios of partially integrated differential branching fractions in the wide high-$q^2$ and low-$q^2$
bins, using the \FerMILC\ form factors~\cite{Lattice:2015tia,Bailey:2015nbd,Bailey:2015dka}, are our main 
results in this section:
\begin{equation}
	\frac{\Delta\BR(B^+\to \pi^+ \mu^+\mu^-)}{\Delta\BR(B^+\to K^+ \mu^+\mu^-)}^\text{SM} \times 10^3 =
        \left\{
        \begin{array}{lrcr}
          26.8\, (0.8)(5.3)(0.4), &  1~\text{GeV}^2 \leq & q^2 \leq &  6~\text{GeV}^2 , \\
          47.2\, (1.3)(3.4)(1.3), & 15~\text{GeV}^2 \leq & q^2 \leq & 22~\text{GeV}^2 ,
        \end{array}
          \right. \label{eq:B2pill_B2Kll_ratio}
\end{equation}
where the errors are from the CKM matrix elements, hadronic form factors, and all others added in
quadrature, respectively.
Binned Standard-Model values for additional $q^2$ intervals and for both the $B^+$ and $B^0$ decay modes are
provided in Table~\ref{tab:dBdq2-ratio}.
Figure~\ref{fig:B2pi_B2K_ratio} compares the above Standard-Model values with recent measurements from LHCb
for the same $q^2$ bins~\cite{Aaij:2015nea}:
 \begin{equation}
	\frac{\Delta\BR(B^+\to \pi^+ \mu^+\mu^-)}{\Delta\BR(B^+\to K^+ \mu^+\mu^-)}^\text{exp} \times 10^3 =
        \left\{
        \begin{array}{lrcr}
              38 (9)(1),&  1~\text{GeV}^2 \leq & q^2 \leq &  6~\text{GeV}^2 , \\
              37 (8)(1),& 15~\text{GeV}^2 \leq & q^2 \leq & 22~\text{GeV}^2 ,
        \end{array} \right.
    \label{eq:LHCb_B2pill_B2Kll_ratio}
\end{equation}
where the quoted errors are statistical and systematic, respectively.
The Standard-Model result for each individual bin is consistent with its experimental
measurement~\cite{Aaij:2015nea}---within 1.1$\sigma$---but the theory band lies below experiment for the
$1~\text{GeV}^2\leq q^2 \leq 6~\text{GeV}^2$ bin, while it lies above for the
$15~\text{GeV}^2\leq q^2\leq22~\text{GeV}^2$ bin.
Combining the two bins, including the theoretical correlations from
Tables~\ref{tab:B2P+_ratio_dBdq2_FF_correlations} and~\ref{tab:B2P+_ratio_dBdq2_Other_correlations}, and
treating the experimental bins as uncorrelated, shows that the LHCb measurement is compatible with the
Standard Model within 1.1$\sigma$ ($\chi^2/\text{dof} = 2.7/2$ and $p=0.26$).
Given the present uncertainties, however, ample room remains for new-physics contributions that may be
observable with improved measurements after the LHCb upgrade.

\subsubsection{Lepton-universality-violating observables} 
\label{subsec:results_lepton}


Lepton-universality-violating effects may give rise to observable deviations in ratios of rare $B$
decays to final states with different charged leptons~\cite{Altmannshofer:2013foa,Gauld:2013qba,%
Buras:2013qja,Datta:2013kja,Gauld:2013qja,Buras:2013dea,Altmannshofer:2014cfa,Hiller:2014yaa,Buras:2014fpa,%
Biswas:2014gga,Glashow:2014iga}, and would constitute a clear sign of physics beyond the Standard Model.
A useful observable to look for such effects is the ratio of partially integrated decay rates to different
charged-lepton final states with the same $q^2$ cuts~\cite{Hiller:2003js}:
\begin{equation}
    R_P^{\ell_1 \ell_2} (q_\text{min}^2, q_\text{max}^2) =
        \frac{\int_{q_\text{min}^2}^{q_\text{max}^2} \,dq^2 \,d\BR(B\to P \ell_1^+\ell_1^-)/dq^2}
        {\int_{q_\text{min}^2}^{q_\text{max}^2} \,dq^2  \,d\BR(B \to P \ell_2^+\ell_2^-)/dq^2},
\end{equation}
where $P = \pi, K$ and $\ell_1, \ell_2 = e, \mu, \tau$.
The quantities $R_{\pi}^{\mu e}$ and $R_{K}^{\mu e}$ are predicted to be unity in the Standard Model, up to
corrections of order $(m_\ell^2/M_B^2, m_\ell^4/q^4)$~\cite{Hiller:2003js,Bobeth:2007dw}, which are tiny
for $\ell = e, \mu$.
Thus any observed deviation of $R_{K(\pi)}^{\mu e}$ from unity would indicate the presence of physics
beyond the Standard Model.

Measurements of $R_K$ at $e^+e^-$ colliders by BaBar~\cite{Lees:2012tva} and Belle~\cite{Wei:2009zv} are
consistent with Standard-Model expectations within large experimental uncertainties of about 20--30\%.
The LHCb Collaboration, however, recently reported a measurement of the ratio %
$R_{K^+}^{\mu e}(1~\text{GeV}^2,6~\text{GeV}^2) = 0.745\left(^{+97}_{-82}\right)$~\cite{Aaij:2014ora} that is
2.6$\sigma$ lower than Standard-Model expectations.
Here we calculate lepton-universality-violating ratios in the Standard Model using the \FerMILC\ \BtoK\ and
\Btopi\ form factors~\cite{Lattice:2015tia,Bailey:2015nbd,Bailey:2015dka}.
Our predictions for $R_{\pi}^{\mu \ell}$ ($\ell = e,\tau$) are the first to use only {\it ab-initio} QCD
information for the hadronic physics, while results for $R_K^{\ell_1\ell_2}$ ($\ell_1,\ell_2=e,\mu,\tau$)
were previously presented by the HPQCD Collaboration using their own lattice-QCD form-factor
determinations~\cite{Bouchard:2013mia}.

\begin{figure}
    \includegraphics[width=0.49\textwidth]{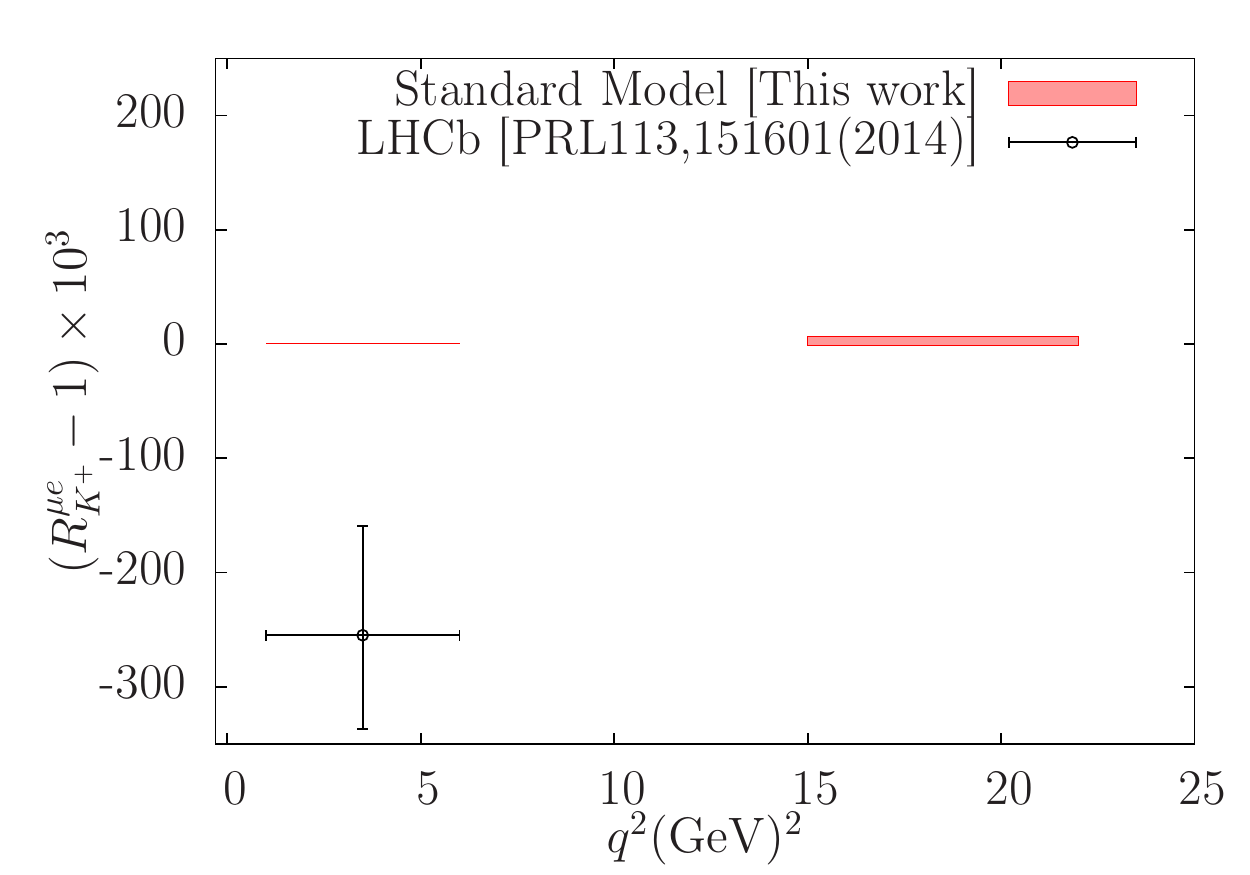} \hfill
    \includegraphics[width=0.49\textwidth]{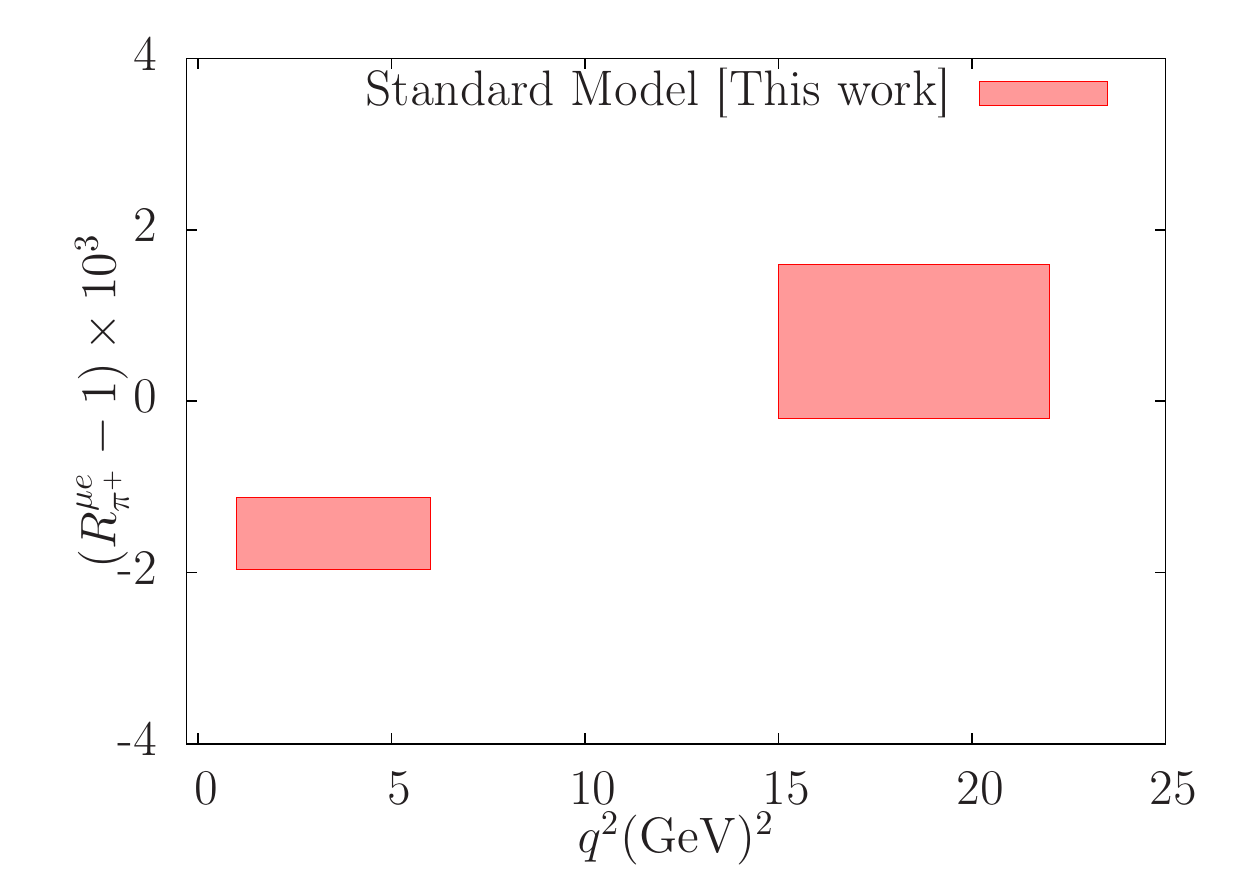}   
    \caption{Standard-Model lepton-universality-violating ratios $R_{K^+}^{\mu e} - 1$ (left) and
        $R_{\pi^+}^{\mu e}-1$ (right) for
        $(q_\text{min}^2, q_\text{max}^2)=(1~\text{GeV}^2,6~\text{GeV}^2)$ and 
        $(15~\text{GeV}^2,22~\text{GeV}^2)$ using the \FerMILC\ form 
        factors~\cite{Lattice:2015tia,Bailey:2015nbd,Bailey:2015dka}.
        The errors in the Standard-Model results are dominated by the form-factor uncertainties; the 
        remaining contributions are too small to be visible.
        The left plot also shows LHCb's measurement for the low-$q^2$ bin~\cite{Aaij:2014ora}.}
    \label{fig:B2P_RPemu}
\end{figure}

Tables~\ref{tab:B2P_RPi} and~\ref{tab:B2P_RK} show $R_{\pi}^{\mu \ell}$ and $R_{K}^{\mu \ell}$ for
$\ell=e,\tau$, respectively, using the same $q^2$ bins employed by LHCb~\cite{Aaij:2015nea}.
Figure~\ref{fig:B2P_RPemu} plots the difference from unity of $R_{K^+}^{\mu e}$ (left) and
$R_{\pi^+}^{\mu e}$ (right) for the wide $q^2$ bins below and above the charmonium resonances.
For the same $q^2$ cuts as LHCb's measurement~\cite{Aaij:2014ora}, we obtain
\begin{equation}
	\left[ R_{K^+}^{\mu e}(1~\text{GeV}^2, 6~\text{GeV}^2) -1 \right] \times 10^{3} = 0.50(43),
\end{equation}
where the error is predominantly from the form-factor uncertainties.
This agrees with the earlier isospin-averaged Standard-Model value $(R_K^{\mu e}-1)\times10^3=0.74(35)$ from
HPQCD for the same $q^2$ interval~\cite{Bouchard:2013mia} with a similar error.
Thus, explicit calculation with lattice QCD confirms the intuitively significant deviation between 
experiment and the Standard Model, observed by LHCb~\cite{Aaij:2014ora}.

\subsection{Rare \texorpdfstring{\boldmath$b\to q\nu\bar{\nu}\; (q=d,s)$}{b to q nu nubar} decay observables}
\label{subsec:results_nunubar}


Rare $B$ decays into neutrino-pair final states have not yet been observed.
The most recent bounds on $\BR(B\to K \nu \bar{\nu})$ from Babar~\cite{Lees:2013kla} and
Belle~\cite{Lutz:2013ftz} are, however, only about a factor of ten larger than Standard-Model expectations,
so prospects are good for its observation by Belle~II~\cite{Aushev:2010bq}.
The Standard-Model decay rate for $\Btopi\nu\bar\nu$ is further suppressed below $\BtoK\nu\bar\nu$ by the
relative CKM factor $|V_{td}/V_{ts}|^2 \approx 0.04$, except for $\BR(B^+\to \pi^+\nu_\tau\bar{\nu_\tau})$,
which is enhanced by long-distance contributions.
Indeed, $B^+\to \pi^+\nu_\tau\bar{\nu_\tau}$ events are included in measurements of the leptonic decay rate
$\BR(B^+ \to \tau^+ \nu_\tau)$, where the $\tau$ subsequently decays as $\tau \to \pi^+
\bar{\nu}_\tau$~\cite{Aubert:2009wt,Lees:2012ju,Adachi:2012mm,Kronenbitter:2015kls}.

In anticipation of such measurements, we provide Standard-Model predictions for $B\to K\nu\bar{\nu}$ and
$B\to\pi\nu\bar{\nu}$ observables using the \FerMILC\ form factors~\cite{Lattice:2015tia,Bailey:2015dka}.
Previous analyses of $B\to K (\pi) \nu \bar{\nu}$ used form factors from light-cone sum rules
(LCSR)~\cite{Altmannshofer:2008dz,Altmannshofer:2009ma} or perturbative QCD (pQCD)~\cite{Wang:2012ab}.
One recent study of $B\to K (\pi) \nu \bar{\nu}$~\cite{Buras:2014fpa} combined lattice-QCD form factors at
high $q^2$~\cite{Bouchard:2013eph} with LCSR form factors~\cite{Straub:2015ica} at low $q^2$ in a
simultaneous $z$-expansion fit.
Our calculations of $B\to K (\pi) \nu \bar{\nu}$ observables are the first to use only \emph{ab-initio} QCD
information for the hadronic physics.
Because experiments cannot identify the outgoing neutrino flavor, we present results for the sum of
contributions from $\nu_e, \nu_\mu$, and $\nu_\tau$.

\begin{figure}
    \includegraphics[width=0.49\textwidth]{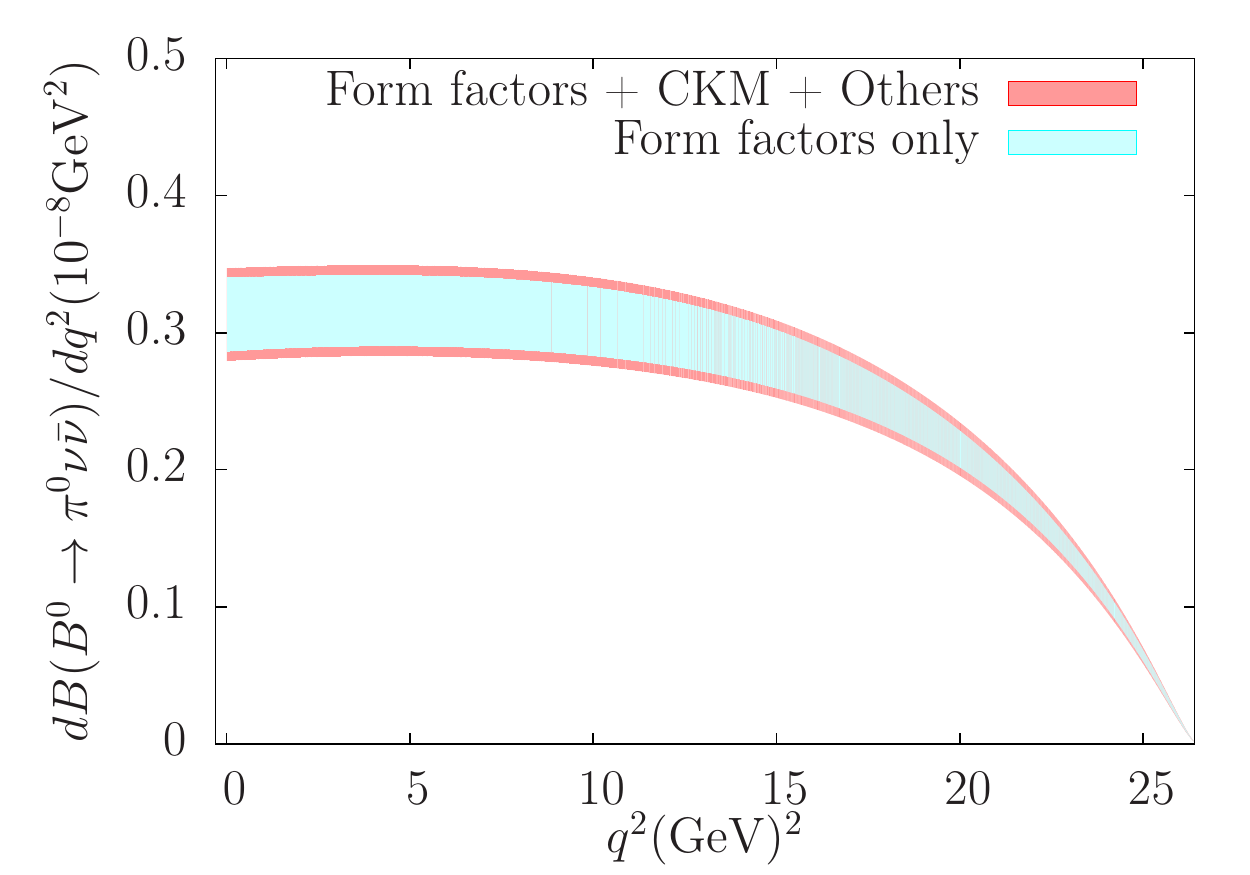}\hfill
    \includegraphics[width=0.49\textwidth]{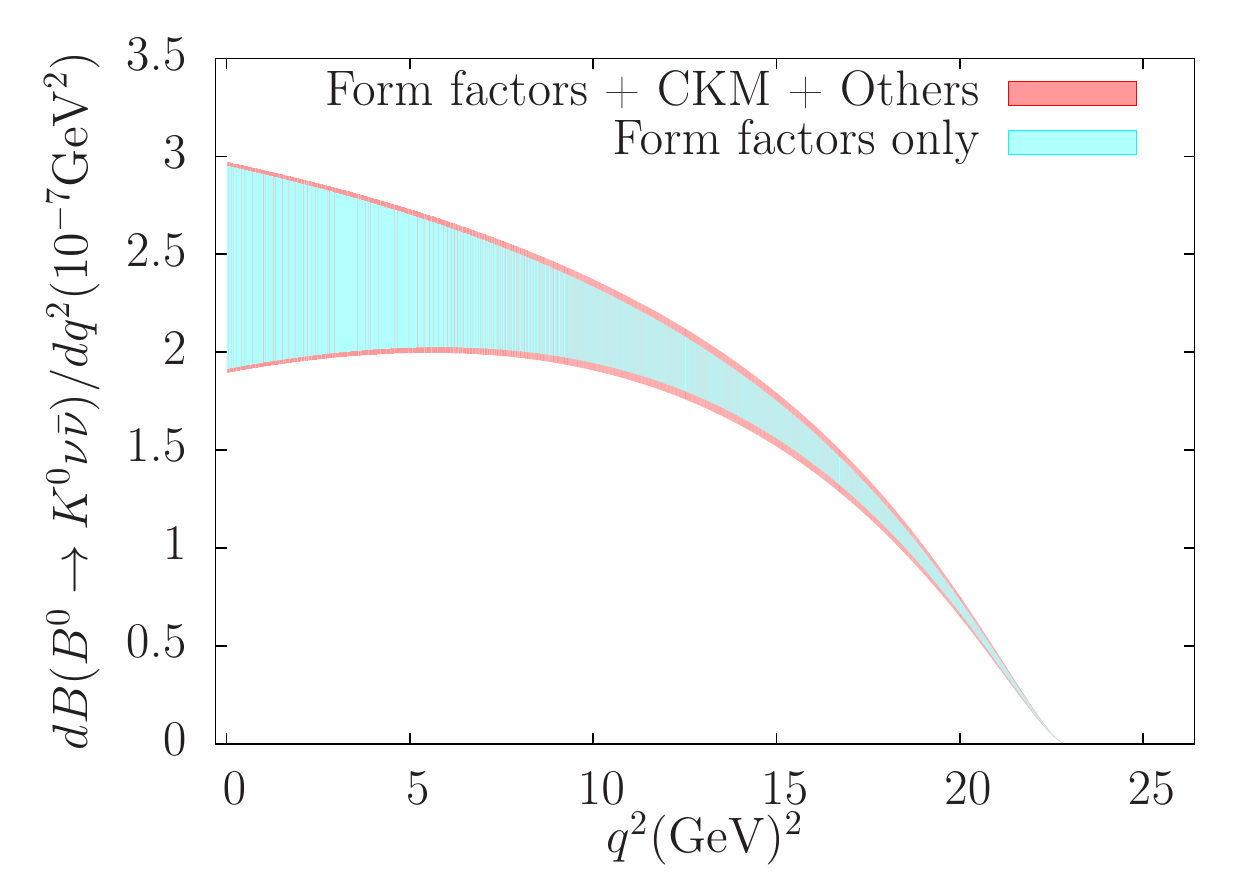} 
    \caption{Standard-Model differential branching fraction for $B^0\to \pi^0 \nu \bar{\nu}$ decay (left) 
        and $B^0\to K^0 \nu \bar{\nu}$ (right) using the \FerMILC\ form
        factors~\cite{Lattice:2015tia,Bailey:2015dka}.}
 \label{fig:B2Knunu_dBdq2}
\end{figure}

For the neutral decay modes $B^0\to K^0 (\pi^0) \nu \bar{\nu}$, the dominant contributions to the
Standard-Model decay rate are from FCNC transitions [see
Eq.~(\ref{eq:Btonunubar_rate})].
Figure~\ref{fig:B2Knunu_dBdq2} shows our Standard-Model prediction for the differential branching fractions
for $B^0\to \pi^0 \nu \bar{\nu}$ and $B^0\to K^0 \nu \bar{\nu}$.
We obtain the total branching fractions
%
\begin{align}
    \BR(B^0\to\pi^0\nu\bar{\nu}) \times 10^7 &= 0.668(41)(49)(16) , \label{eq:B0toPi0_nunubar} \\
    \BR(B^0\to  K^0\nu\bar{\nu}) \times 10^7 &= 40.1(2.2)(4.3)(0.9) , 
\end{align}
where the errors are from the CKM elements, form factors, and the quadrature sum of all other contributions,
respectively.
The ``other" errors in Eqs.~(4.11) and~(4.12) are so much smaller than those from the CKM elements and form
factors because $u\bar{u}$ and $c\bar{c}$ resonances do not contribute to the rate in Eq.~(2.2), and because
the perturbative contributions lumped into $X_t$ are known to about a percent.

For the charged decay modes $B^+\to K^+ (\pi^+) \nu \bar{\nu}$, we obtain the following contributions to the
Standard-Model branching fractions from FCNC transitions:
\begin{align}
    \BR(B^+\to\pi^+\nu\bar{\nu})_{\mathrm{SD}} \times 10^7 &= 1.456(89)(106)(34) ,
    \label{eq:B+toPi+_nunubar_SD} \\
    \BR(B^+\to  K^+\nu\bar{\nu})_{\mathrm{SD}}  \times 10^7 &= 43.2(2.3)(4.6)(1.0) ,
    \label{eq:B+toK+_nunubar_SD}
\end{align}
where again the errors are from the CKM elements, form factors, and the quadrature sum of all other
contributions.
Our result for $\BR(B^+\to K^+ \nu \bar{\nu})_{\mathrm{SD}}$ is consistent with that in
Ref.~\cite{Buras:2014fpa}, albeit with a slightly larger error.
The smaller error in Ref.~\cite{Buras:2014fpa} stems from their use of an additional input at low-$q^2$ from
LCSR~\cite{Straub:2015ica}.
The results quoted in Ref.~\cite{Wang:2012ab} for the total $B\to K(\pi) \nu \bar{\nu}$ branching fractions
for both the charged and neutral modes agree with ours, but they have significantly larger errors.

As discussed in Sec.~\ref{subsec:nunubar}, the decay rates for
$B^+\to K^+ (\pi^+) \nu_\tau \bar{\nu_\tau}$ also receive substantial long-distance contributions from
intermediate tree-level $\tau$ decays.
The numerical values for the long-distance contributions are given in Eqs.~(\ref{eq:BtoPinunubar_LD_result})
and~(\ref{eq:BtoKnunubar_LD_result}).
We add them to the short-distance contributions in Eqs.~(\ref{eq:B+toPi+_nunubar_SD})
and~(\ref{eq:B+toK+_nunubar_SD}) to obtain the full branching ratios:
%
\begin{align}
    \BR(B^+\to\pi^+\nu\bar{\nu}) \times 10^6 &= 9.62(1)(92) ,  \\
    \BR(B^+\to  K^+\nu\bar{\nu}) \times 10^6 &= 4.94(52)(6) ,  
\end{align}
where here the errors are from the short-distance and long-distance contributions respectively.
For $B^+\to\pi^+\nu\bar{\nu}$, the intermediate $\tau$-decay channel increases the Standard-Model rate by an
order of magnitude, whereas for $B^+\to K^+\nu\bar{\nu}$ it only generates about a 10\% enhancement.

\subsection{Tree-level \texorpdfstring{\boldmath$b\to u\ell\nu$}{b to u l nu} decay observables}
\label{subsec:results_ellnu} 

The decay $B\to\pi\tau\nu$ has not yet been observed experimentally.
Recently, however, the Belle Collaboration reported their first search for $B^0\to\pi^-\tau^+\nu_\tau$
decay~\cite{Hamer:2015jsa}, obtaining a result for the total branching fraction with a 2.4$\sigma$
significance, corresponding to an upper limit not far from the Standard-Model prediction.
The upcoming Belle~II experiment is therefore well positioned to measure the branching fraction as well as
the $q^2$ spectrum for this process.
 
Most previous Standard-Model predictions for $B\to\pi\tau\nu$ have relied on estimates of the hadronic form
factors from LCSR~\cite{Khodjamirian:2011ub,Dutta:2013qaa,Wang:2015vgv} or pQCD~\cite{Wang:2012ab}.
Reference~\cite{Fajfer:2012jt} employs form factors from lattice QCD~\cite{Dalgic:2006dt,Bailey:2008wp}.
The scalar form factor, however, was calculated only in Ref.~\cite{Dalgic:2006dt}, and disagrees with more
recent continuum-limit results~\cite{Flynn:2015mha,Lattice:2015tia}.
Because the total uncertainty on $\BR(B\to\pi\tau\nu)$ in Ref.~\cite{Fajfer:2012jt} is dominated by the
error on the scalar form factor, it is now possible to improve the Standard-Model estimate.
Here we use the form factors and value of $|V_{ub}|$, $3.72(16) \times 10^{-3}$, from
\FerMILC~\cite{Lattice:2015tia}, obtained from a simultaneous $z$-expansion fit of lattice-QCD results and
the measured $B\to\pi\ell\nu$ partial branching fractions.
These form factors and $|V_{ub}|$ carry significant correlations, which we incorporate below.
We reiterate that with this choice of inputs we assume that there are no significant new-physics
contributions to $B\to\pi\ell\nu$ decays with light charged leptons.

\begin{figure}
    \includegraphics[width=0.48\textwidth]{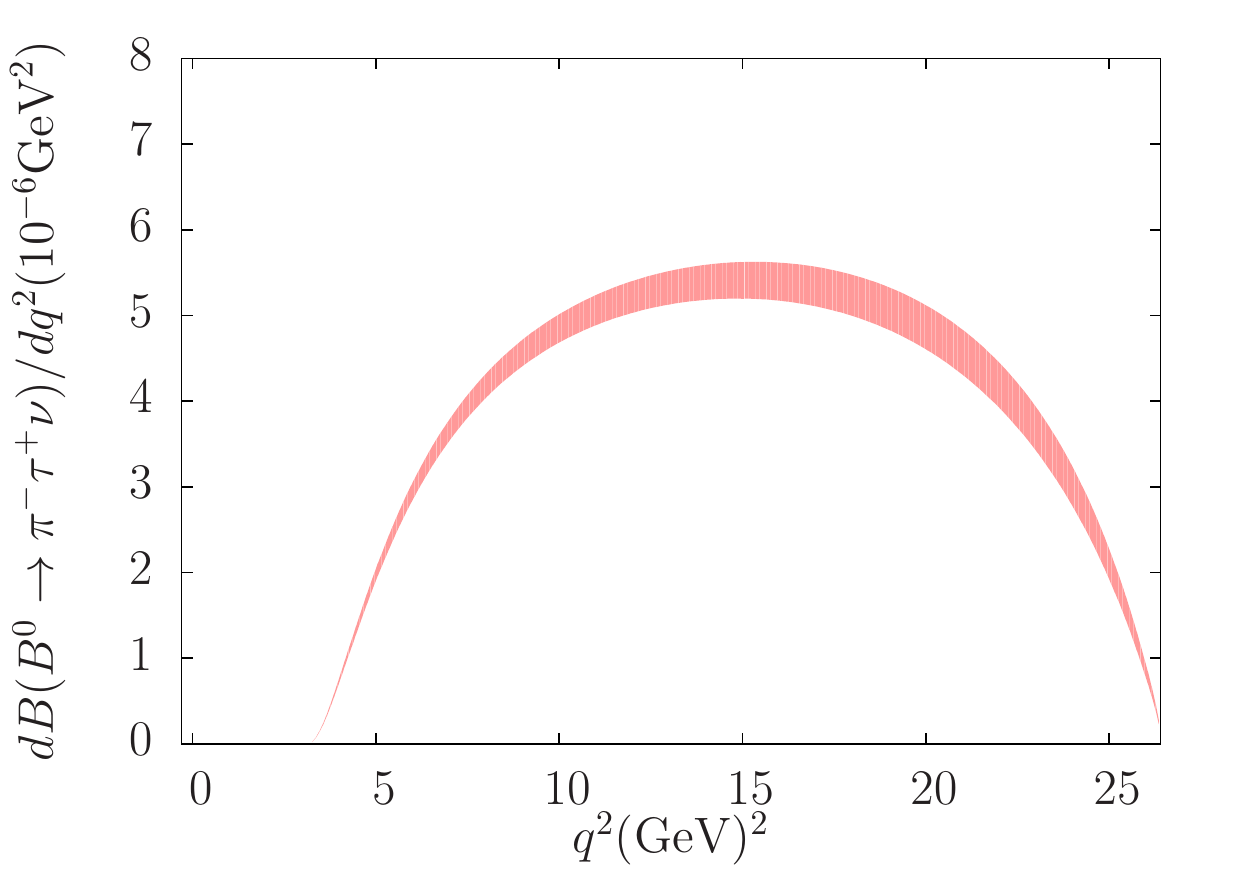} 
    \caption{Standard-Model differential branching fraction for $B^0\to\pi^-\tau^+ \nu_\tau$ decay using 
        the \FerMILC\ form factors and determination of $|V_{ub}|$ from Ref.~\cite{Lattice:2015tia}.}
    \label{fig:B2tau_dBdq2}
\end{figure}

Figure~\ref{fig:B2tau_dBdq2} shows our Standard-Model prediction for the $B^0\to \pi^- \tau^+ \nu_\tau$
differential branching fraction.
For the total integrated branching fractions we find
\begin{align}
    \BR (B^0\to \pi^- \tau^+ \nu_\tau)  =  9.35(38) \times 10^{-5}, \\
    \BR (B^+\to \pi^0 \tau^+ \nu_\tau)  =  4.99(20) \times 10^{-5},
\end{align}
where the error includes the correlated uncertainties from the form factors and $|V_{ub}|$.
Because of the correlations between the form factors and $|V_{ub}|$, it is not possible to quote their
errors individually.
The uncertainties stemming from the parametric inputs are negligible.
Our results are consistent with those quoted in Refs.~\cite{Dutta:2013qaa,Wang:2012ab} within their much
larger uncertainties.
Because the Standard-Model branching fraction for $B\to\pi\tau\nu$ is of the same order of magnitude as
for $B\to\pi\mu\nu$, the Belle II experiment should be in a good position to test our prediction of
its differential decay rate.

Deviations from Standard-Model expectations have been observed for semileptonic $B$-meson decays to
$\tau\nu_\tau$ final states involving tree-level $b\to c$ charged-current interactions.
Given the combined 3.9$\sigma$ excess quoted by HFAG for their averages of $R(D)$ and
$R(D^*)$~\cite{Amhis:2014hma}, it is interesting to consider the analogous ratio for $B\to\pi\ell\nu$ decay,
which instead proceeds through a tree-level $b\to u$ transition.
Assuming, again, that there are no new-physics contributions to decays to light charged leptons,
$B\to\pi\ell\nu$ with $\ell = e, \mu$, the \FerMILC\ form factors~\cite{Lattice:2015tia} yield the 
Standard-Model prediction
\begin{equation}
    R(\pi) \equiv \frac{\BR(\Btopi \tau \nu_\tau)}{\BR(\Btopi \ell \nu_\ell)} = 0.641(17), \label{eq:RPiDef}
\end{equation}
for both the charged and neutral $B$-meson decay modes.
Because $|V_{ub}|$ cancels in the ratio, $R(\pi)$ provides an especially clean probe of new physics,
particularly charged Higgs bosons, independent of the currently observed tension between determinations of
$|V_{ub}|$ from inclusive and exclusive semileptonic $B$-meson
decays~\cite{Amhis:2014hma,Agashe:2014kda,Bevan:2014iga}.

\section{CKM matrix elements and Wilson coefficients}
\label{sec:Implications}

We now illustrate the broader utility of the \FerMILC\ form
factors~\cite{Lattice:2015tia,Bailey:2015nbd,Bailey:2015dka} for Standard-Model and
beyond-the-Standard-Model phenomenology with two concrete examples.
First, in Sec.~\ref{subsec:Vtx}, starting with the assumption that the Standard Model is a complete
description of Nature, we combine our predicted branching fractions with the recent LHCb measurements to
determine the CKM matrix elements $|V_{td}|$, $|V_{ts}|$, and their ratio.
We then compare them with results from other processes.
Second, in Sec.~\ref{subsec:WilsonConstraints}, we make no such assumption but take the CKM matrix
elements from unitarity and combine our theoretical branching fractions with the experimental measurements
to constrain the Wilson coefficients of the $b\to q\ell\ell \; (q=d,s)$ effective Hamiltonian.
We then compare them with Standard-Model values.
 
These analyses are possible because, as stressed above, the \Btopi\ and \BtoK\ form factors are decoupled,
via the effective Hamiltonian, from physics at energy scales above the electroweak scale.

\subsection{Constraints on \texorpdfstring{\boldmath$V_{ts}$}{Vts}, \texorpdfstring{\boldmath$V_{td}$}{Vtd}, and \texorpdfstring{\boldmath$|V_{td}/V_{ts}|$}{Vtd/Vts}}
\label{subsec:Vtx}


In the Standard Model, the ratios of differential branching fractions
$$
    \frac{\Delta\BR(B\to\pi\ell^+\ell^-)}{\Delta\BR(B\to K\ell^+\ell^-)} \quad {\rm{and}} \quad
    \frac{\Delta\BR(B\to\pi\nu\bar{\nu})}{\Delta\BR(B\to K\nu\bar{\nu})}
$$
are both proportional to the ratio of CKM matrix elements $|V_{td}/V_{ts}|^2$.
Thus, they enable determinations of $|V_{td}/V_{ts}|$, independent of that from the ratio of
$B_d$-to-$B_s$-meson oscillation frequencies~\cite{Agashe:2014kda}, which is currently the most precise.

The LHCb experiment's initial observation of $B \to \pi \mu^+\mu^-$~\cite{LHCb:2012de} enabled the first
determination of this ratio of CKM matrix elements from rare semileptonic $B$ decays.
In that work, they obtained $|V_{td}/V_{ts}|=0.266(35)$ using form factors from light-cone sum
rules~\cite{Ball:2004ye} and neglecting the theoretical uncertainty.
More recently, LHCb measured the differential decay rate for $B \to \pi \mu^+\mu^-$ and the ratio
$\Delta\BR(B \to \pi \mu^+\mu^- )/ \Delta\BR(B\to K \mu^+\mu^-)$ in two bins of $q^2$ below and above the
charmonium resonances~\cite{Aaij:2015nea}.
Although the measurement errors decreased, the quoted error in $|V_{td}/V_{ts}|=0.23(^{+5}_{-4})$
increased from including now the theory uncertainty.
Here we obtain the first determination of $|V_{td}/V_{ts}|$ from rare $b \to d(s) \ell^+ \ell^-$ decay
processes using only \emph{ab-initio} lattice-QCD information for the hadronic form factors.

Following Refs.~\cite{LHCb:2012de,Ali:2013zfa} we calculate
\begin{align}
    F^{\pi/K} &\equiv \left| \frac{V_{ts}}{V_{td}} \right|^2 
        \frac{\Delta\BR(B^+\to \pi^+ \mu^+\mu^-)}{\Delta\BR(B^+\to K^+ \mu^+\mu^-)} ,
    \label{eq:FpiKDef}
\end{align}
which removes the CKM matrix elements.
Taking Eq.~(\ref{eq:B2pill_B2Kll_ratio}) for the second factor in Eq.~(\ref{eq:FpiKDef}) and
removing the CKM ratio used there (from Table~\ref{tab:inputs}), we obtain
\begin{equation}
    F^{\pi/K} = \left\{
        \begin{array}{lrcr}
           0.60(12), &  1~\text{GeV}^2 \leq & q^2 \leq &  6~\text{GeV}^2 \\
          1.055(81), & 15~\text{GeV}^2 \leq & q^2 \leq & 22~\text{GeV}^2
        \end{array} \right.,
    \label{eq:FpiKResults}
\end{equation}
where the errors stem predominantly from the form-factor uncertainties.
Combining this Standard-Model calculation of $F^{\pi/K}$ with LHCb's recent measurement~\cite{Aaij:2015nea},
Eq.~(\ref{eq:LHCb_B2pill_B2Kll_ratio}), yields
\begin{equation}
    |V_{td}/V_{ts}| = \left\{
        \begin{array}{lrcr}
            0.252(25)(30), &  1~\text{GeV}^2\leq & q^2 \leq &  6~\text{GeV}^2 \\
            0.187(7)(20),  & 15~\text{GeV}^2\leq & q^2 \leq & 22~\text{GeV}^2
        \end{array} \right. ,
    \label{eq:Vtd_Vts_LowHigh}
\end{equation}
where the errors are from theory and experiment, respectively.
A joint fit over both bins including theoretical correlations (which in practice are negligible) yields our
final result for the ratio of CKM matrix elements:
\begin{equation}
	|V_{td}/V_{ts}| = 0.201(20),
    \label{eq:Vtd_Vts}
\end{equation}
where the error includes both experimental and theoretical uncertainties, and the combined $\chi^2/{\rm dof}
= 2.3/1$ ($p=0.13$).
Equation~(\ref{eq:Vtd_Vts}) agrees with the more precise determination from the oscillation frequencies of
neutral $B_{d,s}$ mesons, $|V_{td}/V_{ts}|=0.216(1)(11)$, as well as that from CKM unitarity,
$|V_{td}/V_{ts}|=0.2115(30)$~\cite{Charles:2004jd}.

The error on $|V_{td}/V_{ts}|$ in Eq.~(\ref{eq:Vtd_Vts}) is more than two times smaller than that obtained
by LHCb in Ref.~\cite{Aaij:2015nea} using the same experimental information.
This improvement stems entirely from the more precise form factors.
Because the error on $|V_{td}/V_{ts}|$ in Eq.~(\ref{eq:Vtd_Vts}) is dominated by the experimental
uncertainty, especially for the high-$q^2$ bin, it will be reduced as measurements of $B\to K(\pi)
\ell^+\ell^-$ decays improve.
Better form-factor calculations will also aid the determination of $|V_{td}/V_{ts}|$ from the low-$q^2$
region.
Future observations of $B \to K(\pi) \nu\bar{\nu}$ in combination with our Standard-Model predictions in
Sec.~\ref{subsec:nunubar} will enable yet another way to determine $|V_{td}/V_{ts}|$.

We can also determine the products of CKM elements $|V_{tb} V_{td}^*|$ and $|V_{tb} V_{ts}^*|$ that appear
in the individual decay rates for $\Btopi \ell^+ \ell^-$ and $\BtoK\ell^+ \ell^-$ decay, respectively.
In analogy with the analysis above, we combine our calculations of the CKM-independent quantities
$\Delta\BR(B^+ \to \pi^+ \mu^+ \mu^-)/|V_{tb} V_{td}^*|^2$ and $\Delta\BR(B^+ \to K^+ \mu^+ \mu^-)/|V_{tb}
V_{ts}^*|^2$ with experimental measurements of the partial branching fractions for the same $q^2$ intervals.
Using the $\Btopi \mu^+\mu^-$ partial branching fractions measured by LHCb~\cite{Aaij:2015nea}, and quoted
in Eq.~(\ref{eq:LHCb_B2pi}), we obtain
\begin{align}
    |V_{tb} V_{td}^*| \times 10^3 &= \left\{
        \begin{array}{lrcr}
        8.34(49)(95), &  1~\text{GeV}^2\leq & q^2 \leq &  6~\text{GeV}^2 , \\
        6.90(26)(81), & 15~\text{GeV}^2\leq & q^2 \leq & 22~\text{GeV}^2 .
        \end{array} \right.
    \label{eq:Vtd_Vtb_LowHigh}
\end{align}
Similarly, using the $\BtoK\mu^+\mu^-$ measurement from Ref.~\cite{Aaij:2014pli}, quoted in
Eq.~(\ref{eq:LHCb_B2K+}), we obtain
\begin{eqnarray}              
    |V_{tb} V_{ts}^*| \times 10^3 & = & \left\{
        \begin{array}{lrcr}
            33.3(2.8)(1.0), & 1.1~\text{GeV}^2 \leq & q^2 \leq &  6~\text{GeV}^2 , \\
            36.0(1.1)(1.1), &  15~\text{GeV}^2 \leq & q^2 \leq & 22~\text{GeV}^2 .
        \end{array} \right.
    \label{eq:Vtds_Vtb_LowHigh}
\end{eqnarray}
The errors given in Eqs.~(\ref{eq:Vtd_Vtb_LowHigh}) and~(\ref{eq:Vtds_Vtb_LowHigh}) are from theory
and experiment, respectively.
Combining the values from the individual $q^2$ bins above including correlations gives
\begin{eqnarray}
	|V_{tb} V_{td}^*| \times 10^3 & =& 7.45(69) , \label{eq:Vtd_Vtb} \\
	|V_{tb} V_{ts}^*| \times 10^3 & =& 35.7(15)  , \label{eq:Vts_Vtb} 
\end{eqnarray}
for our final results, where the errors include both the experimental and theoretical uncertainties.

Taking $|V_{tb}| = 0.9991$ from CKM unitarity~\cite{Charles:2004jd}, where the error is of
order $10^{-5}$ and hence negligible, we can infer values for the magnitudes of the individual CKM
elements $|V_{td}|$ and $|V_{ts}|$.
We find
\begin{eqnarray}
	|V_{td}| & =&  7.45(69) \times 10^{-3} , \label{eq:Vtd} \\
	|V_{ts}| & =& 35.7(1.5) \times 10^{-3} , \label{eq:Vts} 
\end{eqnarray}
where the errors include both the experimental and theoretical uncertainties.
This determination of $|V_{td}|$ agrees with the Particle Data Group (PDG) value
$|V_{td}|=8.4(6)\times10^{-3}$~\cite{Agashe:2014kda} obtained from the oscillation frequency of neutral
$B_d$ mesons, with commensurate precision.
Our $|V_{ts}|$ is 1.4$\sigma$ lower than $|V_{ts}|=40.0(2.7)\times10^{-3}$ from $B_s$-meson oscillations
with an error that is almost two times smaller.
Compared with the determinations $|V_{td}|=7.2(^{+9}_{-8})\times10^{-3}$ and $|V_{ts}|=32(4)\times10^{-3}$
by LHCb~\cite{Aaij:2015nea}---using the same experimental inputs but older form factors---the uncertainties
in Eqs.~(\ref{eq:Vtd}) and~(\ref{eq:Vts}) are 1.2 and 2.7 times  smaller, respectively.
Again, this illustrates the value added from using the more precise hadronic form factors.
It is worth noting that the errors on the $B$-mixing results~\cite{Agashe:2014kda} are dominated by the
uncertainties on the corresponding hadronic matrix elements~\cite{Gamiz:2009ku,Bazavov:2012zs}.
Therefore the errors on $|V_{td}|$ and $|V_{ts}|$ from both neutral $B$-meson oscillations and semileptonic
$B$ decays will decrease with anticipated lattice-QCD improvements (see
Refs.~\cite{Dowdall:2014qka,Bouchard:2014eea} and our discussion in Sec.~\ref{sec:Conclusions}).

Finally, assuming CKM unitarity, our result for $|V_{ts}|$ implies a value for $|V_{cb}|$ via
$|V_{cb}|=|V_{ts}|=A\lambda^2 + O(\lambda^4)$, where the explicit expression for the correction can be found
in Ref.~\cite{Charles:2004jd}.
Taking numerical values for $\{A, \lambda, \bar{\rho}, \bar{\eta} \}$ from Table~\ref{tab:inputs} to
estimate the correction term, we obtain $|V_{cb}| = 36.5 (1.5) \times 10^{-3}$, where the error stems from
the uncertainty in $|V_{ts}|$ in Eq.~(\ref{eq:Vtd}) and the parametric uncertainty due to higher-order
corrections in $\lambda$ is negligible.
This alternate result for $|V_{cb}|$ is $1.6\sigma$ below the exclusive $|V_{cb}|$ determination
from $B \to D^* \ell \nu$~\cite{Bailey:2014tva}, 2.6$\sigma$ below that from
$B\to D\ell\nu$~\cite{GambinoEPS2015}, and $3.5\sigma$ below the inclusive $|V_{cb}|$
determination~\cite{Alberti:2014yda}.
This tension is simply another perspective on the differences we found between the experimental measurements
for the $B \to K \mu^+\mu^-$ partially integrated branching fractions and our Standard-Model predictions,
discussed in Sec.~\ref{subsec:B2Kobs}.

\subsection{Constraints on Wilson coefficients} \label{subsec:WilsonConstraints}


In this section, we investigate the constraints on the Wilson coefficients of the effective Hamiltonian
implied by present $B\to(K,\pi)\mu^+\mu^-$ measurements combined with the \FerMILC\ form
factors~\cite{Lattice:2015tia,Bailey:2015nbd,Bailey:2015dka}.
We focus on high-scale ($\mu_0\simeq 120~\text{GeV}$) contributions to the Wilson coefficients $C_9$ and
$C_{10}$:
\begin{align}
    C_9 (\mu_0) = C_9^\text{SM} (\mu_0) + C_9^\text{NP} (\mu_0) \; , \\
    C_{10} (\mu_0) = C_{10}^\text{SM} (\mu_0) + C_{10}^\text{NP} (\mu_0) \; , 
\end{align}
where the Standard-Model matching conditions are given in Eqs.~(25) and~(26) of Ref.~\cite{Huber:2005ig}
and, for our choice of inputs, correspond to $C_9^\text{SM}(\mu_0)=1.614$ and
$C_{10}^\text{SM}(\mu_0)=-4.255$.
The excellent agreement between the experimental and theoretical determinations of $B\to X_s\gamma$ (see,
for instance, Ref.~\cite{Misiak:2015xwa} and references therein) suggests that any new-physics contributions
to $C_7$ and $C_8$ are small, so we do not consider them here.
We also assume that the Wilson coefficients for $b\to s\ell\ell$ and $b\to d\ell\ell$ transitions are
identical, as they would be in minimal flavor violation, where new-physics contributions to the semileptonic
operators for $b\to q \ell\ell$ are proportional to $V_{tb}^{} V_{tq}^*$.
We further assume that there are no new $CP$-violating phases and take $C_9^\text{NP}(\mu_0)$ and
$C_{10}^\text{NP}(\mu_0)$ to be real.
\begin{figure}
    \includegraphics[width=0.45\textwidth]{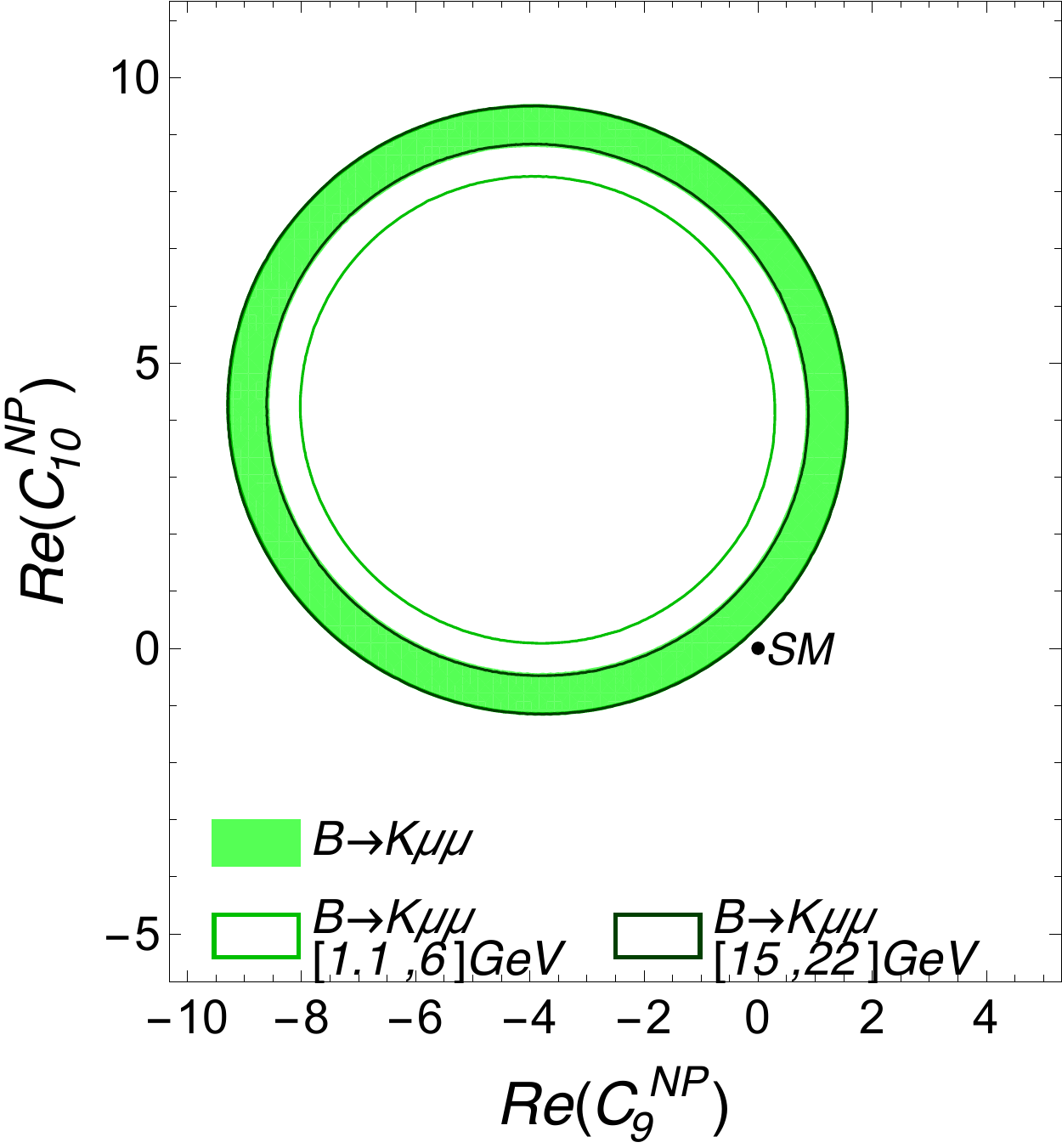}   \hfill
    \includegraphics[width=0.45\textwidth]{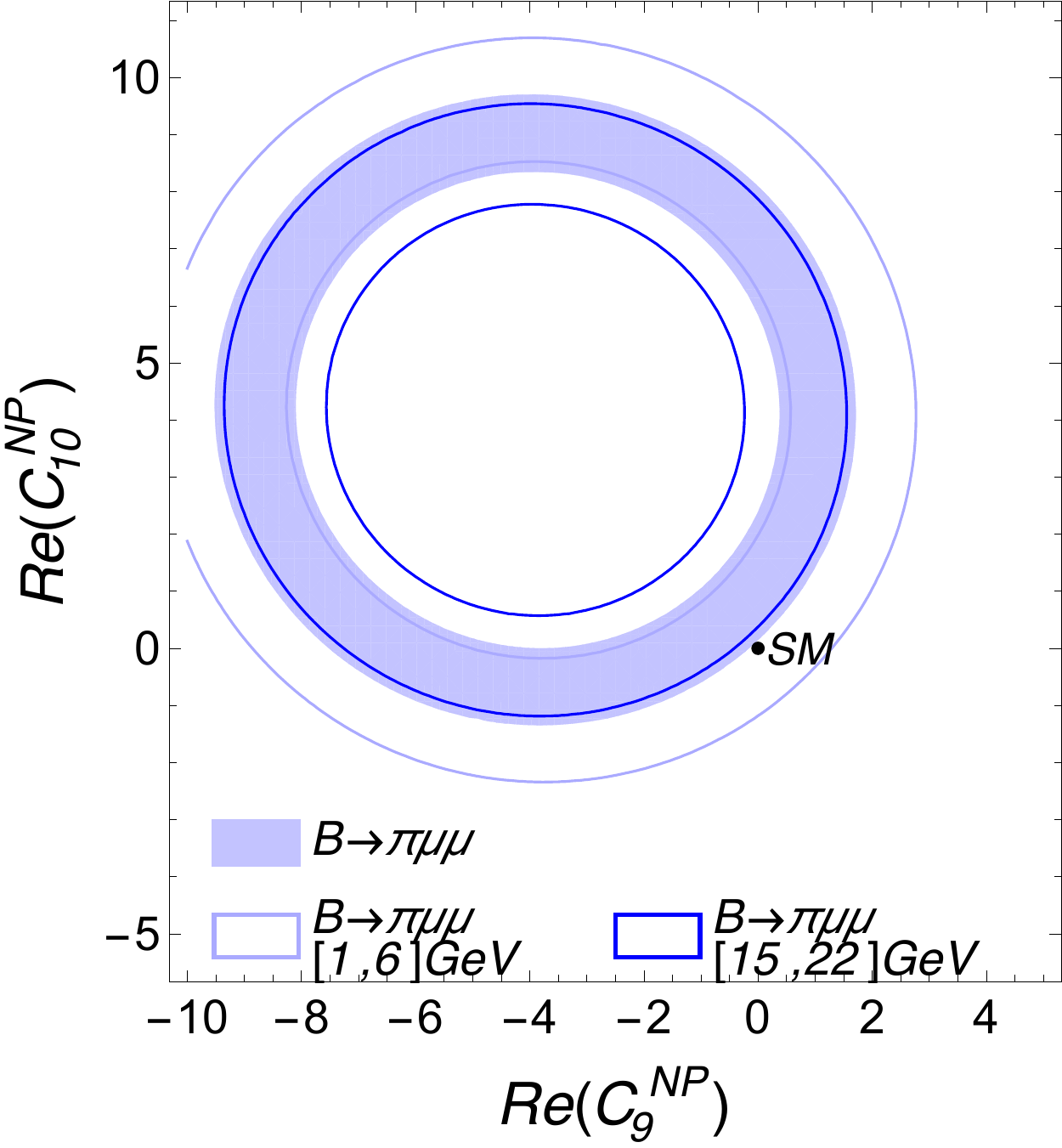}  \\[1.5em]
    \includegraphics[width=0.45\textwidth]{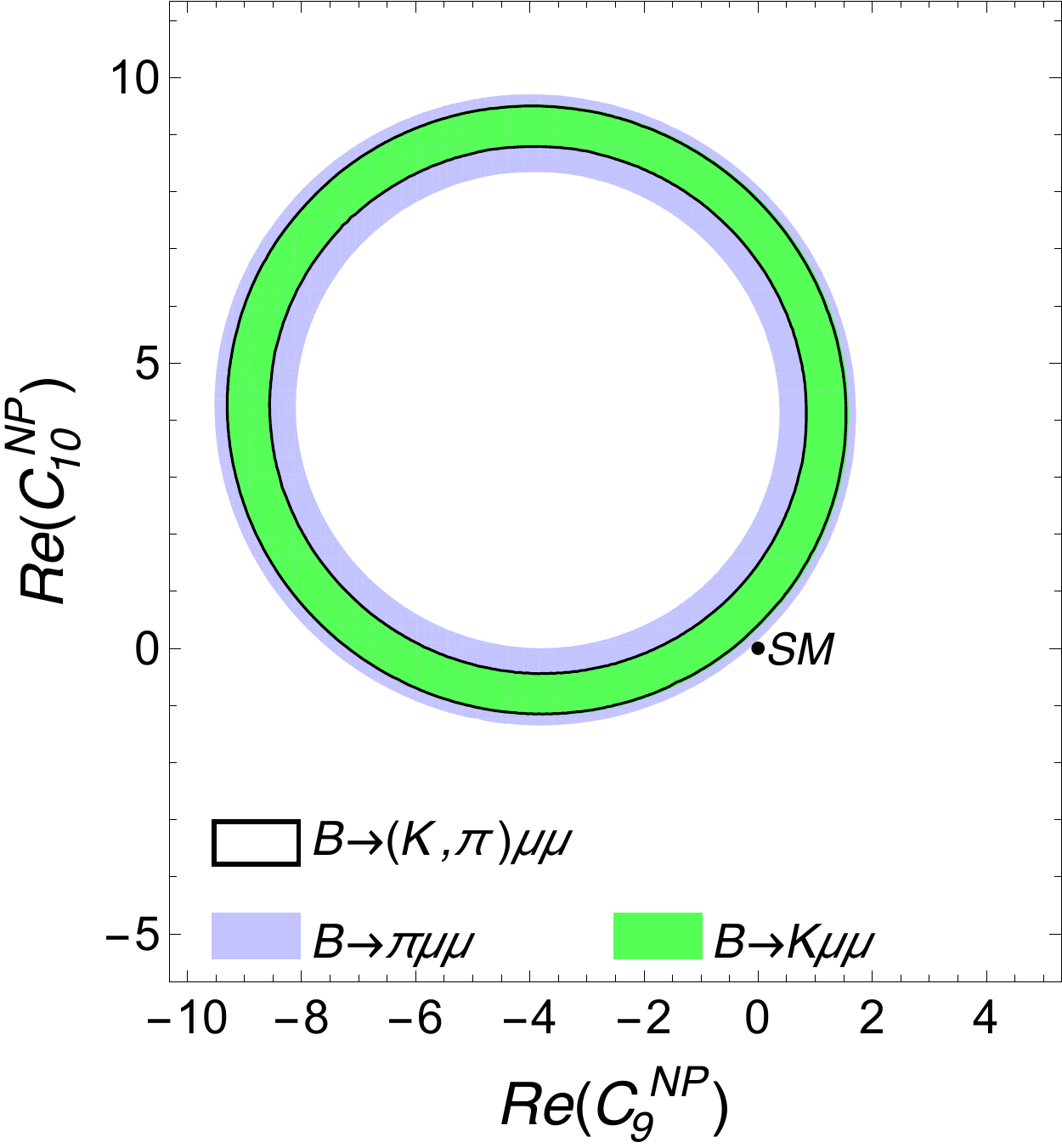} \hfill
    \includegraphics[width=0.45\textwidth]{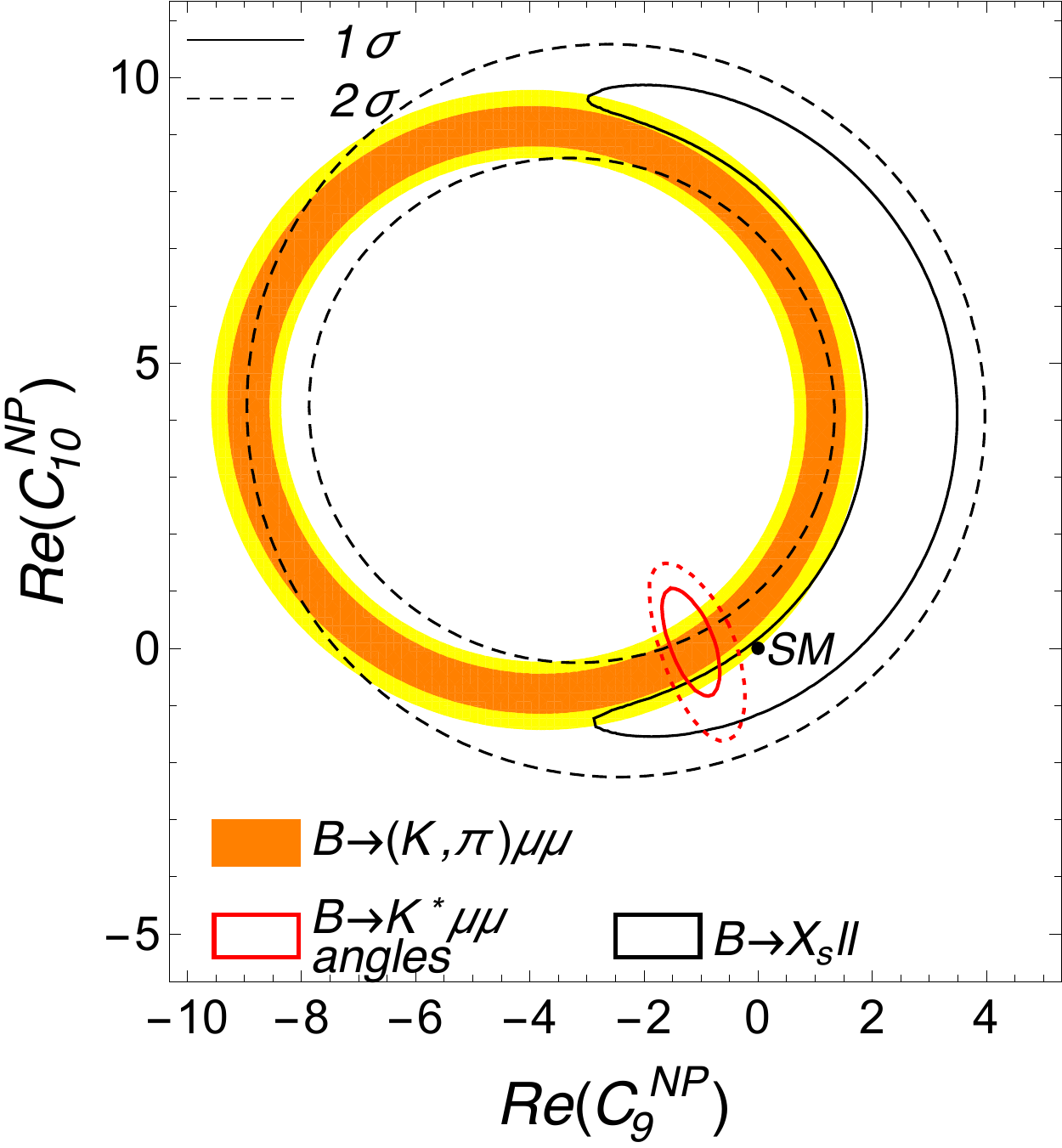}
    \caption{Constraints on the $\Re\left(C_9^\text{NP}\right)$-$\Re\left(C_{10}^\text{NP}\right)$ plane 
        implied by $B\to (K,\pi)\mu^+\mu^-$ data.
        In the top two panels, the light and dark unfilled bands show the $1\sigma$ constraints from 
        the low-$q^2$ ($[1(1.1),6]~\gev^2$) and high-$q^2$ ($[15,22]~\gev^2$) bins, respectively, for
        $B^+\to K^+\mu^+\mu^-$ (left) and $B^+\to \pi^+ \mu^+\mu^-$ (right).
        The filled bands show the $1\sigma$ allowed regions when the two bins are combined.
        Note that the outer low- and high-$q^2$ $B\to K$ contours almost completely overlap.
        The lower left panel shows these two $1\sigma$ regions with an unfilled band obtained from 
        combining the two constraints (almost coincident with the filled \BtoKll\ band).
        The lower right panel compares the $1\sigma$ and $2\sigma$ bands [in orange (gray) and yellow 
        (light gray), respectively] from the $B\to(K,\pi)\mu^+\mu^-$ data  with the constraints from
        $B\to  K^*\ell\ell$ angular observables (red unfilled contours, dotted for 
        $2\sigma$)~\cite{Altmannshofer:2015sma} and  
        inclusive processes (black open contours, dashed for $2\sigma$)~\cite{Huber:2015sra}.}
    \label{fig:c9c10}
\end{figure}
\begin{figure}
    \includegraphics[width=0.45\textwidth]{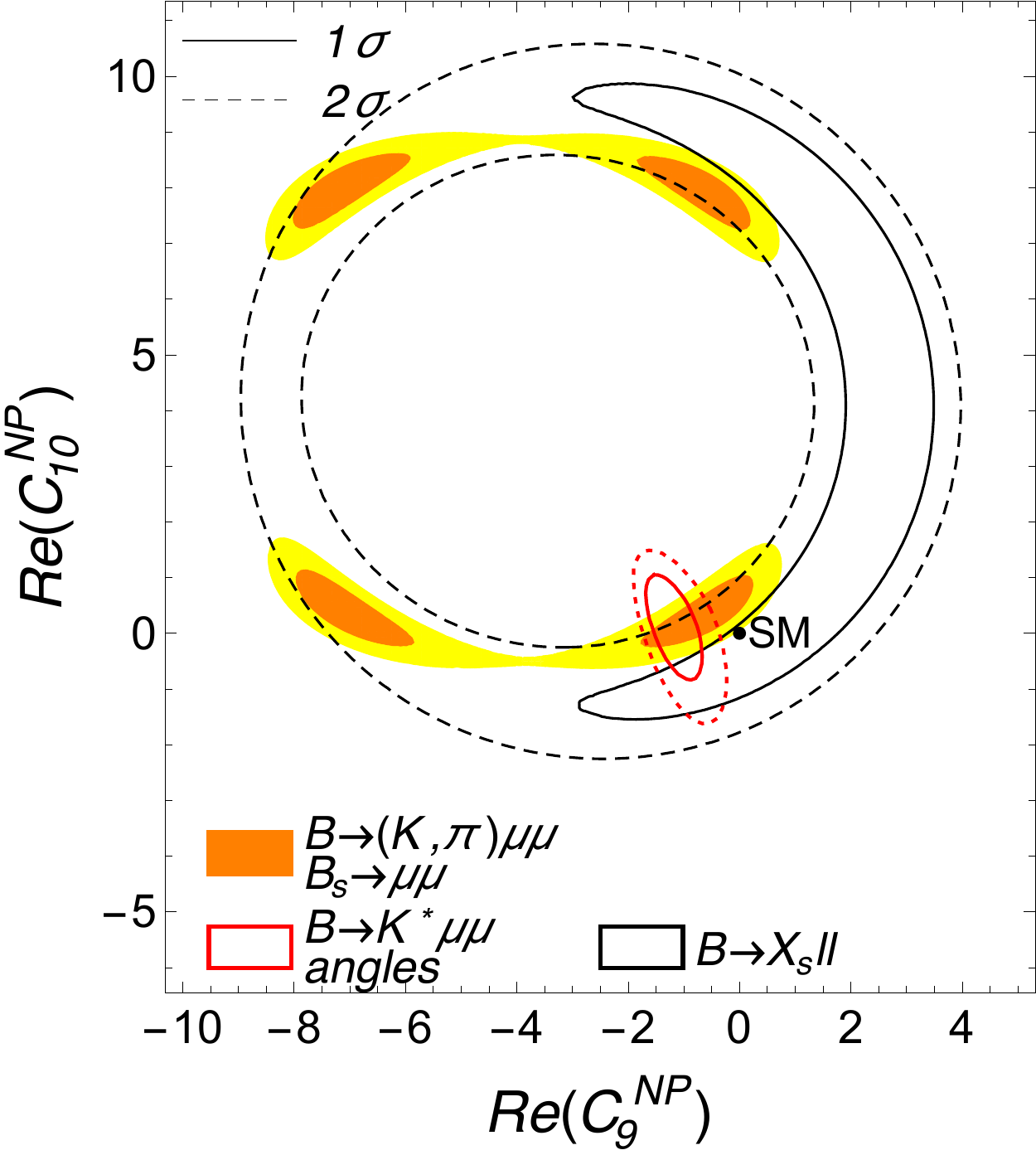} \hfill
    \includegraphics[width=0.45\textwidth]{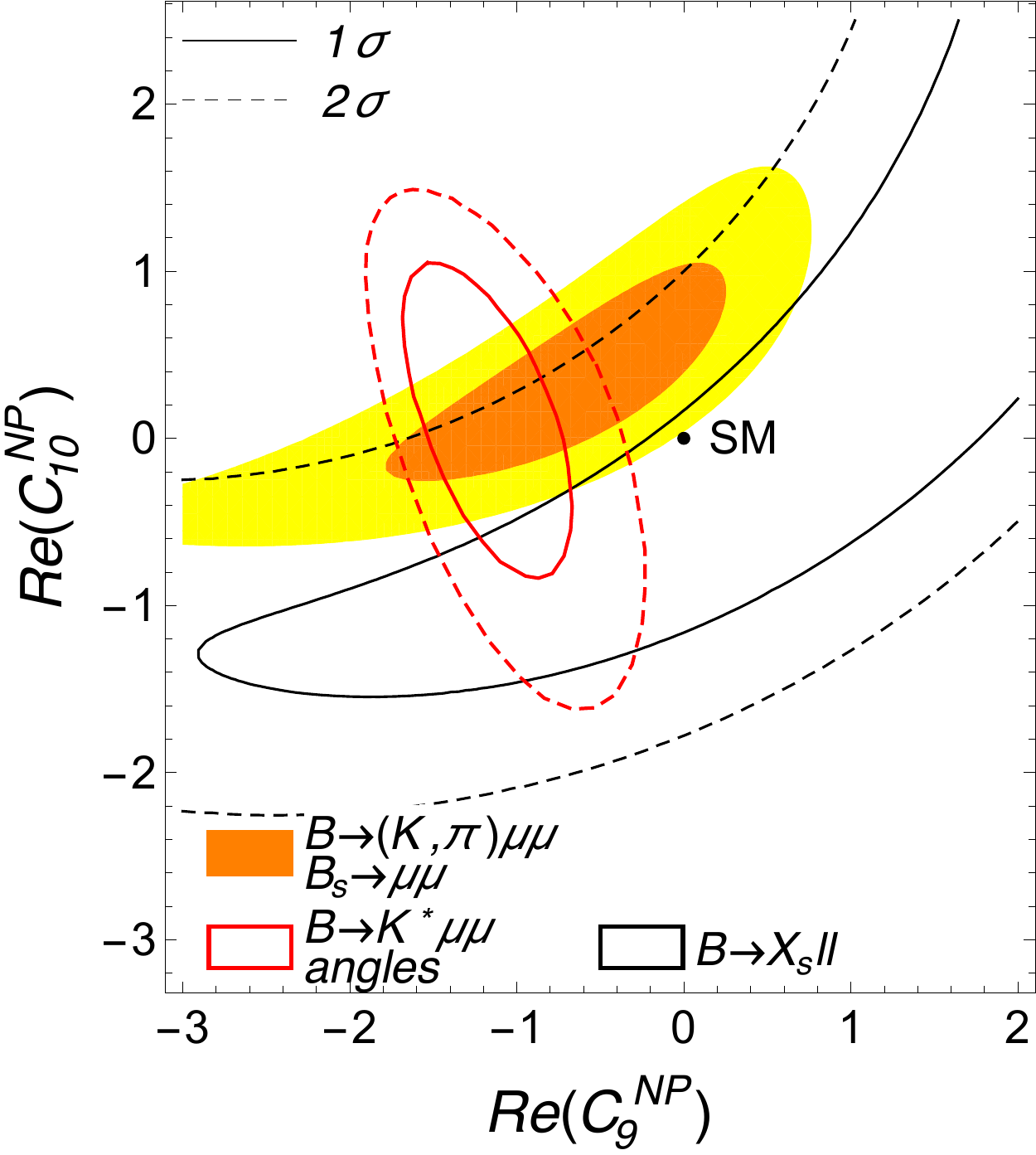} 
    \caption{Constraints on the $\Re(C_9^\text{NP})$-$\Re(C_{10}^\text{NP})$ plane implied by $B\to
        (K,\pi)\mu^+\mu^-$ and $B_s\to \mu^+\mu^-$ data.
        The right panel shows the same contours as the left panel, but focuses on the region near the 
        Standard-Model value.
        The color and styling is the same as in Fig.~\ref{fig:c9c10} (lower right).}
    \label{fig:c9c10Bsmm}
\end{figure}

To obtain the constraints shown here, we employ the measured $B^+\to \pi^+ \mu^+\mu^-$ and $B^+\to K^+
\mu^+\mu^-$ branching ratios in the wide $q^2$ intervals $[1(1.1),6]~\text{GeV}^2$ and
$[15,22]~\text{GeV}^2$ from LHCb~\cite{Aaij:2014pli,Aaij:2015nea}, which are quoted in
Eqs.~(\ref{eq:LHCb_B2pi}) and~(\ref{eq:LHCb_B2K+}).
We adopt a frequentist approach, and construct a $\chi^2$ statistic for these four measurements using a
covariance matrix constructed from the correlation matrices given in
Tables~\ref{tab:B2pi+_B2K+_dBdq2_FF_correlations} and \ref{tab:B2pi+_B2K+_dBdq2_Other_correlations}, and
from the errors quoted in Table IV of Ref.~\cite{Bailey:2015nbd} and in Table~\ref{tab:B+2K_dBdq2}.
We obtain the experimental contribution to the covariance matrix by assuming that the four LHCb measurements
in Eqs.~(\ref{eq:LHCb_B2pi}) and (\ref{eq:LHCb_B2K+}) are uncorrelated, which should be a good approximation
because the high- and low-$q^2$ bins are statistically independent, and the $B^+\to\pi^+\mu^+\mu^-$
measurement is dominated by statistical errors, while that for $B^+\to K^+ \mu^+\mu^-$ is limited by
systematics.

The resulting allowed regions in the $\Re\left(C_9^\text{NP}\right)$-$\Re\left(C_{10}^\text{NP}\right)$
plane are shown in Fig.~\ref{fig:c9c10}.
In the top two panels we present the $1\sigma$ constraints from $B^+\to K^+ \mu^+\mu^-$ (left) and
$B^+\to\pi^+\mu^+\mu^-$ (right), where we show the allowed regions implied by each of the two bins
separately (unfilled bands) as well as their combination (solid bands).
The lower left panel shows the constraint from combining $B^+\to K^+ \mu^+\mu^-$ and $B^+\to\pi^+
\mu^+\mu^-$ branching ratios.
Comparing the top and bottom left panels, we see that the combined $B\to (K,\pi) \mu^+\mu^-$ constraint is
currently controlled by the high-$q^2$ $B^+\to K^+ \mu^+\mu^-$ $[15,22]\; \text{GeV}^2$ bin.
In the lower right panel, the orange and yellow solid bands are the 1$\sigma$ and 2$\sigma$ regions allowed
by $B\to (K,\pi)\mu^+\mu^-$ data.
Allowing for new-physics contributions to $C_9$ and $C_{10}$ yields a best fit with 
$\chi^2_\text{min}/\text{dof}=1.8/2$, corresponding to $p=0.41$.
We find a 2.0$\sigma$ tension between the Standard-Model values for $C_9^\text{NP}=C_{10}^\text{NP}=0$
and those favored by the $B\to (K,\pi)\mu^+\mu^-$ branching ratios. 

In the lower right panel of Fig.~\ref{fig:c9c10}, we compare our allowed region in the
$\Re(C_9^\text{NP})$-$\Re(C_{10}^\text{NP})$ plane obtained from $B\to (K,\pi)\mu^+\mu^-$ branching
fractions alone with the constraints from inclusive $B\to X_s \ell^+\ell^-$ and exclusive $B\to
K^*\mu^+\mu^-$ measurements.
The region favored by inclusive observables (black contours) is taken from Ref.~\cite{Huber:2015sra}, where
the most recent experimental results from BaBar~\cite{Lees:2013nxa, Aubert:2004it} and
Belle~\cite{Iwasaki:2005sy} were used.
Similarly, the region favored by $B\to K^*\mu^+\mu^-$ angular observables (red contours) is taken from
Ref.~\cite{Altmannshofer:2015sma}.
Global analyses of $b\to s$ data similar to that performed in Ref.~\cite{Altmannshofer:2015sma} have also
been presented in Refs.~\cite{Descotes-Genon:2013wba,Altmannshofer:2013foa,Beaujean:2013soa,%
Descotes-Genon:2013zva,Hurth:2013ssa,Descotes-Genon:2014uoa,Hiller:2014yaa,Hurth:2014vma,%
Descotes-Genon:2014joa,Hiller:2014ula,Descotes-Genon:2015xqa}.

In Fig.~\ref{fig:c9c10Bsmm} we add the constraint from the leptonic decay rate $\BR(B_s\to\mu^+\mu^-)$, for 
which lattice QCD also gives a reliable input for the hadronic matrix element $f_{B_s}$~%
\cite{Bazavov:2011aa,McNeile:2011ng,Na:2012kp,Dowdall:2013tga,Carrasco:2013naa,Christ:2014uea}.
The expression for the Standard-Model rate is given in Eqs.~(3) and~(6) of Ref.~\cite{Bobeth:2013uxa}, and
is proportional to $f_{B_s}^2$ and to the CKM combination $|V_{tb}^{} V_{ts}^*|^{2}$. 
For $f_{B_s}$ we use the recent PDG value $f_{B_s}=226.0(2.2)~\text{GeV}$~\cite{Rosner:2015wva}, which was
obtained by averaging the lattice QCD results of
Refs.~\cite{Bazavov:2011aa,McNeile:2011ng,Dowdall:2013tga,Christ:2014uea,Aoki:2014nga}.
We take the remaining parametric inputs from Table~\ref{tab:inputs} to obtain the Standard-Model total
branching ratio
\begin{equation}
	\BR(B_s \to \mu^+ \mu^-)^\text{SM} = 3.39(18)(7)(8) \times 10^{-9} \,,
\end{equation}
where the errors are from the CKM elements, decay constant, and the quadrature sum of all other
contributions, respectively.
We take the nonparametric uncertainties to be 1.5\%~\cite{Bobeth:2013uxa}.

In the most general Standard-Model extension, new-physics contributions to $B_s\to\mu^+\mu^-$ decay can
arise from six operators in the effective Hamiltonian: in addition to $Q_{10}$, there are operators with
lepton currents $\bar\ell\ell$ and $\bar\ell\gamma_5\ell$, and three additional operators obtained by
flipping the chirality of the quark current~\cite{Fleischer:2014jaa}.
In order to combine information from $B_s\to\mu^+\mu^-$ with the constraints on the Wilson coefficients
$C_9$ and $C_{10}$ from $\BtoK(\pi)\mu^+\mu^-$ branching ratios presented above, we assume that only
$C_{10}$ is affected by new physics.
Under this assumption, the $B_s\to \mu^+\mu^-$ rate is proportional to $|C_{10}|^2$, so its inclusion
reduces constraint from a ring to a smaller, roughly elliptical, region.
Taking the measured branching ratio $\BR(B_s \to \mu^+ \mu^-)^\text{exp} = 2.8(^{+0.7}_{-0.6})\times10^{-9}$
from CMS and LHCb~\cite{CMS:2014xfa}, we obtain the yellow and orange shaded bands in
Fig.~\ref{fig:c9c10Bsmm}.
As shown in the left panel, at the 1$\sigma$ level there are four distinct allowed regions in the
$\Re(C_9^\text{NP})$-$\Re(C_{10}^\text{NP})$ plane, which merge into two larger nearly horizontal bands at
2$\sigma$.
The lower-right orange contour is close to the Standard-Model value, and we zoom in on this region in
Fig.~\ref{fig:c9c10Bsmm}, right.
The region allowed by $\BtoK(\pi)\mu^+\mu^-$ and $B_s\to \mu^+\mu^-$ branching ratios is compatible with the
constraint from $B\to K^*\mu^+\mu^-$ angular observables, but is in slight tension with the constraint from
inclusive $B\to X_s \ell^+\ell^-$ decays.
Because the Standard-Model $B_s\to \mu^+\mu^-$ total branching ratio is compatible with experiment,
including this information slightly decreases the tension between the Standard-Model prediction and
experimental measurements of $B\to(K,\pi)\mu\mu$ branching ratios.
The compatibility with the Standard-Model hypothesis increases from $p=0.10$ to $p=0.13$, and the
significance of the tension decreases from 1.7$\sigma$ to 1.5$\sigma$.
Allowing for new physics in $C_9$ and $C_{10}$ yields a best fit with $\chi^2_\text{min}/\text{dof}=1.8/3$,
corresponding to $p=0.61$, and values of $\Re(C_9^\text{NP})$ and $\Re(C_{10}^\text{NP})$ that differ by 
2.1$\sigma$ from the Standard Model.

Because the constraints from $\Btopi(K)\mu^+\mu^-$ branching ratios on the 
$\Re(C_9^\text{NP})$-$\Re(C_{10}^\text{NP})$ plane are limited by the experimental uncertainties, the 
widths of the 
corresponding bands in Figs.~\ref{fig:c9c10} and~\ref{fig:c9c10Bsmm} will be reduced with new measurements
by LHCb from the recently started LHC run and by the upcoming Belle~II experiment.
Therefore, these decays will continue to squeeze the allowed region in the
$\Re(C_9^\text{NP})$-$\Re(C_{10}^\text{NP})$ plane.

\section{Summary of Main Results}
\label{sec:Summary}


We now summarize our main results to help the reader digest the large quantity of information presented in
the previous two sections.
We present them in the same order in which they appeared above.

We begin with tests of heavy-quark and SU(3) symmetries.
By and large, the results given in Sec.~\ref{sec:SymTests} show marginal-to-excellent agreement with the
symmetry limits, but the sizable $q^2$ dependence we observe is an obstacle to providing a simple rule of
thumb.

Our findings for the Standard-Model observables for decays with a charged-lepton pair in the final state,
$\BtoK(\pi)\ell^+\ell^-$ with $\ell=e,\mu,\tau$ are more interesting.
Here, the most reliable results are those for the wide $q^2$ regions above and below the charmonium
resonances.
For decays with a muon pair, we compare our results for the partially integrated branching fractions with
the latest experimental measurements from the LHCb experiment~\cite{Aaij:2014pli,Aaij:2015nea}.
For the wide $q^2$ bins below and above the charmonium resonances, we find that the Standard-Model
expectations for both $\Delta\BR(\Btopi \mu^+\mu^-)$ and $\Delta\BR(\BtoK \mu^+\mu^-)$ are in slight tension
with the experimental measurements (see Fig.~\ref{fig:B2K_dBdq2_integrated}), with the Standard-Model values
being 1--2$\sigma$ higher.
The Standard-Model expectations for the ratio $\Delta\BR(\Btopi \mu^+\mu^-)/\Delta\BR(\BtoK \mu^+\mu^-)$ are
compatible with experiment, however, within 1.1$\sigma$ (see Fig.~\ref{fig:B2pi_B2K_ratio}).
We also provide Standard-Model values for the flat terms and lepton-universality-violating ratios for all
lepton final states $\ell = e, \mu, \tau$.
We confirm the 2.6$\sigma$ discrepancy observed for $R^{\mu e}_{K^+}$ by LHCb~\cite{Aaij:2014ora}.
Semileptonic $B$ decays with $\tau$ pairs in the final state have yet to be observed, so our results for the
associated observables are theoretical predictions that will be tested by experiment.

For decays with a neutrino pair in the final state, $\BtoK(\pi)\nu\bar\nu$, we provide Standard-Model
predictions for the total branching fractions.
We do not present partially integrated branching fractions because there are not yet experimental
measurements of these processes to guide our choice of $q^2$ intervals.
Like $\BtoK(\pi)\ell^+\ell^-$, these processes involve a $b \to d(s)$ FCNC transition, and thus probe some
of the same underlying physics.
The decay rates for $\BtoK(\pi)\nu\bar\nu$, however, depend on only a single operator in the Standard-Model
effective Hamiltonian.
Hence, once measured, they will provide complementary information.
Moreover, the $\BtoK(\pi)\nu\bar\nu$ decay rates do not receive contributions from $u\bar{u}, c\bar{c}$
resonances or nonfactorizable terms, making the theoretical predictions particularly clean.
Because the current bounds on $\BR(\BtoK\nu\bar\nu)$ from Babar~\cite{Lees:2013kla} and
Belle~\cite{Lutz:2013ftz} are only about a factor of ten larger than Standard-Model expectations, one may
anticipate that this process will be observed by the forthcoming Belle~II experiment~\cite{Aushev:2010bq}.
Once experimental analyses of the $\BtoK(\pi)\nu\bar\nu$ differential decay rates have settled on bin sizes,
we can provide predictions for the partially integrated branching fractions matched to the $q^2$ bins
employed.

We also predict the total branching ratio for the tree-level decay
$\Btopi\tau\nu$.
Although this is not a FCNC process, the large $\tau$-lepton mass makes this process sensitive to
contributions from charged Higgs or other scalar bosons.
The ratio of the decay rate for $\Btopi\tau\nu$ over the decay rate for $\Btopi\ell\nu\; (\ell = e, \mu)$ is
of particular interest given the combined 3.9$\sigma$ deviation from the Standard Model for the analogous
ratios for $B\to D \ell \nu$ and $B\to D^* \ell\nu$ semileptonic decays~\cite{Amhis:2014hma}.
We obtain
\begin{equation}
    R(\pi) \equiv \frac{\BR(\Btopi \tau \nu_\tau)}{\BR(\Btopi \ell \nu_\ell)}  = 0.641(17) ,
\end{equation}
where the error quoted includes statistical and systematic uncertainties.
Because the CKM element $|V_{ub}|$ cancels in the ratio, $R(\pi)$ provides an especially clean test of the
Standard Model independent of the tension between inclusive and exclusive
determinations~\cite{Antonelli:2009ws,Butler:2013kdw,Amhis:2014hma,Agashe:2014kda,Bevan:2014iga}.
The Belle experiment recently presented a preliminary 2.4$\sigma$ measurement of the $B^0 \to \pi^- \tau^+
\nu_\tau$ total branching fraction, setting an upper limit only about five times greater than the Standard
Model prediction, so Belle~II may be able to measure not only the total branching fraction but also the
$q^2$ spectrum for this process~\cite{Aushev:2010bq}.  Again, we can provide partially integrated branching 
ratios corresponding to the bins used in future experimental analyses once they are needed.

The differential decay rates for $\Btopi\ell^+\ell^-(\nu\bar\nu)$ and $\BtoK \ell^+\ell^-(\nu\bar\nu)$
decay are proportional to the combinations of CKM elements $|V_{td}V_{tb}^*|$ and $|V_{ts}V_{tb}^*|$,
respectively.
Thus they enable determinations of $|V_{td}|$, $|V_{ts}|$, and their ratio that can be compared with the
current most precise results obtained from the oscillation frequencies of neutral $B_d$ and $B_s$ mesons.
Assuming the Standard Model, we combine the theoretical values for $\Delta\BR(\Btopi \ell^+\ell^-)$,
$\Delta\BR(\BtoK \ell^+\ell^-)$, and their ratio with the recent experimental measurements from
LHCb~\cite{Aaij:2014pli,Aaij:2015nea} to obtain the CKM matrix elements
\begin{equation}
	|V_{td}| = 7.45(69) \times 10^{-3}, \quad |V_{ts}| = 35.7(1.5) \times 10^{-3}, \quad
    \left| \frac{V_{td}}{V_{ts}} \right| = 0.201(20),
\end{equation}
where we take $|V_{tb}|$ from CKM unitarity~\cite{Charles:2004jd}, and the errors include both experimental
and theoretical uncertainties.
These results are compatible with the PDG values from neutral $B$-meson oscillations~\cite{Agashe:2014kda},
with a commensurate uncertainty for $|V_{td}|$, and an error on $|V_{ts}|$ that is almost two times smaller.
Compared with the determinations $|V_{td}| = 7.2(^{+9}_{-8}) \times 10^{-3}$, $|V_{ts}| = 32(4) \times
10^{-3}$, and $|V_{td}/V_{ts}| = 0.24(^{+5}_{-4})$ by LHCb~\cite{Aaij:2015nea} using the same experimental
inputs but older form factors, the uncertainties above are 1.2, 2.7, and 2.3 times smaller, respectively.
This illustrates the impact of using the precise hadronic form factors from \emph{ab-initio} lattice
QCD~\cite{Lattice:2015tia,Bailey:2015nbd,Bailey:2015dka}.
Further, our predictions for $\BR(\Btopi\nu\bar\nu)$, $\BR(\BtoK\nu\bar\nu)$, and their ratio will
facilitate new, independent determinations of $|V_{td}|$, $|V_{ts}|$, and their ratio once these processes
have been observed experimentally.

Finally, we also explore the constraints on possible new-physics contributions to the Wilson coefficients
$C_9$ and $C_{10}$ implied by the LHCb data for the partially integrated wide-bin $\BtoK(\pi) \mu^+\mu^-$
branching ratios when combined with the new lattice-QCD form factors.
We find a $1.7\sigma$ tension between the Standard Model and the values for $\Re(C_9)$ and
$\Re(C_{10})$ preferred by semileptonic $B$-meson decay data alone, as shown in Fig.~\ref{fig:c9c10}.
Including the constraint from $B_s \to \mu\mu$, for which reliable hadronic input from lattice QCD is also
available, shrinks the allowed region and also slightly decreases the significance of the tension with the
Standard Model.
The region allowed by these theoretically clean decay modes is consistent with the constraints obtained from
$B \to K^* \ell \ell$ observables, and the widths of the bands are comparable in size.
The $B \to K^*$ constraints, however, make use of additional theoretical assumptions, which are not needed
in lattice-QCD calculations of $B$-meson leptonic decays or semileptonic decays with a pseudoscalar meson in
the final state.
Hence, using the \FerMILC\ form factors from \emph{ab-initio} QCD~\cite{Lattice:2015tia,Bailey:2015nbd,%
Bailey:2015dka}, we obtain theoretically clean constraints on the Wilson coefficients with uncertainties
similar to previous analyses.

\section{Conclusions and outlook}
\label{sec:Conclusions}

Rare semileptonic $B$-meson decays provide a wealth of processes and observables with which to test the
Standard Model and search for new physics.
Exploiting this wealth, however, requires that both the experimental measurements and the corresponding
theoretical calculations are sufficiently precise and reliable.
Recent progress in both areas has been significant, in particular, on the form factors that parametrize the
momentum-dependent hadronic contributions to $B$-meson semileptonic decays with a pseudoscalar meson in the
final state.
In this paper, we explore the phenomenological implications of new calculations of the
$\Btopi$~\cite{Lattice:2015tia,Bailey:2015nbd} and $\BtoK$~\cite{Bailey:2015dka} transition form factors by
the Fermilab Lattice and MILC Collaborations.
With the new \emph{ab-initio} QCD information on the hadronic matrix elements, we are able to calculate the
observables with fewer assumptions than previously possible.
Indeed, the comparison of our Standard-Model results for the $\BtoK(\pi)\mu^+\mu^-$ partial branching
fractions with experimental measurements reveals that the theoretical uncertainties are now commensurate
with the experimental errors, especially in the high-$q^2$ region.
As a result, these decays are already providing theoretically clean and quantitatively meaningful tests of
the Standard Model and constraints on new physics.
Once the rare decays $\BtoK(\pi)\nu\bar\nu$ and $\Btopi\tau\nu$ are observed, our predictions for these
processes will enable further Standard-Model tests, and, if deviations are seen, provide complementary
information on the underlying new physics.

Our work reveals 1--2$\sigma$ deviations between the Standard Model and experiment for both
$\Btopi\mu^+\mu^-$ and $\BtoK\mu^+\mu^-$ decays, where the theory values lie systematically above the
measurements for all four wide $q^2$ bins outside the charmonium resonance region.
Although the combined tension is less than 2$\sigma$, at the current level of uncertainty, there is still
ample room for new physics.
Sharpening this and other tests and potentially revealing evidence for new physics will require improvements
in both experiment and theory, both of which are expected.
On the experimental side, measurements will continue to improve at the currently running LHCb experiment.
Further, the soon-to-start Belle II experiment expects a great increase in luminosity compared to the
previous Belle experiment.
It may therefore observe heretofore unseen decays such as $\BtoK(\pi)\nu\bar\nu$ and $\Btopi\tau\nu$, for
which the Standard-Model predictions are particularly clean.
On the theoretical side, more precise $\BtoK$ and $\Btopi$ form factors from lattice QCD are anticipated.
A dominant uncertainty in the form factors from Refs.~\cite{Lattice:2015tia,Bailey:2015nbd,Bailey:2015dka}
employed in this work, and for similar efforts using different light- and $b$-quark
actions~\cite{Bouchard:2013pna,Bouchard:2013zda,Flynn:2015mha}, is the combined statistical plus
chiral-extrapolation error.
Fortunately, three- and four-flavor lattice gauge-field ensembles with the average light-quark mass
$(m_u+m_d)/2$ tuned to the physical value are becoming increasingly
available~\cite{Durr:2010aw,Bruno:2014jqa,Blum:2014tka,Bazavov:2015yea}, the use of which will essentially
eliminate this source of error.
Further, the form-factor uncertainties at $q^2=0$ are quite large due to the extrapolation from the range of
simulated lattice momenta $q^2 \gtrsim 16~\text{GeV}^2$ to the low-$q^2$ region using the model-independent
$z$ expansion.
Reducing the form factor uncertainties at low-$q^2$ is necessary in order to make better use of experimental
data at low-$q^2$, and to sharpen comparisons of $q^2$ spectra between theory and experiment.
Lattice-QCD ensembles with finer lattice spacings but similar spatial volumes will enable simulations with
larger pion and kaon momenta, thereby shortening the extrapolation range and reducing the associated error.
In particular, the form-factor calculations of Refs.~\cite{Lattice:2015tia,Bailey:2015nbd,Bailey:2015dka}
will be repeated by the Fermilab Lattice and MILC Collaborations on a new set of ensembles recently
generated by the MILC Collaboration~\cite{Bazavov:2010ru,Bazavov:2015yea} using the highly improved
staggered quark action~\cite{Follana:2006rc}.
These four-flavor ensembles include dynamical up, down, strange, and charm quarks; physical-mass light
quarks; and planned lattice spacings as small as approximately 0.03~fm.

We note that semileptonic $B$-meson decays with a vector meson in the final state, such as $B\to
K^*\ell\ell$, provide an even richer set of observables with which to test the Standard Model, many of 
which are already measured experimentally.
For example, for the analysis of the Wilson coefficients, the constraint from $\BtoK^*\mu^+\mu^-$ angular
observables in the $\Re\left(C_9^{\rm NP}\right)$-$\Re\left(C_{10}^{\rm NP}\right)$ plane is approximately
perpendicular to that from $\Btopi(K)\mu^+\mu^-$ branching ratios and provides complementary information.
The presence of an unstable hadron in the final state, however, makes {\em ab initio} calculations of their
form factors much more complicated.
In fact, finite-volume methods for properly including the width of an unstable final state hadron in
semileptonic $B$-meson decays are still being developed~\cite{Briceno:2014uqa}.
Once such methods are fully established, they will bring the hadronic uncertainties for $\BtoK^*\mu^+\mu^-$
observables under equally good theoretical control as for $\Btopi(K)\mu^+\mu^-$.

$B$-meson leptonic and semileptonic decays are already testing the Standard Model in the quark-flavor
sector, in some cases yielding tantalizing discrepancies at the 2--3$\sigma$ level.
As discussed above, even more experimental and theoretical progress is anticipated.
We are therefore optimistic that the rare semileptonic $B$-meson decays studied in this work---%
$\BtoK(\pi)\mu^+\mu^-$, $\BtoK(\pi)\nu\bar\nu$, or $\Btopi\tau\nu$---may eventually reveal the presence of
new flavor-changing interactions or sources of $CP$-violation in the quark sector.

\acknowledgments

We thank our colleagues in the Fermilab Lattice and MILC Collaborations for an enjoyable collaboration that
helped spur this work.
We also thank Ulrik Egede and Tobias Tekampe for useful correspondence about LHCb's recent
$\BtoK(\pi)\mu^+\mu^-$ measurements, and Wolfgang Altmannshofer, Gudrun Hiller, Jernej Kamenik, Alexander
Khodjamirian, David Straub, and Yuming Wang for valuable discussions about the theory.
This work was supported in part by the U.S.\ Department of Energy under Grants No.~DE-SC0010120 (S.G.)
and No.~DE-FG02-13ER42001 (A.X.K.); %
by the National Science Foundation under Grant No.~PHY-1417805 (D.D., J.L.); %
and by the German Excellence Initiative, the European Union Seventh Framework Programme under grant
agreement No.~291763, and the European Union's Marie Curie COFUND program (A.S.K).
Fermilab is operated by Fermi Research Alliance, LLC, under Contract No.~DE-AC02-07CH11359 with the U.S.\
Department of Energy.
S.G., A.S.K., E.L., and R.Z.\ thank the Kavli Institute for Theoretical Physics, which is supported by the
National Science Foundation under Grant No.~PHY11-25915, for its hospitality while this paper was being
written.
A.S.K., A.X.K., and D.D.\ thank the Mainz Institute for Theoretical Physics for its hospitality while part
of this paper was written.

\appendix

\section{Numerical results for \texorpdfstring{\boldmath$B\to K (\pi) \ell^+ \ell^-$}{B to K(pi)ll} observables}
\label{app:Results}

Here we tabulate the numerical values of $B\to K(\pi)\ell^+\ell^-$ observables in the Standard Model
integrated over different $q^2$ intervals.
We select the same ranges of momentum transfer as the most recent experimental measurements from
LHCb~\cite{Aaij:2015nea,Aaij:2014pli}.


\begin{table}[ht]
    \caption{
        Standard-Model binned flat term $F_H^\ell(q^2_\text{min},q^2_\text{max})$ for \Bptopipll\ decay.
        Errors shown are from form factors and the quadrature sum of all other contributions, respectively.}
    \label{tab:B2pi_FH_charged}
  \begin{tabular}{l@{\hskip 3mm}c@{\hskip 3mm}c@{\hskip 3mm}c}
  \hline\hline
       $[q^2_{\rm min}, q^2_{\rm max}]$ (GeV$^2$)        & $10^8\, F_H^e$ &   $10^3\, F_H^\mu$  & $10^1\, F_H^\tau$    \\
  \hline
$[0.10, 2.00]$ & 242.0(2.2,0.6)  & 96.1(0.8,0.2) & \\
$[2.00, 4.00]$ & 50.6(1.1,0.1)   & 21.5(0.5,0.1) & \\
$[4.00, 6.00]$ & 28.6(1.0,0.1)   & 12.2(0.4,0.0) & \\
$[6.00, 8.00]$ & 19.9(0.9,0.1)   & 8.5(0.4,0.0)  & \\
$[15.00, 17.00]$ & 10.2(0.6,0.3) & 4.3(0.3,0.1)  & 8.5(0.1,0.2) \\
$[17.00, 19.00]$ & 10.1(0.6,0.3) & 4.3(0.3,0.1)  & 8.0(0.1,0.2) \\
$[19.00, 22.00]$ & 10.9(0.6,0.3) & 4.7(0.3,0.1)  & 7.8(0.1,0.2) \\
\hline
$[1.00, 6.00]$ & 52.7(1.3,0.2)   & 22.3(0.5,0.1) & \\
$[15.00, 22.00]$ & 10.4(0.6,0.3) & 4.4(0.3,0.1)  & 8.0(0.1,0.2) \\
$[4m_\ell^2, (M_{B^+}-M_{\pi^+})^2]$ & 55.6(3.6,1.8)  & 17.1(1.0,0.5) & 8.2(0.1,0.2) \\
  \hline
  \hline
  \end{tabular}
\end{table}

\begin{table} 
  \caption{
      Standard-Model binned flat term $F_H^\ell(q^2_\text{min},q^2_\text{max})$ for \Bntopinll\ decay.
      Errors shown are from form factors and the quadrature sum of all other contributions, respectively.}
  \label{tab:B2pi_FH_neutral}
  \begin{tabular}{l@{\hskip 3mm}c@{\hskip 3mm}c@{\hskip 3mm}c}
  \hline\hline
       $[q^2_{\rm min}, q^2_{\rm max}]$ (GeV$^2$)        & $10^8\, F_H^e$ &   $10^3\, F_H^\mu$  & $10^1\, F_H^\tau$    \\
  \hline
$[0.10, 2.00]$ & 249.6(1.8,0.8) & 98.9(0.7,0.3) &  \\
$[2.00, 4.00]$ & 51.0(1.0,0.1)  & 21.7(0.4,0.1) &  \\
$[4.00, 6.00]$ & 28.7(0.9,0.1)  & 12.2(0.4,0.0) &  \\
$[6.00, 8.00]$ & 19.9(0.8,0.1)  & 8.5(0.4,0.0)  &  \\
$[15.00, 17.00]$ & 10.1(0.6,0.3) & 4.3(0.3,0.1) & 8.5(0.1,0.2) \\
$[17.00, 19.00]$ & 10.0(0.6,0.3) & 4.3(0.3,0.1) & 8.0(0.1,0.2) \\
$[19.00, 22.00]$ & 10.8(0.6,0.3) & 4.6(0.3,0.1) & 7.7(0.1,0.2) \\
\hline
$[1.00, 6.00]$ & 53.1(1.2,0.2)  & 22.5(0.5,0.1) & \\
$[15.00, 22.00]$ & 10.3(0.6,0.3) &  4.4(0.3,0.1) & 8.0(0.1,0.2) \\
$[4m_\ell^2, (M_{B^0}-M_{\pi^0})^2]$ & 58.1(3.9,2.0) & 17.4(1.0,0.6) & 8.2(0.1,0.2) \\
  \hline
  \hline
  \end{tabular}
\end{table}

\begin{table} 
    \caption{
        Standard-Model partially integrated branching fractions for $B^+\to \pi^+ \mu^+\mu^-$ decay.
        Results for $B^+\to\pi^+e^+e^-$ are nearly identical.
        Errors shown are from the CKM elements, form factors, and the quadrature sum of all other
        contributions, respectively.
        Results are from Ref.~\cite{Bailey:2015nbd}, but additional digits are presented and the scale error
        has been included in the ``other" error quoted here, to facilitate use with the correlation matrices
        in Tables~\ref{tab:B2pi+_B2K+_dBdq2_FF_correlations} 
        and~\ref{tab:B2pi+_B2K+_dBdq2_Other_correlations}.}
  \label{tab:B+2pi_dBdq2}
  \begin{tabular}{l@{\hskip 3mm}c}
  \hline\hline
     $[q^2_{\rm min}, q^2_{\rm max}]$ (GeV$^2$) &  $10^9\, \Delta\BR(B^+ \to \pi^+ \mu^+  \mu^-)$   \\
\hline        
$[1.00, 6.00]$   & 4.781(0.286,0.541,0.165) \\ 
$[15.00, 22.00]$ & 5.046(0.303,0.338,0.162)\\ 
  \hline
  \hline
  \end{tabular}
\end{table}


\begin{table} 
  \caption{
      Standard-Model partially integrated branching fractions for $B^+\to K^+ \ell^+\ell^-$ decay.
      Results for \BptoKpee\ are nearly the same as for \BptoKpmumu.
      Errors shown are from the CKM elements, form factors, and the quadrature sum of all other 
      contributions, respectively.
      Results for the electron and muon final states are indistinguishable at the current level of 
      precision.
      At low $q^2$, we present two wide bins $[1~\gev^2, 6~\gev^2]$ and $[1.1~\gev^2,6~\gev^2]$ to enable 
      comparison with the LHCb measurements in Refs.~\cite{Aaij:2015nea} and~\cite{Aaij:2014pli}, 
      respectively.}
  \label{tab:B+2K_dBdq2}
  \begin{tabular}{l@{\hskip 3mm}c@{\hskip 3mm}c}
  \hline\hline
     $[q^2_{\rm min}, q^2_{\rm max}]$ (GeV$^2$)           &  $10^9\, \Delta\BR(B^+ \to K^+ \mu^+  \mu^-)$    &    $10^9\, \Delta\BR(B^+ \to K^+ \tau^+  \tau^-)$     \\
\hline        
$[0.10, 2.00]$ & 68.03(3.70,13.72,1.55) & \\ 
$[2.00, 4.00]$ & 71.72(3.91,12.44,1.63) & \\ 
$[4.00, 6.00]$ & 70.59(3.84,10.36,1.54) & \\ 
$[6.00, 8.00]$ & 68.94(3.75,8.47,1.46) & \\ 
$[15.00, 17.00]$ & 46.15(2.51,2.48,1.62) &  39.92(2.17,2.22,1.40) \\ 
$[17.00, 19.00]$ & 34.91(1.90,1.68,1.13) &  39.31(2.14,1.81,1.33) \\ 
$[19.00, 22.00]$ & 25.73(1.40,1.17,0.86) &  43.23(2.35,1.80,1.57) \\ 
\hline
$[1.00, 6.00]$ & 178.35(9.71,29.80,4.00) & \\ 
$[1.10, 6.00]$ & 174.75(9.52,29.07,3.92) & \\ 
$[15.00, 22.00]$ & 106.79(5.82,5.21,3.49) &  122.46(6.67,5.63,4.17) \\ 
$[4m_\ell^2, (M_{B^+}-M_{K^+})^2]$ & 605.33(32.96,65.14,17.03) &  160.36(8.73,7.87,5.46) \\ 
  \hline
  \hline
  \end{tabular}
\end{table}

\begin{table} 
  \caption{
      Standard-Model partially integrated branching fractions for $B^0\to K^0 \ell^+\ell^-$ decay.
      Results for \BntoKnee\ are nearly the same as for \BntoKnmumu.
      Errors shown are from the CKM elements, form factors, and the quadrature sum of all other 
      contributions, respectively.  Results for the electron and muon final states are indistinguishable at 
      the current level of precision.
      At low $q^2$, we present two wide bins $[1~{\rm GeV}^2, 6~{\rm GeV}^2]$ and 
      $[1.1~{\rm GeV}^2,6~{\rm GeV}^2]$ to enable comparison with the LHCb measurements in 
      Refs.~\cite{Aaij:2015nea} and~\cite{Aaij:2014pli}, respectively.}
  \label{tab:B02K_dBdq2}
  \begin{tabular}{l@{\hskip 3mm}c@{\hskip 3mm}c}
  \hline\hline
      $[q^2_{\rm min}, q^2_{\rm max}]$ (GeV$^2$)      &  $10^9\, \Delta\BR(B^0 \to K^0 \mu^+  \mu^-)$    &    $10^9\, \Delta\BR(B^0 \to K^0 \tau^+  \tau^-)$     \\
  \hline        
$[0.10, 2.00]$ & 63.38(3.45,12.70,1.51) & \\ 
$[2.00, 4.00]$ & 65.88(3.59,11.35,1.46) & \\ 
$[4.00, 6.00]$ & 64.94(3.54,9.47,1.37) & \\ 
$[6.00, 8.00]$ & 63.60(3.46,7.76,1.32) & \\ 
$[15.00, 17.00]$ & 42.76(2.33,2.30,1.51) &  36.96(2.01,2.04,1.30) \\ 
$[17.00, 19.00]$ & 32.25(1.76,1.55,1.04) &  36.34(1.98,1.67,1.23) \\ 
$[19.00, 22.00]$ & 23.53(1.28,1.07,0.79) &  39.74(2.16,1.65,1.44) \\ 
\hline
$[1.00, 6.00]$ & 164.09(8.94,27.23,3.59) & \\ 
$[1.10, 6.00]$ & 160.75(8.75,26.56,3.51) & \\ 
$[15.00, 22.00]$ & 98.54(5.37,4.80,3.22) &  113.05(6.16,5.19,3.85) \\ 
$[4m_\ell^2, (M_{B^0}-M_{K^0})^2]$ & 558.80(30.43,59.72,15.73) &  147.45(8.03,7.22,5.02) \\ 
  \hline
  \hline
  \end{tabular}
\end{table}

\begin{table} 
  \caption{
  Correlations between the form-factor contributions to the errors in the Standard-Model 
  partially integrated branching fractions for $B^+\to\pi^+ \ell^+ \ell^-$ decay and $B^+\to K^+ \ell^+ 
  \ell^-$ decay.
  These should be combined with the central values and form-factor errors in the bottom panels of Table~IV 
  from Ref.~\cite{Bailey:2015nbd} and Table~\ref{tab:B+2K_dBdq2} above. 
  The results for the neutral decay modes $B^0 \to \pi^0 (K^0) \ell^+ \ell^-$ should be taken as 100\% 
  correlated with those for the charged decays.}
  \label{tab:B2pi+_B2K+_dBdq2_FF_correlations}
  \begin{tabular}{l@{\hskip 3mm}c@{\hskip 3mm}c@{\hskip 3mm}c@{\hskip 3mm}c@{\hskip 3mm}c}
  \hline\hline
       $[q^2_{\rm min}, q^2_{\rm max}]$ (GeV$^2$) & $[1, 6]_{\pi^+}$ &   $[15, 22]_{\pi^+}$ &   
$[1, 6]_{K^+}$ &   $[1.1, 6]_{K^+}$ &   $[15, 22]_{K^+}$ \\
  \hline
$[1, 6]_{\pi^+}$   & 1.0000 & 0.6071 & 0.0426 & 0.0428 & 0.1190 \\
$[15, 22]_{\pi^+}$ & 0.6071 & 1.0000 & 0.1020 & 0.1023 & 0.2631 \\
$[1, 6]_{K^+}$     & 0.0426 & 0.1020 & 1.0000 & 1.0000 & 0.5099 \\
$[1.1, 6]_{K^+}$   & 0.0428 & 0.1023 & 1.0000 & 1.0000 & 0.5112 \\
$[15, 22]_{K^+}$   & 0.1190 & 0.2631 & 0.5099 & 0.5112 & 1.0000 \\
  \hline\hline
  \end{tabular}
\end{table}

\begin{table} 
  \caption{
    Correlations between the ``other'' contributions to the errors in the Standard-Model partially integrated
    branching fractions for $B^+\to\pi^+ \ell^+ \ell^-$ decay and $B^+\to K^+ \ell^+ \ell^-$ decay.
    These should be combined with the central values and ``other" errors in the bottom panels of Table~IV
    from Ref.~\cite{Bailey:2015nbd} and Table~\ref{tab:B+2K_dBdq2} above.
    The correlation between the combinations of CKM elements that enter the $B^+\to\pi^+ \ell^+ \ell^-$ and
    $B^+\to K^+ \ell^+ \ell^-$ decay rates ($|V_{td}V_{tb}^*|$ and $|V_{ts}V_{tb}^*|$) is 0.878.
    The results for the neutral decay modes $B^0 \to \pi^0 (K^0) \ell^+ \ell^-$ should be taken as 100\%
    correlated with those for the charged decays.}
  \label{tab:B2pi+_B2K+_dBdq2_Other_correlations}
  \begin{tabular}{l@{\hskip 3mm}c@{\hskip 3mm}c@{\hskip 3mm}c@{\hskip 3mm}c@{\hskip 3mm}c}
  	\hline\hline
  	$[q^2_{\rm min}, q^2_{\rm max}]$ (GeV$^2$) & $[1, 6]_{\pi^+}$ &   $[15, 22]_{\pi^+}$ &   
  	$[1, 6]_{K^+}$ &   $[1.1, 6]_{K^+}$ &   $[15, 22]_{K^+}$ \\
  	\hline
        $[1, 6]_{\pi^+}$   & 1.0000 & 0.4504 & 0.9730 & 0.9728 & 0.4860 \\
        $[15, 22]_{\pi^+}$ & 0.4504 & 1.0000 & 0.4212 & 0.4207 & 0.6098 \\
        $[1, 6]_{K^+}$     & 0.9730 & 0.4212 & 1.0000 & 1.0000 & 0.4510 \\
        $[1.1, 6]_{K^+}$   & 0.9728 & 0.4207 & 1.0000 & 1.0000 & 0.4504 \\
        $[15, 22]_{K^+}$   & 0.4860 & 0.6098 & 0.4510 & 0.4504 & 1.0000 \\
  	\hline\hline
  \end{tabular}	
\end{table}

\begin{table} 
  \caption{
    Standard-Model binned flat term $F_H^\ell(q^2_\text{min},q^2_\text{max})$ for \BptoKpll\ decay.
    Errors shown are from form factors and the quadrature sum of all other contributions, respectively.}
  \label{tab:B2K_FH_charged}
  \begin{tabular}{l@{\hskip 3mm}c@{\hskip 3mm}c@{\hskip 3mm}c}
  \hline\hline
       $[q^2_{\rm min}, q^2_{\rm max}]$ (GeV$^2$)        & $10^8\, F_H^e$ &   $10^3\, F_H^\mu$  & $10^1\, F_H^\tau$    \\
  \hline
$[0.10, 2.00]$ & 248.0(2.2,0.5)  & 98.3(0.8,0.2) & \\
$[2.00, 4.00]$ & 55.7(0.7,0.0)   & 23.6(0.3,0.0) & \\
$[4.00, 6.00]$ & 33.3(0.6,0.0)   & 14.2(0.2,0.0) & \\
$[6.00, 8.00]$ & 24.3(0.5,0.0)   & 10.3(0.2,0.0) & \\
$[15.00, 17.00]$ & 14.0(0.3,0.4) & 6.0(0.1,0.2)  & 8.9(0.0,0.3) \\
$[17.00, 19.00]$ & 14.7(0.3,0.5) & 6.3(0.1,0.2)  & 8.6(0.0,0.2) \\
$[19.00, 22.00]$ & 19.7(0.4,0.7) & 8.4(0.2,0.3)  & 8.7(0.0,0.2) \\
\hline
$[1.00, 6.00]$ & 57.8(1.1,0.1)   & 24.5(0.5,0.0) & \\
$[15.00, 22.00]$ & 15.6(0.3,0.5) & 6.6(0.1,0.2)  & 8.7(0.0,0.2) \\
$[4m_\ell^2, (M_{B^+}-M_{K^+})^2]$ & 71.5(6.1,2.1)  & 22.2(1.7,0.7) & 8.9(0.0,0.3) \\
\hline
\hline
  \end{tabular}
\end{table}

\begin{table} 
  \caption{
    Standard-Model binned flat term $F_H^\ell(q^2_\text{min},q^2_\text{max})$ for \BntoKnll\ decay.
    Errors shown are from the form factors and the quadrature sum of all other contributions, respectively.}
  \label{tab:B2K_FH_neutral}
  \begin{tabular}{l@{\hskip 3mm}c@{\hskip 3mm}c@{\hskip 3mm}c}
  \hline\hline
       $[q^2_{\rm min}, q^2_{\rm max}]$ (GeV$^2$)        & $10^8\, F_H^e$ &   $10^3\, F_H^\mu$  & $10^1\, F_H^\tau$    \\
  \hline
$[0.10, 2.00]$ & 258.6(2.4,0.4) &  101.8(0.9,0.2) & \\
$[2.00, 4.00]$ & 55.8(0.7,0.0)  & 23.6(0.3,0.0)   & \\
$[4.00, 6.00]$ & 33.3(0.6,0.0)  & 14.2(0.2,0.0)   & \\
$[6.00, 8.00]$ & 24.3(0.5,0.0)  & 10.3(0.2,0.0)   & \\
$[15.00, 17.00]$ & 14.0(0.3,0.4)  & 6.0(0.1,0.2)  & 8.9(0.0,0.3) \\
$[17.00, 19.00]$ & 14.7(0.3,0.5)  & 6.3(0.1,0.2)  & 8.6(0.0,0.2) \\
$[19.00, 22.00]$ & 19.8(0.4,0.7)  & 8.4(0.2,0.3)  & 8.7(0.0,0.2) \\
\hline
$[1.00, 6.00]$ & 58.0(1.1,0.1)    & 24.5(0.5,0.0) & \\
$[15.00, 22.00]$ & 15.6(0.3,0.5)  & 6.7(0.1,0.2)  & 8.7(0.0,0.2) \\
$[4m_\ell^2, (M_{B^0}-M_{K^0})^2]$ & 73.5(6.4,2.2)   & 22.4(1.7,0.7) & 8.9(0.0,0.3) \\
\hline
\hline
  \end{tabular}
\end{table}


\begin{table} 
    \caption{
        Standard-Model lepton-universality-violating ratios for $\Btopi\ell^+\ell^-$ decay.
        Results are shown for both the charged ($R_{\pi^+}$, left) and neutral ($R_{\pi^0}$, right) modes.
        Errors shown are from the form factors and the quadrature sum of all other contributions, 
        respectively.}
  \label{tab:B2P_RPi}
  \begin{tabular}{l@{\hskip 3mm}c@{\hskip 3mm}c@{\hskip 3mm}c@{\hskip 3mm}c@{\hskip 3mm}c@{\hskip 3mm}c}
  \hline\hline
  $[q^2_{\rm min}, q^2_{\rm max}]$ (GeV$^2$)& $10^3\, (R_{\pi^+}^{\mu e}-1)$     &  $R_{\pi^+}^{\mu \tau}$ & $10^3\, (R_{\pi^0}^{\mu e}-1)$  & $R_{\pi^0}^{\mu \tau}$     \\
  \hline   
$[0.10,   2.00]$ & $-$5.81(0.62 0.07)  &                     &  $-$4.70(0.50 0.02) & \\
$[2.00,   4.00]$ & $-$1.66(0.46 0.06)  &                     &  $-$1.49(0.41 0.05) & \\
$[4.00,   6.00]$ & $-$1.38(0.42 0.05)  &                     &  $-$1.33(0.40 0.04) & \\
$[6.00,   8.00]$ & $-$1.14(0.39 0.04)  &                     &  $-$1.13(0.39 0.04) & \\
$[15.00, 17.00]$ & \mph0.14(0.64 0.01) & \mph0.52(0.08 0.00) & \mph0.11(0.63 0.00) & \mph0.54(0.08 0.00) \\
$[17.00, 19.00]$ & \mph0.58(0.82 0.02) & \mph0.21(0.06 0.00) & \mph0.54(0.82 0.02) & \mph0.22(0.06 0.00) \\
$[19.00, 22.00]$ & \mph1.38(1.21 0.05) &  $-$0.05(0.04 0.01) & \mph1.33(1.21 0.05) &  $-$0.04(0.04 0.01) \\
\hline
$[1.00, 6.00]$ & $-$1.64(0.45 0.06)  &                  & $-$1.45(0.40 0.05) & \\
$[15.00, 22.00]$ & 0.72(0.90 0.02)   & 0.18(0.06 0.01)  & 0.68(0.90 0.02)& 0.19(0.06 0.01) \\
  \hline
  \hline
  \end{tabular}
\end{table}

\begin{table} 
    \caption{
        Standard-Model lepton-universality-violating ratios for $\BtoK\ell^+\ell^-$ decay.
        Results are shown for both the charged ($R_{K^+}$, left) and neutral ($R_{K^0}$, right) modes.
        Errors shown are from the form factors and the quadrature sum of all other contributions, 
        respectively.} 
  \label{tab:B2P_RK}
  \begin{tabular}{l@{\hskip 3mm}c@{\hskip 3mm}c@{\hskip 3mm}c@{\hskip 3mm}c@{\hskip 3mm}c@{\hskip 3mm}c}
  \hline\hline
  $[q^2_{\rm min}, q^2_{\rm max}]$ (GeV$^2$)& $10^3\, (R_{K^+}^{\mu e}-1)$     &  $R_{K^+}^{\mu \tau}$ & $10^3\, (R_{K^0}^{\mu e}-1)$  & $R_{K^0}^{\mu \tau}$     \\
  \hline   
$[0.10, 2.00]$   &  $-$3.30(0.20 0.02) &                     &  $-$4.27(0.22 0.03) & \\
$[2.00, 4.00]$   & \mph0.50(0.38 0.02) &                     & \mph0.44(0.39 0.02) & \\
$[4.00, 6.00]$   & \mph0.62(0.59 0.02) &                     & \mph0.59(0.59 0.02) & \\
$[6.00, 8.00]$   & \mph0.72(0.86 0.02) &                     & \mph0.70(0.85 0.02) & \\
$[15.00, 17.00]$ & \mph1.79(3.20 0.06) & \mph0.16(0.02 0.01) & \mph1.78(3.20 0.06) & \mph0.16(0.02 0.01) \\
$[17.00, 19.00]$ & \mph2.55(4.23 0.09) &  $-$0.11(0.02 0.01) & \mph2.56(4.23 0.09) &  $-$0.11(0.02 0.01) \\
$[19.00, 22.00]$ & \mph5.08(5.95 0.19) &  $-$0.40(0.01 0.01) & \mph5.13(5.94 0.19) &  $-$0.41(0.01 0.01) \\
\hline
$[1.00, 6.00]$   & 0.50(0.43 0.02)  &                  & 0.43(0.44 0.02) & \\
$[15.00, 22.00]$ & 2.83(4.20 0.10)  & $-$0.13(0.02 0.01) & 2.84(4.19 0.10) & $-$0.13(0.02 0.01) \\
  \hline
  \hline
  \end{tabular}
\end{table}


\begin{table} 
    \caption{
        Standard-Model ratio of partially integrated branching ratios
        $\Delta\BR(\Btopi\ell^+\ell^-)$/$\Delta\BR(\BtoK \ell^+\ell^-)$.
        Errors shown are from the CKM elements, form factors, and the quadrature sum of all other 
        contributions, respectively.}
	\label{tab:dBdq2-ratio}
	\begin{tabular}{l@{\hskip 3mm}c@{\hskip 3mm}c@{\hskip 3mm}c}
		\hline\hline
		$[q^2_{\rm min}, q^2_{\rm max}]$ (GeV$^2$)       &  $10^3\, \frac{\Delta\BR(B^+\to \pi^+\mu^+\mu^-)}{\Delta\BR(B^+ \to K^+ \mu^+\mu^-)}$  & $10^3\, \frac{\Delta\BR(B^0\to \pi^0\mu^+\mu^-)}{\Delta\BR(B^0 \to K^0 \mu^+\mu^-)}$        \\
		\hline        
$[0.10, 2.00]$ & 26.63(0.76,6.31,0.42) & 13.07(0.37,3.12,0.20) \\
$[2.0, 4.0]$   & 26.73(0.76,5.48,0.37) & 13.04(0.37,2.68,0.17) \\
$[4.0, 6.0]$   & 27.01(0.77,4.74,0.35) & 13.20(0.37,2.32,0.16) \\
$[6.0, 8.0]$   & 27.47(0.78,4.11,0.34) & 13.46(0.38,2.02,0.16) \\
$[15.0, 17.0]$ & 36.64(1.04,2.95,1.05) & 18.18(0.52,1.47,0.52) \\
$[17.0, 19.0]$ & 43.37(1.23,3.15,1.24) & 21.63(0.61,1.58,0.62) \\
$[19.0, 22.0]$ & 71.37(2.02,4.69,2.03) & 36.11(1.02,2.38,1.03) \\
\hline
$[1.0, 6.0]$   & 26.82(0.76,5.30,0.37) & 13.09(0.37,2.60,0.17) \\
$[15.0, 22.0]$ & 47.21(1.34,3.38,1.34) & 23.59(0.67,1.70,0.67) \\
		\hline
		\hline
	\end{tabular}
\end{table}

\begin{table} 
    \caption{
        Correlations between the form-factor contributions to the errors in the Standard-Model ratio of
        partially integrated branching ratios $\Delta\BR(\Btopi\ell^+\ell^-)$/$\Delta\BR(\BtoK\ell^+\ell^-)$.
        These should be combined with the central values and form-factor errors in the bottom panel of
        Table~\ref{tab:dBdq2-ratio} above.
        The results for the ratio of neutral decay modes should be taken as 100\% correlated with those for 
        the charged decays.}
    \label{tab:B2P+_ratio_dBdq2_FF_correlations}
  \begin{tabular}{l@{\hskip 3mm}c@{\hskip 3mm}c@{\hskip 3mm}c@{\hskip 3mm}c@{\hskip 3mm}c}
  \hline\hline
       $[q^2_{\rm min}, q^2_{\rm max}]$ (GeV$^2$) & $[1, 6]$ &   $[15, 22]$  \\
  \hline
$[1, 6]$   & 1.0000 & 0.4905 \\
$[15, 22]$ & 0.4905 & 1.0000 \\
  \hline\hline
  \end{tabular}
\end{table}

\begin{table} 
    \caption{
        Correlations between the ``other" contributions to the errors in the Standard-Model ratio of 
        partially integrated branching ratios $\Delta\BR(\Btopi\ell^+\ell^-)$/$\Delta\BR(\BtoK 
        \ell^+\ell^-)$.
        These should be combined with the central values and ``other" errors in the bottom panel of 
        Table~\ref{tab:dBdq2-ratio} above.
        The CKM errors are 100\% correlated between the two bins.
        The results for the ratio of neutral decay modes should be taken as 100\% correlated 
        with those for the charged decays.}
  \label{tab:B2P+_ratio_dBdq2_Other_correlations}
  \begin{tabular}{l@{\hskip 3mm}c@{\hskip 3mm}c@{\hskip 3mm}c@{\hskip 3mm}c@{\hskip 3mm}c}
  	\hline\hline
  	$[q^2_{\rm min}, q^2_{\rm max}]$ (GeV$^2$) & $[1, 6]$ &   $[15, 22]$  \\
  	\hline
  	$[1, 6]$   & 1.0000 & 0.0917 \\
  	$[15, 22]$ & 0.0917 & 1.0000 \\
  	\hline\hline
  \end{tabular}	
\end{table}

\clearpage
\section{\texorpdfstring{\boldmath$B\to K (\pi) \ell^+ \ell^-$}{B to K(pi)ll} differential decay rates}
\label{app:Formulae}

Here we summarize the theoretical expressions for the $B\to K (\pi) \ell^+\ell^-$ differential decay rates
in the Standard Model, including the complete dependence on the charged-lepton mass $m_\ell$.
We encourage the users of these formulae to cite explicitly the original papers~\cite{Beneke:2000wa,
Asatrian:2001de, Beneke:2001at, Asatryan:2001zw, Asatrian:2003vq, Seidel:2004jh, Grinstein:2004vb,
Beneke:2004dp, Greub:2008cy, Bobeth:2010wg, Beylich:2011aq, Bobeth:2011gi, Bobeth:2011nj, Bobeth:2012vn},
in which the results collected below were first derived.
In Appendix~\ref{app:main} we present a complete set of expressions needed to describe the high-$q^2$
region.
The discussion of the running of the tensor form factor $f_T$ surrounding Eq.~(\ref{fTrunning}) has
not been discussed in reference to exclusive $b\to s\ell\ell$ decays elsewhere in the literature.
The additional nonfactorizable terms of the type $\phi_B \star T \star \Phi_P$ required at low-$q^2$ are
collected in Appendix~\ref{app:lowq2}.
For the $\ell=\tau$ case, the lower boundary of kinematic range, $q^2_{\rm min} = 4 m_\tau^2$, is larger
than the $\psi^\prime$ mass implying that the high-$q^2$ OPE is sufficient to completely describe this mode.
Relations between the form factors $f_T$ and $f_+$ valid at low and high-$q^2$ are presented in
Appendix~\ref{app:ffrelations}.
A discussion of scale and power-correction uncertainties is given in Appendix~\ref{app:errors}.

\subsection{Main formulas}
\label{app:main}

The double differential $B\to P \ell\ell \; (P = K, \pi \, ; \, \ell = e, \mu, \tau)$ decay rate is given by
(see, for instance, Ref.~\cite{Bobeth:2011nj})
\begin{align}
    \frac{d^2\Gamma}{dq^2\,d\cos\theta} &= a + b \cos\theta + c \cos^2\theta ,
    \label{eq:dGdqdu} \\
    a &= \Gamma_0 \lambda_0^{1/2} \beta_\ell \left[\frac{\lambda_0}{4} |G|^2 + 
        |C_{10}|^2 \left(\frac{\lambda_0}{4}\beta_l^2 |f_+|^2
        +  \frac{m_\ell^2}{q^2}(M_B^2-M_P^2)^2 |f_0|^2 \right) \right]
    \label{eq:aSM} \\
    b &=  0 , \\
    c &= -\frac{1}{4} \Gamma_0 \lambda_0^{3/2} \beta_\ell^3 \left( |G|^2 + |C_{10}f_+|^2 \right),
    \label{eq:cSM} \\
    G &= C_9^\text{eff} f_+ + \frac{2m_b^{\MSbar}(\mu)}{M_B+M_P} C_7^\text{eff} f_T,
    \label{eq:fv} \\ 
    \Gamma_0 &=  C_P \frac{G_F^2 \alpha_e^2 \left|V_{tb}^{} V_{tq}^*\right|^2}{512 \pi^5 M_B^3},
    \label{eq:Gamma0}  \\
    \lambda_0 &=  4M_B^2|\bm{p}_P|^2 = M_B^4 +M_P^4 + q^4 -2 (M_B^2 M_P^2 +M_B^2 q^2 +M_P^2 q^2), 
    \label{eq:lambda0} \\
    \beta_\ell &=  \sqrt{1-\frac{4 m_\ell^2}{q^2}} ,
\end{align}
where $\theta$ is the angle between the negative lepton direction and the $B$ incoming direction in the
dilepton center of mass frame, and $|\bm{p}_P| =\sqrt{E_P^2-M_P^2}$ is the three-momentum of the final-state
meson in the $B$-meson rest frame.
The isospin factor in Eq.~(\ref{eq:Gamma0}) $C_P = 1$ for decays to kaons and charged pions ($\pi^{\pm}$),
while $C_P = 1/2$ for decays to neutral pions ($\pi^0$).
When $m_\ell=0$, $c=-a$.

The form factor $f_+$ is scale independent while the scale dependence of $f_T$ is simply
controlled by the anomalous dimension of the tensor current:
\begin{align}
    f_T (\mu_2) &= f_T (\mu_1)
        \left( \frac{\alpha_s(\mu_2)}{\alpha_s(\mu_1)} \right)^{-\gamma^{(0)}_T/2\beta_0},
\label{fTrunning}
\end{align}
where $\beta_0 = (11 N_c - 2 N_f)/3 = 23/3$ (with $N_f=5$ active flavors) is the leading order QCD beta
function, and the leading order anomalous dimension of the tensor current is given by the anomalous
dimension of the operator $Q_7$ minus the contribution of the explicit bottom mass that appears in $Q_7$,
$\gamma_T^{(0)} = \gamma_7^{(0)} - \gamma_m^{(0)} = 8 C_F - 6 C_F = 2 C_F = 8/3$ (see, for instance,
Ref.~\cite{Buras:1998raa}).

We obtain the expressions for the effective Wilson coefficients $C^{\rm eff}_7$ and $C_9^{\rm eff}$ starting
from Eq.~(47) of Ref.~\cite{Asatrian:2003vq} where the notation $\tilde C_{7,9}^{\rm eff}$ is used; we also
use results from Refs.~\cite{Beneke:2001at, Greub:2008cy, Bobeth:2011gi}.
We have made a number of changes in notation, however, and also removed some terms, as described below.
Reference~\cite{Asatrian:2003vq} uses the notation $\xi_i=V_{ib} V_{id}^*$ with $i=u,c,t$, which can be
generalized to replace $d$ by $q$ so that the final state quark can be either $d$ or~$s$.
In either case, third-column and $q^\text{th}$-column unitarity implies $\xi_u +\xi_c +\xi_t =0$, or
equivalently, $\xi_c/\xi_t = - 1 - \xi_u/\xi_t = -1 - \lambda_u^{(q)}$, defining $\lambda_u^{(q)} = V_{uq}^*
V_{ub}^{}/(V_{tq}^* V_{tb}^{})$.
Also, the extra factor $\alpha_s/(4\pi)$ in the definition of the operators $Q_{7,9}$ has to be taken into
account by replacing $4\pi/\alpha_s C_{7,9} \to C_{7,9}$ in Eq.~(48) of Ref.~\cite{Asatrian:2003vq}.
Our notation simplifies that of Ref.~\cite{Asatrian:2003vq} by not being explicit about the $\mu$ dependence
of the $C_i$.
We set $\omega_{7,9} \to 0$ to remove bremsstrahlung contributions.
We absorb the terms proportional to $\ln(m_b/\mu)$ in Eq.~(48) of Ref.~\cite{Asatrian:2003vq} into the
definitions of the functions $h(m,q^2)$ that we adopt here in Eqs.~(\ref{hmc})-(\ref{hmb}), so that they are
scale dependent.
We use the exact relation $F_{2,c}^{(7)}= -6 F_{1,c}^{(7)}$ to simplify Eq.~(\ref{c7eff}).
Finally, we keep only the first term for $A_8$ in Eq.~(48) of Ref~\cite{Asatrian:2003vq} since the other
terms are higher order, and take $A_8 =C_8$ because of different operator normalization.

With the above choices, our expressions for $C^{\rm eff}_7$ and $C_9^{\rm eff}$ become
\begin{align}
  C^{\rm eff}_7 = \; & 
  C_{7} - 
  \frac{1}{3} \left[ C_{3} + \frac{4}{3}\,C_{4} + 20\,C_{5} 
    + \frac{80}{3} C_{6} \right]
    - \frac{\alpha_s}{4 \pi} \left[ \left(C_{1} - 6\,C_{2}\right) F_{1,c}^{(7)} + C_{8} F_8^{(7)}\right]
    \nonumber \\
& -\frac{\alpha_s}{4\pi} \lambda_u^{(q)}  (C_1 - 6 C_2) \left( F_{1,c}^{(7)} - F_{1,u}^{(7)} \right) ,
\label{c7eff} \\
  C_9^{\rm eff} = \; &    
    C_{9} + \frac{4}{3} C_3 + \frac{64}{9} C_5 + \frac{64}{27} C_6 
    + h(0,q^2) \left( -\frac{1}{2} C_3 - \frac{2}{3} C_4 - 8 C_5 - \frac{32}{3} C_6 \right)
 \nonumber  \\
 & + h(m_b,q^2) \left( -\frac{7}{2} C_3 - \frac{2}{3} C_4 - 38 C_5 - \frac{32}{3} C_6 \right)  \nonumber \\
&+h(m_c,q^2) \left( \frac{4}{3} C_1 + C_2 + 6 C_3 + 60 C_5 \right) 
 + \lambda_u^{(q)}  \Big[ h(m_c,q^2)-h(0,q^2) \Big] \left( \frac{4}{3} C_1 + C_2 \right) \nonumber \\ 
&  - \frac{\alpha_s}{4 \pi} \left[ C_1 F_{1,c}^{(9)} + C_2 F_{2,c}^{(9)} + C_{8} F_8^{(9)}\right] \nonumber \\
& -\frac{\alpha_s}{4\pi}  \lambda_u^{(q)} \left[ C_1 \left( F_{1,c}^{(9)} - F_{1,u}^{(9)} \right) 
+C_2 \left( F_{2,c}^{(9)} - F_{2,u}^{(9)} \right) \right] . \label{c9eff} 
\end{align}
Numerically, the CKM factor $|\lambda_u^{(s)}| \simeq 0.02 \ll |\lambda_u^{(d)}| \simeq 0.4$; thus terms
proportional to $\lambda_u^{(q)}$ are significant only for the $B\to \pi$ mode.
Reference~\cite{Greub:2008cy} provides a \textsl{Mathematica} notebook for the functions $F_{1,2}^{(7,9)}$ 
in the charm-pole-mass scheme.
Our expressions for $C^{\rm eff}_7$ and $C^{\rm eff}_9$ are similar to Eqs.~(2.6) and (2.7) in
Ref.~\cite{Bobeth:2011gi}; however, we have included terms proportional to $\lambda_u^{(q)}$ for $C^{\rm
eff}_7$, and for $C^{\rm eff}_9$ we have not expanded $h(m_c,q^2)$ and $F_{i,c}^{(7,9)}$ in powers of
$m_c^2/q^2$ as in Eq.~(2.7) of Ref.~\cite{Bobeth:2011gi}.

The function $h(m_q,q^2)$ is given in Eq.~(11) of Ref.~\cite{Beneke:2001at}.
The function $h(m_q,q^2)$ for $m_q = 0, m_b$ is also given explicitly in Eqs.~(3.11) and~(3.12) of
Ref.~\cite{Bobeth:2010wg}.
The functions $F_8^{(7,9)}$, $B_0$, and $C_0$ are given in Eqs.~(B.1)--(B.3), and (29) of
Ref.~\cite{Beneke:2001at}.
The functions $F_{1,u}^{(7,9)}$, $F_{2,u}^{(7,9)}$ are given in Eqs.~(22)--(31) of Ref.~\cite{Seidel:2004jh}
but with an extra minus sign compared with the convention adopted here and in
Refs.~\cite{Asatryan:2001zw, Asatrian:2003vq, Greub:2008cy}).
For the convenience of the reader, we present the explicit expressions for all required functions here:
\begin{align}
h(m_c,q^2) =\;  & \frac{4}{9} \left(\ln \frac{\mu^2}{m_c^2} + \frac{2}{3} + z\right) \nonumber \\
 & -\frac{4}{9} (2+z) \sqrt{|z-1|} 
	\begin{cases}
		\arctan \frac{1}{\sqrt{z-1}} & z=\; \frac{4m_c^2}{q^2} > 1 \cr
		\ln \frac{1+ \sqrt{1-z}}{\sqrt{z}} -\frac{i \pi}{2} & z =\; \frac{4m_c^2}{q^2}\leq 1 \cr
	\end{cases} , \label{hmc}\\
h(0,q^2) =\; &  \frac{8}{27} + \frac{4}{9} \left(\ln \frac{\mu^2}{q^2} + i\; \pi\right) , \label{h0}\\
h(m_b,q^2) =\; & \frac{4}{9} \left(\ln \frac{\mu^2}{m_b^2} + \frac{2}{3} + z\right)
-\frac{4}{9} (2+z) \sqrt{z-1} \arctan \frac{1}{\sqrt{z-1}} \;, z=\frac{4m_b^2}{q^2} , \label{hmb}\\
F_8^{(7)}  =\; &  -\frac{32}{9} \, \ln \frac{\mu}{m_b} 
  - \frac 89 \, \frac{\hat s}{1- \hat s} \, \ln \hat s- \frac 89 \, i\pi 
    - \frac 49 \, \frac{ 11 - 16 \hat s + 8 \hat s^2 }{(1-\hat s)^2} \nonumber \\
 & + \frac 49 \, \frac{1}{(1-\hat s)^3} \,
   \left[ 
     (9 \hat s - 5 \hat s^2 + 2 \hat s^3) \, B_0(\hat s)
    - (4 + 2 \hat s) \, C_0(\hat s) \right],\label{F87} \\
F_8^{(9)} =\; & 
    \frac {16}{9} \, \frac{1}{1- \hat s}\, \ln \hat s
             + \frac{8}{9}  \, \frac{5 - 2 \hat s}{(1-\hat s)^2}
 -
    \frac 89 \, \frac{4 - \hat s}{(1-\hat s)^3} 
   \left[  (1 + \hat s) \, B_0(\hat s)
         - 2 \, C_0(\hat s)
   \right], \label{F89}\\
B_0(\hat s)  =\; & -2\,\sqrt{4 /\hat s-1}\,\arctan\frac{1}{\sqrt{4/\hat s-1}}  ,     \\
C_0(\hat s)  =\; & \int_0^1 dx \, \frac{1}{x \, (1-\hat s) + 1} \, 
                 \ln \frac{x^2}{1- x \, (1-x) \, \hat s},  \\
F_{1,u}^{(7)} =\;  &   F_{1,c}^{(7)}|_{m_c\to 0} = -A(\hat s) , \\
F_{2,u}^{(7)} =\;  &   F_{2,c}^{(7)}|_{m_c\to 0} = -6 F_{1,c}^{(7)}|_{m_c\to 0} = 6 A(\hat s) , \\ 
F_{1,u}^{(9)} =\;  &   F_{1,c}^{(9)}|_{m_c\to 0} = -B(\hat s) - 4 C(\hat s) , \\
F_{2,u}^{(9)} =\;  &  F_{2,c}^{(9)}|_{m_c\to 0} = -3 C(\hat s) + 6 B(\hat s) , \\ 
A(\hat s)  =\; &   
-\frac{104}{243}\,\ln(\frac{m_b^2}{\mu^2})
+\frac{4\sh}{27(1-\sh)}\,
  \Big[{\rm Li}_2(\sh)
  +\ln(\sh)\ln(1-\sh)\Big]\CR
&\hspace{-.8cm}+\frac{1}{729(1-\sh)^2}\,
  \Big[6\sh\Big(29-47\sh\Big)\ln(\sh)
  +785-1600\sh+833\sh^2
  +6\pi i\Big(20-49\sh+47\sh^2\Big)\Big]\CR
&\hspace{-.8cm}-\frac{2}{243(1-\sh)^3}\,
  \Big[2\sqrt{z-1}\Big(-4+9\sh-15\sh^2+4\sh^3\Big){\rm arccot}(\sqrt{z-1})
  +9\sh^3\ln^2(\sh)\CR
  &+18\pi i \sh\Big(1-2\sh\Big)\ln(\sh)\Big]\CR
&\hspace{-.8cm}+\frac{2\sh}{243(1-\sh)^4}\,
  \Big[36\,{\rm arccot}^2(\sqrt{z-1})
  +\pi^2\Big(-4+9\sh-9\sh^2+3\sh^3\Big)\Big] , \\
B(\hat s) = \; &    
\frac{8}{243\sh}\Big[
  (4-34\sh-17\pi i\sh)\ln(\frac{m_b^2}{\mu^2})
  +8\sh\ln^2(\frac{m_b^2}{\mu^2})
  +17\sh\ln(\sh)\ln(\frac{m_b^2}{\mu^2})\Big]\CR
&+\frac{(2+\sh)\sqrt{z-1}}{729\sh}\,
  \Big[-48\ln(\frac{m_b^2}{\mu^2})\,{\rm arccot}(\sqrt{z-1})
  -18\pi\ln(z-1)
  +3i\ln^2(z-1)\CR
  &-24i\,{\rm Li}_2(-x_2/x_1)
  -5\pi^2 i
  +6i\Big(-9\ln^2(x_1)+\ln^2(x_2)-2\ln^2(x_4)\CR
    &\hspace{1cm}+6\ln(x_1)\ln(x_2)-4\ln(x_1)\ln(x_3)
    +8\ln(x_1)\ln(x_4)\Big)\CR
  &-12\pi\Big(2\ln(x_1)+\ln(x_3)+\ln(x_4)\Big)\Big]\CR
&\hspace{-.8cm}-\frac{2}{243\sh(1-\sh)}\,
  \Big[4\sh\Big(-8+17\sh\Big)\Big({\rm Li}_2(\sh)+\ln(\sh)\ln(1-\sh)\Big)\CR
  &+3\Big(2+\sh\Big)\Big(3-\sh\Big)\ln^2(x_2/x_1)
  +12\pi\Big(-6-\sh+\sh^2\Big){\rm arccot}(\sqrt{z-1})\Big]\CR
&\hspace{-.8cm}+\frac{2}{2187\sh(1-\sh)^2}\,
  \Big[-18\sh\Big(120-211\sh+73\sh^2\Big)\ln(\sh)\CR
  &-288-8\sh+934\sh^2-692\sh^3
  +18\pi i\sh\Big(82-173\sh+73\sh^2\Big)\Big]\CR
&\hspace{-.8cm}-\frac{4}{243\sh(1-\sh)^3}\,
  \Big[-2\sqrt{z-1}\Big(4-3\sh-18\sh^2+16\sh^3-5\sh^4\Big){\rm arccot}(\sqrt{z-1})\CR
  &-9\sh^3\ln^2(\sh)
  +2\pi i\sh\Big(8-33\sh+51\sh^2-17\sh^3\Big)\ln(\sh)\Big]\CR
&\hspace{-.8cm}+\frac{2}{729\sh(1-\sh)^4}\,
  \Big[72\Big(3-8\sh+2\sh^2\Big){\rm arccot}^2(\sqrt{z-1})\CR
  &-\pi^2\Big(54-53\sh-286\sh^2+612\sh^3-446\sh^4+113\sh^5\Big)\Big] , \\
C(\hat s)  = \; &   -\frac{16}{81}\,\ln(\hat s \frac{m_b^2}{\mu^2})
+\frac{428}{243}-\frac{64}{27}\,\zeta(3)+\frac{16}{81}\,\pi i , 
\end{align}
where $\hat s = q^2/m_b^2$, $x_1 = (1 + i \sqrt{z-1})/2$, $x_2 = (1- i \sqrt{z-1})/2$, $x_3 = (1+
i/\sqrt{z-1})/2$, $x_4 = (1- i/ \sqrt{z-1})/2$ and $z=4/\hat s$.
Note that our functions $F_i^{(9)}$ follow the conventions used in Ref.~\cite{Greub:2008cy}.
All charm- and bottom-quark masses that appear in these formulae are in the pole scheme, with the exception
of the explicit instance of the bottom \MSbar\ mass in Eq.~(\ref{eq:fv}).

\subsection{Additional nonfactorizable contributions required at \boldmath low~\texorpdfstring{$q^2$}{q-squared}}
\label{app:lowq2}
The additional contributions of the form $\phi_B \star T \star \phi_P$ that arise in the SCET expansion at
low $q^2$ have been calculated in the $m_\ell = 0$ limit in Ref.~\cite{Beneke:2001at}.
Some of these terms are truly nonfactorizable effects while others appear when $f_T$ is expressed in terms
of $f_+$ [see Eq.~(\ref{fTscet}) below]; the latter have to be removed because, in this work, we use
the tensor form factor $f_T$ computed directly from lattice QCD.

Following the notation of Ref.~\cite{Bobeth:2007dw}, all of these terms are included in the quantity
$\tau_P$ defined in Eq.~(B.1) of that work.
Compared to that work,%
\footnote{Note that in Eq.~(B.2) of Ref.~\cite{Bobeth:2007dw} there is a typo in the definition of 
$C_P^{(\text{f})}$ and its overall sign has to be reversed.} %
we set the terms $C_P^{(0,{\rm f},{\rm nf})}$ and $T_{P,\pm}^{({\rm f})}$ to zero in our formulae (where the
superscripts ``f" and ``nf" denote factorizable and nonfactorizable, respectively).
Some of these terms correspond to contributions that we have already included in $C_{7,9}^{\rm eff}$.
Others appear in the SCET expansion of $f_T$, which we do not use here because we have the tensor form
factor from lattice QCD, or in the perturbative expansion of the \MSbar\ $b$-quark mass in terms of the
potential subtracted one, which is not relevant because we adopt the \MSbar\ scheme.

The remaining contributions $T_{P,\pm}^{(0,{\rm nf})}$ are genuine nonfactorizable corrections.
From the expression for $F_V=G/f_+$ given in Eq.~(3.2) of Ref.~\cite{Bobeth:2007dw}, we see that
these terms all can be included in a shift in the effective coefficient $C_9^{\rm eff}$:
\begin{align}
\Delta C_9^{\rm eff}  =\; & \frac{2 m_b}{M_B} \frac{\Delta \tau_P}{f_+} , \label{deltac9} \\
\Delta \tau_P  =\; & \frac{\pi^2}{\Nc} \frac{f_B f_P}{M_B} \sum_\pm \int \frac{d\omega}{\omega}
      \Phi_{B,\pm}(\omega) \int_0^1 du\, \Phi_P(u) 
      \left[ T_{P,\pm}^{(0)} + \tilde \alpha_s C_F \;  T_{P,\pm}^{(\rm nf)} \right] . \label{eq:DeltaTauP}
\end{align}
The functions $T_{P,\pm}^{(0,{\rm nf})} = - T_{\parallel,\pm}^{(0,{\rm nf})}$ are given in Eqs.~(17)
and~(18), (25) and~(26), and (28)--(32) of Ref.~\cite{Beneke:2001at} and read
\begin{align}
T_{P,+}^{(0)}   = \;  & 0 , \\
T_{P,-}^{(0)}  =\;  & e_q \frac{M_B \omega}{M_B \omega -q^2 - i \epsilon } \frac{4 M_B}{m_b} \left(
C_3 + \frac{4}{3} C_4 + 16 C_5 + \frac{64}{3} C_6 \right) , \label{TP-0}\\
T^\text{(nf)}_{P,\,+} = \;    & -\frac{M_B}{m_b}\,
\Big[e_u t_\parallel(u,m_c) \, 
(-C_1/6 + C_2 + 6 C_6) \nonumber \\
& + e_d \,t_\parallel(u,m_b)\, (C_3 - C_4/6 + 16 C_5 + 10 C_6/3),
\nonumber \\ 
&
 + \,e_d \,t_\parallel(u,0) \,(C_3 - C_4/6 + 16 C_5 - 8 C_6/3)\Big]
\\
t_\parallel(u,m_q) =\;   & \frac{2 M_B}{\bar{u} E}\,I_1(m_q) + 
\frac{\bar{u} M_B^2+u q^2}{\bar{u}^2 E^2}\,
\left(B_0(\bar{u} M_B^2+u q^2,m_q)-B_0(q^2,m_q)\right) , \\
E = \; & \frac{M_B^2 + M_P^2-q^2}{2 M_B} , \\
I_1(m_q) = \; & 1+\frac{2 m_q^2}{\bar{u} (M_B^2-q^2)}\, \Big[L_1(x_+)+L_1(x_-)-L_1(y_+)-L_1(y_-)\Big], \\
L_1(x) = \; &\ln\frac{x-1}{x}\,\ln(1-x)-\frac{\pi^2}{6}+\mbox{Li}_2\left(
\frac{x}{x-1}\right) , \\
x_\pm = \; &\frac{1}{2}\pm\left(\frac{1}{4}-\frac{m_q^2}{\bar{u} M_B^2+u q^2} \right)^{\!1/2} , \\
y_\pm = \; &\frac{1}{2}\pm\left(\frac{1}{4}-\frac{m_q^2}{q^2}\right)^{\!1/2}, \\
B_0(q^2,m_q) = \; &-2\,\sqrt{4 m_q^2/q^2-1}\,\arctan\frac{1}{\sqrt{4 m_q^2/q^2-1}} , \\
T^{(\rm nf)}_{P,\,-} = \; & -e_q\,\frac{M_B\omega}{M_B \omega
-q^2-i\epsilon}\, \bigg[\frac{8 \,C_8^{\,\rm eff}}{\bar{u}+u q^2/M_B^2} 
\nonumber\\
&
+\,\frac{6 M_B}{m_b}\, \Big(h(m_c,\bar{u}M_B^2+u q^2) 
\,(-C_1/6 + C_2 + C_4 + 10 C_6)  \nonumber \\
&+ h(m_b,\bar{u}M_B^2+u q^2)\,
(C_3 + 5C_4/6  + 16 C_5 + 22 C_6/3)
\nonumber\\
& 
+ \,h(0,\bar{u}M_B^2+u q^2)\,
(C_3 + 17C_4/6  + 16 C_5 + 82C_6/3 )
\nonumber\\
& 
-\frac{8}{27}\,(-15 C_4/2 + 12 C_5 - 32 C_6)\Big)\bigg] , \label{TP-nf} \\
C_8^{\rm eff} = \;  & C_8 + C_3 - \frac{1}{6} C_4 + 20 C_5-\frac{10}{3} C_6, 
\end{align}
where $\tilde \alpha_s = \alpha_s/(4\pi)$, $\bar u = 1-u$, $h(m_q,q^2)$ is defined in Eq.~(\ref{hmc}), and
$e_q$, which appears in Eq.~(\ref{TP-0}), is the charge of the spectator quark ({\em i.e.}, $e_q= -1/3$ for
neutral $B$ and $e_q= 2/3$ for $B^\pm$).
The correct imaginary parts are obtained by replacing $m_q^2\to m_q^2 - i \varepsilon$.
Note that in Ref.~\cite{Beneke:2001at} the functions repeated above are given in terms of barred
coefficients $\overline C_i$ that are simple linear combinations of the coefficients $C_i$ (explicit
expressions that relate the two sets of coefficients are given in Appendix~A of Ref.~\cite{Beneke:2001at}).
The nonfactorizable contribution $\Delta\tau_P$ in Eq.~(\ref{eq:DeltaTauP}) depends upon the light-cone
distribution amplitudes (LCDA) of the kaon, $\Phi_K(u)$, of the pion, $\Phi_\pi(u)$, and of the
$B$ meson, $\Phi_{B,\pm} (\omega)$.
It is customary to expand the kaon and pion LCDAs in terms of Gegenbauer polynomials and keep only the
first few terms~\cite{Braun:2015axa} (see Eqs.~(48)--(54) in Ref.~\cite{Beneke:2001at}):
\begin{align}
\Phi_K (u) &= 6 u (1-u) \left[ 1 + a_1^K \; C_1^{(3/2)} (2u-1) + a_2^K \; C_2^{(3/2)} (2u -1) + \cdots 
\right] , \\ 
\Phi_\pi (u) &= 6 u (1-u) \left[ 1 + a_2^\pi \; C_2^{(3/2)} (2u-1) + a_4^\pi \; C_4^{(3/2)} (2u -1) + \cdots \right] .
\end{align}
Note that $a_1^\pi$ vanishes due to $G$-parity.
The $u$ dependence of $\Phi_P (u)$ is needed because the convolutions involve nontrivial functions
of~$u$.
The first few coefficients $a_i^{\pi,K}$ have been computed in lattice
QCD~\cite{Braun:2006dg,Boyle:2006pw,Arthur:2010xf,Braun:2015axa}.
The $B$-meson LCDAs are known less precisely, but enter only through the first inverse moments:
\begin{align}
    \lambda_{B,+}^{-1} &= \int_0^\infty d\omega \frac{\Phi_{B,+} (\omega)}{\omega} ,\\
    \lambda_{B,-}^{-1} (q^2) &= \int_0^\infty d\omega \frac{\Phi_{B,-} (\omega)}{\omega-q^2/M_B-i \epsilon}.
\end{align}
Following Ref.~\cite{Beneke:2001at}, we model $\Phi_{B,+}$ and $\Phi_{B,-}$ as
\begin{align}
    \Phi_{B,+}(\omega) &= \frac{\omega}{\omega_0^2} e^{-\omega/\omega_0} , \\
    \Phi_{B,-}(\omega) &= \frac{1}{\omega_0} e^{-\omega/\omega_0} .
\end{align}
The value of $\omega_0$ can be fixed using $\lambda_{B,+}$, giving
\begin{align}
\omega_0 &=  \lambda_{B,+} , \\
\lambda_{B,-}^{-1} (q^2) &= \frac{e^{-q^2/(M_B \lambda_{B,+})}}{\lambda_{B,+}} 
\left[- \text{Ei}\left(\frac{q^2}{M_B} \lambda_{B,+}\right) + i\; \pi \right] . 
\end{align}

\subsection{Form-factor relations}
\label{app:ffrelations}
We now present details on the relations between the form factors $f_T$ and $f_+$, tested in
Sec.~\ref{subsec:HQSTests}.
At low-$q^2$ SCET gives~\cite{Beneke:2000wa}
\begin{align}
    \frac{M_B}{M_B+M_P} f_T &= f_+ 
        \left[1 + \frac{\alpha_s}{4\pi}C_F\left(\ln\frac{m_b^2}{\mu^2} + 2L\right) \right]
    \nonumber \\
    &
    -\frac{\pi}{\Nc} \frac{f_Bf_P}{E} \alpha_s C_F \int \frac{d\omega}{\omega} \Phi_{B,+}(\omega)
        \int_0^1 \frac{du}{\bar u}\, \Phi_P(u) ,
    \label{fTscet}
\end{align}
where $L =-[2 E/(M_B - 2 E)] \ln(2E/M_B)$.
The extension of this relation to order $\alpha_s^2$ is also available~\cite{Bell:2010mg}.

At high-$q^2$ the corresponding relation obtained from the high-$q^2$ OPE reads~\cite{Grinstein:2004vb,
Bobeth:2011nj}
\begin{align}
\frac{M_B}{M_B+M_P}  f_T  = \; & \frac{M_B^2}{q^2} f_+ \kappa(\mu) =
\frac{M_B^2}{q^2} f_+ \left( 1 + 2 \frac{D_0^{(v)}}{C_0^{(v)}} \right)
\frac{m_b}{M_B} , \label{eq:kappa}\\
D_0^{(v)} = \; &  \frac{\alpha_s}{4\pi} C_F \left( 2 \ln \frac{\mu}{m_b} +2 \right) , \\
C_0^{(v)} = \; &  1- \frac{\alpha_s}{4\pi} C_F \left( 3 \ln \frac{\mu}{m_b} +4 \right).
\end{align}
Note that we do not use these expressions, except to test them, because we take the form factors directly
from lattice QCD.

\subsection{Scale and power-correction uncertainties}
\label{app:errors}
Scale uncertainties are intended to account for errors introduced by truncating a perturbative expansion,
and should reflect the size of omitted higher-order perturbative corrections.
The standard approach for estimating such missing terms is to vary the unphysical scales: in our case we
vary the two scales $\mu_b \sim m_b$ and $\mu_0 \sim m_W,m_t$.
Higher-order corrections will cancel exactly the explicit dependence in the expressions for the branching
ratios that we use, while the residual scale dependence will be suppressed by one more power of the strong
coupling $\alpha_s$.
In this work we adopt the standard choices $\mu_b=5~\text{GeV}$ and $\mu_0=\sqrt{m_Wm_t}=120~\text{GeV}$.
The scale uncertainty is then obtained by varying \emph{simultaneously} these scales by a factor of two and
taking half the difference between the maximum and minimum observed values.

Uncertainties associated with power corrections are more difficult to estimate because higher-order terms in
the OPE are dynamically suppressed; \emph{i.e.}, the suppression is expected to appear in the
nonperturbative calculation of the matrix elements of higher-dimensional operators.
We estimate this uncertainty by varying by 10\% all terms in the amplitude that are not directly
proportional to $C_{9,10}f_+$ or $C_7f_T$.

\section{Numerical inputs}
\label{app:Inputs}

Here we tabulate the numerical inputs used for the Standard-Model predictions in Secs.~\ref{sec:SMPheno}
and~\ref{sec:Implications}.
Table~\ref{tab:WilsonCoeffs} provides the Wilson coefficients, while Table~\ref{tab:inputs} provides the
other inputs.

\begin{table}[ht]
    \caption{Numerical values of the Standard-Model Wilson coefficients used for the calculations in 
        this work, taken from Ref.~\cite{Huber:2005ig}.
        The dominant source of error in the coefficients is from the variation of the scale 
        $\mu_\text{low}\in[2.5,10]~\text{GeV}$.
        The scale dependencies of the coefficients in some cases are somewhat large, but are meant to 
        cancel against the corresponding scale dependence of the matrix elements. }
    \label{tab:WilsonCoeffs}
    \begin{tabular}{l@{\hskip 3mm}r@{\hskip 3mm}r@{\hskip 3mm}r}
    \hline\hline
           $C_i  (\mu_b)$     &  value    \\
    \hline        
        $C_1$    &  $-$0.29(16)   \\
        $C_2$    &   1.009(10)   \\
        $C_3$    &  $-$0.0047(42)   \\
        $C_4$    &  $-$0.081(39)   \\
        $C_5$    &   0.00036(31)  \\
        $C_6$    &   0.00082(97)  \\
        $C_7$    &  $-$0.297(26)  \\
        $C_8$    &  $-$0.152(15)  \\
        $C_9$    &   4.04(33) \\
        $C_{10}$ &  $-$4.292(73) \\
    \hline\hline
    \end{tabular}
\end{table}

\begin{table} 
\centering
\caption{Numerical inputs used in the phenomenological analysis of this paper.
The CKM combinations are obtained using the determinations of the Wolfenstein parameters $\{A, \lambda,
\bar\rho, \bar\eta \} = \left\{ 0.810(^{+18}_{-24}), 0.22548(^{+68}_{-34}), 0.1453(^{+133}_{-73}),
0.343(^{+11}_{-12}) \right\}$ from the CKMfitter group's global analysis including results through CKM
2014~\cite{Charles:2004jd}.
The hadronic parameters (decay constants and light-cone distribution amplitudes) are taken from unquenched
lattice-QCD calculations except for $\lambda_{B,+}$, for which lattice results are unavailable.
Coupling constants, masses, and lifetimes are taken from the PDG~\cite{Agashe:2014kda}
unless otherwise specified.}
\label{tab:inputs}
\begin{displaymath}
\begin{tabular}{l@{\hskip 10mm}l}
\hline 
\hline
\spp 
$\alpha_s (m_Z) = 0.1185(6)$ 
& 
$\alpha_e (m_Z) =  1/127.940(18)$ 
\\  
\spp
$s_W^2 \equiv \sin^2\theta_W = 0.2312{6(5)}$ 
& 
$G_F = 1.1663787{(6)} \times 10^{-5} \;  \text{GeV}^{-2}$
\\
\hline
\spp
$m_W = 80.385{(15)}\;\gev$ 
& 
$m_Z = 91.1876{(21)}\;\gev$ 
\\ 
\spp
$m_e = 510.99892{8(11)} \;\text{keV}$ 
&
$m_\mu = 105.6583{715(35)} \;\mev$ 
\\ 
\spp 
$m_\tau = 1.776{82(16)} \;\gev$ 
& 
$m_{t,\text{pole}}= 173.{21(87)} \;\gev$ 
\\
\spp
$M_{B^\pm} = 5.27926{(17)} \;\gev$ 
&
$M_{B^0} = 5.27958{(17)} \; \gev$
\\
\spp
$\tau_{B^\pm} = 1.638{(4)} \; \text{ps}$ 
&
$\tau_{B^0} = 1.519{(5)} \; \text{ps}$
\\
\spp
$M_{\pi^\pm} = 139.57018{(35)} \; \mev$
&
$M_{\pi^0} = 134.9766{(6)} \; \mev$
\\
\spp
$M_{K^\pm} = 493.677{(16)} \; \mev$
&
$M_{K^0} = 497.614{(24)} \; \mev$ 
\\
\spp
$m_b^\text{pole}= 4.{91(12)}\;\gev$~\cite{Ali:2013zfa}
&
$m_c^\text{pole} = 1.77(14) \;\gev$~\cite{Ali:2013zfa}
\\
\spp
$m_b^{\overline{\rm MS}}(m_b)=4.1{8(3)} \; \gev$ 
&
$m_c^{\overline{\rm MS}}(m_c)=1.275(25) \; \gev$ 
\\
\hline
\spp
$|V_{ts}^* V_{tb}^{}| = 4.04(11) \times 10^{-2} $
& 
$|V_{td}^* V_{tb}^{}| = 8.55(26) \times 10^{-3} $ 
\\
\spp 
$|\lambda_u^{(s)}| = 0.01980(62) $
& 
$\arg(\lambda_u^{(s)}) = 114.0(1.9)^\circ $ 
\\ \spp
$|\lambda_u^{(d)}| = 0.404(14) $
& 
$\arg(\lambda_u^{(d)}) = -88.9(2.0)^\circ $ 
\\
\spp
$|V_{td}/V_{ts}| = 0.2115(30)$
& 
\\
\hline
\spp
$f_B = 190.5(4.2)\; \mev$~\cite{Aoki:2013ldr}
&
$f_K = 156.3(9) \; \mev$~\cite{Aoki:2013ldr} 
\\
\spp
$a_1^K = 0.061(4) $~\cite{Arthur:2010xf}
&
$a_2^K = 0.18(7) $~\cite{Arthur:2010xf} 
\\
\spp
$a_2^\pi = 0.23(7)$~\cite{Arthur:2010xf}
&
$\lambda_{B,+} (1.5 \; \gev) = 0.51(12) \; \gev$~\cite{Braun:2003wx,Lee:2005gza,Ball:2006nr}
\\
\hline
\spp
$\mu_b = 5\left(^{+5}_{-2.5}\right) \; \gev$ 
&
$\mu_0 = 120\left(^{+120}_{-60}\right) \; \gev $
\\
\spp
$\mu_f = 2\; \gev$
&
\\
\hline
\hline
\end{tabular}
\end{displaymath}
\end{table}

\clearpage

\bibliography{B2Kpheno}

\begin{thebibliography}{214}%
\makeatletter
\providecommand \@ifxundefined [1]{%
 \@ifx{#1\undefined}
}%
\providecommand \@ifnum [1]{%
 \ifnum #1\expandafter \@firstoftwo
 \else \expandafter \@secondoftwo
 \fi
}%
\providecommand \@ifx [1]{%
 \ifx #1\expandafter \@firstoftwo
 \else \expandafter \@secondoftwo
 \fi
}%
\providecommand \natexlab [1]{#1}%
\providecommand \enquote  [1]{``#1''}%
\providecommand \bibnamefont  [1]{#1}%
\providecommand \bibfnamefont [1]{#1}%
\providecommand \citenamefont [1]{#1}%
\providecommand \href@noop [0]{\@secondoftwo}%
\providecommand \href [0]{\begingroup \@sanitize@url \@href}%
\providecommand \@href[1]{\@@startlink{#1}\@@href}%
\providecommand \@@href[1]{\endgroup#1\@@endlink}%
\providecommand \@sanitize@url [0]{\catcode `\\12\catcode `\$12\catcode
  `\&12\catcode `\#12\catcode `\^12\catcode `\_12\catcode `\%12\relax}%
\providecommand \@@startlink[1]{}%
\providecommand \@@endlink[0]{}%
\providecommand \url  [0]{\begingroup\@sanitize@url \@url }%
\providecommand \@url [1]{\endgroup\@href {#1}{\urlprefix }}%
\providecommand \urlprefix  [0]{URL }%
\providecommand \Eprint [0]{\href }%
\providecommand \doibase [0]{http://dx.doi.org/}%
\providecommand \selectlanguage [0]{\@gobble}%
\providecommand \bibinfo  [0]{\@secondoftwo}%
\providecommand \bibfield  [0]{\@secondoftwo}%
\providecommand \translation [1]{[#1]}%
\providecommand \BibitemOpen [0]{}%
\providecommand \bibitemStop [0]{}%
\providecommand \bibitemNoStop [0]{.\EOS\space}%
\providecommand \EOS [0]{\spacefactor3000\relax}%
\providecommand \BibitemShut  [1]{\csname bibitem#1\endcsname}%
\let\auto@bib@innerbib\@empty
\bibitem [{\citenamefont {Hewett}\ \emph {et~al.}(2012)\citenamefont {Hewett},
  \citenamefont {Weerts} \emph {et~al.}}]{Hewett:2012ns}%
  \BibitemOpen
  \bibfield  {author} {\bibinfo {author} {\bibfnamefont {J.~L.}\ \bibnamefont
  {Hewett}}, \bibinfo {author} {\bibfnamefont {H.}~\bibnamefont {Weerts}},
  \emph {et~al.},\ }\href@noop {} {\emph {\bibinfo {title} {{Fundamental
  Physics at the Intensity Frontier}}}}\ (\bibinfo  {publisher} {{U.S.
  Department of Energy}},\ \bibinfo {year} {2012})\ \Eprint
  {http://arxiv.org/abs/1205.2671} {arXiv:1205.2671 [hep-ex]} \BibitemShut
  {NoStop}%
\bibitem [{\citenamefont {Albrecht}\ \emph {et~al.}(2013)\citenamefont
  {Albrecht} \emph {et~al.}}]{Albrecht:2013wet}%
  \BibitemOpen
  \bibfield  {author} {\bibinfo {author} {\bibfnamefont {J.}~\bibnamefont
  {Albrecht}} \emph {et~al.} (\bibinfo {collaboration} {Charged Leptons Working
  Group}),\ }\href@noop {} {\  (\bibinfo {year} {2013})},\ \Eprint
  {http://arxiv.org/abs/1311.5278} {arXiv:1311.5278 [hep-ex]} \BibitemShut
  {NoStop}%
\bibitem [{\citenamefont {Dawson}\ \emph {et~al.}(2013)\citenamefont {Dawson}
  \emph {et~al.}}]{Dawson:2013bba}%
  \BibitemOpen
  \bibfield  {author} {\bibinfo {author} {\bibfnamefont {S.}~\bibnamefont
  {Dawson}} \emph {et~al.} (\bibinfo {collaboration} {Higgs Boson Working
  Group}),\ }\href@noop {} {\  (\bibinfo {year} {2013})},\ \Eprint
  {http://arxiv.org/abs/1310.8361} {arXiv:1310.8361 [hep-ex]} \BibitemShut
  {NoStop}%
\bibitem [{\citenamefont {Butler}\ \emph {et~al.}(2013)\citenamefont {Butler}
  \emph {et~al.}}]{Butler:2013kdw}%
  \BibitemOpen
  \bibfield  {author} {\bibinfo {author} {\bibfnamefont {J.~N.}\ \bibnamefont
  {Butler}} \emph {et~al.} (\bibinfo {collaboration} {Quark Flavor Physics
  Working Group}),\ }\href@noop {} {\  (\bibinfo {year} {2013})},\ \Eprint
  {http://arxiv.org/abs/1311.1076} {arXiv:1311.1076 [hep-ex]} \BibitemShut
  {NoStop}%
\bibitem [{\citenamefont {Bobeth}\ \emph {et~al.}(2001)\citenamefont {Bobeth},
  \citenamefont {Ewerth}, \citenamefont {{Kr\"uger}},\ and\ \citenamefont
  {Urban}}]{Bobeth:2001sq}%
  \BibitemOpen
  \bibfield  {author} {\bibinfo {author} {\bibfnamefont {C.}~\bibnamefont
  {Bobeth}}, \bibinfo {author} {\bibfnamefont {T.}~\bibnamefont {Ewerth}},
  \bibinfo {author} {\bibfnamefont {F.}~\bibnamefont {{Kr\"uger}}}, \ and\
  \bibinfo {author} {\bibfnamefont {J.}~\bibnamefont {Urban}},\ }\href
  {\doibase 10.1103/PhysRevD.64.074014} {\bibfield  {journal} {\bibinfo
  {journal} {Phys. Rev.}\ }\textbf {\bibinfo {volume} {D64}},\ \bibinfo {pages}
  {074014} (\bibinfo {year} {2001})},\ \Eprint
  {http://arxiv.org/abs/hep-ph/0104284} {arXiv:hep-ph/0104284} \BibitemShut
  {NoStop}%
\bibitem [{\citenamefont {Demir}\ \emph {et~al.}(2002)\citenamefont {Demir},
  \citenamefont {Olive},\ and\ \citenamefont {Voloshin}}]{Demir:2002cj}%
  \BibitemOpen
  \bibfield  {author} {\bibinfo {author} {\bibfnamefont {D.~A.}\ \bibnamefont
  {Demir}}, \bibinfo {author} {\bibfnamefont {K.~A.}\ \bibnamefont {Olive}}, \
  and\ \bibinfo {author} {\bibfnamefont {M.~B.}\ \bibnamefont {Voloshin}},\
  }\href {\doibase 10.1103/PhysRevD.66.034015} {\bibfield  {journal} {\bibinfo
  {journal} {Phys. Rev.}\ }\textbf {\bibinfo {volume} {D66}},\ \bibinfo {pages}
  {034015} (\bibinfo {year} {2002})},\ \Eprint
  {http://arxiv.org/abs/hep-ph/0204119} {arXiv:hep-ph/0204119} \BibitemShut
  {NoStop}%
\bibitem [{\citenamefont {Choudhury}\ and\ \citenamefont
  {Gaur}(2002)}]{Choudhury:2002fk}%
  \BibitemOpen
  \bibfield  {author} {\bibinfo {author} {\bibfnamefont {S.~R.}\ \bibnamefont
  {Choudhury}}\ and\ \bibinfo {author} {\bibfnamefont {N.}~\bibnamefont
  {Gaur}},\ }\href {\doibase 10.1103/PhysRevD.66.094015} {\bibfield  {journal}
  {\bibinfo  {journal} {Phys. Rev.}\ }\textbf {\bibinfo {volume} {D66}},\
  \bibinfo {pages} {094015} (\bibinfo {year} {2002})},\ \Eprint
  {http://arxiv.org/abs/hep-ph/0206128} {arXiv:hep-ph/0206128} \BibitemShut
  {NoStop}%
\bibitem [{\citenamefont {Wang}\ \emph {et~al.}(2008)\citenamefont {Wang},
  \citenamefont {Wang}, \citenamefont {Xu},\ and\ \citenamefont
  {Yang}}]{Wang:2007sp}%
  \BibitemOpen
  \bibfield  {author} {\bibinfo {author} {\bibfnamefont {J.-J.}\ \bibnamefont
  {Wang}}, \bibinfo {author} {\bibfnamefont {R.-M.}\ \bibnamefont {Wang}},
  \bibinfo {author} {\bibfnamefont {Y.-G.}\ \bibnamefont {Xu}}, \ and\ \bibinfo
  {author} {\bibfnamefont {Y.-D.}\ \bibnamefont {Yang}},\ }\href {\doibase
  10.1103/PhysRevD.77.014017} {\bibfield  {journal} {\bibinfo  {journal} {Phys.
  Rev.}\ }\textbf {\bibinfo {volume} {D77}},\ \bibinfo {pages} {014017}
  (\bibinfo {year} {2008})},\ \Eprint {http://arxiv.org/abs/0711.0321}
  {arXiv:0711.0321 [hep-ph]} \BibitemShut {NoStop}%
\bibitem [{\citenamefont {Hiller}\ and\ \citenamefont
  {Schmaltz}(2014)}]{Hiller:2014yaa}%
  \BibitemOpen
  \bibfield  {author} {\bibinfo {author} {\bibfnamefont {G.}~\bibnamefont
  {Hiller}}\ and\ \bibinfo {author} {\bibfnamefont {M.}~\bibnamefont
  {Schmaltz}},\ }\href {\doibase 10.1103/PhysRevD.90.054014} {\bibfield
  {journal} {\bibinfo  {journal} {Phys. Rev.}\ }\textbf {\bibinfo {volume}
  {D90}},\ \bibinfo {pages} {054014} (\bibinfo {year} {2014})},\ \Eprint
  {http://arxiv.org/abs/1408.1627} {arXiv:1408.1627 [hep-ph]} \BibitemShut
  {NoStop}%
\bibitem [{\citenamefont {Gripaios}\ \emph {et~al.}(2015)\citenamefont
  {Gripaios}, \citenamefont {Nardecchia},\ and\ \citenamefont
  {Renner}}]{Gripaios:2014tna}%
  \BibitemOpen
  \bibfield  {author} {\bibinfo {author} {\bibfnamefont {B.}~\bibnamefont
  {Gripaios}}, \bibinfo {author} {\bibfnamefont {M.}~\bibnamefont
  {Nardecchia}}, \ and\ \bibinfo {author} {\bibfnamefont {S.~A.}\ \bibnamefont
  {Renner}},\ }\href {\doibase 10.1007/JHEP05(2015)006} {\bibfield  {journal}
  {\bibinfo  {journal} {JHEP}\ }\textbf {\bibinfo {volume} {05}},\ \bibinfo
  {pages} {006} (\bibinfo {year} {2015})},\ \Eprint
  {http://arxiv.org/abs/1412.1791} {arXiv:1412.1791 [hep-ph]} \BibitemShut
  {NoStop}%
\bibitem [{\citenamefont {Sahoo}\ and\ \citenamefont
  {Mohanta}(2015)}]{Sahoo:2015wya}%
  \BibitemOpen
  \bibfield  {author} {\bibinfo {author} {\bibfnamefont {S.}~\bibnamefont
  {Sahoo}}\ and\ \bibinfo {author} {\bibfnamefont {R.}~\bibnamefont
  {Mohanta}},\ }\href {\doibase 10.1103/PhysRevD.91.094019} {\bibfield
  {journal} {\bibinfo  {journal} {Phys. Rev.}\ }\textbf {\bibinfo {volume}
  {D91}},\ \bibinfo {pages} {094019} (\bibinfo {year} {2015})},\ \Eprint
  {http://arxiv.org/abs/1501.05193} {arXiv:1501.05193 [hep-ph]} \BibitemShut
  {NoStop}%
\bibitem [{\citenamefont {Hou}\ \emph {et~al.}(2013)\citenamefont {Hou},
  \citenamefont {Kohda},\ and\ \citenamefont {Xu}}]{Hou:2013btm}%
  \BibitemOpen
  \bibfield  {author} {\bibinfo {author} {\bibfnamefont {W.-S.}\ \bibnamefont
  {Hou}}, \bibinfo {author} {\bibfnamefont {M.}~\bibnamefont {Kohda}}, \ and\
  \bibinfo {author} {\bibfnamefont {F.}~\bibnamefont {Xu}},\ }\href {\doibase
  10.1103/PhysRevD.87.094005} {\bibfield  {journal} {\bibinfo  {journal} {Phys.
  Rev.}\ }\textbf {\bibinfo {volume} {D87}},\ \bibinfo {pages} {094005}
  (\bibinfo {year} {2013})},\ \Eprint {http://arxiv.org/abs/1302.1471}
  {arXiv:1302.1471 [hep-ph]} \BibitemShut {NoStop}%
\bibitem [{\citenamefont {Gauld}\ \emph
  {et~al.}(2014{\natexlab{a}})\citenamefont {Gauld}, \citenamefont {Goertz},\
  and\ \citenamefont {Haisch}}]{Gauld:2013qba}%
  \BibitemOpen
  \bibfield  {author} {\bibinfo {author} {\bibfnamefont {R.}~\bibnamefont
  {Gauld}}, \bibinfo {author} {\bibfnamefont {F.}~\bibnamefont {Goertz}}, \
  and\ \bibinfo {author} {\bibfnamefont {U.}~\bibnamefont {Haisch}},\ }\href
  {\doibase 10.1103/PhysRevD.89.015005} {\bibfield  {journal} {\bibinfo
  {journal} {Phys. Rev.}\ }\textbf {\bibinfo {volume} {D89}},\ \bibinfo {pages}
  {015005} (\bibinfo {year} {2014}{\natexlab{a}})},\ \Eprint
  {http://arxiv.org/abs/1308.1959} {arXiv:1308.1959 [hep-ph]} \BibitemShut
  {NoStop}%
\bibitem [{\citenamefont {Buras}\ and\ \citenamefont
  {Girrbach}(2013)}]{Buras:2013qja}%
  \BibitemOpen
  \bibfield  {author} {\bibinfo {author} {\bibfnamefont {A.~J.}\ \bibnamefont
  {Buras}}\ and\ \bibinfo {author} {\bibfnamefont {J.}~\bibnamefont
  {Girrbach}},\ }\href {\doibase 10.1007/JHEP12(2013)009} {\bibfield  {journal}
  {\bibinfo  {journal} {JHEP}\ }\textbf {\bibinfo {volume} {1312}},\ \bibinfo
  {pages} {009} (\bibinfo {year} {2013})},\ \Eprint
  {http://arxiv.org/abs/1309.2466} {arXiv:1309.2466 [hep-ph]} \BibitemShut
  {NoStop}%
\bibitem [{\citenamefont {Gauld}\ \emph
  {et~al.}(2014{\natexlab{b}})\citenamefont {Gauld}, \citenamefont {Goertz},\
  and\ \citenamefont {Haisch}}]{Gauld:2013qja}%
  \BibitemOpen
  \bibfield  {author} {\bibinfo {author} {\bibfnamefont {R.}~\bibnamefont
  {Gauld}}, \bibinfo {author} {\bibfnamefont {F.}~\bibnamefont {Goertz}}, \
  and\ \bibinfo {author} {\bibfnamefont {U.}~\bibnamefont {Haisch}},\ }\href
  {\doibase 10.1007/JHEP01(2014)069} {\bibfield  {journal} {\bibinfo  {journal}
  {JHEP}\ }\textbf {\bibinfo {volume} {1401}},\ \bibinfo {pages} {069}
  (\bibinfo {year} {2014}{\natexlab{b}})},\ \Eprint
  {http://arxiv.org/abs/1310.1082} {arXiv:1310.1082 [hep-ph]} \BibitemShut
  {NoStop}%
\bibitem [{\citenamefont {Buras}\ \emph {et~al.}(2014)\citenamefont {Buras},
  \citenamefont {De~Fazio},\ and\ \citenamefont {Girrbach}}]{Buras:2013dea}%
  \BibitemOpen
  \bibfield  {author} {\bibinfo {author} {\bibfnamefont {A.~J.}\ \bibnamefont
  {Buras}}, \bibinfo {author} {\bibfnamefont {F.}~\bibnamefont {De~Fazio}}, \
  and\ \bibinfo {author} {\bibfnamefont {J.}~\bibnamefont {Girrbach}},\ }\href
  {\doibase 10.1007/JHEP02(2014)112} {\bibfield  {journal} {\bibinfo  {journal}
  {JHEP}\ }\textbf {\bibinfo {volume} {1402}},\ \bibinfo {pages} {112}
  (\bibinfo {year} {2014})},\ \Eprint {http://arxiv.org/abs/1311.6729}
  {arXiv:1311.6729 [hep-ph]} \BibitemShut {NoStop}%
\bibitem [{\citenamefont {Altmannshofer}\ \emph {et~al.}(2014)\citenamefont
  {Altmannshofer}, \citenamefont {Gori}, \citenamefont {Pospelov},\ and\
  \citenamefont {Yavin}}]{Altmannshofer:2014cfa}%
  \BibitemOpen
  \bibfield  {author} {\bibinfo {author} {\bibfnamefont {W.}~\bibnamefont
  {Altmannshofer}}, \bibinfo {author} {\bibfnamefont {S.}~\bibnamefont {Gori}},
  \bibinfo {author} {\bibfnamefont {M.}~\bibnamefont {Pospelov}}, \ and\
  \bibinfo {author} {\bibfnamefont {I.}~\bibnamefont {Yavin}},\ }\href
  {\doibase 10.1103/PhysRevD.89.095033} {\bibfield  {journal} {\bibinfo
  {journal} {Phys. Rev.}\ }\textbf {\bibinfo {volume} {D89}},\ \bibinfo {pages}
  {095033} (\bibinfo {year} {2014})},\ \Eprint {http://arxiv.org/abs/1403.1269}
  {arXiv:1403.1269 [hep-ph]} \BibitemShut {NoStop}%
\bibitem [{\citenamefont {Buras}\ \emph {et~al.}(2015)\citenamefont {Buras},
  \citenamefont {Girrbach-Noe}, \citenamefont {Niehoff},\ and\ \citenamefont
  {Straub}}]{Buras:2014fpa}%
  \BibitemOpen
  \bibfield  {author} {\bibinfo {author} {\bibfnamefont {A.~J.}\ \bibnamefont
  {Buras}}, \bibinfo {author} {\bibfnamefont {J.}~\bibnamefont {Girrbach-Noe}},
  \bibinfo {author} {\bibfnamefont {C.}~\bibnamefont {Niehoff}}, \ and\
  \bibinfo {author} {\bibfnamefont {D.~M.}\ \bibnamefont {Straub}},\ }\href
  {\doibase 10.1007/JHEP02(2015)184} {\bibfield  {journal} {\bibinfo  {journal}
  {JHEP}\ }\textbf {\bibinfo {volume} {1502}},\ \bibinfo {pages} {184}
  (\bibinfo {year} {2015})},\ \Eprint {http://arxiv.org/abs/1409.4557}
  {arXiv:1409.4557 [hep-ph]} \BibitemShut {NoStop}%
\bibitem [{\citenamefont {Crivellin}\ \emph
  {et~al.}(2015{\natexlab{a}})\citenamefont {Crivellin}, \citenamefont
  {D'Ambrosio},\ and\ \citenamefont {Heeck}}]{Crivellin:2015mga}%
  \BibitemOpen
  \bibfield  {author} {\bibinfo {author} {\bibfnamefont {A.}~\bibnamefont
  {Crivellin}}, \bibinfo {author} {\bibfnamefont {G.}~\bibnamefont
  {D'Ambrosio}}, \ and\ \bibinfo {author} {\bibfnamefont {J.}~\bibnamefont
  {Heeck}},\ }\href {\doibase 10.1103/PhysRevLett.114.151801} {\bibfield
  {journal} {\bibinfo  {journal} {Phys. Rev. Lett.}\ }\textbf {\bibinfo
  {volume} {114}},\ \bibinfo {pages} {151801} (\bibinfo {year}
  {2015}{\natexlab{a}})},\ \Eprint {http://arxiv.org/abs/1501.00993}
  {arXiv:1501.00993 [hep-ph]} \BibitemShut {NoStop}%
\bibitem [{\citenamefont {Crivellin}\ \emph
  {et~al.}(2015{\natexlab{b}})\citenamefont {Crivellin}, \citenamefont
  {D'Ambrosio},\ and\ \citenamefont {Heeck}}]{Crivellin:2015lwa}%
  \BibitemOpen
  \bibfield  {author} {\bibinfo {author} {\bibfnamefont {A.}~\bibnamefont
  {Crivellin}}, \bibinfo {author} {\bibfnamefont {G.}~\bibnamefont
  {D'Ambrosio}}, \ and\ \bibinfo {author} {\bibfnamefont {J.}~\bibnamefont
  {Heeck}},\ }\href {\doibase 10.1103/PhysRevD.91.075006} {\bibfield  {journal}
  {\bibinfo  {journal} {Phys. Rev.}\ }\textbf {\bibinfo {volume} {D91}},\
  \bibinfo {pages} {075006} (\bibinfo {year} {2015}{\natexlab{b}})},\ \Eprint
  {http://arxiv.org/abs/1503.03477} {arXiv:1503.03477 [hep-ph]} \BibitemShut
  {NoStop}%
\bibitem [{\citenamefont {Aliev}\ and\ \citenamefont
  {Savci}(1999)}]{Aliev:1998sk}%
  \BibitemOpen
  \bibfield  {author} {\bibinfo {author} {\bibfnamefont {T.~M.}\ \bibnamefont
  {Aliev}}\ and\ \bibinfo {author} {\bibfnamefont {M.}~\bibnamefont {Savci}},\
  }\href {\doibase 10.1103/PhysRevD.60.014005} {\bibfield  {journal} {\bibinfo
  {journal} {Phys. Rev.}\ }\textbf {\bibinfo {volume} {D60}},\ \bibinfo {pages}
  {014005} (\bibinfo {year} {1999})},\ \Eprint
  {http://arxiv.org/abs/hep-ph/9812272} {arXiv:hep-ph/9812272} \BibitemShut
  {NoStop}%
\bibitem [{\citenamefont {Iltan}(1999)}]{Iltan:1998ra}%
  \BibitemOpen
  \bibfield  {author} {\bibinfo {author} {\bibfnamefont {E.~O.}\ \bibnamefont
  {Iltan}},\ }\href {\doibase 10.1142/S0217751X99002062} {\bibfield  {journal}
  {\bibinfo  {journal} {Int. J. Mod. Phys.}\ }\textbf {\bibinfo {volume}
  {A14}},\ \bibinfo {pages} {4365} (\bibinfo {year} {1999})},\ \Eprint
  {http://arxiv.org/abs/hep-ph/9807256} {arXiv:hep-ph/9807256} \BibitemShut
  {NoStop}%
\bibitem [{\citenamefont {Erkol}\ and\ \citenamefont
  {Turan}(2002)}]{Erkol:2002nw}%
  \BibitemOpen
  \bibfield  {author} {\bibinfo {author} {\bibfnamefont {G.}~\bibnamefont
  {Erkol}}\ and\ \bibinfo {author} {\bibfnamefont {G.}~\bibnamefont {Turan}},\
  }\href {\doibase 10.1088/1126-6708/2002/02/015} {\bibfield  {journal}
  {\bibinfo  {journal} {JHEP}\ }\textbf {\bibinfo {volume} {0202}},\ \bibinfo
  {pages} {015} (\bibinfo {year} {2002})},\ \Eprint
  {http://arxiv.org/abs/hep-ph/0201055} {arXiv:hep-ph/0201055} \BibitemShut
  {NoStop}%
\bibitem [{\citenamefont {Erkol}\ \emph {et~al.}(2005)\citenamefont {Erkol},
  \citenamefont {Wagenaar},\ and\ \citenamefont {Turan}}]{Erkol:2004me}%
  \BibitemOpen
  \bibfield  {author} {\bibinfo {author} {\bibfnamefont {G.}~\bibnamefont
  {Erkol}}, \bibinfo {author} {\bibfnamefont {J.~W.}\ \bibnamefont {Wagenaar}},
  \ and\ \bibinfo {author} {\bibfnamefont {G.}~\bibnamefont {Turan}},\ }\href
  {\doibase 10.1140/epjc/s2005-02197-9} {\bibfield  {journal} {\bibinfo
  {journal} {Eur. Phys. J.}\ }\textbf {\bibinfo {volume} {C41}},\ \bibinfo
  {pages} {189} (\bibinfo {year} {2005})},\ \Eprint
  {http://arxiv.org/abs/hep-ph/0408186} {arXiv:hep-ph/0408186} \BibitemShut
  {NoStop}%
\bibitem [{\citenamefont {Song}\ \emph {et~al.}(2008)\citenamefont {Song},
  \citenamefont {{L\"u}},\ and\ \citenamefont {Lu}}]{Song:2008zzc}%
  \BibitemOpen
  \bibfield  {author} {\bibinfo {author} {\bibfnamefont {H.-Z.}\ \bibnamefont
  {Song}}, \bibinfo {author} {\bibfnamefont {L.-X.}\ \bibnamefont {{L\"u}}}, \
  and\ \bibinfo {author} {\bibfnamefont {G.-R.}\ \bibnamefont {Lu}},\ }\href
  {\doibase 10.1088/0253-6102/50/3/35} {\bibfield  {journal} {\bibinfo
  {journal} {Commun. Theor. Phys.}\ }\textbf {\bibinfo {volume} {50}},\
  \bibinfo {pages} {696} (\bibinfo {year} {2008})}\BibitemShut {NoStop}%
\bibitem [{\citenamefont {Chen}\ and\ \citenamefont
  {Geng}(2006)}]{Chen:2006nua}%
  \BibitemOpen
  \bibfield  {author} {\bibinfo {author} {\bibfnamefont {C.-H.}\ \bibnamefont
  {Chen}}\ and\ \bibinfo {author} {\bibfnamefont {C.-Q.}\ \bibnamefont
  {Geng}},\ }\href {\doibase 10.1088/1126-6708/2006/10/053} {\bibfield
  {journal} {\bibinfo  {journal} {JHEP}\ }\textbf {\bibinfo {volume} {0610}},\
  \bibinfo {pages} {053} (\bibinfo {year} {2006})},\ \Eprint
  {http://arxiv.org/abs/hep-ph/0608166} {arXiv:hep-ph/0608166} \BibitemShut
  {NoStop}%
\bibitem [{\citenamefont {Nierste}\ \emph {et~al.}(2008)\citenamefont
  {Nierste}, \citenamefont {Trine},\ and\ \citenamefont
  {Westhoff}}]{Nierste:2008qe}%
  \BibitemOpen
  \bibfield  {author} {\bibinfo {author} {\bibfnamefont {U.}~\bibnamefont
  {Nierste}}, \bibinfo {author} {\bibfnamefont {S.}~\bibnamefont {Trine}}, \
  and\ \bibinfo {author} {\bibfnamefont {S.}~\bibnamefont {Westhoff}},\ }\href
  {\doibase 10.1103/PhysRevD.78.015006} {\bibfield  {journal} {\bibinfo
  {journal} {Phys. Rev.}\ }\textbf {\bibinfo {volume} {D78}},\ \bibinfo {pages}
  {015006} (\bibinfo {year} {2008})},\ \Eprint {http://arxiv.org/abs/0801.4938}
  {arXiv:0801.4938 [hep-ph]} \BibitemShut {NoStop}%
\bibitem [{\citenamefont {Tanaka}\ and\ \citenamefont
  {Watanabe}(2010)}]{Tanaka:2010se}%
  \BibitemOpen
  \bibfield  {author} {\bibinfo {author} {\bibfnamefont {M.}~\bibnamefont
  {Tanaka}}\ and\ \bibinfo {author} {\bibfnamefont {R.}~\bibnamefont
  {Watanabe}},\ }\href {\doibase 10.1103/PhysRevD.82.034027} {\bibfield
  {journal} {\bibinfo  {journal} {Phys. Rev.}\ }\textbf {\bibinfo {volume}
  {D82}},\ \bibinfo {pages} {034027} (\bibinfo {year} {2010})},\ \Eprint
  {http://arxiv.org/abs/1005.4306} {arXiv:1005.4306 [hep-ph]} \BibitemShut
  {NoStop}%
\bibitem [{\citenamefont {Crivellin}\ \emph {et~al.}(2012)\citenamefont
  {Crivellin}, \citenamefont {Greub},\ and\ \citenamefont
  {Kokulu}}]{Crivellin:2012ye}%
  \BibitemOpen
  \bibfield  {author} {\bibinfo {author} {\bibfnamefont {A.}~\bibnamefont
  {Crivellin}}, \bibinfo {author} {\bibfnamefont {C.}~\bibnamefont {Greub}}, \
  and\ \bibinfo {author} {\bibfnamefont {A.}~\bibnamefont {Kokulu}},\ }\href
  {\doibase 10.1103/PhysRevD.86.054014} {\bibfield  {journal} {\bibinfo
  {journal} {Phys. Rev.}\ }\textbf {\bibinfo {volume} {D86}},\ \bibinfo {pages}
  {054014} (\bibinfo {year} {2012})},\ \Eprint {http://arxiv.org/abs/1206.2634}
  {arXiv:1206.2634 [hep-ph]} \BibitemShut {NoStop}%
\bibitem [{\citenamefont {Sakaki}\ and\ \citenamefont
  {Tanaka}(2013)}]{Sakaki:2012ft}%
  \BibitemOpen
  \bibfield  {author} {\bibinfo {author} {\bibfnamefont {Y.}~\bibnamefont
  {Sakaki}}\ and\ \bibinfo {author} {\bibfnamefont {H.}~\bibnamefont
  {Tanaka}},\ }\href {\doibase 10.1103/PhysRevD.87.054002} {\bibfield
  {journal} {\bibinfo  {journal} {Phys. Rev.}\ }\textbf {\bibinfo {volume}
  {D87}},\ \bibinfo {pages} {054002} (\bibinfo {year} {2013})},\ \Eprint
  {http://arxiv.org/abs/1205.4908} {arXiv:1205.4908 [hep-ph]} \BibitemShut
  {NoStop}%
\bibitem [{\citenamefont {Celis}\ \emph {et~al.}(2013)\citenamefont {Celis},
  \citenamefont {Jung}, \citenamefont {Li},\ and\ \citenamefont
  {Pich}}]{Celis:2012dk}%
  \BibitemOpen
  \bibfield  {author} {\bibinfo {author} {\bibfnamefont {A.}~\bibnamefont
  {Celis}}, \bibinfo {author} {\bibfnamefont {M.}~\bibnamefont {Jung}},
  \bibinfo {author} {\bibfnamefont {X.-Q.}\ \bibnamefont {Li}}, \ and\ \bibinfo
  {author} {\bibfnamefont {A.}~\bibnamefont {Pich}},\ }\href {\doibase
  10.1007/JHEP01(2013)054} {\bibfield  {journal} {\bibinfo  {journal} {JHEP}\
  }\textbf {\bibinfo {volume} {1301}},\ \bibinfo {pages} {054} (\bibinfo {year}
  {2013})},\ \Eprint {http://arxiv.org/abs/1210.8443} {arXiv:1210.8443
  [hep-ph]} \BibitemShut {NoStop}%
\bibitem [{\citenamefont {Crivellin}(2010)}]{Crivellin:2009sd}%
  \BibitemOpen
  \bibfield  {author} {\bibinfo {author} {\bibfnamefont {A.}~\bibnamefont
  {Crivellin}},\ }\href {\doibase 10.1103/PhysRevD.81.031301} {\bibfield
  {journal} {\bibinfo  {journal} {Phys. Rev.}\ }\textbf {\bibinfo {volume}
  {D81}},\ \bibinfo {pages} {031301} (\bibinfo {year} {2010})},\ \Eprint
  {http://arxiv.org/abs/0907.2461} {arXiv:0907.2461 [hep-ph]} \BibitemShut
  {NoStop}%
\bibitem [{\citenamefont {Feger}\ \emph {et~al.}(2010)\citenamefont {Feger},
  \citenamefont {Mannel}, \citenamefont {Klose}, \citenamefont {Lacker},\ and\
  \citenamefont {{L\"uck}}}]{Feger:2010qc}%
  \BibitemOpen
  \bibfield  {author} {\bibinfo {author} {\bibfnamefont {R.}~\bibnamefont
  {Feger}}, \bibinfo {author} {\bibfnamefont {T.}~\bibnamefont {Mannel}},
  \bibinfo {author} {\bibfnamefont {V.}~\bibnamefont {Klose}}, \bibinfo
  {author} {\bibfnamefont {H.}~\bibnamefont {Lacker}}, \ and\ \bibinfo {author}
  {\bibfnamefont {T.}~\bibnamefont {{L\"uck}}},\ }\href {\doibase
  10.1103/PhysRevD.82.073002} {\bibfield  {journal} {\bibinfo  {journal} {Phys.
  Rev.}\ }\textbf {\bibinfo {volume} {D82}},\ \bibinfo {pages} {073002}
  (\bibinfo {year} {2010})},\ \Eprint {http://arxiv.org/abs/1003.4022}
  {arXiv:1003.4022 [hep-ph]} \BibitemShut {NoStop}%
\bibitem [{\citenamefont {Bernlochner}\ \emph {et~al.}(2014)\citenamefont
  {Bernlochner}, \citenamefont {Ligeti},\ and\ \citenamefont
  {Turczyk}}]{Bernlochner:2014ova}%
  \BibitemOpen
  \bibfield  {author} {\bibinfo {author} {\bibfnamefont {F.~U.}\ \bibnamefont
  {Bernlochner}}, \bibinfo {author} {\bibfnamefont {Z.}~\bibnamefont {Ligeti}},
  \ and\ \bibinfo {author} {\bibfnamefont {S.}~\bibnamefont {Turczyk}},\ }\href
  {\doibase 10.1103/PhysRevD.90.094003} {\bibfield  {journal} {\bibinfo
  {journal} {Phys. Rev.}\ }\textbf {\bibinfo {volume} {D90}},\ \bibinfo {pages}
  {094003} (\bibinfo {year} {2014})},\ \Eprint {http://arxiv.org/abs/1408.2516}
  {arXiv:1408.2516 [hep-ph]} \BibitemShut {NoStop}%
\bibitem [{\citenamefont {Buras}\ and\ \citenamefont
  {Girrbach}(2014)}]{Buras:2013ooa}%
  \BibitemOpen
  \bibfield  {author} {\bibinfo {author} {\bibfnamefont {A.~J.}\ \bibnamefont
  {Buras}}\ and\ \bibinfo {author} {\bibfnamefont {J.}~\bibnamefont
  {Girrbach}},\ }\href {\doibase 10.1088/0034-4885/77/8/086201} {\bibfield
  {journal} {\bibinfo  {journal} {Rept. Prog. Phys.}\ }\textbf {\bibinfo
  {volume} {77}},\ \bibinfo {pages} {086201} (\bibinfo {year} {2014})},\
  \Eprint {http://arxiv.org/abs/1306.3775} {arXiv:1306.3775 [hep-ph]}
  \BibitemShut {NoStop}%
\bibitem [{\citenamefont {Altmannshofer}\ and\ \citenamefont
  {Straub}(2015)}]{Altmannshofer:2015sma}%
  \BibitemOpen
  \bibfield  {author} {\bibinfo {author} {\bibfnamefont {W.}~\bibnamefont
  {Altmannshofer}}\ and\ \bibinfo {author} {\bibfnamefont {D.~M.}\ \bibnamefont
  {Straub}},\ }in\ \href@noop {} {\emph {\bibinfo {booktitle} {{50th Rencontres
  de Moriond on Electroweak Interactions and Unified Theories}}}}\ (\bibinfo
  {year} {2015})\ \Eprint {http://arxiv.org/abs/1503.06199} {arXiv:1503.06199
  [hep-ph]} \BibitemShut {NoStop}%
\bibitem [{\citenamefont {Lees}\ \emph
  {et~al.}(2012{\natexlab{a}})\citenamefont {Lees} \emph
  {et~al.}}]{Lees:2012xj}%
  \BibitemOpen
  \bibfield  {author} {\bibinfo {author} {\bibfnamefont {J.~P.}\ \bibnamefont
  {Lees}} \emph {et~al.} (\bibinfo {collaboration} {BaBar Collaboration}),\
  }\href {\doibase 10.1103/PhysRevLett.109.101802} {\bibfield  {journal}
  {\bibinfo  {journal} {Phys. Rev. Lett.}\ }\textbf {\bibinfo {volume} {109}},\
  \bibinfo {pages} {101802} (\bibinfo {year} {2012}{\natexlab{a}})},\ \Eprint
  {http://arxiv.org/abs/1205.5442} {arXiv:1205.5442 [hep-ex]} \BibitemShut
  {NoStop}%
\bibitem [{\citenamefont {Bailey}\ \emph {et~al.}(2012)\citenamefont {Bailey}
  \emph {et~al.}}]{Bailey:2012jg}%
  \BibitemOpen
  \bibfield  {author} {\bibinfo {author} {\bibfnamefont {J.~A.}\ \bibnamefont
  {Bailey}} \emph {et~al.} (\bibinfo {collaboration} {Fermilab Lattice and MILC
  Collaborations}),\ }\href {\doibase 10.1103/PhysRevLett.109.071802}
  {\bibfield  {journal} {\bibinfo  {journal} {Phys. Rev. Lett.}\ }\textbf
  {\bibinfo {volume} {109}},\ \bibinfo {pages} {071802} (\bibinfo {year}
  {2012})},\ \Eprint {http://arxiv.org/abs/1206.4992} {arXiv:1206.4992
  [hep-ph]} \BibitemShut {NoStop}%
\bibitem [{\citenamefont {Huschle}\ \emph {et~al.}(2015)\citenamefont {Huschle}
  \emph {et~al.}}]{Huschle:2015rga}%
  \BibitemOpen
  \bibfield  {author} {\bibinfo {author} {\bibfnamefont {M.}~\bibnamefont
  {Huschle}} \emph {et~al.} (\bibinfo {collaboration} {Belle Collaboration}),\
  }\href {\doibase 10.1103/PhysRevD.92.072014} {\bibfield  {journal} {\bibinfo
  {journal} {Phys. Rev.}\ }\textbf {\bibinfo {volume} {D92}},\ \bibinfo {pages}
  {072014} (\bibinfo {year} {2015})},\ \Eprint
  {http://arxiv.org/abs/1507.03233} {arXiv:1507.03233 [hep-ex]} \BibitemShut
  {NoStop}%
\bibitem [{\citenamefont {Aaij}\ \emph
  {et~al.}(2015{\natexlab{a}})\citenamefont {Aaij} \emph
  {et~al.}}]{Aaij:2015yra}%
  \BibitemOpen
  \bibfield  {author} {\bibinfo {author} {\bibfnamefont {R.}~\bibnamefont
  {Aaij}} \emph {et~al.} (\bibinfo {collaboration} {LHCb Collaboration}),\
  }\href {\doibase 10.1103/PhysRevLett.115.111803} {\bibfield  {journal}
  {\bibinfo  {journal} {Phys. Rev. Lett.}\ }\textbf {\bibinfo {volume} {115}},\
  \bibinfo {pages} {111803} (\bibinfo {year} {2015}{\natexlab{a}})},\ \Eprint
  {http://arxiv.org/abs/1506.08614} {arXiv:1506.08614 [hep-ex]} \BibitemShut
  {NoStop}%
\bibitem [{\citenamefont {Amhis}\ \emph {et~al.}(2014)\citenamefont {Amhis}
  \emph {et~al.}}]{Amhis:2014hma}%
  \BibitemOpen
  \bibfield  {author} {\bibinfo {author} {\bibfnamefont {Y.}~\bibnamefont
  {Amhis}} \emph {et~al.} (\bibinfo {collaboration} {Heavy Flavor Averaging
  Group}),\ }\href {http://www.slac.stanford.edu/xorg/hfag} {\  (\bibinfo
  {year} {2014})},\ \Eprint {http://arxiv.org/abs/1412.7515} {arXiv:1412.7515
  [hep-ex]} \BibitemShut {NoStop}%
\bibitem [{\citenamefont {Aaij}\ \emph
  {et~al.}(2014{\natexlab{a}})\citenamefont {Aaij} \emph
  {et~al.}}]{Aaij:2014ora}%
  \BibitemOpen
  \bibfield  {author} {\bibinfo {author} {\bibfnamefont {R.}~\bibnamefont
  {Aaij}} \emph {et~al.} (\bibinfo {collaboration} {LHCb Collaboration}),\
  }\href {\doibase 10.1103/PhysRevLett.113.151601} {\bibfield  {journal}
  {\bibinfo  {journal} {Phys. Rev. Lett.}\ }\textbf {\bibinfo {volume} {113}},\
  \bibinfo {pages} {151601} (\bibinfo {year} {2014}{\natexlab{a}})},\ \Eprint
  {http://arxiv.org/abs/1406.6482} {arXiv:1406.6482 [hep-ex]} \BibitemShut
  {NoStop}%
\bibitem [{\citenamefont {Bouchard}\ \emph
  {et~al.}(2013{\natexlab{a}})\citenamefont {Bouchard}, \citenamefont {Lepage},
  \citenamefont {Monahan}, \citenamefont {Na},\ and\ \citenamefont
  {Shigemitsu}}]{Bouchard:2013mia}%
  \BibitemOpen
  \bibfield  {author} {\bibinfo {author} {\bibfnamefont {C.}~\bibnamefont
  {Bouchard}}, \bibinfo {author} {\bibfnamefont {G.~P.}\ \bibnamefont
  {Lepage}}, \bibinfo {author} {\bibfnamefont {C.}~\bibnamefont {Monahan}},
  \bibinfo {author} {\bibfnamefont {H.}~\bibnamefont {Na}}, \ and\ \bibinfo
  {author} {\bibfnamefont {J.}~\bibnamefont {Shigemitsu}} (\bibinfo
  {collaboration} {HPQCD Collaboration}),\ }\href {\doibase
  10.1103/PhysRevLett.111.162002} {\bibfield  {journal} {\bibinfo  {journal}
  {Phys. Rev. Lett.}\ }\textbf {\bibinfo {volume} {111}},\ \bibinfo {pages}
  {162002} (\bibinfo {year} {2013}{\natexlab{a}})},\ \bibinfo {note} {(E)
  \href{http://dx.doi.org/10.1103/PhysRevLett.112.149902}{Phys.\ Rev.\ Lett.}\
  \textbf{112}, 149902 (2013)},\ \Eprint {http://arxiv.org/abs/1306.0434}
  {arXiv:1306.0434 [hep-ph]} \BibitemShut {NoStop}%
\bibitem [{\citenamefont {Horgan}\ \emph
  {et~al.}(2014{\natexlab{a}})\citenamefont {Horgan}, \citenamefont {Liu},
  \citenamefont {Meinel},\ and\ \citenamefont {Wingate}}]{Horgan:2013pva}%
  \BibitemOpen
  \bibfield  {author} {\bibinfo {author} {\bibfnamefont {R.~R.}\ \bibnamefont
  {Horgan}}, \bibinfo {author} {\bibfnamefont {Z.}~\bibnamefont {Liu}},
  \bibinfo {author} {\bibfnamefont {S.}~\bibnamefont {Meinel}}, \ and\ \bibinfo
  {author} {\bibfnamefont {M.}~\bibnamefont {Wingate}},\ }\href {\doibase
  10.1103/PhysRevLett.112.212003} {\bibfield  {journal} {\bibinfo  {journal}
  {Phys. Rev. Lett.}\ }\textbf {\bibinfo {volume} {112}},\ \bibinfo {pages}
  {212003} (\bibinfo {year} {2014}{\natexlab{a}})},\ \Eprint
  {http://arxiv.org/abs/1310.3887} {arXiv:1310.3887 [hep-ph]} \BibitemShut
  {NoStop}%
\bibitem [{\citenamefont {Aaij}\ \emph
  {et~al.}(2014{\natexlab{b}})\citenamefont {Aaij} \emph
  {et~al.}}]{Aaij:2014pli}%
  \BibitemOpen
  \bibfield  {author} {\bibinfo {author} {\bibfnamefont {R.}~\bibnamefont
  {Aaij}} \emph {et~al.} (\bibinfo {collaboration} {LHCb Collaboration}),\
  }\href {\doibase 10.1007/JHEP06(2014)133} {\bibfield  {journal} {\bibinfo
  {journal} {JHEP}\ }\textbf {\bibinfo {volume} {1406}},\ \bibinfo {pages}
  {133} (\bibinfo {year} {2014}{\natexlab{b}})},\ \Eprint
  {http://arxiv.org/abs/1403.8044} {arXiv:1403.8044 [hep-ex]} \BibitemShut
  {NoStop}%
\bibitem [{\citenamefont {Aaij}\ \emph
  {et~al.}(2013{\natexlab{a}})\citenamefont {Aaij} \emph
  {et~al.}}]{Aaij:2013qta}%
  \BibitemOpen
  \bibfield  {author} {\bibinfo {author} {\bibfnamefont {R.}~\bibnamefont
  {Aaij}} \emph {et~al.} (\bibinfo {collaboration} {LHCb Collaboration}),\
  }\href {\doibase 10.1103/PhysRevLett.111.191801} {\bibfield  {journal}
  {\bibinfo  {journal} {Phys. Rev. Lett.}\ }\textbf {\bibinfo {volume} {111}},\
  \bibinfo {pages} {191801} (\bibinfo {year} {2013}{\natexlab{a}})},\ \Eprint
  {http://arxiv.org/abs/1308.1707} {arXiv:1308.1707 [hep-ex]} \BibitemShut
  {NoStop}%
\bibitem [{\citenamefont
  {Langenbruch}(2015)}]{ChristophLangenbruchonbehalfoftheLHCb:2015iha}%
  \BibitemOpen
  \bibfield  {author} {\bibinfo {author} {\bibfnamefont {C.}~\bibnamefont
  {Langenbruch}} (\bibinfo {collaboration} {LHCb Collaboration}),\ }in\
  \href@noop {} {\emph {\bibinfo {booktitle} {{50th Rencontres de Moriond on
  Electroweak Interactions and Unified Theories}}}}\ (\bibinfo {year} {2015})\
  \Eprint {http://arxiv.org/abs/1505.04160} {arXiv:1505.04160 [hep-ex]}
  \BibitemShut {NoStop}%
\bibitem [{\citenamefont {Bailey}\ \emph
  {et~al.}(2015{\natexlab{a}})\citenamefont {Bailey} \emph
  {et~al.}}]{Lattice:2015tia}%
  \BibitemOpen
  \bibfield  {author} {\bibinfo {author} {\bibfnamefont {J.~A.}\ \bibnamefont
  {Bailey}} \emph {et~al.} (\bibinfo {collaboration} {Fermilab Lattice and MILC
  Collaborations}),\ }\href {\doibase 10.1103/PhysRevD.92.014024} {\bibfield
  {journal} {\bibinfo  {journal} {Phys. Rev.}\ }\textbf {\bibinfo {volume}
  {D92}},\ \bibinfo {pages} {014024} (\bibinfo {year} {2015}{\natexlab{a}})},\
  \Eprint {http://arxiv.org/abs/1503.07839} {arXiv:1503.07839 [hep-lat]}
  \BibitemShut {NoStop}%
\bibitem [{\citenamefont {Bailey}\ \emph
  {et~al.}(2015{\natexlab{b}})\citenamefont {Bailey} \emph
  {et~al.}}]{Lattice:2015rga}%
  \BibitemOpen
  \bibfield  {author} {\bibinfo {author} {\bibfnamefont {J.~A.}\ \bibnamefont
  {Bailey}} \emph {et~al.} (\bibinfo {collaboration} {Fermilab Lattice and MILC
  Collaborations}),\ }\href {\doibase 10.1103/PhysRevD.92.034506} {\bibfield
  {journal} {\bibinfo  {journal} {Phys. Rev.}\ }\textbf {\bibinfo {volume}
  {D92}},\ \bibinfo {pages} {034506} (\bibinfo {year} {2015}{\natexlab{b}})},\
  \Eprint {http://arxiv.org/abs/1503.07237} {arXiv:1503.07237 [hep-lat]}
  \BibitemShut {NoStop}%
\bibitem [{\citenamefont {Na}\ \emph {et~al.}(2015)\citenamefont {Na},
  \citenamefont {Bouchard}, \citenamefont {Lepage}, \citenamefont {Monahan},\
  and\ \citenamefont {Shigemitsu}}]{Na:2015kha}%
  \BibitemOpen
  \bibfield  {author} {\bibinfo {author} {\bibfnamefont {H.}~\bibnamefont
  {Na}}, \bibinfo {author} {\bibfnamefont {C.~M.}\ \bibnamefont {Bouchard}},
  \bibinfo {author} {\bibfnamefont {G.~P.}\ \bibnamefont {Lepage}}, \bibinfo
  {author} {\bibfnamefont {C.}~\bibnamefont {Monahan}}, \ and\ \bibinfo
  {author} {\bibfnamefont {J.}~\bibnamefont {Shigemitsu}} (\bibinfo
  {collaboration} {HPQCD Collaboration}),\ }\href {\doibase
  10.1103/PhysRevD.92.054510} {\bibfield  {journal} {\bibinfo  {journal} {Phys.
  Rev.}\ }\textbf {\bibinfo {volume} {D92}},\ \bibinfo {pages} {054510}
  (\bibinfo {year} {2015})},\ \Eprint {http://arxiv.org/abs/1505.03925}
  {arXiv:1505.03925 [hep-lat]} \BibitemShut {NoStop}%
\bibitem [{\citenamefont {Detmold}\ \emph {et~al.}(2015)\citenamefont
  {Detmold}, \citenamefont {Lehner},\ and\ \citenamefont
  {Meinel}}]{Detmold:2015aaa}%
  \BibitemOpen
  \bibfield  {author} {\bibinfo {author} {\bibfnamefont {W.}~\bibnamefont
  {Detmold}}, \bibinfo {author} {\bibfnamefont {C.}~\bibnamefont {Lehner}}, \
  and\ \bibinfo {author} {\bibfnamefont {S.}~\bibnamefont {Meinel}},\ }\href
  {\doibase 10.1103/PhysRevD.92.034503} {\bibfield  {journal} {\bibinfo
  {journal} {Phys. Rev.}\ }\textbf {\bibinfo {volume} {D92}},\ \bibinfo {pages}
  {034503} (\bibinfo {year} {2015})},\ \Eprint
  {http://arxiv.org/abs/1503.01421} {arXiv:1503.01421 [hep-lat]} \BibitemShut
  {NoStop}%
\bibitem [{\citenamefont {Glattauer}(2015)}]{GlattauerEPS2015}%
  \BibitemOpen
  \bibfield  {author} {\bibinfo {author} {\bibfnamefont {R.}~\bibnamefont
  {Glattauer}} (\bibinfo {collaboration} {Belle Collaboration}),\ }\href@noop
  {} {\enquote {\bibinfo {title} {{Semileptonic $B$ and $B_s$ decays at
  Belle}},}\ }\bibinfo {howpublished}
  {\href{https://indico.cern.ch/event/356420/session/3/contribution/306}{talk
  presented at EPS-HEP 2015}} (\bibinfo {year} {2015})\BibitemShut {NoStop}%
\bibitem [{\citenamefont {Gambino}(2015)}]{GambinoEPS2015}%
  \BibitemOpen
  \bibfield  {author} {\bibinfo {author} {\bibfnamefont {P.}~\bibnamefont
  {Gambino}},\ }\href@noop {} {\enquote {\bibinfo {title} {{$V_{cb}$ and
  $V_{ub}$: Where do we stand?}}}\ }\bibinfo {howpublished}
  {\href{https://indico.cern.ch/event/356420/session/3/contribution/824}{talk
  presented at EPS-HEP 2015}} (\bibinfo {year} {2015})\BibitemShut {NoStop}%
\bibitem [{\citenamefont {Alberti}\ \emph {et~al.}(2015)\citenamefont
  {Alberti}, \citenamefont {Gambino}, \citenamefont {Healey},\ and\
  \citenamefont {Nandi}}]{Alberti:2014yda}%
  \BibitemOpen
  \bibfield  {author} {\bibinfo {author} {\bibfnamefont {A.}~\bibnamefont
  {Alberti}}, \bibinfo {author} {\bibfnamefont {P.}~\bibnamefont {Gambino}},
  \bibinfo {author} {\bibfnamefont {K.~J.}\ \bibnamefont {Healey}}, \ and\
  \bibinfo {author} {\bibfnamefont {S.}~\bibnamefont {Nandi}},\ }\href
  {\doibase 10.1103/PhysRevLett.114.061802} {\bibfield  {journal} {\bibinfo
  {journal} {Phys. Rev. Lett.}\ }\textbf {\bibinfo {volume} {114}},\ \bibinfo
  {pages} {061802} (\bibinfo {year} {2015})},\ \Eprint
  {http://arxiv.org/abs/1411.6560} {arXiv:1411.6560 [hep-ph]} \BibitemShut
  {NoStop}%
\bibitem [{\citenamefont {Aaij}\ \emph
  {et~al.}(2015{\natexlab{b}})\citenamefont {Aaij} \emph
  {et~al.}}]{Aaij:2015nea}%
  \BibitemOpen
  \bibfield  {author} {\bibinfo {author} {\bibfnamefont {R.}~\bibnamefont
  {Aaij}} \emph {et~al.} (\bibinfo {collaboration} {LHCb Collaboration}),\
  }\href {\doibase 10.1007/JHEP10(2015)034} {\bibfield  {journal} {\bibinfo
  {journal} {JHEP}\ }\textbf {\bibinfo {volume} {10}},\ \bibinfo {pages} {034}
  (\bibinfo {year} {2015}{\natexlab{b}})},\ \Eprint
  {http://arxiv.org/abs/1509.00414} {arXiv:1509.00414 [hep-ex]} \BibitemShut
  {NoStop}%
\bibitem [{\citenamefont {Hamer}\ \emph {et~al.}(2015)\citenamefont {Hamer}
  \emph {et~al.}}]{Hamer:2015jsa}%
  \BibitemOpen
  \bibfield  {author} {\bibinfo {author} {\bibfnamefont {P.}~\bibnamefont
  {Hamer}} \emph {et~al.} (\bibinfo {collaboration} {Belle Collaboration}),\
  }\href@noop {} {\  (\bibinfo {year} {2015})},\ \Eprint
  {http://arxiv.org/abs/1509.06521} {arXiv:1509.06521 [hep-ex]} \BibitemShut
  {NoStop}%
\bibitem [{\citenamefont {Aushev}\ \emph {et~al.}(2010)\citenamefont {Aushev}
  \emph {et~al.}}]{Aushev:2010bq}%
  \BibitemOpen
  \bibfield  {author} {\bibinfo {author} {\bibfnamefont {T.}~\bibnamefont
  {Aushev}} \emph {et~al.},\ }\href@noop {} {\enquote {\bibinfo {title}
  {{Physics at Super $B$ Factory}},}\ }\bibinfo {howpublished} {KEK report}
  (\bibinfo {year} {2010}),\ \Eprint {http://arxiv.org/abs/1002.5012}
  {arXiv:1002.5012 [hep-ex]} \BibitemShut {NoStop}%
\bibitem [{\citenamefont {Aubert}\ \emph {et~al.}(2010)\citenamefont {Aubert}
  \emph {et~al.}}]{Aubert:2009wt}%
  \BibitemOpen
  \bibfield  {author} {\bibinfo {author} {\bibfnamefont {B.}~\bibnamefont
  {Aubert}} \emph {et~al.} (\bibinfo {collaboration} {BaBar}),\ }\href
  {\doibase 10.1103/PhysRevD.81.051101} {\bibfield  {journal} {\bibinfo
  {journal} {Phys. Rev.}\ }\textbf {\bibinfo {volume} {D81}},\ \bibinfo {pages}
  {051101} (\bibinfo {year} {2010})},\ \Eprint {http://arxiv.org/abs/0912.2453}
  {arXiv:0912.2453 [hep-ex]} \BibitemShut {NoStop}%
\bibitem [{\citenamefont {Lees}\ \emph
  {et~al.}(2013{\natexlab{a}})\citenamefont {Lees} \emph
  {et~al.}}]{Lees:2012ju}%
  \BibitemOpen
  \bibfield  {author} {\bibinfo {author} {\bibfnamefont {J.~P.}\ \bibnamefont
  {Lees}} \emph {et~al.} (\bibinfo {collaboration} {BaBar}),\ }\href {\doibase
  10.1103/PhysRevD.88.031102} {\bibfield  {journal} {\bibinfo  {journal} {Phys.
  Rev.}\ }\textbf {\bibinfo {volume} {D88}},\ \bibinfo {pages} {031102}
  (\bibinfo {year} {2013}{\natexlab{a}})},\ \Eprint
  {http://arxiv.org/abs/1207.0698} {arXiv:1207.0698 [hep-ex]} \BibitemShut
  {NoStop}%
\bibitem [{\citenamefont {Adachi}\ \emph {et~al.}(2013)\citenamefont {Adachi}
  \emph {et~al.}}]{Adachi:2012mm}%
  \BibitemOpen
  \bibfield  {author} {\bibinfo {author} {\bibfnamefont {I.}~\bibnamefont
  {Adachi}} \emph {et~al.} (\bibinfo {collaboration} {Belle}),\ }\href
  {\doibase 10.1103/PhysRevLett.110.131801} {\bibfield  {journal} {\bibinfo
  {journal} {Phys. Rev. Lett.}\ }\textbf {\bibinfo {volume} {110}},\ \bibinfo
  {pages} {131801} (\bibinfo {year} {2013})},\ \Eprint
  {http://arxiv.org/abs/1208.4678} {arXiv:1208.4678 [hep-ex]} \BibitemShut
  {NoStop}%
\bibitem [{\citenamefont {Kronenbitter}\ \emph {et~al.}(2015)\citenamefont
  {Kronenbitter} \emph {et~al.}}]{Kronenbitter:2015kls}%
  \BibitemOpen
  \bibfield  {author} {\bibinfo {author} {\bibfnamefont {B.}~\bibnamefont
  {Kronenbitter}} \emph {et~al.} (\bibinfo {collaboration} {Belle}),\ }\href
  {\doibase 10.1103/PhysRevD.92.051102} {\bibfield  {journal} {\bibinfo
  {journal} {Phys. Rev.}\ }\textbf {\bibinfo {volume} {D92}},\ \bibinfo {pages}
  {051102} (\bibinfo {year} {2015})},\ \Eprint
  {http://arxiv.org/abs/1503.05613} {arXiv:1503.05613 [hep-ex]} \BibitemShut
  {NoStop}%
\bibitem [{\citenamefont {Bailey}\ \emph
  {et~al.}(2015{\natexlab{c}})\citenamefont {Bailey} \emph
  {et~al.}}]{Bailey:2015dka}%
  \BibitemOpen
  \bibfield  {author} {\bibinfo {author} {\bibfnamefont {J.~A.}\ \bibnamefont
  {Bailey}} \emph {et~al.} (\bibinfo {collaboration} {Fermilab Lattice and MILC
  Collaborations}),\ }\href@noop {} {\  (\bibinfo {year}
  {2015}{\natexlab{c}})},\ \Eprint {http://arxiv.org/abs/1509.06235}
  {arXiv:1509.06235 [hep-lat]} \BibitemShut {NoStop}%
\bibitem [{\citenamefont {Bailey}\ \emph
  {et~al.}(2015{\natexlab{d}})\citenamefont {Bailey} \emph
  {et~al.}}]{Bailey:2015nbd}%
  \BibitemOpen
  \bibfield  {author} {\bibinfo {author} {\bibfnamefont {J.~A.}\ \bibnamefont
  {Bailey}} \emph {et~al.} (\bibinfo {collaboration} {Fermilab Lattice and MILC
  Collaborations}),\ }\href {\doibase 10.1103/PhysRevLett.115.152002}
  {\bibfield  {journal} {\bibinfo  {journal} {Phys. Rev. Lett.}\ }\textbf
  {\bibinfo {volume} {115}},\ \bibinfo {pages} {152002} (\bibinfo {year}
  {2015}{\natexlab{d}})},\ \Eprint {http://arxiv.org/abs/1507.01618}
  {arXiv:1507.01618 [hep-ph]} \BibitemShut {NoStop}%
\bibitem [{\citenamefont {Bouchard}\ \emph
  {et~al.}(2013{\natexlab{b}})\citenamefont {Bouchard}, \citenamefont {Lepage},
  \citenamefont {Monahan}, \citenamefont {Na},\ and\ \citenamefont
  {Shigemitsu}}]{Bouchard:2013eph}%
  \BibitemOpen
  \bibfield  {author} {\bibinfo {author} {\bibfnamefont {C.}~\bibnamefont
  {Bouchard}}, \bibinfo {author} {\bibfnamefont {G.~P.}\ \bibnamefont
  {Lepage}}, \bibinfo {author} {\bibfnamefont {C.}~\bibnamefont {Monahan}},
  \bibinfo {author} {\bibfnamefont {H.}~\bibnamefont {Na}}, \ and\ \bibinfo
  {author} {\bibfnamefont {J.}~\bibnamefont {Shigemitsu}} (\bibinfo
  {collaboration} {HPQCD Collaboration}),\ }\href {\doibase
  10.1103/PhysRevD.88.079901, 10.1103/PhysRevD.88.054509} {\bibfield  {journal}
  {\bibinfo  {journal} {Phys. Rev.}\ }\textbf {\bibinfo {volume} {D88}},\
  \bibinfo {pages} {054509} (\bibinfo {year} {2013}{\natexlab{b}})},\ \Eprint
  {http://arxiv.org/abs/1306.2384} {arXiv:1306.2384 [hep-lat]} \BibitemShut
  {NoStop}%
\bibitem [{\citenamefont {Bailey}\ \emph {et~al.}(2009)\citenamefont {Bailey}
  \emph {et~al.}}]{Bailey:2008wp}%
  \BibitemOpen
  \bibfield  {author} {\bibinfo {author} {\bibfnamefont {J.~A.}\ \bibnamefont
  {Bailey}} \emph {et~al.} (\bibinfo {collaboration} {Fermilab Lattice and MILC
  Collaborations}),\ }\href {\doibase 10.1103/PhysRevD.79.054507} {\bibfield
  {journal} {\bibinfo  {journal} {Phys. Rev.}\ }\textbf {\bibinfo {volume}
  {D79}},\ \bibinfo {pages} {054507} (\bibinfo {year} {2009})},\ \Eprint
  {http://arxiv.org/abs/0811.3640} {arXiv:0811.3640 [hep-lat]} \BibitemShut
  {NoStop}%
\bibitem [{\citenamefont {Flynn}\ \emph
  {et~al.}(2015{\natexlab{a}})\citenamefont {Flynn} \emph
  {et~al.}}]{Flynn:2015mha}%
  \BibitemOpen
  \bibfield  {author} {\bibinfo {author} {\bibfnamefont {J.~M.}\ \bibnamefont
  {Flynn}} \emph {et~al.} (\bibinfo {collaboration} {RBC and UKQCD
  Collaborations}),\ }\href {\doibase 10.1103/PhysRevD.91.074510} {\bibfield
  {journal} {\bibinfo  {journal} {Phys. Rev.}\ }\textbf {\bibinfo {volume}
  {D91}},\ \bibinfo {pages} {074510} (\bibinfo {year} {2015}{\natexlab{a}})},\
  \Eprint {http://arxiv.org/abs/1501.05373} {arXiv:1501.05373 [hep-lat]}
  \BibitemShut {NoStop}%
\bibitem [{\citenamefont {del Amo~Sanchez}\ \emph {et~al.}(2011)\citenamefont
  {del Amo~Sanchez} \emph {et~al.}}]{delAmoSanchez:2010af}%
  \BibitemOpen
  \bibfield  {author} {\bibinfo {author} {\bibfnamefont {P.}~\bibnamefont {del
  Amo~Sanchez}} \emph {et~al.} (\bibinfo {collaboration} {BaBar
  Collaboration}),\ }\href {\doibase 10.1103/PhysRevD.83.032007} {\bibfield
  {journal} {\bibinfo  {journal} {Phys. Rev.}\ }\textbf {\bibinfo {volume}
  {D83}},\ \bibinfo {pages} {032007} (\bibinfo {year} {2011})},\ \Eprint
  {http://arxiv.org/abs/1005.3288} {arXiv:1005.3288 [hep-ex]} \BibitemShut
  {NoStop}%
\bibitem [{\citenamefont {Ha}\ \emph {et~al.}(2011)\citenamefont {Ha} \emph
  {et~al.}}]{Ha:2010rf}%
  \BibitemOpen
  \bibfield  {author} {\bibinfo {author} {\bibfnamefont {H.}~\bibnamefont {Ha}}
  \emph {et~al.} (\bibinfo {collaboration} {Belle Collaboration}),\ }\href
  {\doibase 10.1103/PhysRevD.83.071101} {\bibfield  {journal} {\bibinfo
  {journal} {Phys. Rev.}\ }\textbf {\bibinfo {volume} {D83}},\ \bibinfo {pages}
  {071101} (\bibinfo {year} {2011})},\ \Eprint {http://arxiv.org/abs/1012.0090}
  {arXiv:1012.0090 [hep-ex]} \BibitemShut {NoStop}%
\bibitem [{\citenamefont {Lees}\ \emph
  {et~al.}(2012{\natexlab{b}})\citenamefont {Lees} \emph
  {et~al.}}]{Lees:2012vv}%
  \BibitemOpen
  \bibfield  {author} {\bibinfo {author} {\bibfnamefont {J.~P.}\ \bibnamefont
  {Lees}} \emph {et~al.} (\bibinfo {collaboration} {BaBar Collaboration}),\
  }\href {\doibase 10.1103/PhysRevD.86.092004} {\bibfield  {journal} {\bibinfo
  {journal} {Phys. Rev.}\ }\textbf {\bibinfo {volume} {D86}},\ \bibinfo {pages}
  {092004} (\bibinfo {year} {2012}{\natexlab{b}})},\ \Eprint
  {http://arxiv.org/abs/1208.1253} {arXiv:1208.1253 [hep-ex]} \BibitemShut
  {NoStop}%
\bibitem [{\citenamefont {Sibidanov}\ \emph {et~al.}(2013)\citenamefont
  {Sibidanov} \emph {et~al.}}]{Sibidanov:2013rkk}%
  \BibitemOpen
  \bibfield  {author} {\bibinfo {author} {\bibfnamefont {A.}~\bibnamefont
  {Sibidanov}} \emph {et~al.} (\bibinfo {collaboration} {Belle
  Collaboration}),\ }\href {\doibase 10.1103/PhysRevD.88.032005} {\bibfield
  {journal} {\bibinfo  {journal} {Phys. Rev.}\ }\textbf {\bibinfo {volume}
  {D88}},\ \bibinfo {pages} {032005} (\bibinfo {year} {2013})},\ \Eprint
  {http://arxiv.org/abs/1306.2781} {arXiv:1306.2781 [hep-ex]} \BibitemShut
  {NoStop}%
\bibitem [{\citenamefont {Horgan}\ \emph
  {et~al.}(2014{\natexlab{b}})\citenamefont {Horgan}, \citenamefont {Liu},
  \citenamefont {Meinel},\ and\ \citenamefont {Wingate}}]{Horgan:2013hoa}%
  \BibitemOpen
  \bibfield  {author} {\bibinfo {author} {\bibfnamefont {R.~R.}\ \bibnamefont
  {Horgan}}, \bibinfo {author} {\bibfnamefont {Z.}~\bibnamefont {Liu}},
  \bibinfo {author} {\bibfnamefont {S.}~\bibnamefont {Meinel}}, \ and\ \bibinfo
  {author} {\bibfnamefont {M.}~\bibnamefont {Wingate}},\ }\href {\doibase
  10.1103/PhysRevD.89.094501} {\bibfield  {journal} {\bibinfo  {journal} {Phys.
  Rev.}\ }\textbf {\bibinfo {volume} {D89}},\ \bibinfo {pages} {094501}
  (\bibinfo {year} {2014}{\natexlab{b}})},\ \Eprint
  {http://arxiv.org/abs/1310.3722} {arXiv:1310.3722 [hep-lat]} \BibitemShut
  {NoStop}%
\bibitem [{\citenamefont {Brice\~no}\ \emph {et~al.}(2015)\citenamefont
  {Brice\~no}, \citenamefont {Hansen},\ and\ \citenamefont
  {Walker-Loud}}]{Briceno:2014uqa}%
  \BibitemOpen
  \bibfield  {author} {\bibinfo {author} {\bibfnamefont {R.~A.}\ \bibnamefont
  {Brice\~no}}, \bibinfo {author} {\bibfnamefont {M.~T.}\ \bibnamefont
  {Hansen}}, \ and\ \bibinfo {author} {\bibfnamefont {A.}~\bibnamefont
  {Walker-Loud}},\ }\href {\doibase 10.1103/PhysRevD.91.034501} {\bibfield
  {journal} {\bibinfo  {journal} {Phys. Rev.}\ }\textbf {\bibinfo {volume}
  {D91}},\ \bibinfo {pages} {034501} (\bibinfo {year} {2015})},\ \Eprint
  {http://arxiv.org/abs/1406.5965} {arXiv:1406.5965 [hep-lat]} \BibitemShut
  {NoStop}%
\bibitem [{\citenamefont {Grinstein}\ and\ \citenamefont
  {Pirjol}(2004)}]{Grinstein:2004vb}%
  \BibitemOpen
  \bibfield  {author} {\bibinfo {author} {\bibfnamefont {B.}~\bibnamefont
  {Grinstein}}\ and\ \bibinfo {author} {\bibfnamefont {D.}~\bibnamefont
  {Pirjol}},\ }\href {\doibase 10.1103/PhysRevD.70.114005} {\bibfield
  {journal} {\bibinfo  {journal} {Phys. Rev.}\ }\textbf {\bibinfo {volume}
  {D70}},\ \bibinfo {pages} {114005} (\bibinfo {year} {2004})},\ \Eprint
  {http://arxiv.org/abs/hep-ph/0404250} {arXiv:hep-ph/0404250} \BibitemShut
  {NoStop}%
\bibitem [{\citenamefont {Becher}\ \emph {et~al.}(2005)\citenamefont {Becher},
  \citenamefont {Hill},\ and\ \citenamefont {Neubert}}]{Becher:2005fg}%
  \BibitemOpen
  \bibfield  {author} {\bibinfo {author} {\bibfnamefont {T.}~\bibnamefont
  {Becher}}, \bibinfo {author} {\bibfnamefont {R.~J.}\ \bibnamefont {Hill}}, \
  and\ \bibinfo {author} {\bibfnamefont {M.}~\bibnamefont {Neubert}},\ }\href
  {\doibase 10.1103/PhysRevD.72.094017} {\bibfield  {journal} {\bibinfo
  {journal} {Phys. Rev.}\ }\textbf {\bibinfo {volume} {D72}},\ \bibinfo {pages}
  {094017} (\bibinfo {year} {2005})},\ \Eprint
  {http://arxiv.org/abs/hep-ph/0503263} {arXiv:hep-ph/0503263} \BibitemShut
  {NoStop}%
\bibitem [{\citenamefont {Ali}\ \emph {et~al.}(2008)\citenamefont {Ali},
  \citenamefont {Pecjak},\ and\ \citenamefont {Greub}}]{Ali:2007sj}%
  \BibitemOpen
  \bibfield  {author} {\bibinfo {author} {\bibfnamefont {A.}~\bibnamefont
  {Ali}}, \bibinfo {author} {\bibfnamefont {B.~D.}\ \bibnamefont {Pecjak}}, \
  and\ \bibinfo {author} {\bibfnamefont {C.}~\bibnamefont {Greub}},\ }\href
  {\doibase 10.1140/epjc/s10052-008-0623-5} {\bibfield  {journal} {\bibinfo
  {journal} {Eur. Phys. J.}\ }\textbf {\bibinfo {volume} {C55}},\ \bibinfo
  {pages} {577} (\bibinfo {year} {2008})},\ \Eprint
  {http://arxiv.org/abs/0709.4422} {arXiv:0709.4422 [hep-ph]} \BibitemShut
  {NoStop}%
\bibitem [{\citenamefont {Bobeth}\ \emph {et~al.}(2010)\citenamefont {Bobeth},
  \citenamefont {Hiller},\ and\ \citenamefont {van Dyk}}]{Bobeth:2010wg}%
  \BibitemOpen
  \bibfield  {author} {\bibinfo {author} {\bibfnamefont {C.}~\bibnamefont
  {Bobeth}}, \bibinfo {author} {\bibfnamefont {G.}~\bibnamefont {Hiller}}, \
  and\ \bibinfo {author} {\bibfnamefont {D.}~\bibnamefont {van Dyk}},\ }\href
  {\doibase 10.1007/JHEP07(2010)098} {\bibfield  {journal} {\bibinfo  {journal}
  {JHEP}\ }\textbf {\bibinfo {volume} {1007}},\ \bibinfo {pages} {098}
  (\bibinfo {year} {2010})},\ \Eprint {http://arxiv.org/abs/1006.5013}
  {arXiv:1006.5013 [hep-ph]} \BibitemShut {NoStop}%
\bibitem [{\citenamefont {Beylich}\ \emph {et~al.}(2011)\citenamefont
  {Beylich}, \citenamefont {Buchalla},\ and\ \citenamefont
  {Feldmann}}]{Beylich:2011aq}%
  \BibitemOpen
  \bibfield  {author} {\bibinfo {author} {\bibfnamefont {M.}~\bibnamefont
  {Beylich}}, \bibinfo {author} {\bibfnamefont {G.}~\bibnamefont {Buchalla}}, \
  and\ \bibinfo {author} {\bibfnamefont {T.}~\bibnamefont {Feldmann}},\ }\href
  {\doibase 10.1140/epjc/s10052-011-1635-0} {\bibfield  {journal} {\bibinfo
  {journal} {Eur. Phys. J.}\ }\textbf {\bibinfo {volume} {C71}},\ \bibinfo
  {pages} {1635} (\bibinfo {year} {2011})},\ \Eprint
  {http://arxiv.org/abs/1101.5118} {arXiv:1101.5118 [hep-ph]} \BibitemShut
  {NoStop}%
\bibitem [{\citenamefont {Bobeth}\ \emph {et~al.}(2012)\citenamefont {Bobeth},
  \citenamefont {Hiller}, \citenamefont {van Dyk},\ and\ \citenamefont
  {Wacker}}]{Bobeth:2011nj}%
  \BibitemOpen
  \bibfield  {author} {\bibinfo {author} {\bibfnamefont {C.}~\bibnamefont
  {Bobeth}}, \bibinfo {author} {\bibfnamefont {G.}~\bibnamefont {Hiller}},
  \bibinfo {author} {\bibfnamefont {D.}~\bibnamefont {van Dyk}}, \ and\
  \bibinfo {author} {\bibfnamefont {C.}~\bibnamefont {Wacker}},\ }\href
  {\doibase 10.1007/JHEP01(2012)107} {\bibfield  {journal} {\bibinfo  {journal}
  {JHEP}\ }\textbf {\bibinfo {volume} {1201}},\ \bibinfo {pages} {107}
  (\bibinfo {year} {2012})},\ \Eprint {http://arxiv.org/abs/1111.2558}
  {arXiv:1111.2558 [hep-ph]} \BibitemShut {NoStop}%
\bibitem [{\citenamefont {Hou}\ \emph {et~al.}(2014)\citenamefont {Hou},
  \citenamefont {Kohda},\ and\ \citenamefont {Xu}}]{Hou:2014dza}%
  \BibitemOpen
  \bibfield  {author} {\bibinfo {author} {\bibfnamefont {W.-S.}\ \bibnamefont
  {Hou}}, \bibinfo {author} {\bibfnamefont {M.}~\bibnamefont {Kohda}}, \ and\
  \bibinfo {author} {\bibfnamefont {F.}~\bibnamefont {Xu}},\ }\href {\doibase
  10.1103/PhysRevD.90.013002} {\bibfield  {journal} {\bibinfo  {journal} {Phys.
  Rev.}\ }\textbf {\bibinfo {volume} {D90}},\ \bibinfo {pages} {013002}
  (\bibinfo {year} {2014})},\ \Eprint {http://arxiv.org/abs/1403.7410}
  {arXiv:1403.7410 [hep-ph]} \BibitemShut {NoStop}%
\bibitem [{\citenamefont {Grinstein}\ \emph {et~al.}(1989)\citenamefont
  {Grinstein}, \citenamefont {Savage},\ and\ \citenamefont
  {Wise}}]{Grinstein:1988me}%
  \BibitemOpen
  \bibfield  {author} {\bibinfo {author} {\bibfnamefont {B.}~\bibnamefont
  {Grinstein}}, \bibinfo {author} {\bibfnamefont {M.~J.}\ \bibnamefont
  {Savage}}, \ and\ \bibinfo {author} {\bibfnamefont {M.~B.}\ \bibnamefont
  {Wise}},\ }\href {\doibase 10.1016/0550-3213(89)90078-3} {\bibfield
  {journal} {\bibinfo  {journal} {Nucl. Phys.}\ }\textbf {\bibinfo {volume}
  {B319}},\ \bibinfo {pages} {271} (\bibinfo {year} {1989})}\BibitemShut
  {NoStop}%
\bibitem [{\citenamefont {Buras}\ \emph {et~al.}(1994)\citenamefont {Buras},
  \citenamefont {Misiak}, \citenamefont {{M\"unz}},\ and\ \citenamefont
  {Pokorski}}]{Buras:1993xp}%
  \BibitemOpen
  \bibfield  {author} {\bibinfo {author} {\bibfnamefont {A.~J.}\ \bibnamefont
  {Buras}}, \bibinfo {author} {\bibfnamefont {M.}~\bibnamefont {Misiak}},
  \bibinfo {author} {\bibfnamefont {M.}~\bibnamefont {{M\"unz}}}, \ and\
  \bibinfo {author} {\bibfnamefont {S.}~\bibnamefont {Pokorski}},\ }\href
  {\doibase 10.1016/0550-3213(94)90299-2} {\bibfield  {journal} {\bibinfo
  {journal} {Nucl. Phys.}\ }\textbf {\bibinfo {volume} {B424}},\ \bibinfo
  {pages} {374} (\bibinfo {year} {1994})},\ \Eprint
  {http://arxiv.org/abs/hep-ph/9311345} {arXiv:hep-ph/9311345} \BibitemShut
  {NoStop}%
\bibitem [{\citenamefont {Huber}\ \emph {et~al.}(2006)\citenamefont {Huber},
  \citenamefont {Lunghi}, \citenamefont {Misiak},\ and\ \citenamefont
  {Wyler}}]{Huber:2005ig}%
  \BibitemOpen
  \bibfield  {author} {\bibinfo {author} {\bibfnamefont {T.}~\bibnamefont
  {Huber}}, \bibinfo {author} {\bibfnamefont {E.}~\bibnamefont {Lunghi}},
  \bibinfo {author} {\bibfnamefont {M.}~\bibnamefont {Misiak}}, \ and\ \bibinfo
  {author} {\bibfnamefont {D.}~\bibnamefont {Wyler}},\ }\href {\doibase
  10.1016/j.nuclphysb.2006.01.037} {\bibfield  {journal} {\bibinfo  {journal}
  {Nucl. Phys.}\ }\textbf {\bibinfo {volume} {B740}},\ \bibinfo {pages} {105}
  (\bibinfo {year} {2006})},\ \Eprint {http://arxiv.org/abs/hep-ph/0512066}
  {arXiv:hep-ph/0512066} \BibitemShut {NoStop}%
\bibitem [{\citenamefont {Altmannshofer}\ \emph
  {et~al.}(2009{\natexlab{a}})\citenamefont {Altmannshofer}, \citenamefont
  {Ball}, \citenamefont {Bharucha}, \citenamefont {Buras}, \citenamefont
  {Straub},\ and\ \citenamefont {Wick}}]{Altmannshofer:2008dz}%
  \BibitemOpen
  \bibfield  {author} {\bibinfo {author} {\bibfnamefont {W.}~\bibnamefont
  {Altmannshofer}}, \bibinfo {author} {\bibfnamefont {P.}~\bibnamefont {Ball}},
  \bibinfo {author} {\bibfnamefont {A.}~\bibnamefont {Bharucha}}, \bibinfo
  {author} {\bibfnamefont {A.~J.}\ \bibnamefont {Buras}}, \bibinfo {author}
  {\bibfnamefont {D.~M.}\ \bibnamefont {Straub}}, \ and\ \bibinfo {author}
  {\bibfnamefont {M.}~\bibnamefont {Wick}},\ }\href {\doibase
  10.1088/1126-6708/2009/01/019} {\bibfield  {journal} {\bibinfo  {journal}
  {JHEP}\ }\textbf {\bibinfo {volume} {0901}},\ \bibinfo {pages} {019}
  (\bibinfo {year} {2009}{\natexlab{a}})},\ \Eprint
  {http://arxiv.org/abs/0811.1214} {arXiv:0811.1214 [hep-ph]} \BibitemShut
  {NoStop}%
\bibitem [{\citenamefont {Misiak}(1993)}]{Misiak:1992bc}%
  \BibitemOpen
  \bibfield  {author} {\bibinfo {author} {\bibfnamefont {M.}~\bibnamefont
  {Misiak}},\ }\href {\doibase 10.1016/0550-3213(93)90235-H} {\bibfield
  {journal} {\bibinfo  {journal} {Nucl. Phys.}\ }\textbf {\bibinfo {volume}
  {B393}},\ \bibinfo {pages} {23} (\bibinfo {year} {1993})}\BibitemShut
  {NoStop}%
\bibitem [{\citenamefont {Buras}\ and\ \citenamefont
  {M{\"u}nz}(1995)}]{Buras:1994dj}%
  \BibitemOpen
  \bibfield  {author} {\bibinfo {author} {\bibfnamefont {A.~J.}\ \bibnamefont
  {Buras}}\ and\ \bibinfo {author} {\bibfnamefont {M.}~\bibnamefont
  {M{\"u}nz}},\ }\href {\doibase 10.1103/PhysRevD.52.186} {\bibfield  {journal}
  {\bibinfo  {journal} {Phys. Rev.}\ }\textbf {\bibinfo {volume} {D52}},\
  \bibinfo {pages} {186} (\bibinfo {year} {1995})},\ \Eprint
  {http://arxiv.org/abs/hep-ph/9501281} {arXiv:hep-ph/9501281} \BibitemShut
  {NoStop}%
\bibitem [{\citenamefont {Aaij}\ \emph
  {et~al.}(2013{\natexlab{b}})\citenamefont {Aaij} \emph
  {et~al.}}]{Aaij:2013pta}%
  \BibitemOpen
  \bibfield  {author} {\bibinfo {author} {\bibfnamefont {R.}~\bibnamefont
  {Aaij}} \emph {et~al.} (\bibinfo {collaboration} {LHCb}),\ }\href {\doibase
  10.1103/PhysRevLett.111.112003} {\bibfield  {journal} {\bibinfo  {journal}
  {Phys. Rev. Lett.}\ }\textbf {\bibinfo {volume} {111}},\ \bibinfo {pages}
  {112003} (\bibinfo {year} {2013}{\natexlab{b}})},\ \Eprint
  {http://arxiv.org/abs/1307.7595} {arXiv:1307.7595 [hep-ex]} \BibitemShut
  {NoStop}%
\bibitem [{\citenamefont {Khodjamirian}\ \emph {et~al.}(2013)\citenamefont
  {Khodjamirian}, \citenamefont {Mannel},\ and\ \citenamefont
  {Wang}}]{Khodjamirian:2012rm}%
  \BibitemOpen
  \bibfield  {author} {\bibinfo {author} {\bibfnamefont {A.}~\bibnamefont
  {Khodjamirian}}, \bibinfo {author} {\bibfnamefont {T.}~\bibnamefont
  {Mannel}}, \ and\ \bibinfo {author} {\bibfnamefont {Y.~M.}\ \bibnamefont
  {Wang}},\ }\href {\doibase 10.1007/JHEP02(2013)010} {\bibfield  {journal}
  {\bibinfo  {journal} {JHEP}\ }\textbf {\bibinfo {volume} {02}},\ \bibinfo
  {pages} {010} (\bibinfo {year} {2013})},\ \Eprint
  {http://arxiv.org/abs/1211.0234} {arXiv:1211.0234 [hep-ph]} \BibitemShut
  {NoStop}%
\bibitem [{\citenamefont {Hambrock}\ \emph {et~al.}(2015)\citenamefont
  {Hambrock}, \citenamefont {Khodjamirian},\ and\ \citenamefont
  {Rusov}}]{Hambrock:2015wka}%
  \BibitemOpen
  \bibfield  {author} {\bibinfo {author} {\bibfnamefont {C.}~\bibnamefont
  {Hambrock}}, \bibinfo {author} {\bibfnamefont {A.}~\bibnamefont
  {Khodjamirian}}, \ and\ \bibinfo {author} {\bibfnamefont {A.}~\bibnamefont
  {Rusov}},\ }\href {\doibase 10.1103/PhysRevD.92.074020} {\bibfield  {journal}
  {\bibinfo  {journal} {Phys. Rev.}\ }\textbf {\bibinfo {volume} {D92}},\
  \bibinfo {pages} {074020} (\bibinfo {year} {2015})},\ \Eprint
  {http://arxiv.org/abs/1506.07760} {arXiv:1506.07760 [hep-ph]} \BibitemShut
  {NoStop}%
\bibitem [{\citenamefont {Khodjamirian}\ \emph {et~al.}(2010)\citenamefont
  {Khodjamirian}, \citenamefont {Mannel}, \citenamefont {Pivovarov},\ and\
  \citenamefont {Wang}}]{Khodjamirian:2010vf}%
  \BibitemOpen
  \bibfield  {author} {\bibinfo {author} {\bibfnamefont {A.}~\bibnamefont
  {Khodjamirian}}, \bibinfo {author} {\bibfnamefont {T.}~\bibnamefont
  {Mannel}}, \bibinfo {author} {\bibfnamefont {A.~A.}\ \bibnamefont
  {Pivovarov}}, \ and\ \bibinfo {author} {\bibfnamefont {Y.-M.}\ \bibnamefont
  {Wang}},\ }\href {\doibase 10.1007/JHEP09(2010)089} {\bibfield  {journal}
  {\bibinfo  {journal} {JHEP}\ }\textbf {\bibinfo {volume} {1009}},\ \bibinfo
  {pages} {089} (\bibinfo {year} {2010})},\ \Eprint
  {http://arxiv.org/abs/1006.4945} {arXiv:1006.4945 [hep-ph]} \BibitemShut
  {NoStop}%
\bibitem [{\citenamefont {Bobeth}\ \emph {et~al.}(2011)\citenamefont {Bobeth},
  \citenamefont {Hiller},\ and\ \citenamefont {van Dyk}}]{Bobeth:2011gi}%
  \BibitemOpen
  \bibfield  {author} {\bibinfo {author} {\bibfnamefont {C.}~\bibnamefont
  {Bobeth}}, \bibinfo {author} {\bibfnamefont {G.}~\bibnamefont {Hiller}}, \
  and\ \bibinfo {author} {\bibfnamefont {D.}~\bibnamefont {van Dyk}},\ }\href
  {\doibase 10.1007/JHEP07(2011)067} {\bibfield  {journal} {\bibinfo  {journal}
  {JHEP}\ }\textbf {\bibinfo {volume} {1107}},\ \bibinfo {pages} {067}
  (\bibinfo {year} {2011})},\ \Eprint {http://arxiv.org/abs/1105.0376}
  {arXiv:1105.0376 [hep-ph]} \BibitemShut {NoStop}%
\bibitem [{\citenamefont {Bobeth}\ \emph {et~al.}(2013)\citenamefont {Bobeth},
  \citenamefont {Hiller},\ and\ \citenamefont {van Dyk}}]{Bobeth:2012vn}%
  \BibitemOpen
  \bibfield  {author} {\bibinfo {author} {\bibfnamefont {C.}~\bibnamefont
  {Bobeth}}, \bibinfo {author} {\bibfnamefont {G.}~\bibnamefont {Hiller}}, \
  and\ \bibinfo {author} {\bibfnamefont {D.}~\bibnamefont {van Dyk}},\ }\href
  {\doibase 10.1103/PhysRevD.87.034016} {\bibfield  {journal} {\bibinfo
  {journal} {Phys. Rev.}\ }\textbf {\bibinfo {volume} {D87}},\ \bibinfo {pages}
  {034016} (\bibinfo {year} {2013})},\ \Eprint {http://arxiv.org/abs/1212.2321}
  {arXiv:1212.2321 [hep-ph]} \BibitemShut {NoStop}%
\bibitem [{\citenamefont {Grinstein}\ and\ \citenamefont
  {Pirjol}(2002)}]{Grinstein:2002cz}%
  \BibitemOpen
  \bibfield  {author} {\bibinfo {author} {\bibfnamefont {B.}~\bibnamefont
  {Grinstein}}\ and\ \bibinfo {author} {\bibfnamefont {D.}~\bibnamefont
  {Pirjol}},\ }\href {\doibase 10.1016/S0370-2693(02)01601-5} {\bibfield
  {journal} {\bibinfo  {journal} {Phys. Lett.}\ }\textbf {\bibinfo {volume}
  {B533}},\ \bibinfo {pages} {8} (\bibinfo {year} {2002})},\ \Eprint
  {http://arxiv.org/abs/hep-ph/0201298} {arXiv:hep-ph/0201298} \BibitemShut
  {NoStop}%
\bibitem [{\citenamefont {Hewett}\ \emph {et~al.}(2004)\citenamefont {Hewett}
  \emph {et~al.}}]{Hewett:2004tv}%
  \BibitemOpen
  \bibinfo {editor} {\bibfnamefont {J.~L.}\ \bibnamefont {Hewett}} \emph
  {et~al.},\ eds.,\ \href@noop {} {\emph {\bibinfo {title} {{The Discovery
  potential of a Super $B$ Factory}}}}\ (\bibinfo  {publisher} {SLAC},\
  \bibinfo {address} {Menlo Park},\ \bibinfo {year} {2004})\ \Eprint
  {http://arxiv.org/abs/hep-ph/0503261} {arXiv:hep-ph/0503261} \BibitemShut
  {NoStop}%
\bibitem [{\citenamefont {Beneke}\ \emph {et~al.}(1999)\citenamefont {Beneke},
  \citenamefont {Buchalla}, \citenamefont {Neubert},\ and\ \citenamefont
  {Sachrajda}}]{Beneke:1999br}%
  \BibitemOpen
  \bibfield  {author} {\bibinfo {author} {\bibfnamefont {M.}~\bibnamefont
  {Beneke}}, \bibinfo {author} {\bibfnamefont {G.}~\bibnamefont {Buchalla}},
  \bibinfo {author} {\bibfnamefont {M.}~\bibnamefont {Neubert}}, \ and\
  \bibinfo {author} {\bibfnamefont {C.~T.}\ \bibnamefont {Sachrajda}},\ }\href
  {\doibase 10.1103/PhysRevLett.83.1914} {\bibfield  {journal} {\bibinfo
  {journal} {Phys. Rev. Lett.}\ }\textbf {\bibinfo {volume} {83}},\ \bibinfo
  {pages} {1914} (\bibinfo {year} {1999})},\ \Eprint
  {http://arxiv.org/abs/hep-ph/9905312} {arXiv:hep-ph/9905312} \BibitemShut
  {NoStop}%
\bibitem [{\citenamefont {Beneke}\ \emph {et~al.}(2000)\citenamefont {Beneke},
  \citenamefont {Buchalla}, \citenamefont {Neubert},\ and\ \citenamefont
  {Sachrajda}}]{Beneke:2000ry}%
  \BibitemOpen
  \bibfield  {author} {\bibinfo {author} {\bibfnamefont {M.}~\bibnamefont
  {Beneke}}, \bibinfo {author} {\bibfnamefont {G.}~\bibnamefont {Buchalla}},
  \bibinfo {author} {\bibfnamefont {M.}~\bibnamefont {Neubert}}, \ and\
  \bibinfo {author} {\bibfnamefont {C.~T.}\ \bibnamefont {Sachrajda}},\ }\href
  {\doibase 10.1016/S0550-3213(00)00559-9} {\bibfield  {journal} {\bibinfo
  {journal} {Nucl. Phys.}\ }\textbf {\bibinfo {volume} {B591}},\ \bibinfo
  {pages} {313} (\bibinfo {year} {2000})},\ \Eprint
  {http://arxiv.org/abs/hep-ph/0006124} {arXiv:hep-ph/0006124} \BibitemShut
  {NoStop}%
\bibitem [{\citenamefont {Bauer}\ \emph {et~al.}(2001)\citenamefont {Bauer},
  \citenamefont {Fleming}, \citenamefont {Pirjol},\ and\ \citenamefont
  {Stewart}}]{Bauer:2000yr}%
  \BibitemOpen
  \bibfield  {author} {\bibinfo {author} {\bibfnamefont {C.~W.}\ \bibnamefont
  {Bauer}}, \bibinfo {author} {\bibfnamefont {S.}~\bibnamefont {Fleming}},
  \bibinfo {author} {\bibfnamefont {D.}~\bibnamefont {Pirjol}}, \ and\ \bibinfo
  {author} {\bibfnamefont {I.~W.}\ \bibnamefont {Stewart}},\ }\href {\doibase
  10.1103/PhysRevD.63.114020} {\bibfield  {journal} {\bibinfo  {journal} {Phys.
  Rev.}\ }\textbf {\bibinfo {volume} {D63}},\ \bibinfo {pages} {114020}
  (\bibinfo {year} {2001})},\ \Eprint {http://arxiv.org/abs/hep-ph/0011336}
  {arXiv:hep-ph/0011336} \BibitemShut {NoStop}%
\bibitem [{\citenamefont {Bauer}\ \emph {et~al.}(2002)\citenamefont {Bauer},
  \citenamefont {Pirjol},\ and\ \citenamefont {Stewart}}]{Bauer:2001yt}%
  \BibitemOpen
  \bibfield  {author} {\bibinfo {author} {\bibfnamefont {C.~W.}\ \bibnamefont
  {Bauer}}, \bibinfo {author} {\bibfnamefont {D.}~\bibnamefont {Pirjol}}, \
  and\ \bibinfo {author} {\bibfnamefont {I.~W.}\ \bibnamefont {Stewart}},\
  }\href {\doibase 10.1103/PhysRevD.65.054022} {\bibfield  {journal} {\bibinfo
  {journal} {Phys. Rev.}\ }\textbf {\bibinfo {volume} {D65}},\ \bibinfo {pages}
  {054022} (\bibinfo {year} {2002})},\ \Eprint
  {http://arxiv.org/abs/hep-ph/0109045} {arXiv:hep-ph/0109045} \BibitemShut
  {NoStop}%
\bibitem [{\citenamefont {Beneke}\ and\ \citenamefont
  {Feldmann}(2001)}]{Beneke:2000wa}%
  \BibitemOpen
  \bibfield  {author} {\bibinfo {author} {\bibfnamefont {M.}~\bibnamefont
  {Beneke}}\ and\ \bibinfo {author} {\bibfnamefont {T.}~\bibnamefont
  {Feldmann}},\ }\href {\doibase 10.1016/S0550-3213(00)00585-X} {\bibfield
  {journal} {\bibinfo  {journal} {Nucl. Phys.}\ }\textbf {\bibinfo {volume}
  {B592}},\ \bibinfo {pages} {3} (\bibinfo {year} {2001})},\ \Eprint
  {http://arxiv.org/abs/hep-ph/0008255} {arXiv:hep-ph/0008255} \BibitemShut
  {NoStop}%
\bibitem [{\citenamefont {Beneke}\ \emph {et~al.}(2001)\citenamefont {Beneke},
  \citenamefont {Feldmann},\ and\ \citenamefont {Seidel}}]{Beneke:2001at}%
  \BibitemOpen
  \bibfield  {author} {\bibinfo {author} {\bibfnamefont {M.}~\bibnamefont
  {Beneke}}, \bibinfo {author} {\bibfnamefont {T.}~\bibnamefont {Feldmann}}, \
  and\ \bibinfo {author} {\bibfnamefont {D.}~\bibnamefont {Seidel}},\ }\href
  {\doibase 10.1016/S0550-3213(01)00366-2} {\bibfield  {journal} {\bibinfo
  {journal} {Nucl. Phys.}\ }\textbf {\bibinfo {volume} {B612}},\ \bibinfo
  {pages} {25} (\bibinfo {year} {2001})},\ \Eprint
  {http://arxiv.org/abs/hep-ph/0106067} {arXiv:hep-ph/0106067} \BibitemShut
  {NoStop}%
\bibitem [{\citenamefont {Bobeth}\ \emph {et~al.}(2007)\citenamefont {Bobeth},
  \citenamefont {Hiller},\ and\ \citenamefont {Piranishvili}}]{Bobeth:2007dw}%
  \BibitemOpen
  \bibfield  {author} {\bibinfo {author} {\bibfnamefont {C.}~\bibnamefont
  {Bobeth}}, \bibinfo {author} {\bibfnamefont {G.}~\bibnamefont {Hiller}}, \
  and\ \bibinfo {author} {\bibfnamefont {G.}~\bibnamefont {Piranishvili}},\
  }\href {\doibase 10.1088/1126-6708/2007/12/040} {\bibfield  {journal}
  {\bibinfo  {journal} {JHEP}\ }\textbf {\bibinfo {volume} {0712}},\ \bibinfo
  {pages} {040} (\bibinfo {year} {2007})},\ \Eprint
  {http://arxiv.org/abs/0709.4174} {arXiv:0709.4174 [hep-ph]} \BibitemShut
  {NoStop}%
\bibitem [{\citenamefont {Buchalla}\ and\ \citenamefont
  {Buras}(1993)}]{Buchalla:1993bv}%
  \BibitemOpen
  \bibfield  {author} {\bibinfo {author} {\bibfnamefont {G.}~\bibnamefont
  {Buchalla}}\ and\ \bibinfo {author} {\bibfnamefont {A.~J.}\ \bibnamefont
  {Buras}},\ }\href {\doibase 10.1016/0550-3213(93)90405-E} {\bibfield
  {journal} {\bibinfo  {journal} {Nucl. Phys.}\ }\textbf {\bibinfo {volume}
  {B400}},\ \bibinfo {pages} {225} (\bibinfo {year} {1993})}\BibitemShut
  {NoStop}%
\bibitem [{\citenamefont {Misiak}\ and\ \citenamefont
  {Urban}(1999)}]{Misiak:1999yg}%
  \BibitemOpen
  \bibfield  {author} {\bibinfo {author} {\bibfnamefont {M.}~\bibnamefont
  {Misiak}}\ and\ \bibinfo {author} {\bibfnamefont {J.}~\bibnamefont {Urban}},\
  }\href {\doibase 10.1016/S0370-2693(99)00150-1} {\bibfield  {journal}
  {\bibinfo  {journal} {Phys. Lett.}\ }\textbf {\bibinfo {volume} {B451}},\
  \bibinfo {pages} {161} (\bibinfo {year} {1999})},\ \Eprint
  {http://arxiv.org/abs/hep-ph/9901278} {arXiv:hep-ph/9901278} \BibitemShut
  {NoStop}%
\bibitem [{\citenamefont {Buchalla}\ and\ \citenamefont
  {Buras}(1999)}]{Buchalla:1998ba}%
  \BibitemOpen
  \bibfield  {author} {\bibinfo {author} {\bibfnamefont {G.}~\bibnamefont
  {Buchalla}}\ and\ \bibinfo {author} {\bibfnamefont {A.~J.}\ \bibnamefont
  {Buras}},\ }\href {\doibase 10.1016/S0550-3213(99)00149-2} {\bibfield
  {journal} {\bibinfo  {journal} {Nucl. Phys.}\ }\textbf {\bibinfo {volume}
  {B548}},\ \bibinfo {pages} {309} (\bibinfo {year} {1999})},\ \Eprint
  {http://arxiv.org/abs/hep-ph/9901288} {arXiv:hep-ph/9901288} \BibitemShut
  {NoStop}%
\bibitem [{\citenamefont {Brod}\ \emph {et~al.}(2011)\citenamefont {Brod},
  \citenamefont {Gorbahn},\ and\ \citenamefont {Stamou}}]{Brod:2010hi}%
  \BibitemOpen
  \bibfield  {author} {\bibinfo {author} {\bibfnamefont {J.}~\bibnamefont
  {Brod}}, \bibinfo {author} {\bibfnamefont {M.}~\bibnamefont {Gorbahn}}, \
  and\ \bibinfo {author} {\bibfnamefont {E.}~\bibnamefont {Stamou}},\ }\href
  {\doibase 10.1103/PhysRevD.83.034030} {\bibfield  {journal} {\bibinfo
  {journal} {Phys. Rev.}\ }\textbf {\bibinfo {volume} {D83}},\ \bibinfo {pages}
  {034030} (\bibinfo {year} {2011})},\ \Eprint {http://arxiv.org/abs/1009.0947}
  {arXiv:1009.0947 [hep-ph]} \BibitemShut {NoStop}%
\bibitem [{\citenamefont {Altmannshofer}\ \emph
  {et~al.}(2009{\natexlab{b}})\citenamefont {Altmannshofer}, \citenamefont
  {Buras}, \citenamefont {Straub},\ and\ \citenamefont
  {Wick}}]{Altmannshofer:2009ma}%
  \BibitemOpen
  \bibfield  {author} {\bibinfo {author} {\bibfnamefont {W.}~\bibnamefont
  {Altmannshofer}}, \bibinfo {author} {\bibfnamefont {A.~J.}\ \bibnamefont
  {Buras}}, \bibinfo {author} {\bibfnamefont {D.~M.}\ \bibnamefont {Straub}}, \
  and\ \bibinfo {author} {\bibfnamefont {M.}~\bibnamefont {Wick}},\ }\href
  {\doibase 10.1088/1126-6708/2009/04/022} {\bibfield  {journal} {\bibinfo
  {journal} {JHEP}\ }\textbf {\bibinfo {volume} {0904}},\ \bibinfo {pages}
  {022} (\bibinfo {year} {2009}{\natexlab{b}})},\ \Eprint
  {http://arxiv.org/abs/0902.0160} {arXiv:0902.0160 [hep-ph]} \BibitemShut
  {NoStop}%
\bibitem [{\citenamefont {Kamenik}\ and\ \citenamefont
  {Smith}(2009)}]{Kamenik:2009kc}%
  \BibitemOpen
  \bibfield  {author} {\bibinfo {author} {\bibfnamefont {J.~F.}\ \bibnamefont
  {Kamenik}}\ and\ \bibinfo {author} {\bibfnamefont {C.}~\bibnamefont
  {Smith}},\ }\href {\doibase 10.1016/j.physletb.2009.09.041} {\bibfield
  {journal} {\bibinfo  {journal} {Phys. Lett.}\ }\textbf {\bibinfo {volume}
  {B680}},\ \bibinfo {pages} {471} (\bibinfo {year} {2009})},\ \Eprint
  {http://arxiv.org/abs/0908.1174} {arXiv:0908.1174 [hep-ph]} \BibitemShut
  {NoStop}%
\bibitem [{\citenamefont {Rosner}\ \emph {et~al.}(2015)\citenamefont {Rosner},
  \citenamefont {Stone},\ and\ \citenamefont {Van~de Water}}]{Rosner:2015wva}%
  \BibitemOpen
  \bibfield  {author} {\bibinfo {author} {\bibfnamefont {J.~L.}\ \bibnamefont
  {Rosner}}, \bibinfo {author} {\bibfnamefont {S.}~\bibnamefont {Stone}}, \
  and\ \bibinfo {author} {\bibfnamefont {R.~S.}\ \bibnamefont {Van~de Water}},\
  }\href@noop {} {\  (\bibinfo {year} {2015})},\ \Eprint
  {http://arxiv.org/abs/1509.02220} {arXiv:1509.02220 [hep-ph]} \BibitemShut
  {NoStop}%
\bibitem [{\citenamefont {Bazavov}\ \emph
  {et~al.}(2012{\natexlab{a}})\citenamefont {Bazavov} \emph
  {et~al.}}]{Bazavov:2011aa}%
  \BibitemOpen
  \bibfield  {author} {\bibinfo {author} {\bibfnamefont {A.}~\bibnamefont
  {Bazavov}} \emph {et~al.} (\bibinfo {collaboration} {Fermilab Lattice and
  MILC Collaborations}),\ }\href {\doibase 10.1103/PhysRevD.85.114506}
  {\bibfield  {journal} {\bibinfo  {journal} {Phys. Rev.}\ }\textbf {\bibinfo
  {volume} {D85}},\ \bibinfo {pages} {114506} (\bibinfo {year}
  {2012}{\natexlab{a}})},\ \Eprint {http://arxiv.org/abs/1112.3051}
  {arXiv:1112.3051 [hep-lat]} \BibitemShut {NoStop}%
\bibitem [{\citenamefont {Na}\ \emph {et~al.}(2012)\citenamefont {Na},
  \citenamefont {Monahan}, \citenamefont {Davies}, \citenamefont {Horgan},
  \citenamefont {Lepage},\ and\ \citenamefont {Shigemitsu}}]{Na:2012kp}%
  \BibitemOpen
  \bibfield  {author} {\bibinfo {author} {\bibfnamefont {H.}~\bibnamefont
  {Na}}, \bibinfo {author} {\bibfnamefont {C.~J.}\ \bibnamefont {Monahan}},
  \bibinfo {author} {\bibfnamefont {C.~T.~H.}\ \bibnamefont {Davies}}, \bibinfo
  {author} {\bibfnamefont {R.}~\bibnamefont {Horgan}}, \bibinfo {author}
  {\bibfnamefont {G.~P.}\ \bibnamefont {Lepage}}, \ and\ \bibinfo {author}
  {\bibfnamefont {J.}~\bibnamefont {Shigemitsu}} (\bibinfo {collaboration}
  {HPQCD Collaboration}),\ }\href {\doibase 10.1103/PhysRevD.86.034506}
  {\bibfield  {journal} {\bibinfo  {journal} {Phys. Rev.}\ }\textbf {\bibinfo
  {volume} {D86}},\ \bibinfo {pages} {034506} (\bibinfo {year} {2012})},\
  \Eprint {http://arxiv.org/abs/1202.4914} {arXiv:1202.4914 [hep-lat]}
  \BibitemShut {NoStop}%
\bibitem [{\citenamefont {Dowdall}\ \emph
  {et~al.}(2013{\natexlab{a}})\citenamefont {Dowdall}, \citenamefont {Davies},
  \citenamefont {Horgan}, \citenamefont {Monahan},\ and\ \citenamefont
  {Shigemitsu}}]{Dowdall:2013tga}%
  \BibitemOpen
  \bibfield  {author} {\bibinfo {author} {\bibfnamefont {R.~J.}\ \bibnamefont
  {Dowdall}}, \bibinfo {author} {\bibfnamefont {C.~T.~H.}\ \bibnamefont
  {Davies}}, \bibinfo {author} {\bibfnamefont {R.~R.}\ \bibnamefont {Horgan}},
  \bibinfo {author} {\bibfnamefont {C.~J.}\ \bibnamefont {Monahan}}, \ and\
  \bibinfo {author} {\bibfnamefont {J.}~\bibnamefont {Shigemitsu}} (\bibinfo
  {collaboration} {HPQCD Collaboration}),\ }\href {\doibase
  10.1103/PhysRevLett.110.222003} {\bibfield  {journal} {\bibinfo  {journal}
  {Phys. Rev. Lett.}\ }\textbf {\bibinfo {volume} {110}},\ \bibinfo {pages}
  {222003} (\bibinfo {year} {2013}{\natexlab{a}})},\ \Eprint
  {http://arxiv.org/abs/1302.2644} {arXiv:1302.2644 [hep-lat]} \BibitemShut
  {NoStop}%
\bibitem [{\citenamefont {Carrasco}\ \emph {et~al.}(2014)\citenamefont
  {Carrasco} \emph {et~al.}}]{Carrasco:2013naa}%
  \BibitemOpen
  \bibfield  {author} {\bibinfo {author} {\bibfnamefont {N.}~\bibnamefont
  {Carrasco}} \emph {et~al.} (\bibinfo {collaboration} {ETM Collaboration}),\
  }\href@noop {} {\bibfield  {journal} {\bibinfo  {journal} {PoS}\ }\textbf
  {\bibinfo {volume} {LATTICE2013}},\ \bibinfo {pages} {313} (\bibinfo {year}
  {2014})},\ \Eprint {http://arxiv.org/abs/1311.2837} {arXiv:1311.2837
  [hep-lat]} \BibitemShut {NoStop}%
\bibitem [{\citenamefont {Christ}\ \emph {et~al.}(2015)\citenamefont {Christ},
  \citenamefont {Flynn}, \citenamefont {Izubuchi}, \citenamefont {Kawanai},
  \citenamefont {Lehner}, \citenamefont {Soni}, \citenamefont {Van~de Water},\
  and\ \citenamefont {Witzel}}]{Christ:2014uea}%
  \BibitemOpen
  \bibfield  {author} {\bibinfo {author} {\bibfnamefont {N.~H.}\ \bibnamefont
  {Christ}}, \bibinfo {author} {\bibfnamefont {J.~M.}\ \bibnamefont {Flynn}},
  \bibinfo {author} {\bibfnamefont {T.}~\bibnamefont {Izubuchi}}, \bibinfo
  {author} {\bibfnamefont {T.}~\bibnamefont {Kawanai}}, \bibinfo {author}
  {\bibfnamefont {C.}~\bibnamefont {Lehner}}, \bibinfo {author} {\bibfnamefont
  {A.}~\bibnamefont {Soni}}, \bibinfo {author} {\bibfnamefont {R.~S.}\
  \bibnamefont {Van~de Water}}, \ and\ \bibinfo {author} {\bibfnamefont
  {O.}~\bibnamefont {Witzel}} (\bibinfo {collaboration} {RBC and UKQCD
  Collaborations}),\ }\href {\doibase 10.1103/PhysRevD.91.054502} {\bibfield
  {journal} {\bibinfo  {journal} {Phys. Rev.}\ }\textbf {\bibinfo {volume}
  {D91}},\ \bibinfo {pages} {054502} (\bibinfo {year} {2015})},\ \Eprint
  {http://arxiv.org/abs/1404.4670} {arXiv:1404.4670 [hep-lat]} \BibitemShut
  {NoStop}%
\bibitem [{\citenamefont {Aoki}\ \emph {et~al.}(2015)\citenamefont {Aoki},
  \citenamefont {Ishikawa}, \citenamefont {Izubuchi}, \citenamefont {Lehner},\
  and\ \citenamefont {Soni}}]{Aoki:2014nga}%
  \BibitemOpen
  \bibfield  {author} {\bibinfo {author} {\bibfnamefont {Y.}~\bibnamefont
  {Aoki}}, \bibinfo {author} {\bibfnamefont {T.}~\bibnamefont {Ishikawa}},
  \bibinfo {author} {\bibfnamefont {T.}~\bibnamefont {Izubuchi}}, \bibinfo
  {author} {\bibfnamefont {C.}~\bibnamefont {Lehner}}, \ and\ \bibinfo {author}
  {\bibfnamefont {A.}~\bibnamefont {Soni}},\ }\href {\doibase
  10.1103/PhysRevD.91.114505} {\bibfield  {journal} {\bibinfo  {journal} {Phys.
  Rev.}\ }\textbf {\bibinfo {volume} {D91}},\ \bibinfo {pages} {114505}
  (\bibinfo {year} {2015})},\ \Eprint {http://arxiv.org/abs/1406.6192}
  {arXiv:1406.6192 [hep-lat]} \BibitemShut {NoStop}%
\bibitem [{\citenamefont {Follana}\ \emph {et~al.}(2008)\citenamefont
  {Follana}, \citenamefont {Davies}, \citenamefont {Lepage},\ and\
  \citenamefont {Shigemitsu}}]{Follana:2007uv}%
  \BibitemOpen
  \bibfield  {author} {\bibinfo {author} {\bibfnamefont {E.}~\bibnamefont
  {Follana}}, \bibinfo {author} {\bibfnamefont {C.~T.~H.}\ \bibnamefont
  {Davies}}, \bibinfo {author} {\bibfnamefont {G.~P.}\ \bibnamefont {Lepage}},
  \ and\ \bibinfo {author} {\bibfnamefont {J.}~\bibnamefont {Shigemitsu}}
  (\bibinfo {collaboration} {HPQCD, UKQCD}),\ }\href {\doibase
  10.1103/PhysRevLett.100.062002} {\bibfield  {journal} {\bibinfo  {journal}
  {Phys. Rev. Lett.}\ }\textbf {\bibinfo {volume} {100}},\ \bibinfo {pages}
  {062002} (\bibinfo {year} {2008})},\ \Eprint {http://arxiv.org/abs/0706.1726}
  {arXiv:0706.1726 [hep-lat]} \BibitemShut {NoStop}%
\bibitem [{\citenamefont {{D\"urr}}\ \emph {et~al.}(2010)\citenamefont
  {{D\"urr}}, \citenamefont {Fodor}, \citenamefont {Hoelbling}, \citenamefont
  {Katz}, \citenamefont {Krieg}, \citenamefont {Kurth}, \citenamefont
  {Lellouch}, \citenamefont {Lippert}, \citenamefont {Ramos},\ and\
  \citenamefont {Szabo}}]{Durr:2010hr}%
  \BibitemOpen
  \bibfield  {author} {\bibinfo {author} {\bibfnamefont {S.}~\bibnamefont
  {{D\"urr}}}, \bibinfo {author} {\bibfnamefont {Z.}~\bibnamefont {Fodor}},
  \bibinfo {author} {\bibfnamefont {C.}~\bibnamefont {Hoelbling}}, \bibinfo
  {author} {\bibfnamefont {S.~D.}\ \bibnamefont {Katz}}, \bibinfo {author}
  {\bibfnamefont {S.}~\bibnamefont {Krieg}}, \bibinfo {author} {\bibfnamefont
  {T.}~\bibnamefont {Kurth}}, \bibinfo {author} {\bibfnamefont
  {L.}~\bibnamefont {Lellouch}}, \bibinfo {author} {\bibfnamefont
  {T.}~\bibnamefont {Lippert}}, \bibinfo {author} {\bibfnamefont
  {A.}~\bibnamefont {Ramos}}, \ and\ \bibinfo {author} {\bibfnamefont {K.~K.}\
  \bibnamefont {Szabo}} (\bibinfo {collaboration} {Budapest-Marseille-Wuppertal
  Collaboration}),\ }\href {\doibase 10.1103/PhysRevD.81.054507} {\bibfield
  {journal} {\bibinfo  {journal} {Phys. Rev.}\ }\textbf {\bibinfo {volume}
  {D81}},\ \bibinfo {pages} {054507} (\bibinfo {year} {2010})},\ \Eprint
  {http://arxiv.org/abs/1001.4692} {arXiv:1001.4692 [hep-lat]} \BibitemShut
  {NoStop}%
\bibitem [{\citenamefont {Bazavov}\ \emph
  {et~al.}(2010{\natexlab{a}})\citenamefont {Bazavov} \emph
  {et~al.}}]{Bazavov:2010hj}%
  \BibitemOpen
  \bibfield  {author} {\bibinfo {author} {\bibfnamefont {A.}~\bibnamefont
  {Bazavov}} \emph {et~al.} (\bibinfo {collaboration} {MILC}),\ }\bibfield
  {booktitle} {\emph {\bibinfo {booktitle} {{Proceedings, 28th International
  Symposium on Lattice field theory (Lattice 2010)}}},\ }\href@noop {}
  {\bibfield  {journal} {\bibinfo  {journal} {PoS}\ }\textbf {\bibinfo {volume}
  {LATTICE2010}},\ \bibinfo {pages} {074} (\bibinfo {year}
  {2010}{\natexlab{a}})},\ \Eprint {http://arxiv.org/abs/1012.0868}
  {arXiv:1012.0868 [hep-lat]} \BibitemShut {NoStop}%
\bibitem [{\citenamefont {Laiho}\ and\ \citenamefont {Van~de
  Water}(2011)}]{Laiho:2011np}%
  \BibitemOpen
  \bibfield  {author} {\bibinfo {author} {\bibfnamefont {J.}~\bibnamefont
  {Laiho}}\ and\ \bibinfo {author} {\bibfnamefont {R.~S.}\ \bibnamefont {Van~de
  Water}},\ }\bibfield  {booktitle} {\emph {\bibinfo {booktitle} {{Proceedings,
  29th International Symposium on Lattice field theory (Lattice 2011)}}},\
  }\href@noop {} {\bibfield  {journal} {\bibinfo  {journal} {PoS}\ }\textbf
  {\bibinfo {volume} {LATTICE2011}},\ \bibinfo {pages} {293} (\bibinfo {year}
  {2011})},\ \Eprint {http://arxiv.org/abs/1112.4861} {arXiv:1112.4861
  [hep-lat]} \BibitemShut {NoStop}%
\bibitem [{\citenamefont {Arthur}\ \emph {et~al.}(2013)\citenamefont {Arthur}
  \emph {et~al.}}]{Arthur:2012yc}%
  \BibitemOpen
  \bibfield  {author} {\bibinfo {author} {\bibfnamefont {R.}~\bibnamefont
  {Arthur}} \emph {et~al.} (\bibinfo {collaboration} {RBC/UKQCD}),\ }\href
  {\doibase 10.1103/PhysRevD.87.094514} {\bibfield  {journal} {\bibinfo
  {journal} {Phys. Rev.}\ }\textbf {\bibinfo {volume} {D87}},\ \bibinfo {pages}
  {094514} (\bibinfo {year} {2013})},\ \Eprint {http://arxiv.org/abs/1208.4412}
  {arXiv:1208.4412 [hep-lat]} \BibitemShut {NoStop}%
\bibitem [{\citenamefont {Dowdall}\ \emph
  {et~al.}(2013{\natexlab{b}})\citenamefont {Dowdall}, \citenamefont {Davies},
  \citenamefont {Lepage},\ and\ \citenamefont {McNeile}}]{Dowdall:2013rya}%
  \BibitemOpen
  \bibfield  {author} {\bibinfo {author} {\bibfnamefont {R.}~\bibnamefont
  {Dowdall}}, \bibinfo {author} {\bibfnamefont {C.}~\bibnamefont {Davies}},
  \bibinfo {author} {\bibfnamefont {G.}~\bibnamefont {Lepage}}, \ and\ \bibinfo
  {author} {\bibfnamefont {C.}~\bibnamefont {McNeile}} (\bibinfo
  {collaboration} {HPQCD}),\ }\href {\doibase 10.1103/PhysRevD.88.074504}
  {\bibfield  {journal} {\bibinfo  {journal} {Phys.Rev.}\ }\textbf {\bibinfo
  {volume} {D88}},\ \bibinfo {pages} {074504} (\bibinfo {year}
  {2013}{\natexlab{b}})},\ \Eprint {http://arxiv.org/abs/1303.1670}
  {arXiv:1303.1670 [hep-lat]} \BibitemShut {NoStop}%
\bibitem [{\citenamefont {Bazavov}\ \emph {et~al.}(2014)\citenamefont {Bazavov}
  \emph {et~al.}}]{Bazavov:2014wgs}%
  \BibitemOpen
  \bibfield  {author} {\bibinfo {author} {\bibfnamefont {A.}~\bibnamefont
  {Bazavov}} \emph {et~al.} (\bibinfo {collaboration} {Fermilab Lattice and
  MILC}),\ }\href {\doibase 10.1103/PhysRevD.90.074509} {\bibfield  {journal}
  {\bibinfo  {journal} {Phys. Rev.}\ }\textbf {\bibinfo {volume} {D90}},\
  \bibinfo {pages} {074509} (\bibinfo {year} {2014})},\ \Eprint
  {http://arxiv.org/abs/1407.3772} {arXiv:1407.3772 [hep-lat]} \BibitemShut
  {NoStop}%
\bibitem [{\citenamefont {Carrasco}\ \emph {et~al.}(2015)\citenamefont
  {Carrasco} \emph {et~al.}}]{Carrasco:2014poa}%
  \BibitemOpen
  \bibfield  {author} {\bibinfo {author} {\bibfnamefont {N.}~\bibnamefont
  {Carrasco}} \emph {et~al.} (\bibinfo {collaboration} {ETM}),\ }\href
  {\doibase 10.1103/PhysRevD.91.054507} {\bibfield  {journal} {\bibinfo
  {journal} {Phys. Rev.}\ }\textbf {\bibinfo {volume} {D91}},\ \bibinfo {pages}
  {054507} (\bibinfo {year} {2015})},\ \Eprint {http://arxiv.org/abs/1411.7908}
  {arXiv:1411.7908 [hep-lat]} \BibitemShut {NoStop}%
\bibitem [{\citenamefont {Ali}\ \emph {et~al.}(2014)\citenamefont {Ali},
  \citenamefont {Parkhomenko},\ and\ \citenamefont {Rusov}}]{Ali:2013zfa}%
  \BibitemOpen
  \bibfield  {author} {\bibinfo {author} {\bibfnamefont {A.}~\bibnamefont
  {Ali}}, \bibinfo {author} {\bibfnamefont {A.~Y.}\ \bibnamefont
  {Parkhomenko}}, \ and\ \bibinfo {author} {\bibfnamefont {A.~V.}\ \bibnamefont
  {Rusov}},\ }\href {\doibase 10.1103/PhysRevD.89.094021} {\bibfield  {journal}
  {\bibinfo  {journal} {Phys. Rev.}\ }\textbf {\bibinfo {volume} {D89}},\
  \bibinfo {pages} {094021} (\bibinfo {year} {2014})},\ \Eprint
  {http://arxiv.org/abs/1312.2523} {arXiv:1312.2523 [hep-ph]} \BibitemShut
  {NoStop}%
\bibitem [{\citenamefont {Bernard}\ \emph {et~al.}(2001)\citenamefont {Bernard}
  \emph {et~al.}}]{Bernard:2001av}%
  \BibitemOpen
  \bibfield  {author} {\bibinfo {author} {\bibfnamefont {C.~W.}\ \bibnamefont
  {Bernard}} \emph {et~al.},\ }\href {\doibase 10.1103/PhysRevD.64.054506}
  {\bibfield  {journal} {\bibinfo  {journal} {Phys. Rev.}\ }\textbf {\bibinfo
  {volume} {D64}},\ \bibinfo {pages} {054506} (\bibinfo {year} {2001})},\
  \Eprint {http://arxiv.org/abs/hep-lat/0104002} {arXiv:hep-lat/0104002}
  \BibitemShut {NoStop}%
\bibitem [{\citenamefont {Aubin}\ \emph {et~al.}(2004)\citenamefont {Aubin}
  \emph {et~al.}}]{Aubin:2004wf}%
  \BibitemOpen
  \bibfield  {author} {\bibinfo {author} {\bibfnamefont {C.}~\bibnamefont
  {Aubin}} \emph {et~al.},\ }\href {\doibase 10.1103/PhysRevD.70.094505}
  {\bibfield  {journal} {\bibinfo  {journal} {Phys. Rev.}\ }\textbf {\bibinfo
  {volume} {D70}},\ \bibinfo {pages} {094505} (\bibinfo {year} {2004})},\
  \Eprint {http://arxiv.org/abs/hep-lat/0402030} {arXiv:hep-lat/0402030
  [hep-lat]} \BibitemShut {NoStop}%
\bibitem [{\citenamefont {Bazavov}\ \emph
  {et~al.}(2010{\natexlab{b}})\citenamefont {Bazavov} \emph
  {et~al.}}]{Bazavov:2009bb}%
  \BibitemOpen
  \bibfield  {author} {\bibinfo {author} {\bibfnamefont {A.}~\bibnamefont
  {Bazavov}} \emph {et~al.},\ }\href {\doibase 10.1103/RevModPhys.82.1349}
  {\bibfield  {journal} {\bibinfo  {journal} {Rev. Mod. Phys.}\ }\textbf
  {\bibinfo {volume} {82}},\ \bibinfo {pages} {1349} (\bibinfo {year}
  {2010}{\natexlab{b}})},\ \Eprint {http://arxiv.org/abs/0903.3598}
  {arXiv:0903.3598 [hep-lat]} \BibitemShut {NoStop}%
\bibitem [{\citenamefont {El-Khadra}\ \emph {et~al.}(1997)\citenamefont
  {El-Khadra}, \citenamefont {Kronfeld},\ and\ \citenamefont
  {Mackenzie}}]{ElKhadra:1996mp}%
  \BibitemOpen
  \bibfield  {author} {\bibinfo {author} {\bibfnamefont {A.~X.}\ \bibnamefont
  {El-Khadra}}, \bibinfo {author} {\bibfnamefont {A.~S.}\ \bibnamefont
  {Kronfeld}}, \ and\ \bibinfo {author} {\bibfnamefont {P.~B.}\ \bibnamefont
  {Mackenzie}},\ }\href {\doibase 10.1103/PhysRevD.55.3933} {\bibfield
  {journal} {\bibinfo  {journal} {Phys. Rev.}\ }\textbf {\bibinfo {volume}
  {D55}},\ \bibinfo {pages} {3933} (\bibinfo {year} {1997})},\ \Eprint
  {http://arxiv.org/abs/hep-lat/9604004} {arXiv:hep-lat/9604004} \BibitemShut
  {NoStop}%
\bibitem [{\citenamefont {El-Khadra}\ \emph {et~al.}(2001)\citenamefont
  {El-Khadra}, \citenamefont {Kronfeld}, \citenamefont {Mackenzie},
  \citenamefont {Ryan},\ and\ \citenamefont {Simone}}]{ElKhadra:2001rv}%
  \BibitemOpen
  \bibfield  {author} {\bibinfo {author} {\bibfnamefont {A.~X.}\ \bibnamefont
  {El-Khadra}}, \bibinfo {author} {\bibfnamefont {A.~S.}\ \bibnamefont
  {Kronfeld}}, \bibinfo {author} {\bibfnamefont {P.~B.}\ \bibnamefont
  {Mackenzie}}, \bibinfo {author} {\bibfnamefont {S.~M.}\ \bibnamefont {Ryan}},
  \ and\ \bibinfo {author} {\bibfnamefont {J.~N.}\ \bibnamefont {Simone}},\
  }\href {\doibase 10.1103/PhysRevD.64.014502} {\bibfield  {journal} {\bibinfo
  {journal} {Phys. Rev.}\ }\textbf {\bibinfo {volume} {D64}},\ \bibinfo {pages}
  {014502} (\bibinfo {year} {2001})},\ \Eprint
  {http://arxiv.org/abs/hep-ph/0101023} {arXiv:hep-ph/0101023} \BibitemShut
  {NoStop}%
\bibitem [{\citenamefont {{Be\'cirevi\'c}}\ \emph {et~al.}(2003)\citenamefont
  {{Be\'cirevi\'c}}, \citenamefont {{Prelov\v sek}},\ and\ \citenamefont
  {Zupan}}]{Becirevic:2002sc}%
  \BibitemOpen
  \bibfield  {author} {\bibinfo {author} {\bibfnamefont {D.}~\bibnamefont
  {{Be\'cirevi\'c}}}, \bibinfo {author} {\bibfnamefont {S.}~\bibnamefont
  {{Prelov\v sek}}}, \ and\ \bibinfo {author} {\bibfnamefont {J.}~\bibnamefont
  {Zupan}},\ }\href {\doibase 10.1103/PhysRevD.67.054010} {\bibfield  {journal}
  {\bibinfo  {journal} {Phys. Rev.}\ }\textbf {\bibinfo {volume} {D67}},\
  \bibinfo {pages} {054010} (\bibinfo {year} {2003})},\ \Eprint
  {http://arxiv.org/abs/hep-lat/0210048} {arXiv:hep-lat/0210048} \BibitemShut
  {NoStop}%
\bibitem [{\citenamefont {Aubin}\ and\ \citenamefont
  {Bernard}(2007)}]{Aubin:2007mc}%
  \BibitemOpen
  \bibfield  {author} {\bibinfo {author} {\bibfnamefont {C.}~\bibnamefont
  {Aubin}}\ and\ \bibinfo {author} {\bibfnamefont {C.}~\bibnamefont
  {Bernard}},\ }\href {\doibase 10.1103/PhysRevD.76.014002} {\bibfield
  {journal} {\bibinfo  {journal} {Phys. Rev.}\ }\textbf {\bibinfo {volume}
  {D76}},\ \bibinfo {pages} {014002} (\bibinfo {year} {2007})},\ \Eprint
  {http://arxiv.org/abs/0704.0795} {arXiv:0704.0795 [hep-lat]} \BibitemShut
  {NoStop}%
\bibitem [{\citenamefont {Bijnens}\ and\ \citenamefont
  {Jemos}(2010)}]{Bijnens:2010ws}%
  \BibitemOpen
  \bibfield  {author} {\bibinfo {author} {\bibfnamefont {J.}~\bibnamefont
  {Bijnens}}\ and\ \bibinfo {author} {\bibfnamefont {I.}~\bibnamefont
  {Jemos}},\ }\href {\doibase 10.1016/j.nuclphysb.2010.06.021,
  10.1016/j.nuclphysb.2010.10.024} {\bibfield  {journal} {\bibinfo  {journal}
  {Nucl. Phys.}\ }\textbf {\bibinfo {volume} {B840}},\ \bibinfo {pages} {54}
  (\bibinfo {year} {2010})},\ \Eprint {http://arxiv.org/abs/1006.1197}
  {arXiv:1006.1197 [hep-ph]} \BibitemShut {NoStop}%
\bibitem [{\citenamefont {Kronfeld}(2000)}]{Kronfeld:2000ck}%
  \BibitemOpen
  \bibfield  {author} {\bibinfo {author} {\bibfnamefont {A.~S.}\ \bibnamefont
  {Kronfeld}},\ }\href {\doibase 10.1103/PhysRevD.62.014505} {\bibfield
  {journal} {\bibinfo  {journal} {Phys. Rev.}\ }\textbf {\bibinfo {volume}
  {D62}},\ \bibinfo {pages} {014505} (\bibinfo {year} {2000})},\ \Eprint
  {http://arxiv.org/abs/hep-lat/0002008} {arXiv:hep-lat/0002008} \BibitemShut
  {NoStop}%
\bibitem [{\citenamefont {Harada}\ \emph {et~al.}(2002)\citenamefont {Harada},
  \citenamefont {Hashimoto}, \citenamefont {Ishikawa}, \citenamefont
  {Kronfeld}, \citenamefont {Onogi},\ and\ \citenamefont
  {Yamada}}]{Harada:2001fi}%
  \BibitemOpen
  \bibfield  {author} {\bibinfo {author} {\bibfnamefont {J.}~\bibnamefont
  {Harada}}, \bibinfo {author} {\bibfnamefont {S.}~\bibnamefont {Hashimoto}},
  \bibinfo {author} {\bibfnamefont {K.-I.}\ \bibnamefont {Ishikawa}}, \bibinfo
  {author} {\bibfnamefont {A.~S.}\ \bibnamefont {Kronfeld}}, \bibinfo {author}
  {\bibfnamefont {T.}~\bibnamefont {Onogi}}, \ and\ \bibinfo {author}
  {\bibfnamefont {N.}~\bibnamefont {Yamada}},\ }\href {\doibase
  10.1103/PhysRevD.65.094513, 10.1103/PhysRevD.71.019903} {\bibfield  {journal}
  {\bibinfo  {journal} {Phys. Rev.}\ }\textbf {\bibinfo {volume} {D65}},\
  \bibinfo {pages} {094513} (\bibinfo {year} {2002})},\ \Eprint
  {http://arxiv.org/abs/hep-lat/0112044} {arXiv:hep-lat/0112044} \BibitemShut
  {NoStop}%
\bibitem [{\citenamefont {Oktay}\ and\ \citenamefont
  {Kronfeld}(2008)}]{Oktay:2008ex}%
  \BibitemOpen
  \bibfield  {author} {\bibinfo {author} {\bibfnamefont {M.~B.}\ \bibnamefont
  {Oktay}}\ and\ \bibinfo {author} {\bibfnamefont {A.~S.}\ \bibnamefont
  {Kronfeld}},\ }\href {\doibase 10.1103/PhysRevD.78.014504} {\bibfield
  {journal} {\bibinfo  {journal} {Phys. Rev.}\ }\textbf {\bibinfo {volume}
  {D78}},\ \bibinfo {pages} {014504} (\bibinfo {year} {2008})},\ \Eprint
  {http://arxiv.org/abs/0803.0523} {arXiv:0803.0523 [hep-lat]} \BibitemShut
  {NoStop}%
\bibitem [{\citenamefont {Boyd}\ \emph {et~al.}(1995)\citenamefont {Boyd},
  \citenamefont {Grinstein},\ and\ \citenamefont {Lebed}}]{Boyd:1994tt}%
  \BibitemOpen
  \bibfield  {author} {\bibinfo {author} {\bibfnamefont {C.~G.}\ \bibnamefont
  {Boyd}}, \bibinfo {author} {\bibfnamefont {B.}~\bibnamefont {Grinstein}}, \
  and\ \bibinfo {author} {\bibfnamefont {R.~F.}\ \bibnamefont {Lebed}},\ }\href
  {\doibase 10.1103/PhysRevLett.74.4603} {\bibfield  {journal} {\bibinfo
  {journal} {Phys. Rev. Lett.}\ }\textbf {\bibinfo {volume} {74}},\ \bibinfo
  {pages} {4603} (\bibinfo {year} {1995})},\ \Eprint
  {http://arxiv.org/abs/hep-ph/9412324} {arXiv:hep-ph/9412324} \BibitemShut
  {NoStop}%
\bibitem [{\citenamefont {Bourrely}\ \emph {et~al.}(2009)\citenamefont
  {Bourrely}, \citenamefont {Caprini},\ and\ \citenamefont
  {Lellouch}}]{Bourrely:2008za}%
  \BibitemOpen
  \bibfield  {author} {\bibinfo {author} {\bibfnamefont {C.}~\bibnamefont
  {Bourrely}}, \bibinfo {author} {\bibfnamefont {I.}~\bibnamefont {Caprini}}, \
  and\ \bibinfo {author} {\bibfnamefont {L.}~\bibnamefont {Lellouch}},\ }\href
  {\doibase 10.1103/PhysRevD.82.099902, 10.1103/PhysRevD.79.013008} {\bibfield
  {journal} {\bibinfo  {journal} {Phys. Rev.}\ }\textbf {\bibinfo {volume}
  {D79}},\ \bibinfo {pages} {013008} (\bibinfo {year} {2009})},\ \Eprint
  {http://arxiv.org/abs/0807.2722} {arXiv:0807.2722 [hep-ph]} \BibitemShut
  {NoStop}%
\bibitem [{\citenamefont {Olive}\ \emph {et~al.}(2014)\citenamefont {Olive}
  \emph {et~al.}}]{Agashe:2014kda}%
  \BibitemOpen
  \bibfield  {author} {\bibinfo {author} {\bibfnamefont {K.~A.}\ \bibnamefont
  {Olive}} \emph {et~al.} (\bibinfo {collaboration} {Particle Data Group}),\
  }\href {\doibase 10.1088/1674-1137/38/9/090001} {\bibfield  {journal}
  {\bibinfo  {journal} {Chin. Phys.}\ }\textbf {\bibinfo {volume} {C38}},\
  \bibinfo {pages} {090001} (\bibinfo {year} {2014})}\BibitemShut {NoStop}%
\bibitem [{\citenamefont {Lang}\ \emph {et~al.}(2015)\citenamefont {Lang},
  \citenamefont {Mohler}, \citenamefont {{Prelov\v sek}},\ and\ \citenamefont
  {Woloshyn}}]{Lang:2015hza}%
  \BibitemOpen
  \bibfield  {author} {\bibinfo {author} {\bibfnamefont {C.~B.}\ \bibnamefont
  {Lang}}, \bibinfo {author} {\bibfnamefont {D.}~\bibnamefont {Mohler}},
  \bibinfo {author} {\bibfnamefont {S.}~\bibnamefont {{Prelov\v sek}}}, \ and\
  \bibinfo {author} {\bibfnamefont {R.~M.}\ \bibnamefont {Woloshyn}},\ }\href
  {\doibase 10.1016/j.physletb.2015.08.038} {\bibfield  {journal} {\bibinfo
  {journal} {Phys. Lett.}\ }\textbf {\bibinfo {volume} {B750}},\ \bibinfo
  {pages} {17} (\bibinfo {year} {2015})},\ \Eprint
  {http://arxiv.org/abs/1501.01646} {arXiv:1501.01646 [hep-lat]} \BibitemShut
  {NoStop}%
\bibitem [{\citenamefont {Isgur}\ and\ \citenamefont
  {Wise}(1990)}]{Isgur:1990kf}%
  \BibitemOpen
  \bibfield  {author} {\bibinfo {author} {\bibfnamefont {N.}~\bibnamefont
  {Isgur}}\ and\ \bibinfo {author} {\bibfnamefont {M.~B.}\ \bibnamefont
  {Wise}},\ }\href {\doibase 10.1103/PhysRevD.42.2388} {\bibfield  {journal}
  {\bibinfo  {journal} {Phys. Rev.}\ }\textbf {\bibinfo {volume} {D42}},\
  \bibinfo {pages} {2388} (\bibinfo {year} {1990})}\BibitemShut {NoStop}%
\bibitem [{\citenamefont {Burdman}\ and\ \citenamefont
  {Donoghue}(1991)}]{Burdman:1992hg}%
  \BibitemOpen
  \bibfield  {author} {\bibinfo {author} {\bibfnamefont {G.}~\bibnamefont
  {Burdman}}\ and\ \bibinfo {author} {\bibfnamefont {J.~F.}\ \bibnamefont
  {Donoghue}},\ }\href {\doibase 10.1016/0370-2693(91)91538-7} {\bibfield
  {journal} {\bibinfo  {journal} {Phys. Lett.}\ }\textbf {\bibinfo {volume}
  {B270}},\ \bibinfo {pages} {55} (\bibinfo {year} {1991})}\BibitemShut
  {NoStop}%
\bibitem [{\citenamefont {Charles}\ \emph {et~al.}(1999)\citenamefont
  {Charles}, \citenamefont {Le~Yaouanc}, \citenamefont {Oliver}, \citenamefont
  {Pene},\ and\ \citenamefont {Raynal}}]{Charles:1998dr}%
  \BibitemOpen
  \bibfield  {author} {\bibinfo {author} {\bibfnamefont {J.}~\bibnamefont
  {Charles}}, \bibinfo {author} {\bibfnamefont {A.}~\bibnamefont {Le~Yaouanc}},
  \bibinfo {author} {\bibfnamefont {L.}~\bibnamefont {Oliver}}, \bibinfo
  {author} {\bibfnamefont {O.}~\bibnamefont {Pene}}, \ and\ \bibinfo {author}
  {\bibfnamefont {J.~C.}\ \bibnamefont {Raynal}},\ }\href {\doibase
  10.1103/PhysRevD.60.014001} {\bibfield  {journal} {\bibinfo  {journal} {Phys.
  Rev.}\ }\textbf {\bibinfo {volume} {D60}},\ \bibinfo {pages} {014001}
  (\bibinfo {year} {1999})},\ \Eprint {http://arxiv.org/abs/hep-ph/9812358}
  {arXiv:hep-ph/9812358} \BibitemShut {NoStop}%
\bibitem [{\citenamefont {Bajc}\ \emph {et~al.}(1996)\citenamefont {Bajc},
  \citenamefont {Fajfer},\ and\ \citenamefont {Oakes}}]{Bajc:1996py}%
  \BibitemOpen
  \bibfield  {author} {\bibinfo {author} {\bibfnamefont {B.}~\bibnamefont
  {Bajc}}, \bibinfo {author} {\bibfnamefont {S.}~\bibnamefont {Fajfer}}, \ and\
  \bibinfo {author} {\bibfnamefont {R.~J.}\ \bibnamefont {Oakes}},\ }\href@noop
  {} {\  (\bibinfo {year} {1996})},\ \Eprint
  {http://arxiv.org/abs/hep-ph/9612276} {arXiv:hep-ph/9612276} \BibitemShut
  {NoStop}%
\bibitem [{\citenamefont {Gulez}\ \emph {et~al.}(2006)\citenamefont {Gulez},
  \citenamefont {Gray}, \citenamefont {Wingate}, \citenamefont {Davies},
  \citenamefont {Lepage},\ and\ \citenamefont {Shigemitsu}}]{Dalgic:2006dt}%
  \BibitemOpen
  \bibfield  {author} {\bibinfo {author} {\bibfnamefont {E.}~\bibnamefont
  {Gulez}}, \bibinfo {author} {\bibfnamefont {A.}~\bibnamefont {Gray}},
  \bibinfo {author} {\bibfnamefont {M.}~\bibnamefont {Wingate}}, \bibinfo
  {author} {\bibfnamefont {C.~T.~H.}\ \bibnamefont {Davies}}, \bibinfo {author}
  {\bibfnamefont {G.~P.}\ \bibnamefont {Lepage}}, \ and\ \bibinfo {author}
  {\bibfnamefont {J.}~\bibnamefont {Shigemitsu}} (\bibinfo {collaboration}
  {HPQCD Collaboration}),\ }\href {\doibase 10.1103/PhysRevD.73.074502}
  {\bibfield  {journal} {\bibinfo  {journal} {Phys. Rev.}\ }\textbf {\bibinfo
  {volume} {D73}},\ \bibinfo {pages} {074502} (\bibinfo {year} {2006})},\
  \bibinfo {note} {(E)
  \href{http://dx.doi.org/10.1103/PhysRevD.75.119906}{Phys.\ Rev.}\
  \textbf{D75}, 119906 (2007)},\ \Eprint {http://arxiv.org/abs/hep-lat/0601021}
  {arXiv:hep-lat/0601021} \BibitemShut {NoStop}%
\bibitem [{\citenamefont {Bouchard}\ \emph
  {et~al.}(2013{\natexlab{c}})\citenamefont {Bouchard}, \citenamefont {Lepage},
  \citenamefont {Monahan}, \citenamefont {Na},\ and\ \citenamefont
  {Shigemitsu}}]{Bouchard:2013pna}%
  \BibitemOpen
  \bibfield  {author} {\bibinfo {author} {\bibfnamefont {C.}~\bibnamefont
  {Bouchard}}, \bibinfo {author} {\bibfnamefont {G.~P.}\ \bibnamefont
  {Lepage}}, \bibinfo {author} {\bibfnamefont {C.}~\bibnamefont {Monahan}},
  \bibinfo {author} {\bibfnamefont {H.}~\bibnamefont {Na}}, \ and\ \bibinfo
  {author} {\bibfnamefont {J.}~\bibnamefont {Shigemitsu}} (\bibinfo
  {collaboration} {HPQCD Collaboration}),\ }\href {\doibase
  10.1103/PhysRevD.88.054509} {\bibfield  {journal} {\bibinfo  {journal} {Phys.
  Rev.}\ }\textbf {\bibinfo {volume} {D88}},\ \bibinfo {pages} {054509}
  (\bibinfo {year} {2013}{\natexlab{c}})},\ \bibinfo {note} {(E)
  \href{http://dx.doi.org/10.1103/PhysRevD.88.079901}{Phys.\ Rev.}\
  \textbf{D88}, 079901 (2013)},\ \Eprint {http://arxiv.org/abs/1306.2384}
  {arXiv:1306.2384 [hep-lat]} \BibitemShut {NoStop}%
\bibitem [{\citenamefont {Flynn}\ \emph
  {et~al.}(2015{\natexlab{b}})\citenamefont {Flynn}, \citenamefont {Fritzsch},
  \citenamefont {Kawanai}, \citenamefont {Lehner}, \citenamefont {Samways},
  \citenamefont {Sachrajda}, \citenamefont {Van~de Water},\ and\ \citenamefont
  {Witzel}}]{Flynn:2015xna}%
  \BibitemOpen
  \bibfield  {author} {\bibinfo {author} {\bibfnamefont {J.~M.}\ \bibnamefont
  {Flynn}}, \bibinfo {author} {\bibfnamefont {P.}~\bibnamefont {Fritzsch}},
  \bibinfo {author} {\bibfnamefont {T.}~\bibnamefont {Kawanai}}, \bibinfo
  {author} {\bibfnamefont {C.}~\bibnamefont {Lehner}}, \bibinfo {author}
  {\bibfnamefont {B.}~\bibnamefont {Samways}}, \bibinfo {author} {\bibfnamefont
  {C.~T.}\ \bibnamefont {Sachrajda}}, \bibinfo {author} {\bibfnamefont {R.~S.}\
  \bibnamefont {Van~de Water}}, \ and\ \bibinfo {author} {\bibfnamefont
  {O.}~\bibnamefont {Witzel}} (\bibinfo {collaboration} {RBC and UKQCD
  Collaborations}),\ }\href@noop {} {\  (\bibinfo {year}
  {2015}{\natexlab{b}})},\ \Eprint {http://arxiv.org/abs/1506.06413}
  {arXiv:1506.06413 [hep-lat]} \BibitemShut {NoStop}%
\bibitem [{\citenamefont {Tekampe}(2015)}]{TekampeDPF2015}%
  \BibitemOpen
  \bibfield  {author} {\bibinfo {author} {\bibfnamefont {T.}~\bibnamefont
  {Tekampe}} (\bibinfo {collaboration} {LHCb Collaboration}),\ }\href@noop {}
  {\enquote {\bibinfo {title} {{First measurement of the differential branching
  fraction and CP asymmetry of the $B^+ \to \pi^+ \mu^+ \mu^-$ decay}},}\
  }\bibinfo {howpublished}
  {\href{https://indico.cern.ch/event/361123/session/4/contribution/409}{talk
  presented at DPF 2015}} (\bibinfo {year} {2015})\BibitemShut {NoStop}%
\bibitem [{\citenamefont {Wei}\ \emph {et~al.}(2009)\citenamefont {Wei} \emph
  {et~al.}}]{Wei:2009zv}%
  \BibitemOpen
  \bibfield  {author} {\bibinfo {author} {\bibfnamefont {J.-T.}\ \bibnamefont
  {Wei}} \emph {et~al.} (\bibinfo {collaboration} {Belle Collaboration}),\
  }\href {\doibase 10.1103/PhysRevLett.103.171801} {\bibfield  {journal}
  {\bibinfo  {journal} {Phys. Rev. Lett.}\ }\textbf {\bibinfo {volume} {103}},\
  \bibinfo {pages} {171801} (\bibinfo {year} {2009})},\ \Eprint
  {http://arxiv.org/abs/0904.0770} {arXiv:0904.0770 [hep-ex]} \BibitemShut
  {NoStop}%
\bibitem [{\citenamefont {Aaltonen}\ \emph {et~al.}(2011)\citenamefont
  {Aaltonen} \emph {et~al.}}]{Aaltonen:2011qs}%
  \BibitemOpen
  \bibfield  {author} {\bibinfo {author} {\bibfnamefont {T.}~\bibnamefont
  {Aaltonen}} \emph {et~al.} (\bibinfo {collaboration} {CDF Collaboration}),\
  }\href {\doibase 10.1103/PhysRevLett.107.201802} {\bibfield  {journal}
  {\bibinfo  {journal} {Phys. Rev. Lett.}\ }\textbf {\bibinfo {volume} {107}},\
  \bibinfo {pages} {201802} (\bibinfo {year} {2011})},\ \Eprint
  {http://arxiv.org/abs/1107.3753} {arXiv:1107.3753 [hep-ex]} \BibitemShut
  {NoStop}%
\bibitem [{\citenamefont {Lees}\ \emph
  {et~al.}(2012{\natexlab{c}})\citenamefont {Lees} \emph
  {et~al.}}]{Lees:2012tva}%
  \BibitemOpen
  \bibfield  {author} {\bibinfo {author} {\bibfnamefont {J.~P.}\ \bibnamefont
  {Lees}} \emph {et~al.} (\bibinfo {collaboration} {BaBar Collaboration}),\
  }\href {\doibase 10.1103/PhysRevD.86.032012} {\bibfield  {journal} {\bibinfo
  {journal} {Phys. Rev.}\ }\textbf {\bibinfo {volume} {D86}},\ \bibinfo {pages}
  {032012} (\bibinfo {year} {2012}{\natexlab{c}})},\ \Eprint
  {http://arxiv.org/abs/1204.3933} {arXiv:1204.3933 [hep-ex]} \BibitemShut
  {NoStop}%
\bibitem [{\citenamefont {Lyon}\ and\ \citenamefont
  {Zwicky}(2014)}]{Lyon:2014hpa}%
  \BibitemOpen
  \bibfield  {author} {\bibinfo {author} {\bibfnamefont {J.}~\bibnamefont
  {Lyon}}\ and\ \bibinfo {author} {\bibfnamefont {R.}~\bibnamefont {Zwicky}},\
  }\href@noop {} {\  (\bibinfo {year} {2014})},\ \Eprint
  {http://arxiv.org/abs/1406.0566} {arXiv:1406.0566 [hep-ph]} \BibitemShut
  {NoStop}%
\bibitem [{\citenamefont {Aaij}\ \emph
  {et~al.}(2013{\natexlab{c}})\citenamefont {Aaij} \emph
  {et~al.}}]{Aaij:2012vr}%
  \BibitemOpen
  \bibfield  {author} {\bibinfo {author} {\bibfnamefont {R.}~\bibnamefont
  {Aaij}} \emph {et~al.} (\bibinfo {collaboration} {LHCb Collaboration}),\
  }\href {\doibase 10.1007/JHEP02(2013)105} {\bibfield  {journal} {\bibinfo
  {journal} {JHEP}\ }\textbf {\bibinfo {volume} {1302}},\ \bibinfo {pages}
  {105} (\bibinfo {year} {2013}{\natexlab{c}})},\ \Eprint
  {http://arxiv.org/abs/1209.4284} {arXiv:1209.4284 [hep-ex]} \BibitemShut
  {NoStop}%
\bibitem [{\citenamefont {Aaij}\ \emph {et~al.}(2012)\citenamefont {Aaij} \emph
  {et~al.}}]{LHCb:2012de}%
  \BibitemOpen
  \bibfield  {author} {\bibinfo {author} {\bibfnamefont {R.}~\bibnamefont
  {Aaij}} \emph {et~al.} (\bibinfo {collaboration} {LHCb Collaboration}),\
  }\href {\doibase 10.1007/JHEP12(2012)125} {\bibfield  {journal} {\bibinfo
  {journal} {JHEP}\ }\textbf {\bibinfo {volume} {1212}},\ \bibinfo {pages}
  {125} (\bibinfo {year} {2012})},\ \Eprint {http://arxiv.org/abs/1210.2645}
  {arXiv:1210.2645 [hep-ex]} \BibitemShut {NoStop}%
\bibitem [{\citenamefont {Altmannshofer}\ and\ \citenamefont
  {Straub}(2013)}]{Altmannshofer:2013foa}%
  \BibitemOpen
  \bibfield  {author} {\bibinfo {author} {\bibfnamefont {W.}~\bibnamefont
  {Altmannshofer}}\ and\ \bibinfo {author} {\bibfnamefont {D.~M.}\ \bibnamefont
  {Straub}},\ }\href {\doibase 10.1140/epjc/s10052-013-2646-9} {\bibfield
  {journal} {\bibinfo  {journal} {Eur. Phys. J.}\ }\textbf {\bibinfo {volume}
  {C73}},\ \bibinfo {pages} {2646} (\bibinfo {year} {2013})},\ \Eprint
  {http://arxiv.org/abs/1308.1501} {arXiv:1308.1501 [hep-ph]} \BibitemShut
  {NoStop}%
\bibitem [{\citenamefont {Datta}\ \emph {et~al.}(2014)\citenamefont {Datta},
  \citenamefont {Duraisamy},\ and\ \citenamefont {Ghosh}}]{Datta:2013kja}%
  \BibitemOpen
  \bibfield  {author} {\bibinfo {author} {\bibfnamefont {A.}~\bibnamefont
  {Datta}}, \bibinfo {author} {\bibfnamefont {M.}~\bibnamefont {Duraisamy}}, \
  and\ \bibinfo {author} {\bibfnamefont {D.}~\bibnamefont {Ghosh}},\ }\href
  {\doibase 10.1103/PhysRevD.89.071501} {\bibfield  {journal} {\bibinfo
  {journal} {Phys. Rev.}\ }\textbf {\bibinfo {volume} {D89}},\ \bibinfo {pages}
  {071501} (\bibinfo {year} {2014})},\ \Eprint {http://arxiv.org/abs/1310.1937}
  {arXiv:1310.1937 [hep-ph]} \BibitemShut {NoStop}%
\bibitem [{\citenamefont {Biswas}\ \emph {et~al.}(2015)\citenamefont {Biswas},
  \citenamefont {Chowdhury}, \citenamefont {Han},\ and\ \citenamefont
  {Lee}}]{Biswas:2014gga}%
  \BibitemOpen
  \bibfield  {author} {\bibinfo {author} {\bibfnamefont {S.}~\bibnamefont
  {Biswas}}, \bibinfo {author} {\bibfnamefont {D.}~\bibnamefont {Chowdhury}},
  \bibinfo {author} {\bibfnamefont {S.}~\bibnamefont {Han}}, \ and\ \bibinfo
  {author} {\bibfnamefont {S.~J.}\ \bibnamefont {Lee}},\ }\href {\doibase
  10.1007/JHEP02(2015)142} {\bibfield  {journal} {\bibinfo  {journal} {JHEP}\
  }\textbf {\bibinfo {volume} {02}},\ \bibinfo {pages} {142} (\bibinfo {year}
  {2015})},\ \Eprint {http://arxiv.org/abs/1409.0882} {arXiv:1409.0882
  [hep-ph]} \BibitemShut {NoStop}%
\bibitem [{\citenamefont {Glashow}\ \emph {et~al.}(2015)\citenamefont
  {Glashow}, \citenamefont {Guadagnoli},\ and\ \citenamefont
  {Lane}}]{Glashow:2014iga}%
  \BibitemOpen
  \bibfield  {author} {\bibinfo {author} {\bibfnamefont {S.~L.}\ \bibnamefont
  {Glashow}}, \bibinfo {author} {\bibfnamefont {D.}~\bibnamefont {Guadagnoli}},
  \ and\ \bibinfo {author} {\bibfnamefont {K.}~\bibnamefont {Lane}},\ }\href
  {\doibase 10.1103/PhysRevLett.114.091801} {\bibfield  {journal} {\bibinfo
  {journal} {Phys. Rev. Lett.}\ }\textbf {\bibinfo {volume} {114}},\ \bibinfo
  {pages} {091801} (\bibinfo {year} {2015})},\ \Eprint
  {http://arxiv.org/abs/1411.0565} {arXiv:1411.0565 [hep-ph]} \BibitemShut
  {NoStop}%
\bibitem [{\citenamefont {Hiller}\ and\ \citenamefont
  {Kr{\"u}ger}(2004)}]{Hiller:2003js}%
  \BibitemOpen
  \bibfield  {author} {\bibinfo {author} {\bibfnamefont {G.}~\bibnamefont
  {Hiller}}\ and\ \bibinfo {author} {\bibfnamefont {F.}~\bibnamefont
  {Kr{\"u}ger}},\ }\href {\doibase 10.1103/PhysRevD.69.074020} {\bibfield
  {journal} {\bibinfo  {journal} {Phys. Rev.}\ }\textbf {\bibinfo {volume}
  {D69}},\ \bibinfo {pages} {074020} (\bibinfo {year} {2004})},\ \Eprint
  {http://arxiv.org/abs/hep-ph/0310219} {arXiv:hep-ph/0310219} \BibitemShut
  {NoStop}%
\bibitem [{\citenamefont {Lees}\ \emph
  {et~al.}(2013{\natexlab{b}})\citenamefont {Lees} \emph
  {et~al.}}]{Lees:2013kla}%
  \BibitemOpen
  \bibfield  {author} {\bibinfo {author} {\bibfnamefont {J.~P.}\ \bibnamefont
  {Lees}} \emph {et~al.} (\bibinfo {collaboration} {BaBar Collaboration}),\
  }\href {\doibase 10.1103/PhysRevD.87.112005} {\bibfield  {journal} {\bibinfo
  {journal} {Phys. Rev.}\ }\textbf {\bibinfo {volume} {D87}},\ \bibinfo {pages}
  {112005} (\bibinfo {year} {2013}{\natexlab{b}})},\ \Eprint
  {http://arxiv.org/abs/1303.7465} {arXiv:1303.7465 [hep-ex]} \BibitemShut
  {NoStop}%
\bibitem [{\citenamefont {Lutz}\ \emph {et~al.}(2013)\citenamefont {Lutz} \emph
  {et~al.}}]{Lutz:2013ftz}%
  \BibitemOpen
  \bibfield  {author} {\bibinfo {author} {\bibfnamefont {O.}~\bibnamefont
  {Lutz}} \emph {et~al.} (\bibinfo {collaboration} {Belle Collaboration}),\
  }\href {\doibase 10.1103/PhysRevD.87.111103} {\bibfield  {journal} {\bibinfo
  {journal} {Phys. Rev.}\ }\textbf {\bibinfo {volume} {D87}},\ \bibinfo {pages}
  {111103} (\bibinfo {year} {2013})},\ \Eprint {http://arxiv.org/abs/1303.3719}
  {arXiv:1303.3719 [hep-ex]} \BibitemShut {NoStop}%
\bibitem [{\citenamefont {Wang}\ and\ \citenamefont
  {Xiao}(2012)}]{Wang:2012ab}%
  \BibitemOpen
  \bibfield  {author} {\bibinfo {author} {\bibfnamefont {W.-F.}\ \bibnamefont
  {Wang}}\ and\ \bibinfo {author} {\bibfnamefont {Z.-J.}\ \bibnamefont
  {Xiao}},\ }\href {\doibase 10.1103/PhysRevD.86.114025} {\bibfield  {journal}
  {\bibinfo  {journal} {Phys. Rev.}\ }\textbf {\bibinfo {volume} {D86}},\
  \bibinfo {pages} {114025} (\bibinfo {year} {2012})},\ \Eprint
  {http://arxiv.org/abs/1207.0265} {arXiv:1207.0265 [hep-ph]} \BibitemShut
  {NoStop}%
\bibitem [{\citenamefont {Bharucha}\ \emph {et~al.}(2015)\citenamefont
  {Bharucha}, \citenamefont {Straub},\ and\ \citenamefont
  {Zwicky}}]{Straub:2015ica}%
  \BibitemOpen
  \bibfield  {author} {\bibinfo {author} {\bibfnamefont {A.}~\bibnamefont
  {Bharucha}}, \bibinfo {author} {\bibfnamefont {D.~M.}\ \bibnamefont
  {Straub}}, \ and\ \bibinfo {author} {\bibfnamefont {R.}~\bibnamefont
  {Zwicky}},\ }\href@noop {} {\  (\bibinfo {year} {2015})},\ \Eprint
  {http://arxiv.org/abs/1503.05534} {arXiv:1503.05534 [hep-ph]} \BibitemShut
  {NoStop}%
\bibitem [{\citenamefont {Khodjamirian}\ \emph {et~al.}(2011)\citenamefont
  {Khodjamirian}, \citenamefont {Mannel}, \citenamefont {Offen},\ and\
  \citenamefont {Wang}}]{Khodjamirian:2011ub}%
  \BibitemOpen
  \bibfield  {author} {\bibinfo {author} {\bibfnamefont {A.}~\bibnamefont
  {Khodjamirian}}, \bibinfo {author} {\bibfnamefont {T.}~\bibnamefont
  {Mannel}}, \bibinfo {author} {\bibfnamefont {N.}~\bibnamefont {Offen}}, \
  and\ \bibinfo {author} {\bibfnamefont {Y.~M.}\ \bibnamefont {Wang}},\ }\href
  {\doibase 10.1103/PhysRevD.83.094031} {\bibfield  {journal} {\bibinfo
  {journal} {Phys. Rev.}\ }\textbf {\bibinfo {volume} {D83}},\ \bibinfo {pages}
  {094031} (\bibinfo {year} {2011})},\ \Eprint {http://arxiv.org/abs/1103.2655}
  {arXiv:1103.2655 [hep-ph]} \BibitemShut {NoStop}%
\bibitem [{\citenamefont {Dutta}\ \emph {et~al.}(2013)\citenamefont {Dutta},
  \citenamefont {Bhol},\ and\ \citenamefont {Giri}}]{Dutta:2013qaa}%
  \BibitemOpen
  \bibfield  {author} {\bibinfo {author} {\bibfnamefont {R.}~\bibnamefont
  {Dutta}}, \bibinfo {author} {\bibfnamefont {A.}~\bibnamefont {Bhol}}, \ and\
  \bibinfo {author} {\bibfnamefont {A.~K.}\ \bibnamefont {Giri}},\ }\href
  {\doibase 10.1103/PhysRevD.88.114023} {\bibfield  {journal} {\bibinfo
  {journal} {Phys. Rev.}\ }\textbf {\bibinfo {volume} {D88}},\ \bibinfo {pages}
  {114023} (\bibinfo {year} {2013})},\ \Eprint {http://arxiv.org/abs/1307.6653}
  {arXiv:1307.6653 [hep-ph]} \BibitemShut {NoStop}%
\bibitem [{\citenamefont {Wang}\ and\ \citenamefont
  {Shen}(2015)}]{Wang:2015vgv}%
  \BibitemOpen
  \bibfield  {author} {\bibinfo {author} {\bibfnamefont {Y.-M.}\ \bibnamefont
  {Wang}}\ and\ \bibinfo {author} {\bibfnamefont {Y.-L.}\ \bibnamefont
  {Shen}},\ }\href {\doibase 10.1016/j.nuclphysb.2015.07.016} {\bibfield
  {journal} {\bibinfo  {journal} {Nucl. Phys.}\ }\textbf {\bibinfo {volume}
  {B898}},\ \bibinfo {pages} {563} (\bibinfo {year} {2015})},\ \Eprint
  {http://arxiv.org/abs/1506.00667} {arXiv:1506.00667 [hep-ph]} \BibitemShut
  {NoStop}%
\bibitem [{\citenamefont {Fajfer}\ \emph {et~al.}(2012)\citenamefont {Fajfer},
  \citenamefont {Kamenik}, \citenamefont {Nisandzic},\ and\ \citenamefont
  {Zupan}}]{Fajfer:2012jt}%
  \BibitemOpen
  \bibfield  {author} {\bibinfo {author} {\bibfnamefont {S.}~\bibnamefont
  {Fajfer}}, \bibinfo {author} {\bibfnamefont {J.~F.}\ \bibnamefont {Kamenik}},
  \bibinfo {author} {\bibfnamefont {I.}~\bibnamefont {Nisandzic}}, \ and\
  \bibinfo {author} {\bibfnamefont {J.}~\bibnamefont {Zupan}},\ }\href
  {\doibase 10.1103/PhysRevLett.109.161801} {\bibfield  {journal} {\bibinfo
  {journal} {Phys. Rev. Lett.}\ }\textbf {\bibinfo {volume} {109}},\ \bibinfo
  {pages} {161801} (\bibinfo {year} {2012})},\ \Eprint
  {http://arxiv.org/abs/1206.1872} {arXiv:1206.1872 [hep-ph]} \BibitemShut
  {NoStop}%
\bibitem [{\citenamefont {Bevan}\ \emph {et~al.}(2014)\citenamefont {Bevan}
  \emph {et~al.}}]{Bevan:2014iga}%
  \BibitemOpen
  \bibfield  {author} {\bibinfo {author} {\bibfnamefont {A.~J.}\ \bibnamefont
  {Bevan}} \emph {et~al.} (\bibinfo {collaboration} {Belle and BaBar
  Collaborations}),\ }\href {\doibase 10.1140/epjc/s10052-014-3026-9}
  {\bibfield  {journal} {\bibinfo  {journal} {Eur. Phys. J.}\ }\textbf
  {\bibinfo {volume} {C74}},\ \bibinfo {pages} {3026} (\bibinfo {year}
  {2014})},\ \Eprint {http://arxiv.org/abs/1406.6311} {arXiv:1406.6311
  [hep-ex]} \BibitemShut {NoStop}%
\bibitem [{\citenamefont {Ball}\ and\ \citenamefont
  {Zwicky}(2005)}]{Ball:2004ye}%
  \BibitemOpen
  \bibfield  {author} {\bibinfo {author} {\bibfnamefont {P.}~\bibnamefont
  {Ball}}\ and\ \bibinfo {author} {\bibfnamefont {R.}~\bibnamefont {Zwicky}},\
  }\href {\doibase 10.1103/PhysRevD.71.014015} {\bibfield  {journal} {\bibinfo
  {journal} {Phys. Rev.}\ }\textbf {\bibinfo {volume} {D71}},\ \bibinfo {pages}
  {014015} (\bibinfo {year} {2005})},\ \Eprint
  {http://arxiv.org/abs/hep-ph/0406232} {arXiv:hep-ph/0406232} \BibitemShut
  {NoStop}%
\bibitem [{\citenamefont {Charles}\ \emph {et~al.}(2005)\citenamefont {Charles}
  \emph {et~al.}}]{Charles:2004jd}%
  \BibitemOpen
  \bibfield  {author} {\bibinfo {author} {\bibfnamefont {J.}~\bibnamefont
  {Charles}} \emph {et~al.} (\bibinfo {collaboration} {CKMfitter Group}),\
  }\href {\doibase 10.1140/epjc/s2005-02169-1} {\bibfield  {journal} {\bibinfo
  {journal} {Eur. Phys. J.}\ }\textbf {\bibinfo {volume} {C41}},\ \bibinfo
  {pages} {1} (\bibinfo {year} {2005})},\ \Eprint
  {http://arxiv.org/abs/hep-ph/0406184} {arXiv:hep-ph/0406184} \BibitemShut
  {NoStop}%
\bibitem [{\citenamefont {{G\'amiz}}\ \emph {et~al.}(2009)\citenamefont
  {{G\'amiz}}, \citenamefont {Davies}, \citenamefont {Lepage}, \citenamefont
  {Shigemitsu},\ and\ \citenamefont {Wingate}}]{Gamiz:2009ku}%
  \BibitemOpen
  \bibfield  {author} {\bibinfo {author} {\bibfnamefont {E.}~\bibnamefont
  {{G\'amiz}}}, \bibinfo {author} {\bibfnamefont {C.~T.~H.}\ \bibnamefont
  {Davies}}, \bibinfo {author} {\bibfnamefont {G.~P.}\ \bibnamefont {Lepage}},
  \bibinfo {author} {\bibfnamefont {J.}~\bibnamefont {Shigemitsu}}, \ and\
  \bibinfo {author} {\bibfnamefont {M.}~\bibnamefont {Wingate}} (\bibinfo
  {collaboration} {HPQCD Collaboration}),\ }\href {\doibase
  10.1103/PhysRevD.80.014503} {\bibfield  {journal} {\bibinfo  {journal} {Phys.
  Rev.}\ }\textbf {\bibinfo {volume} {D80}},\ \bibinfo {pages} {014503}
  (\bibinfo {year} {2009})},\ \Eprint {http://arxiv.org/abs/0902.1815}
  {arXiv:0902.1815 [hep-lat]} \BibitemShut {NoStop}%
\bibitem [{\citenamefont {Bazavov}\ \emph
  {et~al.}(2012{\natexlab{b}})\citenamefont {Bazavov} \emph
  {et~al.}}]{Bazavov:2012zs}%
  \BibitemOpen
  \bibfield  {author} {\bibinfo {author} {\bibfnamefont {A.}~\bibnamefont
  {Bazavov}} \emph {et~al.} (\bibinfo {collaboration} {Fermilab Lattice and
  MILC Collaborations}),\ }\href {\doibase 10.1103/PhysRevD.86.034503}
  {\bibfield  {journal} {\bibinfo  {journal} {Phys. Rev.}\ }\textbf {\bibinfo
  {volume} {D86}},\ \bibinfo {pages} {034503} (\bibinfo {year}
  {2012}{\natexlab{b}})},\ \Eprint {http://arxiv.org/abs/1205.7013}
  {arXiv:1205.7013 [hep-lat]} \BibitemShut {NoStop}%
\bibitem [{\citenamefont {Dowdall}\ \emph {et~al.}(2014)\citenamefont
  {Dowdall}, \citenamefont {Davies}, \citenamefont {Horgan}, \citenamefont
  {Lepage}, \citenamefont {Monahan},\ and\ \citenamefont
  {Shigemitsu}}]{Dowdall:2014qka}%
  \BibitemOpen
  \bibfield  {author} {\bibinfo {author} {\bibfnamefont {R.~J.}\ \bibnamefont
  {Dowdall}}, \bibinfo {author} {\bibfnamefont {C.~T.~H.}\ \bibnamefont
  {Davies}}, \bibinfo {author} {\bibfnamefont {R.~R.}\ \bibnamefont {Horgan}},
  \bibinfo {author} {\bibfnamefont {G.~P.}\ \bibnamefont {Lepage}}, \bibinfo
  {author} {\bibfnamefont {C.~J.}\ \bibnamefont {Monahan}}, \ and\ \bibinfo
  {author} {\bibfnamefont {J.}~\bibnamefont {Shigemitsu}} (\bibinfo
  {collaboration} {HPQCD Collaboration}),\ }\href@noop {} {\bibfield  {journal}
  {\bibinfo  {journal} {PoS}\ }\textbf {\bibinfo {volume} {LATTICE2014}},\
  \bibinfo {pages} {373} (\bibinfo {year} {2014})},\ \Eprint
  {http://arxiv.org/abs/1411.6989} {arXiv:1411.6989 [hep-lat]} \BibitemShut
  {NoStop}%
\bibitem [{\citenamefont {Bouchard}\ \emph
  {et~al.}(2014{\natexlab{a}})\citenamefont {Bouchard}, \citenamefont
  {Freeland}, \citenamefont {Bernard}, \citenamefont {Chang}, \citenamefont
  {El-Khadra}, \citenamefont {G‡miz}, \citenamefont {Kronfeld}, \citenamefont
  {Laiho},\ and\ \citenamefont {Van~de Water}}]{Bouchard:2014eea}%
  \BibitemOpen
  \bibfield  {author} {\bibinfo {author} {\bibfnamefont {C.~M.}\ \bibnamefont
  {Bouchard}}, \bibinfo {author} {\bibfnamefont {E.}~\bibnamefont {Freeland}},
  \bibinfo {author} {\bibfnamefont {C.~W.}\ \bibnamefont {Bernard}}, \bibinfo
  {author} {\bibfnamefont {C.~C.}\ \bibnamefont {Chang}}, \bibinfo {author}
  {\bibfnamefont {A.~X.}\ \bibnamefont {El-Khadra}}, \bibinfo {author}
  {\bibfnamefont {M.~E.}\ \bibnamefont {G‡miz}}, \bibinfo {author}
  {\bibfnamefont {A.~S.}\ \bibnamefont {Kronfeld}}, \bibinfo {author}
  {\bibfnamefont {J.}~\bibnamefont {Laiho}}, \ and\ \bibinfo {author}
  {\bibfnamefont {R.~S.}\ \bibnamefont {Van~de Water}} (\bibinfo
  {collaboration} {Fermilab Lattice and MILC Collaborations}),\ }\href@noop {}
  {\bibfield  {journal} {\bibinfo  {journal} {PoS}\ }\textbf {\bibinfo {volume}
  {LATTICE2014}},\ \bibinfo {pages} {378} (\bibinfo {year}
  {2014}{\natexlab{a}})},\ \Eprint {http://arxiv.org/abs/1412.5097}
  {arXiv:1412.5097 [hep-lat]} \BibitemShut {NoStop}%
\bibitem [{\citenamefont {Bailey}\ \emph {et~al.}(2014)\citenamefont {Bailey}
  \emph {et~al.}}]{Bailey:2014tva}%
  \BibitemOpen
  \bibfield  {author} {\bibinfo {author} {\bibfnamefont {J.~A.}\ \bibnamefont
  {Bailey}} \emph {et~al.} (\bibinfo {collaboration} {Fermilab Lattice and MILC
  Collaborations}),\ }\href {\doibase 10.1103/PhysRevD.89.114504} {\bibfield
  {journal} {\bibinfo  {journal} {Phys. Rev.}\ }\textbf {\bibinfo {volume}
  {D89}},\ \bibinfo {pages} {114504} (\bibinfo {year} {2014})},\ \Eprint
  {http://arxiv.org/abs/1403.0635} {arXiv:1403.0635 [hep-lat]} \BibitemShut
  {NoStop}%
\bibitem [{\citenamefont {Misiak}\ \emph {et~al.}(2015)\citenamefont {Misiak}
  \emph {et~al.}}]{Misiak:2015xwa}%
  \BibitemOpen
  \bibfield  {author} {\bibinfo {author} {\bibfnamefont {M.}~\bibnamefont
  {Misiak}} \emph {et~al.},\ }\href {\doibase 10.1103/PhysRevLett.114.221801}
  {\bibfield  {journal} {\bibinfo  {journal} {Phys. Rev. Lett.}\ }\textbf
  {\bibinfo {volume} {114}},\ \bibinfo {pages} {221801} (\bibinfo {year}
  {2015})},\ \Eprint {http://arxiv.org/abs/1503.01789} {arXiv:1503.01789
  [hep-ph]} \BibitemShut {NoStop}%
\bibitem [{\citenamefont {Huber}\ \emph {et~al.}(2015)\citenamefont {Huber},
  \citenamefont {Hurth},\ and\ \citenamefont {Lunghi}}]{Huber:2015sra}%
  \BibitemOpen
  \bibfield  {author} {\bibinfo {author} {\bibfnamefont {T.}~\bibnamefont
  {Huber}}, \bibinfo {author} {\bibfnamefont {T.}~\bibnamefont {Hurth}}, \ and\
  \bibinfo {author} {\bibfnamefont {E.}~\bibnamefont {Lunghi}},\ }\href
  {\doibase 10.1007/JHEP06(2015)176} {\bibfield  {journal} {\bibinfo  {journal}
  {JHEP}\ }\textbf {\bibinfo {volume} {06}},\ \bibinfo {pages} {176} (\bibinfo
  {year} {2015})},\ \Eprint {http://arxiv.org/abs/1503.04849} {arXiv:1503.04849
  [hep-ph]} \BibitemShut {NoStop}%
\bibitem [{\citenamefont {Lees}\ \emph {et~al.}(2014)\citenamefont {Lees} \emph
  {et~al.}}]{Lees:2013nxa}%
  \BibitemOpen
  \bibfield  {author} {\bibinfo {author} {\bibfnamefont {J.~P.}\ \bibnamefont
  {Lees}} \emph {et~al.} (\bibinfo {collaboration} {BaBar Collaboration}),\
  }\href {\doibase 10.1103/PhysRevLett.112.211802} {\bibfield  {journal}
  {\bibinfo  {journal} {Phys. Rev. Lett.}\ }\textbf {\bibinfo {volume} {112}},\
  \bibinfo {pages} {211802} (\bibinfo {year} {2014})},\ \Eprint
  {http://arxiv.org/abs/1312.5364} {arXiv:1312.5364 [hep-ex]} \BibitemShut
  {NoStop}%
\bibitem [{\citenamefont {Aubert}\ \emph {et~al.}(2004)\citenamefont {Aubert}
  \emph {et~al.}}]{Aubert:2004it}%
  \BibitemOpen
  \bibfield  {author} {\bibinfo {author} {\bibfnamefont {B.}~\bibnamefont
  {Aubert}} \emph {et~al.} (\bibinfo {collaboration} {BaBar Collaboration}),\
  }\href {\doibase 10.1103/PhysRevLett.93.081802} {\bibfield  {journal}
  {\bibinfo  {journal} {Phys. Rev. Lett.}\ }\textbf {\bibinfo {volume} {93}},\
  \bibinfo {pages} {081802} (\bibinfo {year} {2004})},\ \Eprint
  {http://arxiv.org/abs/hep-ex/0404006} {arXiv:hep-ex/0404006 [hep-ex]}
  \BibitemShut {NoStop}%
\bibitem [{\citenamefont {Iwasaki}\ \emph {et~al.}(2005)\citenamefont {Iwasaki}
  \emph {et~al.}}]{Iwasaki:2005sy}%
  \BibitemOpen
  \bibfield  {author} {\bibinfo {author} {\bibfnamefont {M.}~\bibnamefont
  {Iwasaki}} \emph {et~al.} (\bibinfo {collaboration} {Belle Collaboration}),\
  }\href {\doibase 10.1103/PhysRevD.72.092005} {\bibfield  {journal} {\bibinfo
  {journal} {Phys. Rev.}\ }\textbf {\bibinfo {volume} {D72}},\ \bibinfo {pages}
  {092005} (\bibinfo {year} {2005})},\ \Eprint
  {http://arxiv.org/abs/hep-ex/0503044} {arXiv:hep-ex/0503044 [hep-ex]}
  \BibitemShut {NoStop}%
\bibitem [{\citenamefont {Descotes-Genon}\ \emph
  {et~al.}(2013{\natexlab{a}})\citenamefont {Descotes-Genon}, \citenamefont
  {Matias},\ and\ \citenamefont {Virto}}]{Descotes-Genon:2013wba}%
  \BibitemOpen
  \bibfield  {author} {\bibinfo {author} {\bibfnamefont {S.}~\bibnamefont
  {Descotes-Genon}}, \bibinfo {author} {\bibfnamefont {J.}~\bibnamefont
  {Matias}}, \ and\ \bibinfo {author} {\bibfnamefont {J.}~\bibnamefont
  {Virto}},\ }\href {\doibase 10.1103/PhysRevD.88.074002} {\bibfield  {journal}
  {\bibinfo  {journal} {Phys. Rev.}\ }\textbf {\bibinfo {volume} {D88}},\
  \bibinfo {pages} {074002} (\bibinfo {year} {2013}{\natexlab{a}})},\ \Eprint
  {http://arxiv.org/abs/1307.5683} {arXiv:1307.5683 [hep-ph]} \BibitemShut
  {NoStop}%
\bibitem [{\citenamefont {Beaujean}\ \emph {et~al.}(2014)\citenamefont
  {Beaujean}, \citenamefont {Bobeth},\ and\ \citenamefont {van
  Dyk}}]{Beaujean:2013soa}%
  \BibitemOpen
  \bibfield  {author} {\bibinfo {author} {\bibfnamefont {F.}~\bibnamefont
  {Beaujean}}, \bibinfo {author} {\bibfnamefont {C.}~\bibnamefont {Bobeth}}, \
  and\ \bibinfo {author} {\bibfnamefont {D.}~\bibnamefont {van Dyk}},\ }\href
  {\doibase 10.1140/epjc/s10052-014-2897-0} {\bibfield  {journal} {\bibinfo
  {journal} {Eur. Phys. J.}\ }\textbf {\bibinfo {volume} {C74}},\ \bibinfo
  {pages} {2897} (\bibinfo {year} {2014})},\ \bibinfo {note} {(E)
  \href{http://dx.doi.org/10.1140/epjc/s10052-014-3179-6}{Eur.\ Phys.\ J.}\
  \textbf{C74}, 3179 (2014)},\ \Eprint {http://arxiv.org/abs/1310.2478}
  {arXiv:1310.2478 [hep-ph]} \BibitemShut {NoStop}%
\bibitem [{\citenamefont {Descotes-Genon}\ \emph
  {et~al.}(2013{\natexlab{b}})\citenamefont {Descotes-Genon}, \citenamefont
  {Matias},\ and\ \citenamefont {Virto}}]{Descotes-Genon:2013zva}%
  \BibitemOpen
  \bibfield  {author} {\bibinfo {author} {\bibfnamefont {S.}~\bibnamefont
  {Descotes-Genon}}, \bibinfo {author} {\bibfnamefont {J.}~\bibnamefont
  {Matias}}, \ and\ \bibinfo {author} {\bibfnamefont {J.}~\bibnamefont
  {Virto}},\ }\href@noop {} {\bibfield  {journal} {\bibinfo  {journal} {PoS}\
  }\textbf {\bibinfo {volume} {EPS-HEP2013}},\ \bibinfo {pages} {361} (\bibinfo
  {year} {2013}{\natexlab{b}})},\ \Eprint {http://arxiv.org/abs/1311.3876}
  {arXiv:1311.3876 [hep-ph]} \BibitemShut {NoStop}%
\bibitem [{\citenamefont {Hurth}\ and\ \citenamefont
  {Mahmoudi}(2014)}]{Hurth:2013ssa}%
  \BibitemOpen
  \bibfield  {author} {\bibinfo {author} {\bibfnamefont {T.}~\bibnamefont
  {Hurth}}\ and\ \bibinfo {author} {\bibfnamefont {F.}~\bibnamefont
  {Mahmoudi}},\ }\href {\doibase 10.1007/JHEP04(2014)097} {\bibfield  {journal}
  {\bibinfo  {journal} {JHEP}\ }\textbf {\bibinfo {volume} {04}},\ \bibinfo
  {pages} {097} (\bibinfo {year} {2014})},\ \Eprint
  {http://arxiv.org/abs/1312.5267} {arXiv:1312.5267 [hep-ph]} \BibitemShut
  {NoStop}%
\bibitem [{\citenamefont {Descotes-Genon}\ \emph
  {et~al.}(2014{\natexlab{a}})\citenamefont {Descotes-Genon}, \citenamefont
  {Hofer}, \citenamefont {Matias},\ and\ \citenamefont
  {Virto}}]{Descotes-Genon:2014uoa}%
  \BibitemOpen
  \bibfield  {author} {\bibinfo {author} {\bibfnamefont {S.}~\bibnamefont
  {Descotes-Genon}}, \bibinfo {author} {\bibfnamefont {L.}~\bibnamefont
  {Hofer}}, \bibinfo {author} {\bibfnamefont {J.}~\bibnamefont {Matias}}, \
  and\ \bibinfo {author} {\bibfnamefont {J.}~\bibnamefont {Virto}},\ }\href
  {\doibase 10.1007/JHEP12(2014)125} {\bibfield  {journal} {\bibinfo  {journal}
  {JHEP}\ }\textbf {\bibinfo {volume} {12}},\ \bibinfo {pages} {125} (\bibinfo
  {year} {2014}{\natexlab{a}})},\ \Eprint {http://arxiv.org/abs/1407.8526}
  {arXiv:1407.8526 [hep-ph]} \BibitemShut {NoStop}%
\bibitem [{\citenamefont {Hurth}\ \emph {et~al.}(2014)\citenamefont {Hurth},
  \citenamefont {Mahmoudi},\ and\ \citenamefont {Neshatpour}}]{Hurth:2014vma}%
  \BibitemOpen
  \bibfield  {author} {\bibinfo {author} {\bibfnamefont {T.}~\bibnamefont
  {Hurth}}, \bibinfo {author} {\bibfnamefont {F.}~\bibnamefont {Mahmoudi}}, \
  and\ \bibinfo {author} {\bibfnamefont {S.}~\bibnamefont {Neshatpour}},\
  }\href {\doibase 10.1007/JHEP12(2014)053} {\bibfield  {journal} {\bibinfo
  {journal} {JHEP}\ }\textbf {\bibinfo {volume} {12}},\ \bibinfo {pages} {053}
  (\bibinfo {year} {2014})},\ \Eprint {http://arxiv.org/abs/1410.4545}
  {arXiv:1410.4545 [hep-ph]} \BibitemShut {NoStop}%
\bibitem [{\citenamefont {Descotes-Genon}\ \emph
  {et~al.}(2014{\natexlab{b}})\citenamefont {Descotes-Genon}, \citenamefont
  {Hofer}, \citenamefont {Matias},\ and\ \citenamefont
  {Virto}}]{Descotes-Genon:2014joa}%
  \BibitemOpen
  \bibfield  {author} {\bibinfo {author} {\bibfnamefont {S.}~\bibnamefont
  {Descotes-Genon}}, \bibinfo {author} {\bibfnamefont {L.}~\bibnamefont
  {Hofer}}, \bibinfo {author} {\bibfnamefont {J.}~\bibnamefont {Matias}}, \
  and\ \bibinfo {author} {\bibfnamefont {J.}~\bibnamefont {Virto}},\
  }\href@noop {} {\  (\bibinfo {year} {2014}{\natexlab{b}})},\ \Eprint
  {http://arxiv.org/abs/1411.0922} {arXiv:1411.0922 [hep-ph]} \BibitemShut
  {NoStop}%
\bibitem [{\citenamefont {Hiller}\ and\ \citenamefont
  {Schmaltz}(2015)}]{Hiller:2014ula}%
  \BibitemOpen
  \bibfield  {author} {\bibinfo {author} {\bibfnamefont {G.}~\bibnamefont
  {Hiller}}\ and\ \bibinfo {author} {\bibfnamefont {M.}~\bibnamefont
  {Schmaltz}},\ }\href {\doibase 10.1007/JHEP02(2015)055} {\bibfield  {journal}
  {\bibinfo  {journal} {JHEP}\ }\textbf {\bibinfo {volume} {02}},\ \bibinfo
  {pages} {055} (\bibinfo {year} {2015})},\ \Eprint
  {http://arxiv.org/abs/1411.4773} {arXiv:1411.4773 [hep-ph]} \BibitemShut
  {NoStop}%
\bibitem [{\citenamefont {Descotes-Genon}\ \emph {et~al.}(2015)\citenamefont
  {Descotes-Genon}, \citenamefont {Hofer}, \citenamefont {Matias},\ and\
  \citenamefont {Virto}}]{Descotes-Genon:2015xqa}%
  \BibitemOpen
  \bibfield  {author} {\bibinfo {author} {\bibfnamefont {S.}~\bibnamefont
  {Descotes-Genon}}, \bibinfo {author} {\bibfnamefont {L.}~\bibnamefont
  {Hofer}}, \bibinfo {author} {\bibfnamefont {J.}~\bibnamefont {Matias}}, \
  and\ \bibinfo {author} {\bibfnamefont {J.}~\bibnamefont {Virto}},\ }\href
  {\doibase 10.1088/1742-6596/631/1/012027} {\bibfield  {journal} {\bibinfo
  {journal} {J. Phys. Conf. Ser.}\ }\textbf {\bibinfo {volume} {631}},\
  \bibinfo {pages} {012027} (\bibinfo {year} {2015})},\ \Eprint
  {http://arxiv.org/abs/1503.03328} {arXiv:1503.03328 [hep-ph]} \BibitemShut
  {NoStop}%
\bibitem [{\citenamefont {McNeile}\ \emph {et~al.}(2012)\citenamefont
  {McNeile}, \citenamefont {Davies}, \citenamefont {Follana}, \citenamefont
  {Hornbostel},\ and\ \citenamefont {Lepage}}]{McNeile:2011ng}%
  \BibitemOpen
  \bibfield  {author} {\bibinfo {author} {\bibfnamefont {C.}~\bibnamefont
  {McNeile}}, \bibinfo {author} {\bibfnamefont {C.~T.~H.}\ \bibnamefont
  {Davies}}, \bibinfo {author} {\bibfnamefont {E.}~\bibnamefont {Follana}},
  \bibinfo {author} {\bibfnamefont {K.}~\bibnamefont {Hornbostel}}, \ and\
  \bibinfo {author} {\bibfnamefont {G.~P.}\ \bibnamefont {Lepage}} (\bibinfo
  {collaboration} {HPQCD Collaboration}),\ }\href {\doibase
  10.1103/PhysRevD.85.031503} {\bibfield  {journal} {\bibinfo  {journal} {Phys.
  Rev.}\ }\textbf {\bibinfo {volume} {D85}},\ \bibinfo {pages} {031503}
  (\bibinfo {year} {2012})},\ \Eprint {http://arxiv.org/abs/1110.4510}
  {arXiv:1110.4510 [hep-lat]} \BibitemShut {NoStop}%
\bibitem [{\citenamefont {Bobeth}\ \emph {et~al.}(2014)\citenamefont {Bobeth},
  \citenamefont {Gorbahn}, \citenamefont {Hermann}, \citenamefont {Misiak},
  \citenamefont {Stamou},\ and\ \citenamefont {Steinhauser}}]{Bobeth:2013uxa}%
  \BibitemOpen
  \bibfield  {author} {\bibinfo {author} {\bibfnamefont {C.}~\bibnamefont
  {Bobeth}}, \bibinfo {author} {\bibfnamefont {M.}~\bibnamefont {Gorbahn}},
  \bibinfo {author} {\bibfnamefont {T.}~\bibnamefont {Hermann}}, \bibinfo
  {author} {\bibfnamefont {M.}~\bibnamefont {Misiak}}, \bibinfo {author}
  {\bibfnamefont {E.}~\bibnamefont {Stamou}}, \ and\ \bibinfo {author}
  {\bibfnamefont {M.}~\bibnamefont {Steinhauser}},\ }\href {\doibase
  10.1103/PhysRevLett.112.101801} {\bibfield  {journal} {\bibinfo  {journal}
  {Phys. Rev. Lett.}\ }\textbf {\bibinfo {volume} {112}},\ \bibinfo {pages}
  {101801} (\bibinfo {year} {2014})},\ \Eprint {http://arxiv.org/abs/1311.0903}
  {arXiv:1311.0903 [hep-ph]} \BibitemShut {NoStop}%
\bibitem [{\citenamefont {Fleischer}(2014)}]{Fleischer:2014jaa}%
  \BibitemOpen
  \bibfield  {author} {\bibinfo {author} {\bibfnamefont {R.}~\bibnamefont
  {Fleischer}},\ }\href {\doibase 10.1142/S0217751X14440047} {\bibfield
  {journal} {\bibinfo  {journal} {Int. J. Mod. Phys.}\ }\textbf {\bibinfo
  {volume} {A29}},\ \bibinfo {pages} {1444004} (\bibinfo {year} {2014})},\
  \Eprint {http://arxiv.org/abs/1407.0916} {arXiv:1407.0916 [hep-ph]}
  \BibitemShut {NoStop}%
\bibitem [{\citenamefont {Khachatryan}\ \emph {et~al.}(2015)\citenamefont
  {Khachatryan} \emph {et~al.}}]{CMS:2014xfa}%
  \BibitemOpen
  \bibfield  {author} {\bibinfo {author} {\bibfnamefont {V.}~\bibnamefont
  {Khachatryan}} \emph {et~al.} (\bibinfo {collaboration} {LHCb and CMS
  Collaborations}),\ }\href {\doibase 10.1038/nature14474} {\bibfield
  {journal} {\bibinfo  {journal} {Nature}\ }\textbf {\bibinfo {volume} {522}},\
  \bibinfo {pages} {68} (\bibinfo {year} {2015})},\ \Eprint
  {http://arxiv.org/abs/1411.4413} {arXiv:1411.4413 [hep-ex]} \BibitemShut
  {NoStop}%
\bibitem [{\citenamefont {Antonelli}\ \emph {et~al.}(2010)\citenamefont
  {Antonelli} \emph {et~al.}}]{Antonelli:2009ws}%
  \BibitemOpen
  \bibfield  {author} {\bibinfo {author} {\bibfnamefont {M.}~\bibnamefont
  {Antonelli}} \emph {et~al.},\ }\href {\doibase 10.1016/j.physrep.2010.05.003}
  {\bibfield  {journal} {\bibinfo  {journal} {Phys. Rept.}\ }\textbf {\bibinfo
  {volume} {494}},\ \bibinfo {pages} {197} (\bibinfo {year} {2010})},\ \Eprint
  {http://arxiv.org/abs/0907.5386} {arXiv:0907.5386 [hep-ph]} \BibitemShut
  {NoStop}%
\bibitem [{\citenamefont {Bouchard}\ \emph
  {et~al.}(2014{\natexlab{b}})\citenamefont {Bouchard}, \citenamefont {Lepage},
  \citenamefont {Monahan}, \citenamefont {Na},\ and\ \citenamefont
  {Shigemitsu}}]{Bouchard:2013zda}%
  \BibitemOpen
  \bibfield  {author} {\bibinfo {author} {\bibfnamefont {C.~M.}\ \bibnamefont
  {Bouchard}}, \bibinfo {author} {\bibfnamefont {G.~P.}\ \bibnamefont
  {Lepage}}, \bibinfo {author} {\bibfnamefont {C.~J.}\ \bibnamefont {Monahan}},
  \bibinfo {author} {\bibfnamefont {H.}~\bibnamefont {Na}}, \ and\ \bibinfo
  {author} {\bibfnamefont {J.}~\bibnamefont {Shigemitsu}} (\bibinfo
  {collaboration} {HPQCD Collaboration}),\ }\href@noop {} {\bibfield  {journal}
  {\bibinfo  {journal} {PoS}\ }\textbf {\bibinfo {volume} {LATTICE2013}},\
  \bibinfo {pages} {387} (\bibinfo {year} {2014}{\natexlab{b}})},\ \Eprint
  {http://arxiv.org/abs/1310.3207} {arXiv:1310.3207 [hep-lat]} \BibitemShut
  {NoStop}%
\bibitem [{\citenamefont {{D\"urr}}\ \emph {et~al.}(2011)\citenamefont
  {{D\"urr}}, \citenamefont {Fodor}, \citenamefont {Hoelbling}, \citenamefont
  {Katz}, \citenamefont {Krieg}, \citenamefont {Kurth}, \citenamefont
  {Lellouch}, \citenamefont {Lippert}, \citenamefont {Szabo},\ and\
  \citenamefont {Vulvert}}]{Durr:2010aw}%
  \BibitemOpen
  \bibfield  {author} {\bibinfo {author} {\bibfnamefont {S.}~\bibnamefont
  {{D\"urr}}}, \bibinfo {author} {\bibfnamefont {Z.}~\bibnamefont {Fodor}},
  \bibinfo {author} {\bibfnamefont {C.}~\bibnamefont {Hoelbling}}, \bibinfo
  {author} {\bibfnamefont {S.~D.}\ \bibnamefont {Katz}}, \bibinfo {author}
  {\bibfnamefont {S.}~\bibnamefont {Krieg}}, \bibinfo {author} {\bibfnamefont
  {T.}~\bibnamefont {Kurth}}, \bibinfo {author} {\bibfnamefont
  {L.}~\bibnamefont {Lellouch}}, \bibinfo {author} {\bibfnamefont
  {T.}~\bibnamefont {Lippert}}, \bibinfo {author} {\bibfnamefont {K.~K.}\
  \bibnamefont {Szabo}}, \ and\ \bibinfo {author} {\bibfnamefont
  {G.}~\bibnamefont {Vulvert}} (\bibinfo {collaboration}
  {Budapest-Marseille-Wuppertal Collaboration}),\ }\href {\doibase
  10.1007/JHEP08(2011)148} {\bibfield  {journal} {\bibinfo  {journal} {JHEP}\
  }\textbf {\bibinfo {volume} {08}},\ \bibinfo {pages} {148} (\bibinfo {year}
  {2011})},\ \Eprint {http://arxiv.org/abs/1011.2711} {arXiv:1011.2711
  [hep-lat]} \BibitemShut {NoStop}%
\bibitem [{\citenamefont {Bruno}\ \emph {et~al.}(2015)\citenamefont {Bruno}
  \emph {et~al.}}]{Bruno:2014jqa}%
  \BibitemOpen
  \bibfield  {author} {\bibinfo {author} {\bibfnamefont {M.}~\bibnamefont
  {Bruno}} \emph {et~al.},\ }\href {\doibase 10.1007/JHEP02(2015)043}
  {\bibfield  {journal} {\bibinfo  {journal} {JHEP}\ }\textbf {\bibinfo
  {volume} {02}},\ \bibinfo {pages} {043} (\bibinfo {year} {2015})},\ \Eprint
  {http://arxiv.org/abs/1411.3982} {arXiv:1411.3982 [hep-lat]} \BibitemShut
  {NoStop}%
\bibitem [{\citenamefont {Blum}\ \emph {et~al.}(2014)\citenamefont {Blum} \emph
  {et~al.}}]{Blum:2014tka}%
  \BibitemOpen
  \bibfield  {author} {\bibinfo {author} {\bibfnamefont {T.}~\bibnamefont
  {Blum}} \emph {et~al.} (\bibinfo {collaboration} {RBC and UKQCD
  Collaborations}),\ }\href@noop {} {\  (\bibinfo {year} {2014})},\ \Eprint
  {http://arxiv.org/abs/1411.7017} {arXiv:1411.7017 [hep-lat]} \BibitemShut
  {NoStop}%
\bibitem [{\citenamefont {Bazavov}\ \emph {et~al.}(2015)\citenamefont {Bazavov}
  \emph {et~al.}}]{Bazavov:2015yea}%
  \BibitemOpen
  \bibfield  {author} {\bibinfo {author} {\bibfnamefont {A.}~\bibnamefont
  {Bazavov}} \emph {et~al.} (\bibinfo {collaboration} {MILC Collaboration}),\
  }\href@noop {} {\  (\bibinfo {year} {2015})},\ \Eprint
  {http://arxiv.org/abs/1503.02769} {arXiv:1503.02769 [hep-lat]} \BibitemShut
  {NoStop}%
\bibitem [{\citenamefont {Bazavov}\ \emph
  {et~al.}(2010{\natexlab{c}})\citenamefont {Bazavov} \emph
  {et~al.}}]{Bazavov:2010ru}%
  \BibitemOpen
  \bibfield  {author} {\bibinfo {author} {\bibfnamefont {A.}~\bibnamefont
  {Bazavov}} \emph {et~al.} (\bibinfo {collaboration} {MILC Collaboration}),\
  }\href {\doibase 10.1103/PhysRevD.82.074501} {\bibfield  {journal} {\bibinfo
  {journal} {Phys. Rev.}\ }\textbf {\bibinfo {volume} {D82}},\ \bibinfo {pages}
  {074501} (\bibinfo {year} {2010}{\natexlab{c}})},\ \Eprint
  {http://arxiv.org/abs/1004.0342} {arXiv:1004.0342 [hep-lat]} \BibitemShut
  {NoStop}%
\bibitem [{\citenamefont {Follana}\ \emph {et~al.}(2007)\citenamefont
  {Follana}, \citenamefont {Mason}, \citenamefont {Davies}, \citenamefont
  {Hornbostel}, \citenamefont {Lepage}, \citenamefont {Shigemitsu},
  \citenamefont {Trottier},\ and\ \citenamefont {Wong}}]{Follana:2006rc}%
  \BibitemOpen
  \bibfield  {author} {\bibinfo {author} {\bibfnamefont {E.}~\bibnamefont
  {Follana}}, \bibinfo {author} {\bibfnamefont {Q.}~\bibnamefont {Mason}},
  \bibinfo {author} {\bibfnamefont {C.}~\bibnamefont {Davies}}, \bibinfo
  {author} {\bibfnamefont {K.}~\bibnamefont {Hornbostel}}, \bibinfo {author}
  {\bibfnamefont {G.~P.}\ \bibnamefont {Lepage}}, \bibinfo {author}
  {\bibfnamefont {J.}~\bibnamefont {Shigemitsu}}, \bibinfo {author}
  {\bibfnamefont {H.}~\bibnamefont {Trottier}}, \ and\ \bibinfo {author}
  {\bibfnamefont {K.}~\bibnamefont {Wong}} (\bibinfo {collaboration} {HPQCD
  Collaboration}),\ }\href {\doibase 10.1103/PhysRevD.75.054502} {\bibfield
  {journal} {\bibinfo  {journal} {Phys. Rev.}\ }\textbf {\bibinfo {volume}
  {D75}},\ \bibinfo {pages} {054502} (\bibinfo {year} {2007})},\ \Eprint
  {http://arxiv.org/abs/hep-lat/0610092} {arXiv:hep-lat/0610092 [hep-lat]}
  \BibitemShut {NoStop}%
\bibitem [{\citenamefont {Asatryan}\ \emph {et~al.}(2001)\citenamefont
  {Asatryan}, \citenamefont {Asatrian}, \citenamefont {Greub},\ and\
  \citenamefont {Walker}}]{Asatrian:2001de}%
  \BibitemOpen
  \bibfield  {author} {\bibinfo {author} {\bibfnamefont {H.~H.}\ \bibnamefont
  {Asatryan}}, \bibinfo {author} {\bibfnamefont {H.~M.}\ \bibnamefont
  {Asatrian}}, \bibinfo {author} {\bibfnamefont {C.}~\bibnamefont {Greub}}, \
  and\ \bibinfo {author} {\bibfnamefont {M.}~\bibnamefont {Walker}},\ }\href
  {\doibase 10.1016/S0370-2693(01)00441-5} {\bibfield  {journal} {\bibinfo
  {journal} {Phys. Lett.}\ }\textbf {\bibinfo {volume} {B507}},\ \bibinfo
  {pages} {162} (\bibinfo {year} {2001})},\ \Eprint
  {http://arxiv.org/abs/hep-ph/0103087} {arXiv:hep-ph/0103087} \BibitemShut
  {NoStop}%
\bibitem [{\citenamefont {Asatryan}\ \emph {et~al.}(2002)\citenamefont
  {Asatryan}, \citenamefont {Asatrian}, \citenamefont {Greub},\ and\
  \citenamefont {Walker}}]{Asatryan:2001zw}%
  \BibitemOpen
  \bibfield  {author} {\bibinfo {author} {\bibfnamefont {H.~H.}\ \bibnamefont
  {Asatryan}}, \bibinfo {author} {\bibfnamefont {H.~M.}\ \bibnamefont
  {Asatrian}}, \bibinfo {author} {\bibfnamefont {C.}~\bibnamefont {Greub}}, \
  and\ \bibinfo {author} {\bibfnamefont {M.}~\bibnamefont {Walker}},\ }\href
  {\doibase 10.1103/PhysRevD.65.074004} {\bibfield  {journal} {\bibinfo
  {journal} {Phys. Rev.}\ }\textbf {\bibinfo {volume} {D65}},\ \bibinfo {pages}
  {074004} (\bibinfo {year} {2002})},\ \Eprint
  {http://arxiv.org/abs/hep-ph/0109140} {arXiv:hep-ph/0109140} \BibitemShut
  {NoStop}%
\bibitem [{\citenamefont {Asatrian}\ \emph {et~al.}(2004)\citenamefont
  {Asatrian}, \citenamefont {Bieri}, \citenamefont {Greub},\ and\ \citenamefont
  {Walker}}]{Asatrian:2003vq}%
  \BibitemOpen
  \bibfield  {author} {\bibinfo {author} {\bibfnamefont {H.~M.}\ \bibnamefont
  {Asatrian}}, \bibinfo {author} {\bibfnamefont {K.}~\bibnamefont {Bieri}},
  \bibinfo {author} {\bibfnamefont {C.}~\bibnamefont {Greub}}, \ and\ \bibinfo
  {author} {\bibfnamefont {M.}~\bibnamefont {Walker}},\ }\href {\doibase
  10.1103/PhysRevD.69.074007} {\bibfield  {journal} {\bibinfo  {journal} {Phys.
  Rev.}\ }\textbf {\bibinfo {volume} {D69}},\ \bibinfo {pages} {074007}
  (\bibinfo {year} {2004})},\ \Eprint {http://arxiv.org/abs/hep-ph/0312063}
  {arXiv:hep-ph/0312063} \BibitemShut {NoStop}%
\bibitem [{\citenamefont {Seidel}(2004)}]{Seidel:2004jh}%
  \BibitemOpen
  \bibfield  {author} {\bibinfo {author} {\bibfnamefont {D.}~\bibnamefont
  {Seidel}},\ }\href {\doibase 10.1103/PhysRevD.70.094038} {\bibfield
  {journal} {\bibinfo  {journal} {Phys. Rev.}\ }\textbf {\bibinfo {volume}
  {D70}},\ \bibinfo {pages} {094038} (\bibinfo {year} {2004})},\ \Eprint
  {http://arxiv.org/abs/hep-ph/0403185} {arXiv:hep-ph/0403185} \BibitemShut
  {NoStop}%
\bibitem [{\citenamefont {Beneke}\ \emph {et~al.}(2005)\citenamefont {Beneke},
  \citenamefont {Feldmann},\ and\ \citenamefont {Seidel}}]{Beneke:2004dp}%
  \BibitemOpen
  \bibfield  {author} {\bibinfo {author} {\bibfnamefont {M.}~\bibnamefont
  {Beneke}}, \bibinfo {author} {\bibfnamefont {T.}~\bibnamefont {Feldmann}}, \
  and\ \bibinfo {author} {\bibfnamefont {D.}~\bibnamefont {Seidel}},\ }\href
  {\doibase 10.1140/epjc/s2005-02181-5} {\bibfield  {journal} {\bibinfo
  {journal} {Eur. Phys. J.}\ }\textbf {\bibinfo {volume} {C41}},\ \bibinfo
  {pages} {173} (\bibinfo {year} {2005})},\ \Eprint
  {http://arxiv.org/abs/hep-ph/0412400} {arXiv:hep-ph/0412400} \BibitemShut
  {NoStop}%
\bibitem [{\citenamefont {Greub}\ \emph {et~al.}(2008)\citenamefont {Greub},
  \citenamefont {Pilipp},\ and\ \citenamefont {Schupbach}}]{Greub:2008cy}%
  \BibitemOpen
  \bibfield  {author} {\bibinfo {author} {\bibfnamefont {C.}~\bibnamefont
  {Greub}}, \bibinfo {author} {\bibfnamefont {V.}~\bibnamefont {Pilipp}}, \
  and\ \bibinfo {author} {\bibfnamefont {C.}~\bibnamefont {Schupbach}},\ }\href
  {\doibase 10.1088/1126-6708/2008/12/040} {\bibfield  {journal} {\bibinfo
  {journal} {JHEP}\ }\textbf {\bibinfo {volume} {0812}},\ \bibinfo {pages}
  {040} (\bibinfo {year} {2008})},\ \Eprint {http://arxiv.org/abs/0810.4077}
  {arXiv:0810.4077 [hep-ph]} \BibitemShut {NoStop}%
\bibitem [{\citenamefont {Buras}(1999)}]{Buras:1998raa}%
  \BibitemOpen
  \bibfield  {author} {\bibinfo {author} {\bibfnamefont {A.~J.}\ \bibnamefont
  {Buras}},\ }in\ \href@noop {} {\emph {\bibinfo {booktitle} {Probing the
  standard model of particle interactions}}},\ \bibinfo {editor} {edited by\
  \bibinfo {editor} {\bibfnamefont {R.}~\bibnamefont {Gupta}}, \bibinfo
  {editor} {\bibfnamefont {A.}~\bibnamefont {Morel}}, \bibinfo {editor}
  {\bibfnamefont {E.}~\bibnamefont {{de Rafael}}}, \ and\ \bibinfo {editor}
  {\bibfnamefont {F.}~\bibnamefont {David}}}\ (\bibinfo  {publisher} {North
  Holland},\ \bibinfo {address} {Amsterdam},\ \bibinfo {year} {1999})\ pp.\
  \bibinfo {pages} {281--539},\ \Eprint {http://arxiv.org/abs/hep-ph/9806471}
  {arXiv:hep-ph/9806471} \BibitemShut {NoStop}%
\bibitem [{\citenamefont {Braun}\ \emph {et~al.}(2015)\citenamefont {Braun},
  \citenamefont {Collins}, \citenamefont {G{\"o}ckeler}, \citenamefont
  {P{\'e}rez-Rubio}, \citenamefont {Sch{\"a}fer}, \citenamefont {Schiel},\ and\
  \citenamefont {Sternbeck}}]{Braun:2015axa}%
  \BibitemOpen
  \bibfield  {author} {\bibinfo {author} {\bibfnamefont {V.~M.}\ \bibnamefont
  {Braun}}, \bibinfo {author} {\bibfnamefont {S.}~\bibnamefont {Collins}},
  \bibinfo {author} {\bibfnamefont {M.}~\bibnamefont {G{\"o}ckeler}}, \bibinfo
  {author} {\bibfnamefont {P.}~\bibnamefont {P{\'e}rez-Rubio}}, \bibinfo
  {author} {\bibfnamefont {A.}~\bibnamefont {Sch{\"a}fer}}, \bibinfo {author}
  {\bibfnamefont {R.~W.}\ \bibnamefont {Schiel}}, \ and\ \bibinfo {author}
  {\bibfnamefont {A.}~\bibnamefont {Sternbeck}} (\bibinfo {collaboration}
  {QCDSF Collaboration}),\ }\href {\doibase 10.1103/PhysRevD.92.014504}
  {\bibfield  {journal} {\bibinfo  {journal} {Phys. Rev.}\ }\textbf {\bibinfo
  {volume} {D92}},\ \bibinfo {pages} {014504} (\bibinfo {year} {2015})},\
  \Eprint {http://arxiv.org/abs/1503.03656} {arXiv:1503.03656 [hep-lat]}
  \BibitemShut {NoStop}%
\bibitem [{\citenamefont {Braun}\ \emph {et~al.}(2006)\citenamefont {Braun}
  \emph {et~al.}}]{Braun:2006dg}%
  \BibitemOpen
  \bibfield  {author} {\bibinfo {author} {\bibfnamefont {V.~M.}\ \bibnamefont
  {Braun}} \emph {et~al.} (\bibinfo {collaboration} {QCDSF and UKQCD
  Collaborations}),\ }\href {\doibase 10.1103/PhysRevD.74.074501} {\bibfield
  {journal} {\bibinfo  {journal} {Phys. Rev.}\ }\textbf {\bibinfo {volume}
  {D74}},\ \bibinfo {pages} {074501} (\bibinfo {year} {2006})},\ \Eprint
  {http://arxiv.org/abs/hep-lat/0606012} {arXiv:hep-lat/0606012} \BibitemShut
  {NoStop}%
\bibitem [{\citenamefont {Boyle}\ \emph {et~al.}(2006)\citenamefont {Boyle},
  \citenamefont {Donnellan}, \citenamefont {Flynn}, \citenamefont
  {{J\"uttner}}, \citenamefont {Noaki}, \citenamefont {Sachrajda},\ and\
  \citenamefont {Tweedie}}]{Boyle:2006pw}%
  \BibitemOpen
  \bibfield  {author} {\bibinfo {author} {\bibfnamefont {P.~A.}\ \bibnamefont
  {Boyle}}, \bibinfo {author} {\bibfnamefont {M.~A.}\ \bibnamefont
  {Donnellan}}, \bibinfo {author} {\bibfnamefont {J.~M.}\ \bibnamefont
  {Flynn}}, \bibinfo {author} {\bibfnamefont {A.}~\bibnamefont {{J\"uttner}}},
  \bibinfo {author} {\bibfnamefont {J.}~\bibnamefont {Noaki}}, \bibinfo
  {author} {\bibfnamefont {C.~T.}\ \bibnamefont {Sachrajda}}, \ and\ \bibinfo
  {author} {\bibfnamefont {R.~J.}\ \bibnamefont {Tweedie}} (\bibinfo
  {collaboration} {UKQCD Collaboration}),\ }\href {\doibase
  10.1016/j.physletb.2006.07.033} {\bibfield  {journal} {\bibinfo  {journal}
  {Phys. Lett.}\ }\textbf {\bibinfo {volume} {B641}},\ \bibinfo {pages} {67}
  (\bibinfo {year} {2006})},\ \Eprint {http://arxiv.org/abs/hep-lat/0607018}
  {arXiv:hep-lat/0607018} \BibitemShut {NoStop}%
\bibitem [{\citenamefont {Arthur}\ \emph {et~al.}(2011)\citenamefont {Arthur}
  \emph {et~al.}}]{Arthur:2010xf}%
  \BibitemOpen
  \bibfield  {author} {\bibinfo {author} {\bibfnamefont {R.}~\bibnamefont
  {Arthur}} \emph {et~al.} (\bibinfo {collaboration} {RBC and UKQCD
  Collaborations}),\ }\href {\doibase 10.1103/PhysRevD.83.074505} {\bibfield
  {journal} {\bibinfo  {journal} {Phys. Rev.}\ }\textbf {\bibinfo {volume}
  {D83}},\ \bibinfo {pages} {074505} (\bibinfo {year} {2011})},\ \Eprint
  {http://arxiv.org/abs/1011.5906} {arXiv:1011.5906 [hep-lat]} \BibitemShut
  {NoStop}%
\bibitem [{\citenamefont {Bell}\ \emph {et~al.}(2011)\citenamefont {Bell},
  \citenamefont {Beneke}, \citenamefont {Huber},\ and\ \citenamefont
  {Li}}]{Bell:2010mg}%
  \BibitemOpen
  \bibfield  {author} {\bibinfo {author} {\bibfnamefont {G.}~\bibnamefont
  {Bell}}, \bibinfo {author} {\bibfnamefont {M.}~\bibnamefont {Beneke}},
  \bibinfo {author} {\bibfnamefont {T.}~\bibnamefont {Huber}}, \ and\ \bibinfo
  {author} {\bibfnamefont {X.-Q.}\ \bibnamefont {Li}},\ }\href {\doibase
  10.1016/j.nuclphysb.2010.09.022} {\bibfield  {journal} {\bibinfo  {journal}
  {Nucl. Phys.}\ }\textbf {\bibinfo {volume} {B843}},\ \bibinfo {pages} {143}
  (\bibinfo {year} {2011})},\ \Eprint {http://arxiv.org/abs/1007.3758}
  {arXiv:1007.3758 [hep-ph]} \BibitemShut {NoStop}%
\bibitem [{\citenamefont {Aoki}\ \emph {et~al.}(2014)\citenamefont {Aoki} \emph
  {et~al.}}]{Aoki:2013ldr}%
  \BibitemOpen
  \bibfield  {author} {\bibinfo {author} {\bibfnamefont {S.}~\bibnamefont
  {Aoki}} \emph {et~al.} (\bibinfo {collaboration} {Flavor Lattice Averaging
  Group}),\ }\href {\doibase 10.1140/epjc/s10052-014-2890-7} {\bibfield
  {journal} {\bibinfo  {journal} {Eur. Phys. J.}\ }\textbf {\bibinfo {volume}
  {C74}},\ \bibinfo {pages} {2890} (\bibinfo {year} {2014})},\ \Eprint
  {http://arxiv.org/abs/1310.8555} {arXiv:1310.8555 [hep-lat]} \BibitemShut
  {NoStop}%
\bibitem [{\citenamefont {Braun}\ \emph {et~al.}(2004)\citenamefont {Braun},
  \citenamefont {Ivanov},\ and\ \citenamefont {Korchemsky}}]{Braun:2003wx}%
  \BibitemOpen
  \bibfield  {author} {\bibinfo {author} {\bibfnamefont {V.~M.}\ \bibnamefont
  {Braun}}, \bibinfo {author} {\bibfnamefont {D.~Y.}\ \bibnamefont {Ivanov}}, \
  and\ \bibinfo {author} {\bibfnamefont {G.~P.}\ \bibnamefont {Korchemsky}},\
  }\href {\doibase 10.1103/PhysRevD.69.034014} {\bibfield  {journal} {\bibinfo
  {journal} {Phys. Rev.}\ }\textbf {\bibinfo {volume} {D69}},\ \bibinfo {pages}
  {034014} (\bibinfo {year} {2004})},\ \Eprint
  {http://arxiv.org/abs/hep-ph/0309330} {arXiv:hep-ph/0309330} \BibitemShut
  {NoStop}%
\bibitem [{\citenamefont {Lee}\ and\ \citenamefont
  {Neubert}(2005)}]{Lee:2005gza}%
  \BibitemOpen
  \bibfield  {author} {\bibinfo {author} {\bibfnamefont {S.~J.}\ \bibnamefont
  {Lee}}\ and\ \bibinfo {author} {\bibfnamefont {M.}~\bibnamefont {Neubert}},\
  }\href {\doibase 10.1103/PhysRevD.72.094028} {\bibfield  {journal} {\bibinfo
  {journal} {Phys. Rev.}\ }\textbf {\bibinfo {volume} {D72}},\ \bibinfo {pages}
  {094028} (\bibinfo {year} {2005})},\ \Eprint
  {http://arxiv.org/abs/hep-ph/0509350} {arXiv:hep-ph/0509350} \BibitemShut
  {NoStop}%
\bibitem [{\citenamefont {Ball}\ and\ \citenamefont
  {Zwicky}(2006)}]{Ball:2006nr}%
  \BibitemOpen
  \bibfield  {author} {\bibinfo {author} {\bibfnamefont {P.}~\bibnamefont
  {Ball}}\ and\ \bibinfo {author} {\bibfnamefont {R.}~\bibnamefont {Zwicky}},\
  }\href {\doibase 10.1088/1126-6708/2006/04/046} {\bibfield  {journal}
  {\bibinfo  {journal} {JHEP}\ }\textbf {\bibinfo {volume} {0604}},\ \bibinfo
  {pages} {046} (\bibinfo {year} {2006})},\ \Eprint
  {http://arxiv.org/abs/hep-ph/0603232} {arXiv:hep-ph/0603232} \BibitemShut
  {NoStop}%
\end{thebibliography}%
\bibliographystyle{apsrev4-1}
\end{document}